\documentclass[11pt, a4paper, twoside, titlepage]{report}

\usepackage[dvipsnames]{xcolor}
\usepackage{graphicx} 
\usepackage{bm}
\usepackage{graphicx}
\usepackage{amssymb}
\usepackage{amsfonts}
\usepackage{float}
\usepackage{cuted}
\usepackage{pdfpages}
\usepackage{booktabs}
\usepackage{placeins}
\usepackage{bm}
\usepackage{graphicx}
\usepackage{amssymb}
\usepackage{amsfonts}
\usepackage{fancybox}
\usepackage{fancyhdr}
\usepackage{float}
\usepackage{lipsum}
\usepackage{mathtools}
\usepackage{cuted}
\usepackage{dsfont}
\usepackage{placeins}
\usepackage{booktabs}
\usepackage{multirow}
\usepackage{multicol}
\usepackage{braket}
\usepackage{caption}
\usepackage{subfig}
\usepackage[margin=1cm]{caption}
\usepackage{chngcntr}
\usepackage[toc,page]{appendix}
\usepackage{amsmath}
\usepackage{dsfont}
\usepackage{amsthm}
\newtheorem{theorem}{Theorem}

\newtheorem{corollary}{Corollary}

\usepackage{epigraph}
\usepackage{tesis_ESI_en}
\usepackage{amsmath}
\usepackage{pstricks}
\usepackage{pdflscape}

\DeclareMathOperator*{\argmax}{arg\,max}

\newcommand{\comment}[1]{}

\newcommand{\overbar}[1]{\mkern 1.5mu\overline{\mkern-1.5mu#1\mkern-1.5mu}\mkern 1.5mu}
\newcommand{\eref}[1]{(\ref{#1})}
\newtheorem{Def}{{\em Definition}}
\newtheorem{Pro}{{\em Proposition}}
\newcommand\ytl[2]{
\parbox[b]{8em}{\hfill{\color{cyan}\bfseries\sffamily #1}~$\cdots\cdots$~}\makebox[0pt][c]{$\bullet$}\vrule\quad \parbox[c]{4.5cm}{\vspace{7pt}\color{black}\raggedright\sffamily #2.\\[7pt]}\\[-3pt]}

\begin{document}

\pagenumbering{roman}
\pagestyle{empty}
\pagestyle{empty}
\vspace*{6cm}
\begin{center}
\begin{minipage}[c][0.7\height][t]{\textwidth}
	\begin{center}
	\LARGE{\bf Error Correction for Reliable Quantum Computing}
	\end{center}
\end{minipage}\end{center}
\cleardoublepage


\begin{center}
\begin{minipage}[c][0.8\height][t]{\textwidth}
	\begin{center}
	\LARGE{\bf{Error Correction for Reliable Quantum Computing}}
	\end{center}
\end{minipage}\vspace*{1cm}
\begin{minipage}[c][0.2\height][t]{\textwidth}
	\begin{center}
	\Large{\bf{by}}
	\end{center}
\end{minipage}\vspace*{1cm}
\begin{minipage}[c][0.8\height][t]{\textwidth}
	\begin{center}
	\Large{\bf{Patricio Fuentes Ugartemendia}}
	\end{center}
\end{minipage}
\begin{minipage}[c]{\textwidth}
\begin{center}
\vspace{0.75cm}
\includegraphics[scale=0.8]{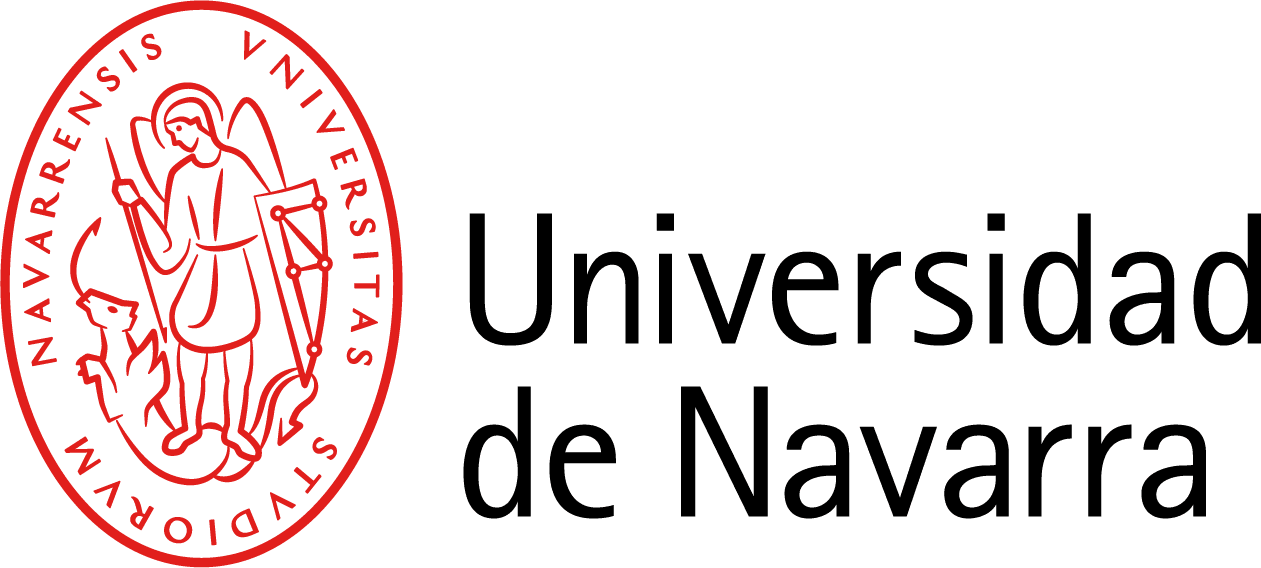}
\vspace{1cm}
\end{center}
\end{minipage}
\vspace{1cm}
\begin{minipage}[t][1.1\height][t]{1.0\textwidth}
	\begin{center}
	\large A dissertation submitted to the \\ \vspace{0.15cm}TECNUN - SCHOOL OF ENGINEERING \\\vspace{0.15cm}
	in partial fulfillment of the requirements for the Degree of\vspace{0.15cm} \\
	DOCTOR OF PHILOSOPHY
	\end{center}
	\vspace{0.24cm}
\end{minipage}

\vspace{0cm}
\begin{minipage}[t][1.1\height][t]{\textwidth}
	\begin{center}
	\large Under the supervision of:\\
	\vspace{0.15cm}
	Professor Pedro M. Crespo Bofill \\
	\end{center}
	\vspace{1.24cm}
\end{minipage}

{\large University of Navarra--TECNUN\\
School of Engineering\\
2021}\end{center}
\clearpage

\normalsize\vspace*{15cm}
\begin{minipage}{13cm}
Doctoral Dissertation by the University of Navarra-TECNUN\\ 

\copyright{Patricio Fuentes Ugartemendia}, Dec 2021\\[2ex] 

Donostia-San Sebasti\'an
\end{minipage}

\cleardoublepage

\vspace*{\fill}
For my parents Armando and Pili, my sister Idoia, and my dog Moose. \newline \newline 
\indent For Maria, who I would never have met had I not pursued this PhD. 
\vspace*{\fill}
\cleardoublepage

\pagestyle{plain}
\pagenumbering{roman}
\setcounter{page}{3}

\pagestyle{fancyplain}

\chapter*{Acknowledgements}

It is with great pride and no small sense of accomplishment that I present to you my PhD dissertation. Although my name is the one that appears on the cover, by no means should the work in this document be attributed only to me. Writing a PhD dissertation is no small feat, in fact I have seen it compared to bearing children (an exaggeration certainly...), and one that cannot possibly be achieved alone. For this reason, it is only appropriate that, prior to diving into the contents of this thesis, all those involved in this process be appropriately thanked and their contributions acknowledged. I must, however, make the disclaimer that I have unchained the literary demon that lives within me to write this section. Readers, proceed at your own peril.

It should come as no surprise that the first person to whom I must express my gratitude is my supervisor Professor Pedro Crespo Bofill. He has been present throughout all my academic career-defining moments; having supervised my bachelor thesis, my master's thesis and now my PhD thesis. I have actually known Pedro since I first set foot in Tecnun, as, by what I now believe to have been fate, he became my academic advisor in my first year as an undergraduate student. In all this time, Pedro has been nothing but a joyful and motivating presence, his support and belief in me never wavering. I also find it impossible to overstate his commitment to his students; even at the most improper of times, he has always been willing to spare a moment to give me guidance and advice. His technical expertise is folklore at our university, but only as a PhD student have I truly come to comprehend the depths of his genius; how one can know so much about such diverse fields of science I know not. For giving me the chance to pursue my academic dreams and for so many other things, thank you Pedro. 

It is only fitting that the second person I thank be my colleague and co-author Josu Etxezarreta Martinez. He is the pioneer of quantum information at our university and the one person that managed to sell me on this fascinating realm. Conducting our research and our dissertations together has been one of the highlights of my PhD journey and one of the primary reasons for my academic success. Besides his scholarly contributions to my work, I must also thank Josu for an innumerable amount of non-technical things. Because there are too many to mention herein, I will simply say the following: May our academic prowess continue to increase as a function of our coffee consumption, and may you never cease to ``\textit{yayeet}'' and ``\textit{yeah buddy!}''. For your invaluable aid and for becoming one of my closest friends, thank you Josu. 

To finish with my fellow co-authors, I must now thank Professor Javier Garcia Frias at the University of Delaware. This dissertation would be nowhere near its current form if not for his involvement. I also feel obliged to thank Xabier Insausti and Marta Zárraga at my own department (department of Mathematical Principles). They have given me the chance to learn concepts completely outside the scope of quantum information and have allowed me to participate and contribute towards a scientific publication on a topic that was completely alien to me a year back. Moreover, it would be unjust to complete this paragraph without mention of Professor Jesús Gutiérrez. Despite not being directly involved in my thesis, it is because of him that I have been able to work within the department of Mathematical Principles. For your help and patience, I thank you all. 

At this point, it is time for me to thank the past and present members of office D15. It is only right that I begin with Imanol Granada, a.k.a, ``\textit{the Prophet}'' or ``\textit{Sir Safemoon}'', the man responsible for bringing forth what can only be described as the biggest personal economic bonanza I have ever seen. Know that you are dearly missed in D15 and that I cannot wait to see what lies ahead in our joint blockchain endeavours. For bringing me into the cryptoverse and lowering my aversion to risk, thank you Imanol. Next we have Fernando Rosety, another member who no longer dwells among us in D15. The office is not the same without your humming of classical music. While some leave, others come, and so I now move to the newest member of our office, Toni de Martí. Although I have only known him for a scant three months, I have never met anyone with such a surprising array of stories and anecdotes. I must also mention Iñigo Barasoain and Fran Velásquez. The former I thank for his respect (despite having ascended to the higher plane of mathematics he continues to engage with us lowling engineers) and the latter for his cheerfulness and his innate ability to take jokes. 

Along these lines, I must now mention all those friends and acquaintances at the university who, perhaps unbeknownst to them, have helped me along the way. First comes Paul Zabalegui. Our friendship began in summer swim school when we were about ten years old and although we lost touch after it finished, fate would have us meet at university once again. There are few people with whom I share so many of my passions and I feel fortunate to count him among my dearest friends. Next we have Unai Ayucar. Although it may border on insanity at times, his untethered imagination has always motivated me to revise my opinion of what is possible in life. This unfettered creativity also extends to his cooking, as only he could possibly think it a good idea to fry \textit{paella} in litres of olive oil. Last among my estranged classmates is Daniel Talan. Despite being in Switzerland, our voice calls and discussions have become a welcome and enjoyable pastime. Thank you.

I continue at the risk of making these acknowledgements longer than the dissertation itself, but it would be an injustice to leave the following people without mention. All my friends and teammates at Txuri-Urdin. Through the rough patches of our first few seasons to winning $3$ league titles in a row, I will forever cherish the memories of our time together. In particular, I want to thank Luis Gimenez, Lucas Serna, Pablo Zaballa, Mikel Mendizabal, and Borja Aizpurua personally. You have done more for this thesis than you know. In similar fashion, I also want to give thanks to my teammates, coaches, and staff at the Spanish national team. Actually, I should just thank the game of hockey itself. Even after twenty two years of playing, nothing makes me feel more alive than stepping on the ice. The game has blunted the edge of many a sharp knife in my life and its capacity to distract me has been invaluable to my success as a researcher. For all that you have given me, thank you hockey. 

The game took me to Saint Andrew's College and so we now turn to that marvelous place. I would not be who I am today had I not spent two years at SAC. True to its motto, the place molded me into a significantly more mature individual and taught me skills that have shined with brilliance during my time as a PhD student. To all my teachers, coaches, and friends at SAC, thank you. Special shoutouts to my friends in 1st Hockey and the so-called ``Dawgz'': Graham, Humza, David, YoungWoo, West, and Andy. Although we have not seen each other in a long time, I know our friendship still runs as deep as it did when we gamed away our days at the Manor. Following this theme of childhood friends, thank you to Isma, Guille, Luken, and Andrey. From the day we first met in kindergarten at Saint Patrick's, our passion for sports and later on fitness kept us united. I pray that our games of \textit{pádel} never stop being so competitive. 

Barring chance or extremely good fortune, those I will thank next will likely never read this dissertation. Thank you to R. A Salvatore, Steven Erikson, Brandon Sanderson, Patrick Rothfuss, and all those other authors from whom I have learned so much. Thank you also to Satoshi Nakamoto. I do not doubt that your creation will change the world for the better. I must also thank the Counting Crows, Rise Against, Machine Gun Kelly, and the Chainsmokers for motivating me in all those instances when my discipline evaporated. In relation to this, I must also thank the people at Youtube and Nespresso (and Iñigo Gutierrez by association); procrastination via the classic video and black coffee combo will never get old. Finally, a special shoutout to all those online meetings, home workouts, and quarantine periods brought to me by the COVID-$19$ pandemic. Somehow, throughout this entire ordeal, my work has not been impeded at any point in time. I have been extremely fortunate.

Having left the best for last, I must now turn to those closest to my heart. After spending days thinking about how to thank my parents appropriately, I found it best to use the following quote by the writer Chuck Palahniuk: ``First your parents, they give you your life, but then they try to give you their life''. Despite this, I am still left with a feeling of insufficiency. I guess my parents have done so much for me that it is not possible to put my gratitude into words. \textit{Gracias attatto y amatxo por todo}. The same goes for my sister Idoia, to whom I owe more than can be expressed. Thank you for always being there, for always having time to talk, and most of all, for always being willing to listen. Through thick and thin, your presence has never failed to remind me that there is always light at the end of the tunnel. Lastly, thank you Moose for being such an energetic furball and for greeting me at the door everyday as if you had not seen me in years. 

Alas, it is now time for me to thank one final person. Maria, you, above anyone else, have lived through the highs and the lows of my PhD. You have seen the extent to which manuscript revisions can frustrate me and how much ideas can consume me. You remained at my side through it all, and only through your intervention have I managed to complete my work. \textit{Maria, tesian gertatu zaidan gauzarik onena zara}. 

That was a lengthy acknowledgements section, I know. Even so, I cannot help but feel like I have not thanked all the people that have helped me along the way. In any case, I think that it is high time for me to stop torturing the reader with my sorry attempts at Shakesperean prose. To all those I have named herein and those who I have surely forgotten, thank you for helping me navigate the trials and tribulations of this voyage, I could not have done it without you.

\cleardoublepage

\chapter*{Abstract}

    Quantum computers herald the arrival of a new era in which previously intractable computational problems will be solved efficiently. However, quantum technology is held down by decoherence, a phenomenon that is omnipresent in the quantum paradigm and that renders quantum information useless when left unchecked. The science of quantum error correction, a discipline that seeks to combine and protect quantum information from the effects of decoherence using structures known as codes, has arisen to meet this challenge. Stabilizer codes, a particular subclass of quantum codes, have enabled fast progress in the field of quantum error correction by allowing parallels to be drawn with the widely studied field of classical error correction. This has resulted in the construction of the quantum counterparts of well-known capacity-approaching classical codes like sparse codes and quantum turbo codes. However, quantum codes obtained in this manner do not entirely evoke the stupendous error correcting abilities of their classical counterparts. This occurs because classical strategies ignore important differences between the quantum and classical paradigms, an issue that needs to be addressed if quantum error correction is to succeed in its battle with decoherence. In this dissertation we study a phenomenon exclusive to the quantum paradigm, known as degeneracy, and its effects on the performance of sparse quantum codes. Furthermore, we also analyze and present methods to improve the performance of a specific family of sparse quantum codes in various different scenarios.

\chapter*{Research Papers}

This thesis is the culmination of two and a half years of work within the \textit{Mathematical Principles group} of the {Department of Biomedical Engineering and Sciences} at the \textit{Tecnun - School of Engineering (University of Navarra)}. Throughout this time, I have published a number of research papers, detailed below in chronological order. In terms of their relationship to this dissertation, this thesis is mostly comprised of the results obtained in those articles that I have first-authored myself (shown in blue). To provide context, I have included a brief summary of the other works that I have co-authored in Chapter 8.

\begin{itemize}
    \item  \textcolor{Blue}{\textbf{P. Fuentes}, J. Etxezarreta Martinez, P. M. Crespo, and J. Garcia-Frias, ``Approach for the construction of non-Calderbank-Steane-Shor low-density-generator-matrix based quantum codes,'' \textit{Phys. Rev. A}, vol. 102, pp. 012423, 2020. doi:10.1103/PhysRevA.102.012423.}
    
    \item \textcolor{Blue}{\textbf{P. Fuentes}, J. Etxezarreta Martinez, P. M. Crespo, and J. Garcia-Fr\'ias, ``Performance of non-CSS LDGM-based quantum codes over the Misidentified Depolarizing Channel,'' \textit{IEEE International Conference on Quantum Computing and Engineering (QCE20)}, 2020. doi:10.1109/QCE49297.2020.00022.}
    
    \item J. Etxezarreta Martinez, \textbf{P. Fuentes}, P. M. Crespo, and J. Garcia-Frias, ``Pauli Channel Online Estimation Protocol for Quantum Turbo Codes,'' \textit{IEEE International Conference on Quantum Computing and Engineering (QCE20)}, 2020. doi: 10.1109/QCE49297.2020.00023.
    
    \item J. Etxezarreta Martinez, \textbf{P. Fuentes}, P. M. Crespo, and J. Garcia-Frias, ``Approximating Decoherence Processes for the Design and Simulation of Quantum Error Correction Codes in Classical Computers,'' \emph{IEEE Access}, vol. 8, pp. 172623-172643, 2020. doi: 10.1109/ACCESS.2020.3025619.
    
    \item \textcolor{Blue}{\textbf{P. Fuentes}, J. Etxezarreta Martinez, P. M. Crespo, and J. Garcia-Frias, ``Design of LDGM-based quantum codes for asymmetric quantum channels,'' \textit{Phys. Rev. A}, vol. 103, pp. 022617, 2021. doi: 10.1103/PhysRevA.103.022617.}
    
    \item \textcolor{Blue}{\textbf{P. Fuentes}, J. Etxezarreta Martinez, P. M. Crespo, and J. Garcia-Frias, {``Degeneracy and its impact on the decoding of sparse quantum codes,''} \emph{IEEE Access}, vol. 9, pp. 89093-89119, 2021. doi: 10.1109/ACCESS.2021.3089829.}
    
    \item J. Etxezarreta Martinez, \textbf{P. Fuentes}, P. M. Crespo, and J. Garcia-Frias, ``Time-varying quantum channel models for superconducting qubits,'' \textit{npj Quantum Information}, vol. 7, no. 115, 2021. doi: 10.1038/s41534-021-00448-5.
    
    \item \textcolor{Blue}{\textbf{P. Fuentes}, J. Etxezarreta Martinez, P. M. Crespo, and J. Garcia-Frias, {``On the logical error rate of sparse quantum codes,''}  submitted to \textit{IEEE Trans. on Quantum Eng.}, 2021. arXiv: 2108.10645v2.}
    
    \item J. Etxezarreta Martinez, \textbf{P. Fuentes}, P. M. Crespo, and J. Garcia-Frias, ``Quantum outage probability for time-varying quantum channels,'' submitted to \textit{Phys. Rev. A}, 2021. arXiv:2108.13701.

\end{itemize}

\cleardoublepage

\chapter*{Glossary}

A list of used acronyms is provided below.\\

\begin{tabular}{ll}
\textbf{BER}  		&       \textit{Bit Error Rate} \\
\textbf{BP}  		&       \textit{Belief Propagation} \\
\textbf{BSC}  		&       \textit{Binary Symmetric Channel} \\
\textbf{BWML}  		&       \textit{Bit-Wise Maximum Likelihood} \\
\textbf{CSS}  		&       \textit{Calderbank-Shor-Steane} \\
\textbf{EFB}  		&       \textit{Enhanced Feedback} \\
\textbf{ECC}  		&       \textit{Elliptic Curve Cryptography} \\
\textbf{LDGM}  		&       \textit{Low Density Generator Matrix} \\
\textbf{LDPC}  		&       \textit{Low Density Parity-Check} \\
\textbf{LLR}  		&       \textit{Log-Likelihood Ratio} \\
\textbf{LUT}  		&       \textit{Look-Up Table} \\
\textbf{NP}  		&       \textit{Non-deterministic Polynomial} \\
\textbf{OSD}  		&       \textit{Ordered Statistics Decoder} \\
\textbf{PCM}  		&       \textit{Parity Check Matrix} \\
\textbf{QCC}  		&       \textit{Quantum Convolutional Code} \\
\textbf{QEC}  		&       \textit{Quantum Error Correction} \\
\textbf{QLDGM}  		&       \textit{Quantum Low Density Generator Matrix} \\
\textbf{QLDPC}  		&       \textit{Quantum Low Density Parity Check} \\
\textbf{QMLD}  		&       \textit{Quantum Maximum Likelihood Decoding} \\
\textbf{QPCM}  		&       \textit{Quantum Parity Check Matrix} \\
\textbf{QSC}  		&       \textit{Quantum Stabilizer Code} \\
\textbf{QTC}  		&       \textit{Quantum Turbo Code} \\
\textbf{RSA}  		&       \textit{Rivest, Shamir and Adleman} \\
\textbf{SP(A)}  		&       \textit{Sum-Product (Algorithm)} \\
\textbf{WER}  		&       \textit{Word Error Rate} \\

\textbf{i.i.d.}  		&       \textit{independent and identically distributed} \\

\end{tabular}



\cleardoublepage
\chapter*{Notation}

Although all symbols are defined at their first appearance, some are repeated throughout the dissertation. A list of the most frequent symbols is provided below. \\

\begin{tabular}{ll}
\( \mathcal{H}_2\) 	&  Complex Hilbert space of dimension 2. \\
\(\otimes\)   	&  Tensor product. \\
\(I\)    	&  Single qubit Pauli matrix (identity). \\
\(X\)   &  Single qubit Pauli matrix (bit flip). \\
\(Z\)  &  Single qubit Pauli matrix (phase flip). \\
\(Y\) & Single qubit Pauli matrix (bit \& phase flip). \\
\(\Pi^{\otimes N}\) 	&  Set of $N$-fold tensor products of single qubit Pauli operators. \\
\(\mathcal{G}_N\)   	&  $N$-fold Pauli group. \\
\(\overbar{\mathcal{G}}_N\)   	&  Effective $N$-fold Pauli group. \\
\(\xi\)   	&  Quantum channel. \\						
\(\xi_P\)   	&  Pauli channel. \\
\(|\cdot|\)   	&  Absolute value of a number or carnality of a set. \\
\(\mathcal{S}\) & Stabilizer set defined over the Pauli group. \\
\(\mathcal{\overbar{S}}\) & Stabilizer set defined over the effective Pauli group. \\
\(k\) & Length of information word. \\
\(n\) & Length of a codeword. \\
\(\mathcal{C(\overbar{S})}\) & Stabilizer code. \\
\(\ket{\psi}\) & Quantum Information state. \\
\(\ket{\overbar{\psi}}\) & Encoded quantum state. \\
\(\mathbb{F}_2^{2N}\) & Set of length $2N$ binary vectors. \\
\(\odot\) & Symplectic product. \\
\(\oplus\) & Modulo 2 sum. \\
\(\circledast\) 	&  Modulo 2 product. \\
\(\mathbf{w}\)   	&  Quantum syndrome. \\
\(\cdot\)   	&  Group operation over the Pauli Group. \\
\(\star\)   	&  Group operation over the effective Pauli group. \\			
\end{tabular}

\newpage

\begin{tabular}{ll}

\(\beta\)   	&  Symplectic isomorphism/map. \\
\(\mathcal{\overbar{Z}(\overbar{S})}\) & Effective centralizer of a stabilizer. \\
\(\mathcal{\overbar{N}(\overbar{S})}\) & Effective normalizer of a stabilizer. \\
\(\mathbf{S}_v\) & Stabilizer operator. \\
\(\mathbf{T}_i\) & Pure error operator \& effective centralizer coset representative. \\
\(\mathbf{L}_j\) & Logical operator \& stabilizer coset representative. \\
\(C^\perp\) & Dual of a an error correction code. \\

\end{tabular}

\cleardoublepage

\clearemptydoublepage
\pagestyle{fancyplain}
\def\contentsname{Contents}
\setcounter{tocdepth}{3}
\tableofcontents
\clearemptydoublepage



\pagenumbering{arabic}

\listparindent=10mm
\parskip=10pt

\setcounter{page}{1}
\def\chaptername{CHAPTER}


\chapter{Introduction} \label{cp1_intro}

\epigraph{\textit{``Veris in numeris''}}{\textbf{Satoshi Nakamoto}.}

\noindent\hrulefill

The behaviour and composition of matter in its most reduced scale has long been pondered by the scientific community. In fact, the concept of the atom dates back to the 5th century BCE, which is when the Greek philosophers Leucippus and Democritus first brought up the idea. Since then, understanding of the topic has progressed immensely, especially with the development of quantum mechanics. Unfortunately, despite our advancements, many areas in the field of physics still defy human understanding. This is, in no small part, due to the incapacity of classical computers to simulate the time evolution of subatomic systems. It was precisely for this reason that, in his revolutionary work \cite{feynman},  Richard Feynman posited that without devices that obeyed the eldritch laws of quantum mechanics (the fundamental theory in physics that describes the behaviour of subatomic particles) it would not be possible to accurately portray the behaviour of matter. Thenceforth, research has shown that quantum constructs have myriads of applications beyond Feynman's original proposal and that they are especially well suited to efficiently solve certain tasks which are computationally unmanageable for classical instruments. The advantages provided by these machines stem from the quantum nature of their most basic component: the quantum bit (qubit). While classical bits can only exist in one of two states, $0$ or $1$, qubits manifest as a superposition of these two states, which means that they are both $0$ and $1$ simultaneously. 

The implications that the superposition property of qubits has on the field of computation are vast. Whereas a classical $N$-bit register stores a single $N$-bit value, superposition allows an $N$-qubit quantum register to store $2^N$ states concurrently. Then, through the devious application of global function optimization techniques, the $2^N$ states can be evaluated \textit{in parallel}\footnote{The analogy with parallel computing is useful to understand the advantages that quantum computers provide. However, it must be stated that quantum computing and parallel computing are not one and the same. Quantum computing exploits the superposition property of qubits to consider the entire solution space of a particular problem simultaneously. Parallel computing makes use of multiple processors to evaluate each possible solution independently on each of them. The former is limited by the amount of qubits it can employ while the latter is limited by the number of processors at its disposal.} for a cost analogous to that of a single classical evaluation \cite{bab1, bab2}. For this reason, specific problems that are computationally hard in classical terms, such as the factorization of large numbers or performing a search through an unstructured database, become significantly less complex on quantum machines capable of running quantum algorithms \cite{shor, grover1, grover2}. For instance, Shor's algorithm for the factorization of prime numbers runs in polynomial time while the best known classical algorithm for this same purpose runs in exponential time \cite{shor}. Other currently known notable tasks that are better addressed using quantum technology are the discrete logarithm problem \cite{shor}, Byzantine agreement \cite{byzantine}, or parallel computation in communication networks \cite{parallel1, parallel2, parallel3}. Aside from the field of computation science, numerous other scientific fields stand to gain from the development of the quantum framework. A good example is the area of communication security, where the journey towards quantum secure cryptographic schemes has already begun with the proposal of the BB84 \cite{BB84} and the E91 \cite{E91} protocols. 

This staggering theoretical potential has transformed quantum technology into the harbinger of a new era in the fields of computation and communications, and its capability to outperform classical methods in the areas of information processing, storage, and transmission can no longer be disputed. Unfortunately, despite the scientific community's unwavering commitment to the construction of a full-scale quantum computer, devices capable of realizing the promise of quantum information science have not yet become a reality. Mostly, this can be attributed to the phenomenon known as decoherence \cite{decoherence1, decoherence2, decoherence3}, which describes the process by which the quantum objects that store quantum information lose coherence as a consequence of their interaction with the environment. The only way to indefinitely maintain coherence requires the perfect isolation of a quantum state, a process that prohibits any interaction or manipulation of said state, which means that decoherence is unavoidable when working with quantum information. In consequence, for quantum computers to be useful, they must guarantee sufficiently long quantum information coherence time periods for practical applications. Satisfying this requirement is no simple feat, as it implies that quantum processors must function correctly even when their elemental information units suffer from decoherence effects. It is to find an answer to this dilemma that the scientific discipline known as the theory of Quantum Error Correction (QEC) has arisen; to find ways to ensure that quantum technology can operate in time intervals that are long enough for the advantages of quantum computing to come to light. In fact, the corruptive power of decoherence is strong enough to make many experts believe that, without appropriate error correction strategies, quantum computing itself hangs in the balance. 

This widespread concern with regard to the achievability of quantum computing in the absence of error correction has caused the field of QEC to experience a drastic surge in popularity since the first quantum code was introduced in \cite{decoherence1}. Significant breakthroughs have been made during this rise to fame, of which (arguably) the most important is the formulation of Quantum Stabilizer Codes (QSCs) in Gottesmans PhD thesis \cite{QSC}. In said work, Gottesman shows how, by casting existing groups of classical codes into the framework of QSCs, quantum counterparts of these classical designs can be derived, effectively allowing the development of quantum coding schemes from existing classical strategies. This formalism has enabled an almost seamless transition from classical error correction to QEC and has led to the construction of many QEC code families like Quantum Reed-Muller codes \cite{QRM}, sparse quantum codes or Quantum Low Density Parity Check
(QLDPC) codes \cite{bicycle, ldpc1, qldpc15, jgf1, jgf2, patrick, patrick-asym}, Quantum Convolutional
Codes (QCC) \cite{QCC1, QCC2, QCC3}, Quantum Turbo Codes (QTC) \cite{QTC, EAQTC, josu1, josu2} and Quantum Topological Codes \cite{toric, top1, top2, surface}. 

It is also pivotal for the mechanisms through which a QEC code bestows quantum information with a robust defence against quantum decoherence to be of reasonable complexity. The encoding and decoding requirements of QEC codes are of paramount importance since the quantum gates which implement error correction operations are also faulty and may induce additional errors in the quantum information. This gives rise to the concept of fault-tolerance or fault-tolerant computing \cite{threshold, fault1, fault2, fault3}, which is a term used to refer to the notion of a quantum apparatus functioning correctly despite the fact that its most basic components may sometimes be faulty themselves. 

Among the aforementioned quantum code families, the quantum counterparts of sparse codes stand out as being especially well suited to implement fault-tolerant error correction methods. The sparsity of their decoding matrices implies that only a few quantum interactions per qubit are necessary in the error correction procedure, and ensures that additional quantum gate errors are avoided. In consequence, this field is evolving rapidly and numerous new construction and design methods for sparse quantum codes are being proposed \cite{ref1, ref2, ref3, ref4, ref5, ref6}. These codes, which are also known as QLDPC codes, can be defined as stabilizer codes with sparse generators and they can be constructed using a variety of different methods. One of the most commonly employed design strategies consists in taking classical LDPC codes as the starting point and adapting them so that they can be used in the quantum paradigm \cite{qldpc15}.

Classical LDPC codes, along with turbo codes \cite{turbo}, represent forms of sparse or random-like codes that can be decoded probabilistically and that are capable of approaching the theoretical communication limits of a communication channel with a reasonable decoding complexity. This stems from the fact that they provide sufficient structure for the decoding  process to function correctly, while, simultaneously, the randomness involved in the design itself guarantees excellent performance. Decoding is performed by means of the Sum Product Algorithm (SPA) \cite{spa}, which is a generic message passing algorithm that computes various marginal functions associated with a global function. Related decoding methodologies for probabilistic codes, such as Belief Propagation (BP) \cite{BP} or the Viterbi algorithm \cite{Viterbi}, have been shown to be specific instances of the SPA \cite{Wiberg}. The SPA operates over tree-like graphs known as factor graphs, which are used to represent a complicated ``global" function of many variables as a product of simpler ``local" functions, each depending on a subset of the variables. Factor graphs express which variables are arguments of which local functions and the SPA derives its computational efficiency by exploiting the way in which the global function factors into those products of ``local" functions. When the factor graph is a tree, the SPA converges to the exact solution in a time bounded by the tree's depth. In scenarios where the algorithm does not converge, i.e., when the factor graph has loops, it still represents a good heuristic method to implement sub-optimal decoding if the loops are long enough. In \cite{qBP1} and \cite{qBP2}, it was shown that decoding stabilizer QEC codes on memoryless quantum channels can be defined as the execution of the SPA over a typically loopy factor graph. 

Being linear block codes, classical LDPC codes are designed by defining a set of parity check equations that involve information bits. To guarantee their low density, each equation involves a small number of bits, and each particular bit is involved in a reduced number of equations. This set of constraints is defined by means of a Parity Check Matrix (PCM) where each row denotes a parity check equation and each column denotes a coded bit. The PCM can also be represented by means of a factor graph, where two types of interconnected nodes, variable nodes and parity check nodes, represent each of the columns and each of the rows of the PCM, respectively. Because of the low density requirements imposed in the design of these codes, the corresponding factor graph will have a small number of loops, ensuring good performance when decoded using the SPA. Given their capacity-approaching performance under SPA decoding, as well as their potential upside to implement effective error correcting strategies, deriving good quantum sparse codes is germane to the field of QEC.


\section{Motivation and Objectives} \label{cp1:objectives}

Two primary issues arise when designing sparse quantum codes. On the one hand, most randomly generated LDPC codes are not suitable for the quantum paradigm and so the number of good classical codes applicable to the quantum domain is reduced. On the other, decoding based on the SPA is affected by a quantum phenomenon known as \textit{degeneracy} \cite{degen1, degen2, degen3, degen4}, which has no classical equivalent. In the literature, the first problem is addressed by using constructions with stringent requirements, like the one simultaneously proposed by Calderbank, Shor, and Steane in \cite{CSS1, CSS2} (CSS codes). Unfortunately, these methods introduce particular drawbacks that further complicate the design of QLDPC codes and that, as of yet, have not been completely resolved. The second issue, which pertains to degeneracy, poses a quandary that is more difficult to give an answer to. This happens because the decoding algorithm \cite{spa, BP} used to decode degenerate QLDPC codes is designed, in principle, for a classical environment in which degeneracy is not present, and so it will be completely blind to this phenomenon\footnote{The effects of degeneracy are mitigated in entanglement assisted schemes \cite{EA1, EA2, EA3}. Such strategies make use of pre-shared Bell states to reduce the degenerate content of a quantum code to the point that, depending on the amount of pre-shared information, the scenario becomes increasingly similar to that of classical decoding \cite{EAQTC}. However, unassisted coding strategies that cannot make use of entanglement experience the full-fledged impact of this phenomenon.}.

In this thesis we attempt to tackle both of these concerns: improving the performance of QLDPC codes and studying degeneracy, by establishing the following key objectives:

\subsection{Performance of QLDPC codes across the landscape of quantum channels}

Due to the ease with which quantum codes can be built from their classical counterparts based on the CSS construction, most of the existing QLDPC codes are built using this methodology. CSS codes are a particular subset of the QSC family that provide a straightforward method to design quantum codes via existing classical codes. Although the construction introduces additional code design challenges (the classical codes must comply with a specific algebraic condition), the method ensures that the resulting quantum code is applicable in any quantum environment. Thus, codes built in this manner can be used across the entire spectrum of quantum channel models, which means that they can be studied and optimized for different practical scenarios. 

Unfortunately, the CSS construction is not without its faults. Classical codes that can be employed in CSS schemes must meet specific requirements that limit the performance that is attainable with the resulting quantum code. This impediment to the performance of CSS codes, known as the \textit{CSS lower bound} \cite{CSSbound}, implies that the best possible theoretical performance over a quantum channel cannot be met when using a CSS code. This has inspired the search for non-CSS constructions, as they are not limited by the CSS lower bound and should be able to outperform CSS codes provided that they are designed optimally. Non-CSS LDPC-based codes were proposed in \cite{nonCSS1} and \cite{nonCSS2}. Despite showing promise, these codes failed to outperform existing CSS QLDPC codes for comparable block lengths.

Against this backdrop, it is clear that there is ample room for scientific growth within the niche of CSS-based design of sparse quantum codes. However, the overarching nature of our first objective requires that we addresses its constituent parts: the design of non-CSS codes and the performance of CSS codes over different quantum channel models, separately. For this reason, in this dissertation we restrict the study of non-CSS QLDPC designs to the framework of the depolarizing channel (the most extended quantum channel model) and we use the widely-established CSS construction technique to build codes and optimize them for less conventional quantum channel models. 
\subsection{Understanding and exploiting degeneracy}

When quantum stabilizer codes built from sparse classical codes are employed in the quantum paradigm, they are impacted by a phenomenon known as degeneracy \cite{degen1, degen2, degen3, degen4}, which has no classical equivalent. This causes stabilizer codes to exhibit a particular coset structure in which multiple different error patterns act identically on the transmitted information \cite{CSS2, symplec1, symplec2}. 
Although the manifestation of degeneracy in the design of sparse quantum codes and its effects on the decoding process has been studied extensively \cite{QSC, degen3, degen4, logical, softVit, neural, Hard, softPoul}, especially for QTCs and quantum topological codes \cite{EAQTC, softVit, toricphd1, toricphd2, sabo}, it remains a somewhat obscure topic in the literature. This can be attributed to the varying and sometimes inconsistent notation and the oft confusing nature of the notion of degeneracy itself. In consequence, although degeneracy should theoretically improve performance, limited research exists on how to quantify and exploit this phenomenon in the framework of QLDPC codes. 

For these reasons, the second objective of this thesis is to accurately characterize the phenomenon of degeneracy as it pertains to sparse quantum codes. In a similar manner to the first objective, we will employ a two pronged strategy to realize this goal: first, we seek to completely describe degeneracy and the mechanisms that govern its behaviour, and then we use this framework to devise methods to diagnose and exploit its manifestation. 

\section{Outline and Contributions of the Thesis}

\begin{figure}[h!]
	\begin{center}
		\includegraphics[width=0.95\columnwidth]{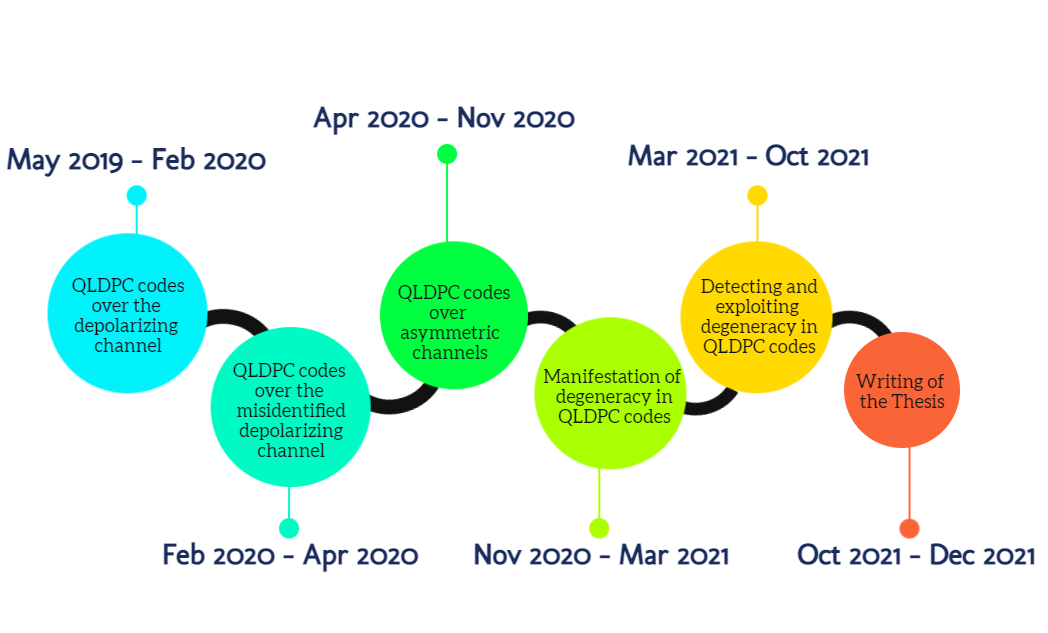}
		\caption{Timeline of the research carried out in this thesis.}
		\label{cp1:chro}
	\end{center}
\end{figure}

The contents of this thesis are structured in a way that, according to the authors view, provides the best possible reading experience. This means that instead of following the chronology of the research, the chapters of the dissertation have been placed so that understanding of each particular chapter is facilitated by those that precede it. A timeline of how the research that comprises this thesis actually evolved can be seen in Figure \ref{cp1:chro}.

The thesis begins with a brief commentary on quantum computers and what they excel at in Chapter 2. This chapter discusses concepts related to complexity theory that shed light on the importance of quantum computing and quantum error correction. In Chapter 3, an introduction to fundamental concepts and preliminaries related to quantum information science and classical error correction is provided. From here on out, the remaining chapters of the dissertation can be grouped into two distinct parts, each one related to the objetives described previously in subsection \ref{cp1:objectives}: Part $1$ (Chapters 4 and 5) discusses the degeneracy phenomenon and its impact on sparse quantum codes, and Part $2$ (Chapters 6 and 7) focuses on the design of non-CSS QLDPC codes and the optimization of CSS QLDPC schemes for different quantum channel models. Then, in the final two chapters of the thesis, a summary of co-authored research (Chapter 8) and possible future work (Chapter 9) is provided.

\subsection{Chapter 2: Why quantum?}

Chapter 2 seeks to clearly present the arguments in favour of quantum computing. For this purpose, it discusses why quantum computers are better suited than their classical counterparts to perform specific tasks like the factorization of prime numbers. It also delves into the technologies that are currently being employed to physically implement quantum computers and provides an overview of the major players in the field of experimental quantum computing.

This chapter is meant as a cursory presentation on the advantages of quantum computing. People that are familiar with the field can skip this section.

\subsection{Chapter 3: Preliminaries on Quantum Information and Classical Error Correction}

Herein we lay the groundwork necessary to follow the discourse in subsequent chapters by introducing basic concepts of quantum information theory and classical error correction. The chapter is divided into two sections: section \ref{cp3:quant} which is devoted to quantum information theory and section \ref{cp3:class} which is dedicated to classical error correction. The chapter commences in section \ref{sec:qubit} with a discussion on the qubit. Then, in Section \ref{sec:entanglement}, we introduce the concept of entanglement, which has become almost folklore in the quantum information community. Following this, we present the notions of unitary operators and gates in the quantum domain in section \ref{sec:noise}. This section on quantum information theory is concluded with an introduction to the Pauli group and a brief overview of the most common quantum channel models. 

Following section \ref{cp3:quant} we turn to the realm of classical error correction. Section \ref{cp3:class} is comprised of subsection \ref{sec:linear-code}, which introduces the concept of linear block codes, and subsection \ref{sec:LDPC}, which presents the family of LDPC codes and other basic communication and graph theory notions like factor graphs and iterative decoding. 

Although this chapter seeks to make the dissertation self-contained, many other important concepts related to quantum information and classical error correction have not been included. If needed, we refer the reader to \cite{NielsenChuang} and \cite{cp1_Gamal} for further detail on quantum information and classical error correction, respectively. Furthermore, because scientists from various different fields of study are involved in quantum computing, it may be that either section $3.1$ or section $3.2$ of this chapter is well known to many readers. In such a case, those familiar with a particular section should skip it and move on to the next section or to Chapter 4.

\subsection{Chapter 4: Degeneracy and its impact on Decoding (Degeneracy I)}

Chapter 4 studies the phenomenon of degeneracy from the perspective of group theory with the purpose of completely characterizing it in the context of sparse quantum codes. The chapter begins by casting the notion of stabilizer codes into a group theoretical framework and using it to perform necessary distinctions between ideas that are sometimes misunderstood in the field of QEC. Following this, the classical and quantum decoding problems are presented, and their similarities and differences are discussed. Then, we proceed by studying the emergence of
degeneracy and the impact that disregarding its existence has on the decoding process. The chapter is closed with a detailed example that serves to illustrate many of these concepts.

The contents of this chapter are based on the following journal paper:

\begin{itemize}
    \item P. Fuentes, J. Etxezarreta Martinez, P. M. Crespo, and J. Garcia-Frias, {``Degeneracy and its impact on the decoding of sparse quantum codes,''} \emph{IEEE Access}, vol. 9, pp. 89093-89119, 2021. doi: 10.1109/ACCESS.2021.3089829.
\end{itemize}

\subsection{Chapter 5: Detecting Degeneracy and Improved Decoding Strategies (Degeneracy II)}

This chapter considers the issue of detecting the presence of degeneracy when using sparse quantum codes. We begin the chapter by discussing the reasons for which limited research exists on how to quantify the true impact that the degeneracy phenomenon has on QLDPC codes. Then, we discuss why two different performance assessment metrics have been used in the literature of sparse quantum codes, and we show how only one of them provides an accurate portrayal of performance. Following this, we devise a method to assess the effects of degeneracy on sparse quantum codes and we explain another previously existing strategy to do so. Finally, we use our strategy to analyze the frequency with which different types of errors occur when using sparse quantum codes and we provide insight on how the design and decoding of these codes can be improved.

The method proposed in this chapter as well as most of its contents have been published in the following journal paper:

\begin{itemize}
    \item P. Fuentes, J. Etxezarreta Martinez, P. M. Crespo, and J. Garcia-Frias, {``On the logical error rate of sparse quantum codes,''} submitted to \textit{IEEE Trans. on Quantum Eng.}, 2021. arXiv: 2108.10645v2.
\end{itemize}

\subsection{Chapter 6: Non-CSS QLDPC codes (QLDPC I)}

Most QLDPC codes are built by casting classical LDPC codes in the framework of stabilizer codes, which enables the design of quantum codes from any arbitrary classical binary and quaternary codes. Some of the best performing QLDPC codes are obtained by combining the CSS construction with Low Density Generator Matrix (LDGM) codes, which are a particular type of LDPC code. However, CSS constructions are limited by an unsurpassable bound, which has inspired the search for non-CSS constructions as they should theoretically be able to outperform CSS codes. In this chapter, we show how the nature of CSS designs and the manner in which they must be decoded limits the performance that they can achieve. Then, we introduce a non-CSS quantum code construction that we derive from the best CSS QLDGM construction that can be found in the literature. We close the chapter by showing how codes designed using this method outperform CSS QLDGM codes and most other QLDPC codes of comparable complexity.

The work that appears in this chapter has been published in the following journal paper:

\begin{itemize}
    \item P. Fuentes, J. Etxezarreta Martinez, P. M. Crespo, and J. Garcia-Frias, ``Approach for the construction of non-Calderbank-Steane-Shor low-density-generator-matrix based quantum codes,'' \textit{Phys. Rev. A}, vol. 102, pp. 012423, 2020. doi:10.1103/PhysRevA.102.012423.
\end{itemize}

\subsection{Chapter 7: Performance of QLDPC codes over Pauli channels (QLDPC II)}

Most of the research related to QLDPC codes has been conducted under the tacit premise that perfect knowledge of the quantum channel in question is available. In reality, such a scenario is highly unlikely, which makes it necessary to analyze the change in the behaviour of these codes as a function of the existing information about the quantum channel. In the first section of this chapter, section \ref{sec:mismatch}, we study the behaviour of the non-CSS QLDGM codes introduced in Chapter 6 under the umbrella of channel mismatch, a term that makes reference to a scenario in which the true channel information and that which is known is different. 

Generally, it has also been the norm in the literature of QEC to consider only the depolarizing channel: the symmetric instance of the generic Pauli channel, that incurs bit-flips, phase-flips, or a combination of both with the same probability. However, because of the behaviour of the materials they are built from, it is not appropriate to employ the depolarizing channel model to represent specific quantum devices. Instead, they must be modelled using a different quantum channel capable of accurately representing asymmetric scenarios in which the likelihood of a phase-flip is higher than that of a bit-flip. Thus, in the second section of this chapter, section \ref{sec:asym}, we study ways in which to adapt the design of the CSS LDGM-based codes discussed in Chapter 6 to asymmetric quantum channels.

The work that comprises this chapter has been published in the following papers:

\begin{itemize}
    \item P. Fuentes, J. Etxezarreta Martinez, P. M. Crespo, and J. Garcia-Fr\'ias, ``Performance of non-CSS LDGM-based quantum codes over the Misidentified Depolarizing Channel,'' \textit{IEEE International Conference on Quantum Computing and Engineering (QCE20)}, 2020. doi:10.1109/QCE49297.2020.00022.
    \item P. Fuentes, J. Etxezarreta Martinez, P. M. Crespo, and J. Garcia-Frias, ``Design of LDGM-based quantum codes for asymmetric quantum channels,'' \textit{Phys. Rev. A}, vol. 103, pp. 022617, 2021. doi: 10.1103/PhysRevA.103.022617.
\end{itemize}

\subsection{Chapter 8: Quantum Turbo Codes and Time-varying quantum channels}

In this chapter, we provide a summary of other research that the author has co-authored and participated in during this PhD dissertation. Although outside the niche of sparse quantum codes, this research is also related to QEC. It is primarily the work of the author's colleague and first author of the journal papers, Josu Etxezarreta Martinez. For this reason, only a succinct overview is contained within this chapter and readers are referred to the first author's own PhD dissertation for discussions that do this work justice. Chapter $8$ is comprised of four sections, each one devoted to a specific topic: Section \ref{sec:turbo} discusses contributions that have been made to the field of QTCs, section \ref{sec:twirl} looks at various mathematical tools that can be used to describe the effects of the decoherence phenomenon, section \ref{sec:TV} goes over the idea of time-varying quantum channels, and finally, section \ref{sec:TV-2} looks at the theoretical limits of error correction in the context of time-varying quantum channels.   

The research that appears in this chapter has been published in the following journal and conference papers:

\begin{itemize}
    \item J. Etxezarreta Martinez, P. Fuentes, P. M. Crespo, and J. Garcia-Frias, ``Pauli Channel Online Estimation Protocol for Quantum Turbo Codes,'' \textit{IEEE International Conference on Quantum Computing and Engineering (QCE20)}, 2020. doi: 10.1109/QCE49297.2020.00023.
    \item J. Etxezarreta Martinez, P. Fuentes, P. M. Crespo, and J. Garcia-Frias, ``Approximating Decoherence Processes for the Design and Simulation of Quantum Error Correction Codes in Classical Computers,'' \emph{IEEE Access}, vol. 8, pp. 172623-172643, 2020. doi: 10.1109/ACCESS.2020.3025619.
    \item J. Etxezarreta Martinez, P. Fuentes, P. M. Crespo, and J. Garcia-Frias, ``Time-varying quantum channel models for superconducting qubits,'' \textit{npj Quantum Information}, vol. 7, no. 115, 2021. doi: 10.1038/s41534-021-00448-5.
    \item J. Etxezarreta Martinez, P. Fuentes, P. M. Crespo, and J. Garcia-Frias, ``Quantum outage probability for time-varying quantum channels,'' submitted to \textit{Phys. Rev. A}, 2021. arXiv:2108.13701.
\end{itemize}

\subsection{Chapter 9: Conclusion and Future Work}

This final chapter concludes our discourse by summarizing the conclusions of our work and analyzing possible routes that the research conducted in this thesis may follow in the future.

\section{How to read this thesis}

\begin{figure}[h!]
	\begin{center}
		\includegraphics[width=\columnwidth, height = 4.5in]{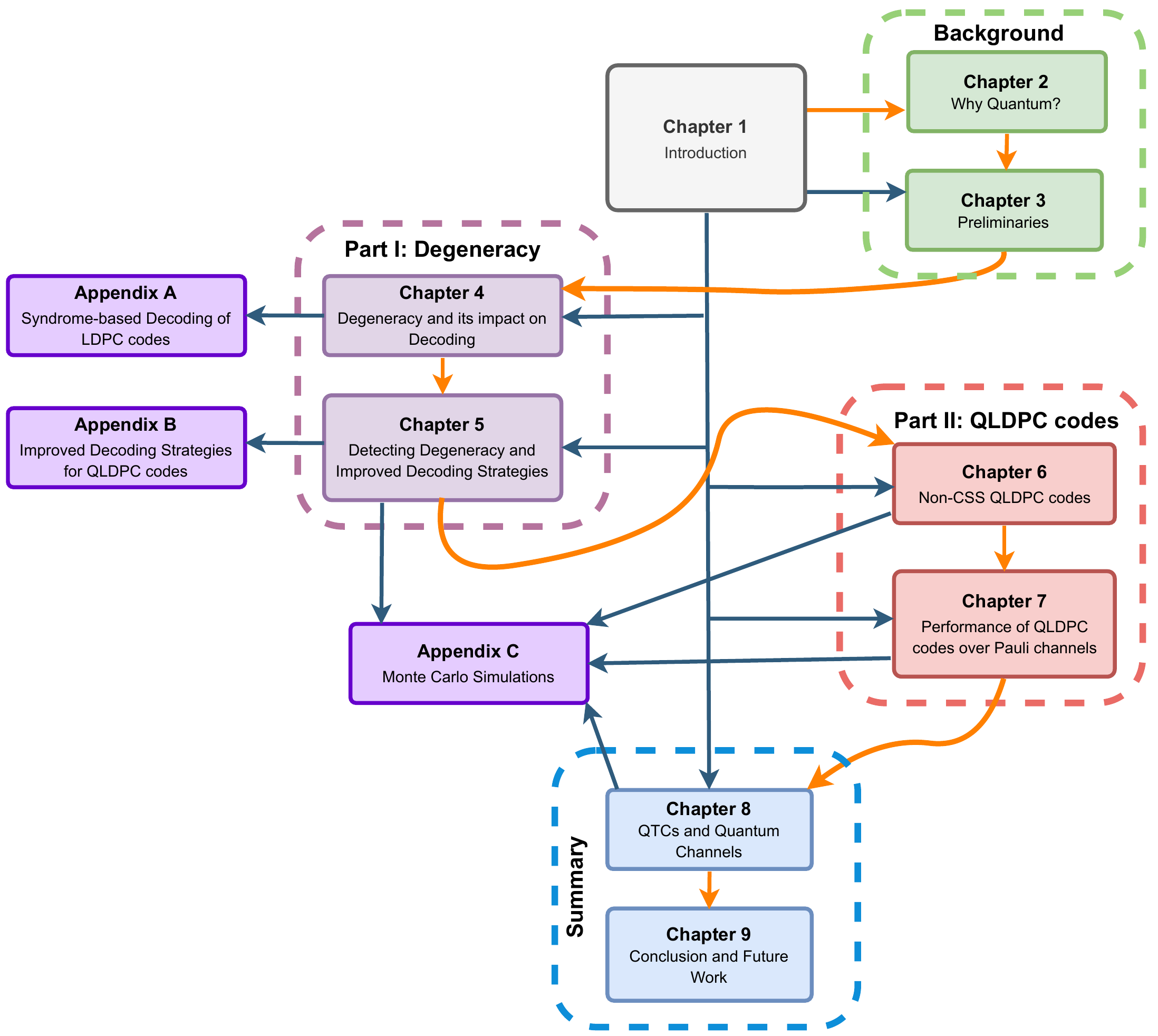}
		\caption{Block diagram detailing the dependencies between chapters.}
		\label{cp1_reading}
	\end{center}
\end{figure}

    An outline of the contents of this thesis is shown in Figure \ref{cp1_reading}. The different parts of this dissertation (enclosed by dotted rectangles in Figure \ref{cp1_reading}) need not be read sequentially, as they are mostly independent from each other. It is worth noting, however, that the notation and particular contents of the chapters of this thesis differ from the original journal articles they are based on. This has been done for the sake of clarity and to maintain the integrity of the notation employed herein. For this reason, it is the author's belief that readers will benefit most from reading each chapter as it is presented (reading the dissertation from start to finish). Thus, we believe that going through this thesis by following the orange path shown in Figure \ref{cp1_reading} will result in the best possible reading experience.

\clearemptydoublepage
\chapter{Why Quantum?} \label{chapter2}

\epigraph{\textit{``Insanity is doing the same thing over and over and expecting different results''}}{\textbf{Albert Einstein}.}

\noindent\hrulefill

The advent of computers in the late twentieth century has profoundly transformed society. Originally designed to facilitate the computation of mathematical calculations, these machines have transcended their primary purpose and are now present in almost all aspects of modern life. Computers are essentially omnipresent, running everything from national monetary frameworks to power grids, serving as the gateway to the vast digital world we refer to as the internet, and being the most valuable asset in the life of many individuals and companies. However, even though computers have enabled mankind to address issues that could not even be conceived prior to their invention, problems that cannot be solved using the worlds most powerful machines still remain. This inability of classical computers to solve specific problems has both negative and positive implications. For instance, in the realm of drug design where complex optimization problems are commonplace, the incapacity of classical computers to solve these problems slows down the process of drug discovery and is perceived as something negative. In contrast, in the field of cryptography, the fact that certain problems can not be solved with classical machines is what makes electronic devices and protocols secure. 

One might assume that based on technological advancement and the constant improvement of electronic components, those complex problems that cannot be currently solved using classical means may potentially be solvable with future classical methods. In other words, what is impossible in today's computers may not be so in those of the future. However, we know this not to be the case thanks to the strong Church-Turing thesis \cite{Turing}. The thesis tells us that every physical implementation of universal computation can simulate any other implementation with only a polynomial slowdown. Essentially, this means that while future classical computers may be better than current ones at attempting to solve these problems, the difficulty of solving the problems scales with the size of the input in the same way on both hardwares, i.e, the complexity of solving the task can be understood as being independent of the computer it is run on. Therefore, we can say that there is a subset of problems that, no matter how advanced electronic computational methods become, will be impossible to solve by classical computers in a reasonable amount of time. To better understand this concept, we need to look at it through the lens of classical complexity theory, which is the scientific discipline tasked with classifying computational problems according to their difficulty.

In complexity theory, so-called easy, or classically tractable, problems can be solved by computer algorithms that run in polynomial time; i.e., for a problem of size $N$, the time or number of steps needed to find the solution is a polynomial function of $N$. Algorithms for solving hard, or intractable, problems, on the other hand, require times that are exponential functions of the problem size $N$. Polynomial-time algorithms are considered to be efficient, while exponential-time algorithms are considered inefficient, because the execution times of the latter grow much more rapidly as the problem size increases. Based on this perspective, complexity theorists refer to the aforementioned clasically-intractable problems as Non-deterministic Polynomial (NP) time problems, a term that represents a class of computational problems for which no efficient solution algorithm has been found. Problems are said to belong to the NP class if their solution can be guessed and verified in polynomial time, and are labelled as non-deterministic because no particular rule is followed to make the guess. Thus, although a solution to an NP problem can be verified ``quickly" (in polynomial time), there is no known way to find a solution rapidly. That is, the time required to solve the problem using any currently known algorithm increases exponentially as the size of the problem grows. It is for this reason that the search for a polynomial time algorithm capable of solving NP problems, called the\textit{ P versus NP problem}, is one of the fundamental unsolved problems in computer science today.

Numerous well-known problems in computer science and mathematics belong to the family of NP problems. Among them, the most popular are the Traveling salesman problem, the factorization of numbers into primes or the discrete logarithm problem. The first problem is common in optimization scenarios and consists in finding the minimum cyclic path connecting $N$ points with specified distances between them. The latter two problems are prevalent in cryptography. For instance, the security of the Rivest Shamir Adleman (RSA) public key cryptography protocol \cite{RSA}, which is widely employed in traditional finance, relies on the fact that factoring numbers into their prime components is an NP problem. Similarly, many public-private key pair generation cryptographic schemes such as Elliptic Curve Cryptography (ECC) \cite{ecc}, prevalent in blockchain technology based protocols like Bitcoin \cite{bitcoin}, are secure due to fact that the discrete logarithm problem also belongs to the class of NP problems. Based on this discussion, it is clear that the development of a technology capable of solving classically-unapproachable NP problems will have a disruptive effect on many scientific fields. In fact, given that NP problems are actually quite frequent, it is likely that such a technology will become the catalyst for a societal upheaval similar to the one that occurred when classical computers first burst onto the scene. 

It is for these reasons that quantum computing has garnered so much attention during the past decade, as these revolutionary computers are the scientific communities best bet to tackle some of the NP problems that have so thwarted all previous classical solution attempts. The concept of quantum computing was first proposed by Feynman \cite{feynman}, a realization that came to him after devoting time to studying a particular NP problem: the simulation of quantum systems. The difficulty of simulating quantum phenomena using classical methods is best explained by Gottesman in \cite{QSC}, but the main takeaway is that keeping track of a quantum state in a classical computer requires exponential classical resources. More specifically, while an $N$-bit classical computer has $2^N$ possible states, its state space is only $N$-dimensional, since a state can be described by a binary vector with $N$ components. In contrast, an $N$-qubit quantum computer has a $2^N$ dimensional state space, since a complex vector with $2^N$ components is necessary to describe any given state. The same reason that makes the simulation of quantum systems an NP problem on classical computers led Feynman to conjecture that computers that obeyed the laws of quantum mechanics would have the capacity to bypass classical computational limits. Although it may not have seemed so at the time, this statement is groundbreaking. It implies that the classification of problems into complexity classes does not apply to quantum computers, which, aside from foreshadowing that some classically intractable problems may actually become tractable on quantum machines, also suggests that the strong Church-Turing thesis is wrong. Since Feynman's proclamation, quantum algorithms capable of solving classical NP problems in polynomial time have been discovered, further cementing the promise of quantum computing. The best known examples are Shor's algorithm \cite{shor}, which can factor numbers into their prime components in polynomial time, and Grover's algorithm \cite{grover1, grover2}, which provides a quadratic speedup when searching for an entry in an unordered database made up of a finite number of objects\footnote{It's complexity is $O(\sqrt{N})$, whereas the best known classical algorithms scale as $O(N)$, where $N$ denotes the number of objects in the database.}.

In light of the astonishing promise and breathtaking potential of operational quantum computers, agents in both academia and the private sector are racing towards the development and construction of quantum computers sophisticated enough to run quantum algorithms. For a quantum information processing system to be useful, it requires long-lived quantum states and a viable way to interact with them. Although there are different ways of doing so, we generally consider these systems to be comprised of a number of two-level subsystems called qubits\footnote{The term comes from ``quantum bit'', since qubits are the quantum analogue of bits in a classical computer.}. For quantum computers to be good, their constituent qubits must exhibit the following traits:

\begin{itemize}
    \item \textbf{Long coherence times} $\rightarrow$ The quantum mechanical properties of qubits can only be leveraged whilst they remain coherent, i.e, while they are in a state of superposition. Qubits lose coherence when they interact with the outside world, an inevitable phenomenon baptized (unsurprisingly) as decoherence. Thus, the coherence time of a qubit is a measure of how long it stays in a workable superposition state, i.e, how much time passes before it ``decoheres''. Clearly, longer coherence times will allow for longer and more complex quantum computation. However, qubit coherence times can only be increased by minimizing the interaction of subatomic quantum particles with the outside world, an extremely complicated task that makes building quantum computers a remarkable feat of engineering.  
    
    \item \textbf{High connectivity} $\rightarrow$ It is desirable for the qubits of a quantum computer to be highly connected, as this allows operations to act on specific qubits simultaneously. Because decoherence arises when qubits interact with their environment (which includes other qubits), achieving high connectivity between qubits while ensuring long coherence times is a complex task.
    \item \textbf{High fidelity gate operations} $\rightarrow$ Quantum computation is achieved through the execution of sequences of operations known as quantum gates. Physical implementations of these gates are not perfect and they are faulty by nature (this varies depending on the technology that is employed), which results in excess ``noise'' being added to quantum information when it is processed with quantum gates. Naturally, higher fidelity gate operations will lead to better quantum computing.
    \item \textbf{High scalability} $\rightarrow$ In broad terms, the power of a quantum computater is determined by the number of qubits that make it up. Thus, it is important for qubits to be scalable so that increasing numbers of them can be employed to construct quantum machines. 
\end{itemize}

Based on the above list, it is easy to see that satisfying these requirements essentially comes down to the capacity of a quantum computer to handle decoherence-induced noise, as this phenomenon manifests with time, connectivity, and when performing operations. As mentioned previously in the introduction, QEC is (arguably) the best and only way to tackle this issue and minimize the negative impact of decoherence. Despite the theoretical nature of most of the current work on QEC, error correcting codes are indispensable for quantum computers to evolve beyond their presently reduced applications and achieve true and indisputable quantum supremacy \cite{supremacy}. Before delving into the realm of QEC in subsequent chapters, it is worthwhile to briefly discuss some of the most relevant technologies that are being explored as possible physical realizations of qubits. At the time of writing, the most advanced technologies for the construction of qubits are:

\begin{enumerate}
    \item \textbf{Superconducting Circuits}: Qubits can be built based on superconductors by placing a resistance-free current in a superposition state, using a microwave signal, and making it oscillate around a circuit loop. The defining trait of superconducting qubits and the primary reason for them being so well-known is that they have to be kept at cryogenic temperatures (below 100mK, or 0.1 degrees above absolute zero), a necessary requirement for the resistance of the superconductor to vanish. This technology is advantageous for mainly two reasons: It has a faster quantum gate time than other technologies, allowing for much faster computation, and the technology behind superconducting qubits can take advantage of proven existing electronic circuit design methods and processes (such as printable circuits) to tackle the scalability issue of quantum computing. Unfortunately, superconducting qubit technology is not without its faults. Qubits built in this manner have short coherence times and low connectivity. Furthermore, this technology requires the achievement and maintenance of cryogenic temperatures, which can be expensive and cumbersome, as well as individual calibration (each superconducting qubit is slightly different). 
    
    \begin{figure}[!htp]
	\centering
	\includegraphics[height=2.5in, width=0.5\columnwidth]{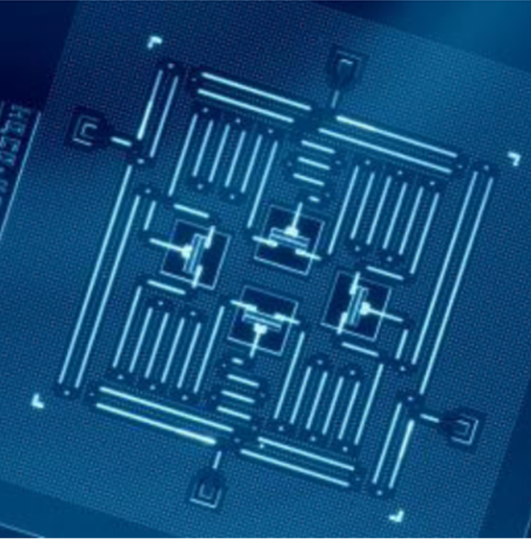}
	\caption{$4$ qubit superconducting processor fabricated at IBM \cite{supercond}.}
	\label{superconductorrrrrr}
\end{figure}

    \item \textbf{Trapped Ions}: Ion Trap quantum computers work by trapping ions (charged atoms) using electric fields and holding them in place. Then, the outermost electron orbiting the nucleus can be put in different states and used as a qubit. The main appeal of trapped ion technology is its stability; trapped ion qubits have much longer coherence times than their superconducting counterparts. Additionally, although ions need to be cooled to perform optimally, the temperature requirement is much less prohibitive than for superconducting technology. Another important advantage is that ion trap qubits can be reconfigured, which allows for high qubit connectivity and avoids some of the issues and computational overhead found with other technologies. The downside to trapped ion quantum computing is its significantly slower operation time in comparison to other implementations. Other important drawbacks are the fact that ions need to be kept in high vacuum and that the technology involved in creating trapped ions, which requires the integration of techniques from a wide range of scientific domains, is not yet mature. 
    
    \begin{figure}[!h]
	\centering
	\includegraphics[height = 2.75in, width=0.7\columnwidth]{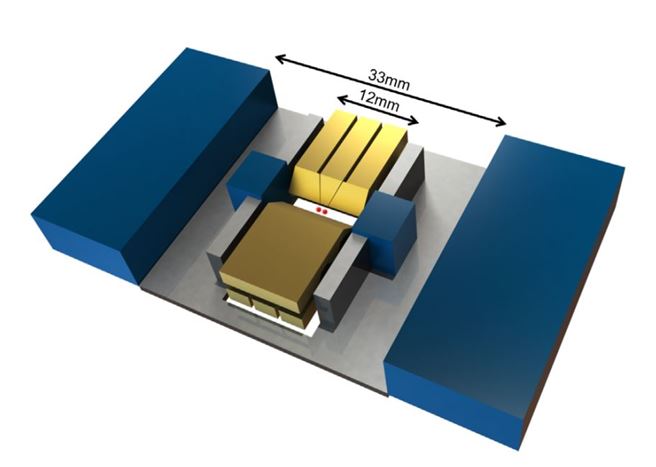}
	\caption{Schematic of the linear Paul ion trap (yellow) fitted with four permanent magnets (blue), arranged to create a strong magnetic field gradient along the trap axis \cite{ion-trap}.}
	\label{superconductorrr}
    \end{figure}
    
    \item \textbf{Photonics}: Photons (particles of light) operating on silicon chip pathways can be used to construct qubits. The primary advantage of this technology is that it does not require extreme cooling, allowing for more energy-efficient and less cumbersome quantum computing \cite{photonic}. In a similar manner to superconducting qubits, because it is based on the use of silicon chips, this approach to quantum computing can exploit existing semiconductor industry infrastructure, which makes it highly scalable. Given that this technology is still nascent, important issues such as qubit connectivity remain to be proven.

    \item \textbf{Neutral Atoms}: The neutral atom approach to quantum computing is similar to that of ion traps but instead of using charged particles, neutral atoms are used as qubits. Aside from exhibiting long coherence times, neutral atoms have the additional advantage of being configurable into arrays of single neutral atoms, which has the potential of becoming a very powerful and scalable technology to build and manipulate thousands of qubits \cite{neutral}. Despite its promise, as is the case with many of these qubit implementation methods, the principal concerns regarding the use of this technology are that it is still in its first stages of development.
    
    \item \textbf{Nitrogen-Vacancy Center}: One of the most recent qubit construction methods is based on the use of an electron spin inside a Nitrogen-Vacancy (NV) centre in a diamond lattice. NV centres are point defects in a diamond lattice characterized by having a nearest-neighbor pair of a nitrogen atom, which substitutes for a carbon atom, and a lattice vacancy (see Figure \ref{NV}). Qubits built in this manner have long coherence times and can work at a large variety of temperatures. Once more, as of yet, there are few experimental results related to this approach to quantum computating. 
    
        \begin{figure}[!h]
	\centering
	\includegraphics[height=2.5in,width=0.5\columnwidth]{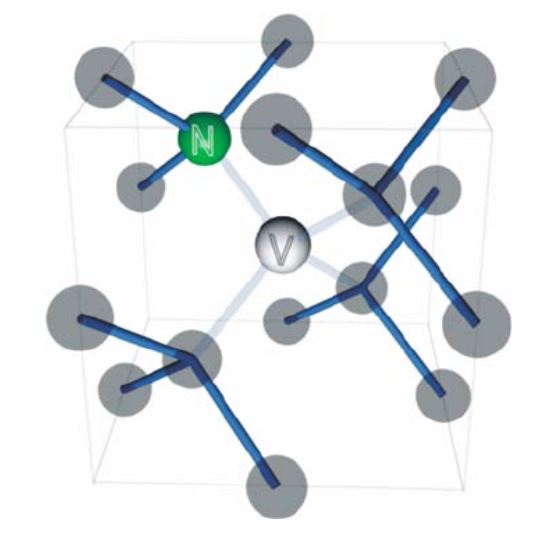}
	\caption{Schematic representation of the nitrogen vacancy (NV) centre structure \cite{NV-cite}.}
	\label{NV}
\end{figure}
    
\end{enumerate}

\begin{landscape}

\begin{figure}[!h]
	\centering
	\includegraphics[height=3.8in,width=\columnwidth]{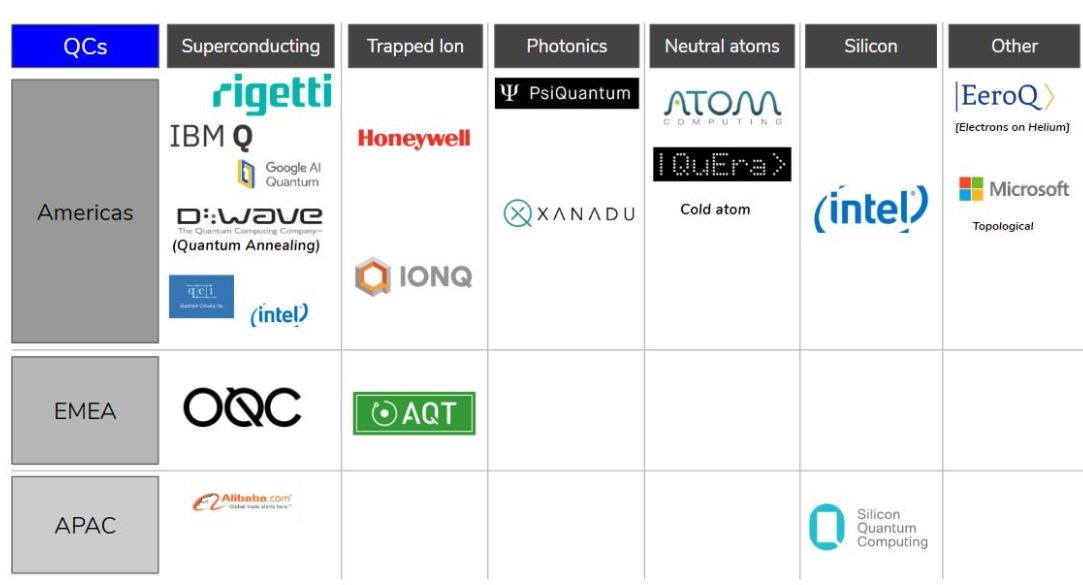}
	\caption{Notable companies involved in quantum computing classified by their chosen technology and geographical location.}
	\label{players}
\end{figure}

\end{landscape}

This wide range of available technologies to physically implement quantum computers presents an unprecedented opportunity for entrepreneurship. In fact, the allure of the field is so strong it has attracted world-renowned companies and inspired the creation of many startups. Figure \ref{players} classifies the biggest players in quantum computing according to the technology they have chosen to implement their quantum processors. Among them, D-Wave Systems, a company based in Canada, was the first to offer commercial access to quantum computers. Currently, IBM, which boasts a 127 qubit superconducting processor, QuEra, who claim to have built a 256 neutral atom qubit computer, and IonQ, with its high fidelity 32 qubit trapped ion chip, appear to be leading the charge towards fully operational universal quantum computers.

\subsubsection*{An important remark}

It is irrefutable that quantum computers have the potential to transform the world as we know it. However, it is important to remain grounded and to understand that they will likely ``only'' deliver tremendous speed-ups for particular types of problems. Fortunately, because some of these problems, like those related to optimization, are present in almost every aspect of society, quantum computing will possibly impact many areas of human life. Nonetheless, it should be stressed that quantum computers are not the \textit{be-all and end-all} of science. We should also remember that, because the field of quantum computing is still in its infancy, we do not yet fully comprehend which problems are suited for quantum speed-ups and how to develop algorithms to demonstrate them. 

\clearemptydoublepage
\chapter{Preliminaries} \label{chapter3}

\epigraph{\textit{``Life before death. Strength before weakness. Journey before destination''}}{\textbf{Brandon Sanderson}.}

\noindent\hrulefill

This chapter serves as a basic introduction to the realms of quantum information theory and classical error correction. It includes background material (terminology and notation) that aims to facilitate the reading and understanding of the rest of this dissertation. The chapter is divided into two major sections: section \ref{cp3:quant} dedicated to quantum information theory and section \ref{cp3:class} devoted to classical error correction. Because the people involved in the field of quantum information come from a wide variety of scientific disciplines, some readers may find one (or both) sections familiar. In such a case, the appropriate sections should be skipped, although we do suggest reading section \ref{sec:N-qubits} as it introduces notation that differs slightly from the one employed in the literature.

\section{Quantum Information} \label{cp3:quant}

This section commences with an overview of the postulates of quantum mechanics and a discussion on the qubit and its various representations. Then, in subsections \ref{sec:entanglement} and \ref{sec:no-cloning}, we go over important aspects such as entanglement and the no-cloning theorem. Following this, we introduce the concept of quantum noise and unitary operators in section \ref{sec:noise}. We conclude this introduction to quantum information by presenting the Pauli group in subsection \ref{sec:N-qubits} and providing a brief overview of the most common quantum channel models in section \ref{sec:channels}. 

\subsection{Postulates of Quantum Mechanics}

The postulates of quantum mechanics provide us with the necessary tools to study the behaviour of subatomic particles. As stated in \cite{NielsenChuang}, they are the result of a long process of trial and (mostly) error, which involved a considerable amount of guessing and fumbling by the originators of the theory. Essentially, these postulates are the axioms on which the theory and mathematical framework of quantum mechanics is built. The motivation and reasoning behind them is not always clear (even to experts), but knowing what they represent can be helpful to understand other quantum mechanical concepts. For the sake of simplicity, in what follows we simply state the basic postulates of quantum mechanics (the notation we employ will be introduced later on). We will then explain and reference them as needed throughout this chapter. Because discussions regarding the origin and physical meaning of these postulates is beyond the scope of this dissertation, the reader is referred to \cite{NielsenChuang, introQIC} for a rigorous discourse on this topic. The basic postulates of quantum mechanics are: 

\begin{itemize}
    \item \textbf{Postulate 1 - State Space}: Associated to any isolated physical system is a complex vector space with inner product (that is, a Hilbert space) known as the state space of the system. The system is completely described by its state vector, which is a unit vector in the system’s state space.
    \item \textbf{Postulate 2 - Evolution}: The evolution of a closed quantum system is described by a unitary transformation. That is, the state $\ket{\psi}$ of the system at time $t_1$ is related to the state $\ket{\psi'}$ of the system at time $t_2$ by a unitary operator $U$ which depends only on the times $t_1$ and $t_2$, $$\ket{\psi'}=U\ket{\psi}.$$
    \item \textbf{Postulate 3 - Measurement}: Quantum measurements are described by a collection $\{M_m\}$ of measurement operators. These are operators acting on the state space of the system being measured. The index $m$ refers to the measurement outcomes that may occur in the experiment. If the state of the quantum system is $\ket{\psi}$ immediately before the measurement then the probability that result $m$ occurs is given by
    $$ \mathrm{P}(m) = \bra{\psi}M_m^\dagger M_m\ket{\psi},$$ and the state of the system after the measurement is $$ \frac{M_m\ket{\psi}}{\sqrt{\bra{\psi}M_m^\dagger M_m\ket{\psi}}}.$$ The measurement operators satisfy the completeness equation,
$$ \sum_m M_m^\dagger M_m = I, $$ where $I$ is the identity matrix.

    \item \textbf{Postulate 4 - Composite systems}: The state space of a composite physical system is the tensor product of the state spaces of the component physical systems. Moreover, if we have systems numbered $1$ through $N$, and system number $i$ is prepared in the state $\ket{\psi_i}$ then the joint state of the total system is $\ket{\psi_1}\otimes\ket{\psi_2}\otimes\ldots\otimes\ket{\psi_N}$, where $\otimes$ denotes the tensor product.
 
\end{itemize}

\subsection{The Qubit} \label{sec:qubit}

The simplest quantum mechanical system and the basic unit in quantum information is known as the qubit. In contrast to classical bits, which can exist in only one of two possible states, $0$ or $1$, qubits exhibit a unique property, known as quantum superposition, that allows them to exist as a linear combination of these states. This means that, while a classical bit is an element of the binary field $\mathbb{F}_2$, a qubit is an element of the two dimensional complex Hilbert space $\mathcal{H}_2$. From the first postulate of quantum mechanics we know that a quantum mechanical system is described using the state vector formulation, also known as Braket or Dirac notation \cite{NielsenChuang, Dirac, introQIC}. Thus, the superposition state of a qubit can be written as

\begin{equation} \label{eq:qubit}
    \ket{\psi} = \alpha\ket{0} + \beta\ket{1},
\end{equation}

where $\alpha$, $\beta \in \mathbb{C}$ and $\mathopen|\alpha\mathclose|^2 + \mathopen|\beta\mathclose|^2 = 1$. The vectors $\ket{0}$ and $\ket{1}$ are orthonormal basis states that span $\mathcal{H}_2$ and are jointly referred to as the computational basis states of a qubit. The complex numbers $\alpha$ and $\beta$ are known as the qubit amplitudes. 

\subsubsection*{Quantum measurement}

The third postulate of quantum mechanics tells us that the probability that a quantum state $\ket{\psi}$ is in the state $\ket{x}$ is given by
\begin{equation} \label{eq:measure}
    \mathrm{P}(\ket{x}) = \bra{\psi}M_x^\dagger M_x\ket{\psi}.
\end{equation}

In quantum mechanics, the symbol $\ket{\cdot}$ represents a column vector known as a ket, and the symbol $\bra{\cdot}$ represents a row vector known as a bra (hence why this is known as Braket notation). For every ket $\ket{a}$ there is a bra $\bra{a}$ and they can be easily obtained from each other by computing the conjugate transpose, i.e, $\ket{a} = \bra{a}^\dagger$. If we choose the measurement operator $M_x = \ket{x}\bra{x}$ and introduce it in \eref{eq:measure}, we obtain 

\begin{align} \label{eq:probs_demo}
\begin{split} 
\mathrm{P}(\ket{x}) =& \bra{\psi}M_x^\dagger M_x\ket{\psi} =  \bra{\psi}(\ket{x}\bra{x})^\dagger (\ket{x}\bra{x})\ket{\psi} \\
=& \bra{\psi}(\ket{x}\bra{x})(\ket{x}\bra{x})\ket{\psi} = \braket{\psi|x}\braket{x|x}\braket{x|\psi} \\
=& \braket{\psi|x}\braket{x|\psi} = \mathopen|\braket{x|\psi}\mathclose|^2, \\
\end{split}
\end{align}

where we have used $\braket{x|x} = 1$ and the notation $\braket{\cdot | \cdot}$ represents the inner product between the vectors $\ket{\cdot}$ and $\bra{\cdot}$. 
If we now write the computational basis using vector notation as $$ \ket{0} = \begin{bmatrix} 1 \\0 \\\end{bmatrix}  , \ \ket{1} = \begin{bmatrix} 0 \\1 \\\end{bmatrix} ,$$ we can use the measurement rule derived in \eref{eq:probs_demo} to see that the probability that the state $\ket{\psi}$ is in either of the base states is related to the amplitudes as:

\begin{align} \label{eq:probs}
\begin{split} 
\mathrm{P}(\ket{0}) = \mathopen|\braket{0|\psi}\mathclose|^2 = \mathopen|\bra{0} (\alpha\ket{0} + \beta\ket{1})\mathclose|^2 \\
= \mathopen|\alpha\braket{0|0} + \beta\braket{0|1}\mathclose|^2 = \mathopen|\alpha\mathclose|^2, \\
\mathrm{P}(\ket{1}) = \mathopen|\braket{1|\psi}\mathclose|^2 = \mathopen|\bra{1} (\alpha\ket{0} + \beta\ket{1})\mathclose|^2 \\
= \mathopen|\alpha\braket{1|0} + \beta\braket{1|1}\mathclose|^2 = \mathopen|\beta\mathclose|^2. \\
\end{split}
\end{align}

Based on \eref{eq:probs}, we can understand why we previously stated that the amplitudes should satisfy $|\alpha|^2 + |\beta|^2 = 1$. This comes from the completeness equation for the measurement operators given in the third postulate of quantum mechanics, which ensures that all probabilities add up to $1$: $$1 = \sum_m\mathrm{P}(m) = \sum_m \bra{\psi}M_m^\dagger M_m\ket{\psi}. $$ This outcome, also known as the normalization condition, guarantees that we will always obtain a measurement outcome when measuring a qubit. Thus, we can re-define the qubit as a continuum of the states of the computational basis, that, upon measurement, will collapse to either one of the base states with a probability $|\alpha|^2$ or $|\beta|^2$, respectively. 

\subsubsection*{Global phase}

Another important consequence of the measurement postulate of quantum mechanics is the fact that the global phase of a quantum state has no observable consequence. Based on what we discussed previously, we can easily compute the probability of measuring a quantum state $\ket{w} = i\ket{0}$ in a specific state $\ket{x}$ as

$$\mathrm{P}(\ket{x}) = \mathopen|\braket{x|w}\mathclose|^2 = \mathopen|\bra{x}(i\ket{0})\mathclose|^2 = \mathopen|i\braket{x|0}\mathclose|^2 = \mathopen|\braket{x|0}\mathclose|^2,$$

where $i^2 = -1$. Notice how the probabilities for the state $i\ket{0}$ are identical to the probabilities for the state $\ket{0}$, i.e, $\mathopen|\bra{x}(i\ket{0})\mathclose|^2 = \mathopen|\braket{x|0}\mathclose|^2$. Because quantum measurement is the only possible way we have to extract information from a qubit, this means that the states $i\ket{0}$ and $\ket{0}$ are equivalent in all relevant physical ways. More generally, it can be said that quantum states that differ only by the overall factor $e^{-i\gamma}$ where $\gamma$ is a real number, which we refer to as the global phase, are physically indistinguishable. This means that, for an arbitrary quantum state $e^{-i\gamma}\ket{a}$, the global phase has no observable consequence
\begin{equation}\label{eq:phase}
 \mathopen|\bra{x}(e^{-i\gamma}\ket{a})\mathclose|^2 = \mathopen|e^{-i\gamma}\braket{x|a}\mathclose|^2 =  \mathopen|\braket{x|a}\mathclose|^2 .
 \end{equation}
 
 \subsubsection*{Pure states and Mixed states}

An important distinction that can be made when studying qubit states is that of \textit{pure} or \textit{mixed} states. A pure qubit state is defined as a coherent superposition of the basis states, meaning that it can be described as a linear combination of $\ket{0}$ and $\ket{1}$. Thus, pure qubit states are completely specified by a single ket and can be written as shown in \eref{eq:qubit}. Mixed qubit states are defined as the statistical combination or incoherent mixture of different pure states that cannot be represented using the Dirac notation (mixed quantum states cannot be written as a single ket). Instead, they are represented in terms of the density matrix formulation of quantum mechanics, which is useful to describe qubits whose state is not completely known in state vector terms. 

\subsubsection{The Bloch Sphere}

A practical way of visualizing the two-dimensional complex Hilbert space that defines a qubit is to represent it using a unit-radius sphere. In the jargon of quantum mechanics, this particular sphere is known as the Bloch sphere\footnote{It is so named as a tribute to the work of Swiss physicist Felix Bloch on the quantum theory of solids.}. The top and bottom of the sphere on the $Z$-axis are generally chosen to correspond to the $\ket{0}$ and $\ket{1}$ base states, which in turn can represent the physical spin-up and spin-down states of an electron. In fact, some literature actually represents the computational basis using the notation $\ket{\uparrow} = \ket{0}$ and $\ket{\downarrow} = \ket{1}$. 

The superposition state given in \eref{eq:qubit} can be rewritten using polar coordinates as 

\begin{equation}
    \ket{\psi} = r_{\alpha}e^{i\phi_{\alpha}}\ket{0} +  r_{\beta}e^{i\phi_{\beta}}\ket{1},
\end{equation}

where the parameters $r_{\alpha}, \phi_\alpha, r_\beta,$ and $\phi_\beta$ are real numbers. Knowing that the global phase $e^{i\gamma}$ has no observable consequence\footnote{In polar coordinates this can be shown as $\mathopen|e^{i\gamma}\alpha\mathclose|^2 = (e^{i\gamma}\alpha)^\dagger e^{i\gamma}\alpha = (e^{-i\gamma}\alpha^\dagger) (e^{i\gamma}\alpha) = \alpha^\dagger\alpha = \mathopen|\alpha\mathclose|^2 $.}, we can multiply our state by $e^{-i\phi_\alpha}$, which yields

\begin{align} 
\begin{split} 
        \ket{\psi'} &= e^{-i\phi_\alpha}\ket{\psi} = r_{\alpha}\ket{0} +  r_{\beta}e^{-i\phi_\alpha}e^{i\phi_{\beta}}\ket{1} \\
    &= r_{\alpha}\ket{0} +  r_{\beta}e^{i(\phi_\beta - \phi_\alpha)}\ket{1} = r_{\alpha}\ket{0} +  r_{\beta}e^{i\phi}\ket{1}  ,
\end{split}
\end{align}

where $\phi = \phi_\beta - \phi_\alpha$. If we write the complex number $r_{\beta}e^{i\phi}$ in cartesian coordinates as $x + iy$ and considering that the normalization condition $\braket{\psi'|\psi'} = 1$ must hold, then

\begin{align} \label{eq:bloch-cart}
\begin{split} 
        \braket{\psi'|\psi'} &= \mathopen|r_\alpha\mathclose|^2 + \mathopen|x+iy\mathclose|^2 \\ &= \mathopen|r_\alpha\mathclose|^2 + (x+iy)^*(x+iy) = r_\alpha^2 + x^2 + y^2 = 1  .
\end{split}
\end{align}

Notice that the expression shown in \eref{eq:bloch-cart} is the equation of a unit radius sphere with cartesian coordinates $r_\alpha, x,$ and $y$. By introducing spherical coordinates we can write the state $\ket{\psi}$ (recall that because the global phase has no observable consequence the states $\ket{\psi}$ and $\ket{\psi'}$ are equivalent in all relevant physical manners) as

\begin{equation}
    \ket{\psi} = \cos{\theta'}\ket{0} +  e^{i\phi}\sin{\theta'}\ket{1}  .
\end{equation}

\begin{figure}[!h]
	\centering
	\includegraphics[width=0.65\columnwidth]{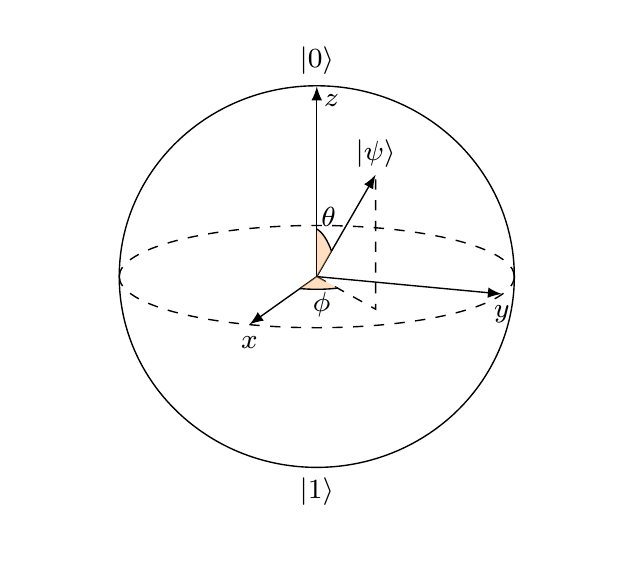}
	\caption{Graphical representation of a qubit $\ket{\psi}$ on the Bloch sphere.}
	\label{bloch-sphere}
\end{figure}

Now, in order for each point on the sphere to be identified by a unique set of spherical coordinates, we must restrict their range. For instance, note how for $\theta' = 0 \rightarrow \ket{\psi} = \ket{0}$ and for $\theta' = \pi \rightarrow \ket{\psi} = -\ket{0}$. This means that the north and south poles of the sphere are physically the same state, since we know the global phase factor $-1 = e^{-i\pi}$ to be irrelevant. Thus, we apply the restriction $0\leq \theta' \leq \frac{\pi}{2}$ to uniquely identify all the points on the sphere. If we introduce $\theta = 2\theta'$, then we can write 

\begin{equation} \label{eq:Bloch-sphere}
    \ket{\psi} = \cos\bigg({\frac{\theta}{2}}\bigg)\ket{0} +  e^{i\phi}\sin\bigg({\frac{\theta}{2}}\bigg)\ket{1} ,
\end{equation}

where $0\leq \theta \leq \pi, 0\leq \phi \leq 2\pi $ are the coordinates of the points of the Bloch sphere. Using $\theta = 2\theta'$ is a useful convention, as it ensures that the basis state $\ket{0}$ corresponds to the north pole of the Bloch sphere and that the basis state $\ket{1}$ corresponds to the south pole of the sphere. A graphical representation of the Bloch sphere is shown in Figure \ref{bloch-sphere}. Aside from providing a visual tool to understand the qubit, the Bloch sphere also makes it easier to interpret the concept of pure quantum states and mixed quantum states: pure states can be understood as points on the surface of the sphere and mixed quantum states can be defined as points within the sphere. 

\subsection{Entanglement} \label{sec:entanglement}

Entanglement is (arguably) the most notorious phenomenon in quantum mechanics. Described by Einstein as ``\textit{spooky action at a distance}'' \cite{Einstein}, it defines a property that links separated qubits and it is the source behind the power of quantum computers. A bipartite quantum system comprised of two single-qubit systems, $\ket{\psi}_\mathrm{A} = \alpha\ket{0}_\mathrm{A} + \beta\ket{1}_\mathrm{A}$ and $\ket{\psi'}_\mathrm{B} = \alpha'\ket{0}_\mathrm{B} + \beta'\ket{1}_\mathrm{B}$, can be formulated as

\begin{align} \label{eq:entanglement}
\begin{split} 
        \ket{\psi}_\mathrm{A} \otimes \ket{\psi'}_\mathrm{B} &= (\alpha\ket{0}_\mathrm{A} + \beta\ket{1}_\mathrm{A}) \otimes (\alpha'\ket{0}_\mathrm{B} + \beta'\ket{1}_\mathrm{B}) \\
        &= \alpha\alpha'\ket{00}_\mathrm{AB} + \alpha\beta'\ket{01}_\mathrm{AB} + \beta\alpha'\ket{10}_\mathrm{AB} + \beta\beta'\ket{11}_\mathrm{AB},
\end{split}
\end{align}

where the notation $\ket{ij} = \ket{i} \otimes \ket{j}$, i.e, it represents the tensor product between two qubits in states $\ket{i}$ and $\ket{j}$, respectively. Using vector notation, the base states of this two qubit system can be written as

$$ \ket{00} = \begin{bmatrix} 1 \\0 \\0 \\0 \\\end{bmatrix} , \ \ket{01} = \begin{bmatrix} 0 \\1 \\0\\0\\\end{bmatrix}  , \ \ket{10} = \begin{bmatrix} 0 \\0 \\1\\0\\\end{bmatrix}  , \ \ket{11} = \begin{bmatrix} 0 \\0 \\0\\1\\\end{bmatrix}  .$$

Now consider a superposition state of just two of these basis states $\ket{\sigma}_\mathrm{AB} = \tau(\ket{00}_\mathrm{AB} + \ket{11}_\mathrm{AB})$. Because this state cannot be written as the tensor product of each of its constituent qubits, i.e, it cannot be obtained from a tensor product $\ket{\psi}_\mathrm{A} \otimes \ket{\psi'}_\mathrm{B}$, it is said to be entangled. Thus, entanglement is defined as the phenomenon by which composite quantum systems can be in states that cannot be written as a product of states of their constituent qubits. 

The intricate connection that entangled qubits are bestowed makes it a useful property in myriads of quantum computing applications, the most prominent of which are Quantum Key Distribution (QKD) \cite{qkd1,qkd2,qkd3}, quantum teleportation \cite{telep1,telep2} and superdense coding \cite{superdense1, superdense2}. Another common way of describing quantum entanglement is by means of the Bell states or Einstein-Podolsky-Rosen (EPR) pairs \cite{NielsenChuang}. Bell states are useful because of the relative simplicity with which they can be generated (only two quantum gates are required to create an entangled Bell state), hence why most $2$-qubit entanglement protocols rely on them. These states are generally denoted as

\begin{align} \label{eq:EPRpairs}
\begin{split} 
        \ket{\Phi^{+}} &= \frac{1}{\sqrt{2}}(\ket{00} + \ket{11}) , \ \ket{\Phi^{-}} = \frac{1}{\sqrt{2}}(\ket{00} - \ket{11}) \\
        \ket{\Psi^{+}} &= \frac{1}{\sqrt{2}}(\ket{01} + \ket{10}) , \ \ket{\Psi^{-}} = \frac{1}{\sqrt{2}}(\ket{01} - \ket{10}).
\end{split}
\end{align}

\subsection{Quantum Noise} \label{sec:noise}

Qubits can lose their coherence and become corrupted through a multitude of different mechanisms. Quantum states can be subjected to noise from a communications channel, they can deteriorate with the passage of time due to undesired interactions with their surroundings, and they can even be exposed to faulty error-inducing operations on a quantum computer. All these phenomena are grouped under a single term, decoherence, which describes the process by which the coherent superposition of the basis states that compose the quantum information state of interest become perturbed \cite{josurev}. This framework is useful because it allows us to model the coherence loss of qubits regardless of the technology used to implement them\footnote{Depending on the specific technology with which qubits are constructed, decoherence will arise due to different physical phenomena.}. 

The effects of decoherence can be conveniently described using the mathematical abstraction of quantum channels, which map an input quantum state to a ``noisy'' output quantum state \cite{reviewPat}. The goal of quantum error correction is to revert the action of these quantum channels and restore the output quantum state into the input quantum state. A pure input quantum state can result in a pure output state, which may be equal to the pure input state (in the extremely unlikely scenario that no channel corruption has occurred), or a mixed output state, if our input qubits have become entangled with the environment. Although it might seem that correcting a mixed output state will be harder than acting on a pure output, this is not the case. The mixed output state can be interpreted as an ensemble of pure states, which implies that if each of the constituent pure states can be corrected back to its original form, we will have recovered the full mixed state. From the perspective of the density matrix formulation of quantum mechanics, the actions of quantum channels can be understood as the application of a so-called \textit{superoperator} on the input density matrix that describes our data qubits \cite{josurev}. This superoperator can be diagonalized and written as the sum of a variety of different matrices that act on the input states with different probability. If a QEC code can correct any of the possible matrices, it will also be capable of correcting the full superoperator \cite{QSC}. Although these individual matrices need not be unitary, the effects of decoherence on qubits are often modelled using combinations of specific unitary matrices.

\subsubsection{Unitary operators} \label{sec:unitaries}

A linear operator whose hermitian conjugate (adjoint) is also its inverse is known as a unitary operator. More explicitly, a unitary operator $U$ acts linearly on a quantum state as $$ U(\ket{\psi}) = U(\alpha\ket{0} + \beta\ket{1}) = U(\alpha\ket{0}) + U(\beta\ket{1}) \ , $$ and it fulfils $$ U^{-1} = U^\dagger \ .$$

From the evolution postulate of quantum mechanics we know that unitary operators play an integral role in this theory. Explicitly, this postulate tells us that a closed quantum system in state $\ket{\psi}$ at time $t_1$ will be related to its state $\ket{\psi'}$ at time $t_2$ by a unitary operator $U$ that depends strictly on the times $t_1$ and $t_2$ as 

\begin{equation} \label{eq:unitary-evo}
    \ket{\psi'} = U\ket{\psi}  .
\end{equation} This means that the evolution of a closed quantum system is described by a unitary transformation. A direct consequence of this is that the quantum analogues of single bit classical logic gates, single-qubit quantum gates, are described by unitary $2\times2$ matrices. These matrices must be unitary because of the normalization condition given in \eref{eq:bloch-cart}. Recall that any quantum state $\ket{\psi} = \alpha\ket{0} + \beta\ket{1}$ must fulfil $\mathopen|\alpha\mathclose|^2 + \mathopen|\beta\mathclose|^2 = 1$, which for quantum state $ \ket{\psi'} = U\ket{\psi}$ holds only if the matrix $U$ is unitary. 

\subsubsection*{The No-cloning Theorem} \label{sec:no-cloning}

A peculiar phenomenon in quantum mechanics, and one that also results from the second postulate of quantum mechanics, is the fact that quantum information cannot be copied or cloned. This concept will be alien to those with a classical information background, since the action of copying or replicating information is a tenet in many classical error correction and information storage strategies. Conventionally referred to as the No-Cloning theorem, the impossibility of copying quantum states arises as a result of the linearity of quantum mechanics. To show this, assume we have an operator $C$ that copies the arbitrary quantum states $\ket{\psi}$ and $\ket{\sigma}$ as

$$ C\ket{\psi} = \psi \otimes \psi , \ C\ket{\sigma} = \sigma \otimes \sigma \ .$$

Then if we wish to copy the sum of both states we perform

$$ C(\ket{\psi}+\ket{\sigma}) = (\ket{\psi}+\ket{\sigma}) \otimes (\ket{\psi}+\ket{\sigma}) \ .$$

Because the transformation $C$ must be linear (all unitary operators are bounded linear operators), then 

$$ C(\ket{\psi}+\ket{\sigma}) =  C(\ket{\psi}) + C(\ket{\sigma}) = \psi \otimes \psi + \sigma \otimes \sigma .$$

However, it is obvious that $ (\ket{\psi}+\ket{\sigma}) \otimes (\ket{\psi}+\ket{\sigma}) \neq \psi \otimes \psi + \sigma \otimes \sigma$, which means that our copying operator $C$ has failed to copy the state $\ket{\psi}+\ket{\sigma}$. Thus, the primary and most relevant consequence of the no cloning theorem is that it is impossible to correctly copy superpositions of the basis states. This means that, in contrast to classical methods, quantum error correction strategies cannot rely on backup copies to preserve information. Instead, they must protect the original quantum state from errors for as long as required\footnote{The longer that qubits are protected the longer their coherence time will be, which should allow the execution of more complex quantum algorithms.}.

\subsubsection*{The Pauli Matrices}

The second postulate of quantum mechanics hides a significant caveat: it does not tell us which unitary operators act on the quantum state, it simply tells us that the closed quantum system will evolve as shown in \eref{eq:unitary-evo}. In fact, it turns out that it is possible for any unitary transformation, i.e, any possible $U \in \mathbb{C}^{2\times2}$, to act on a qubit\footnote{This also means that any unitary matrix can act as a quantum gate.} \cite{NielsenChuang}. Fortunately, the phenomenon of \textit{error discretization}, which will be presented later on, tells us that a QEC code only needs to consider a basis of $\mathbb{C}^{2\times 2}$ to be able to correct all the possible unitary transformations that can act on a qubit. For now, we present the most common single-qubit unitary operators in quantum mechanics. 

The Pauli matrices are primordial elements in the fields of quantum computation and information. They are a set of $2\times 2$ complex matrices that are both Hermitian\footnote{A matrix $A$ is Hermitian if $A^\dagger = A$.} and unitary. Because any complex unitary $2 \times 2$ matrix is also a single-qubit quantum gate, the Pauli matrices can also be referred to as Pauli gates, i.e, the terms are interchangeable in this context. Although notation in the literature varies, they are commonly represented as follows:

\begin{itemize}
    \item \textbf{Pauli Identity Matrix/Gate}: The Pauli identity matrix is defined as $$ I \equiv \sigma_0 \equiv \begin{pmatrix}
1 & 0  \\
0 & 1
\end{pmatrix} \ . $$ The action of the Pauli Identity gate on any quantum state $\ket{\psi} = \alpha\ket{0} + \beta\ket{1}$ is $$ \ket{\psi'} = I\ket{\psi} =  \begin{pmatrix}
1 & 0  \\
0 & 1
\end{pmatrix} \begin{pmatrix}
\alpha   \\
\beta 
\end{pmatrix} = \alpha\ket{0} + \beta\ket{1} =  \ket{\psi} \ .$$

\item \textbf{Pauli $X$ Matrix/Gate}: The Pauli $X$ matrix is defined as $$ X \equiv \sigma_1 \equiv \sigma_x \equiv \begin{pmatrix}
0 & 1  \\
1 & 0
\end{pmatrix} \ . $$ The Pauli $X$ gate, also known as the bit flip gate, acts by swapping the probabilities of the computational basis states. On the Bloch sphere, this can be seen as a $180^\circ$ rotation about the $x$-axis (See Figure \ref{fig:PauliX}). More explicitly, the action of the Pauli $X$ gate on any quantum state $\ket{\psi} = \alpha\ket{0} + \beta\ket{1}$ can be written as $$ \ket{\psi'} = X\ket{\psi} =  \begin{pmatrix}
0 & 1  \\
1 & 0
\end{pmatrix} \begin{pmatrix}
\alpha   \\
\beta 
\end{pmatrix} = \beta\ket{0} + \alpha\ket{1} \ .$$

\begin{figure*}[!h]
	\centering
	\subfloat[]{%
		\includegraphics[width=.4\textwidth]{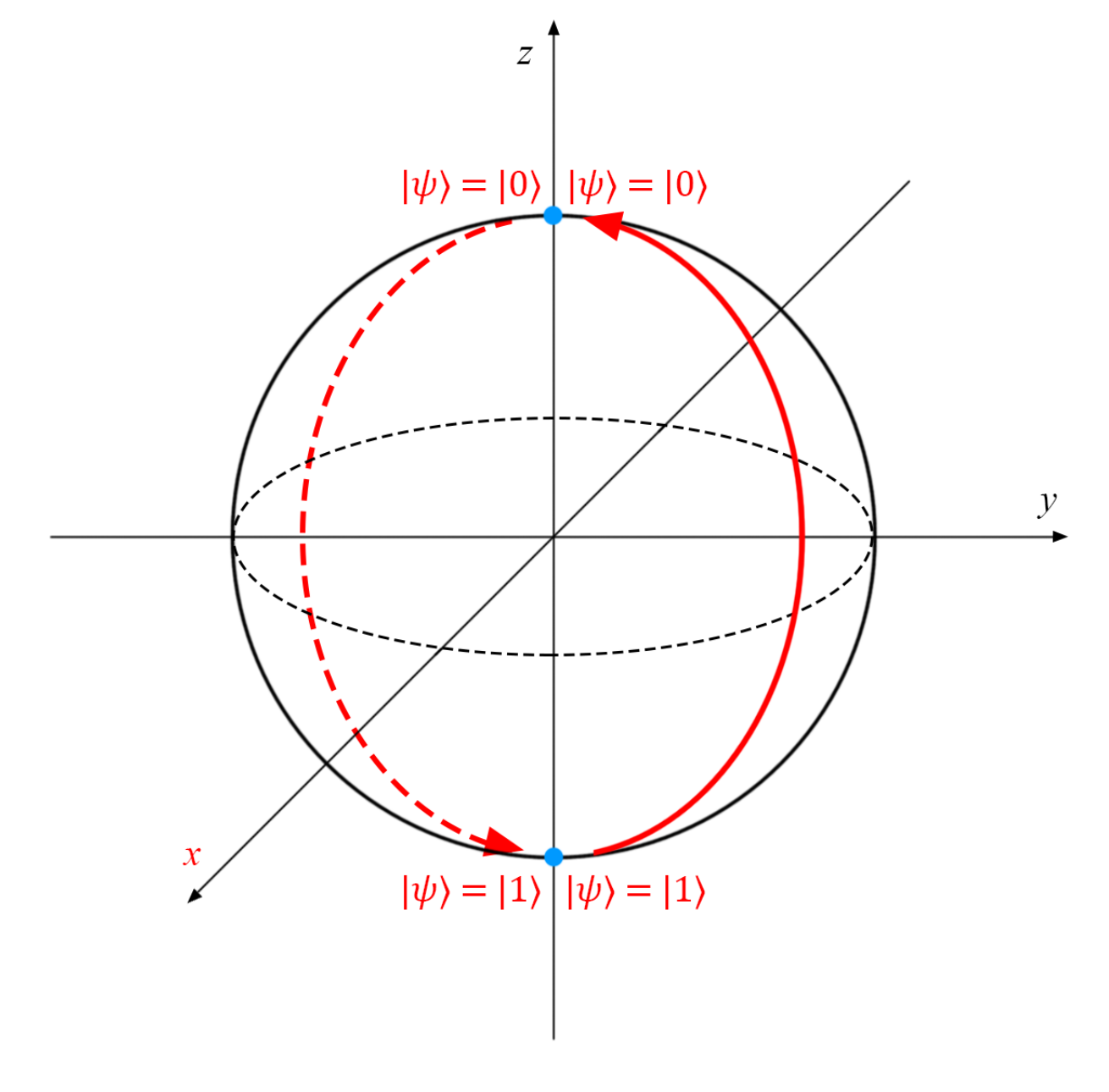}\label{fig:PauliX}
	} 
	\subfloat[]{%
		\includegraphics[width=.4\textwidth]{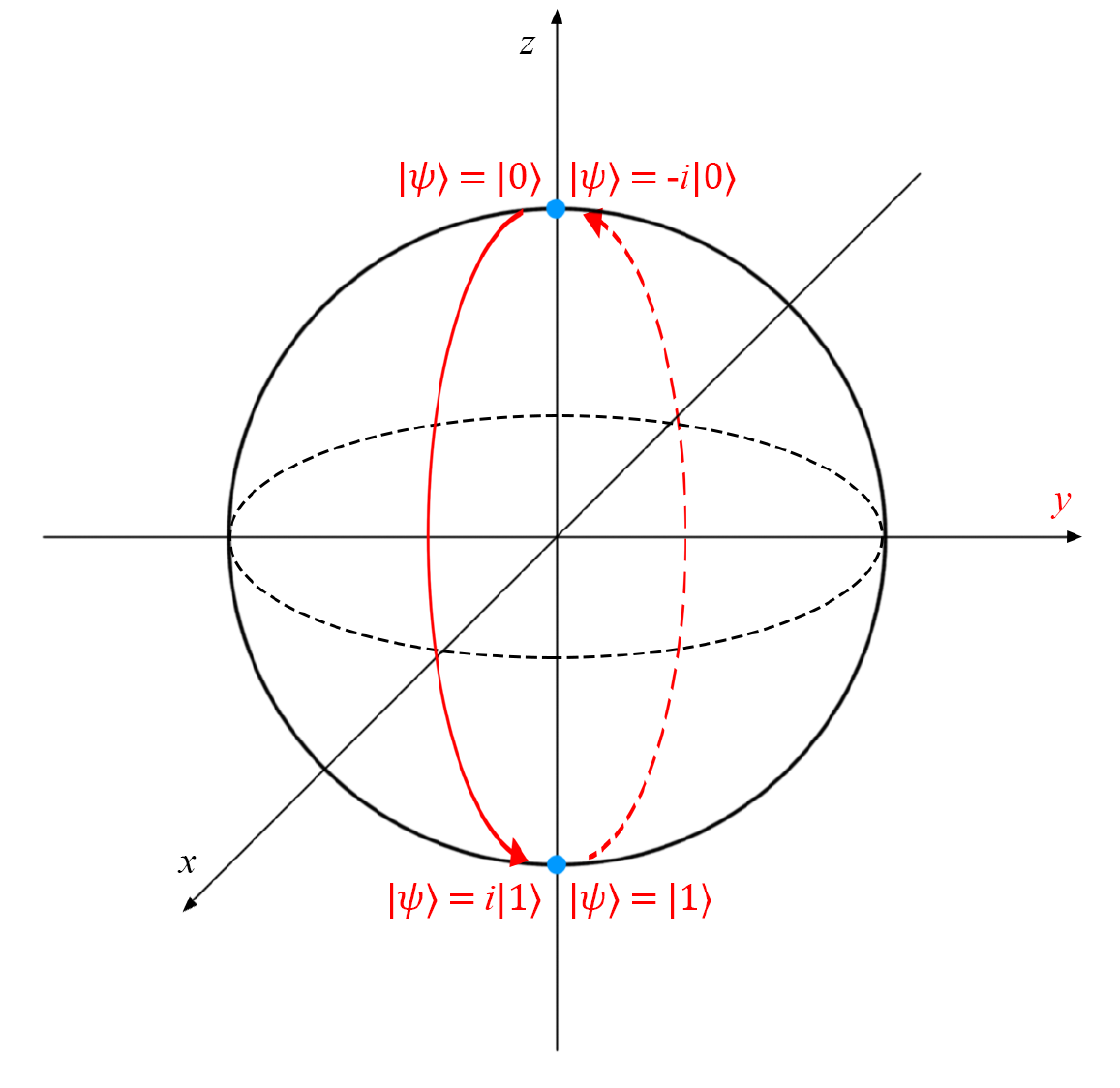} \label{fig:PauliY}
	} \hfill
		\subfloat[]{%
		\includegraphics[width=.4\textwidth]{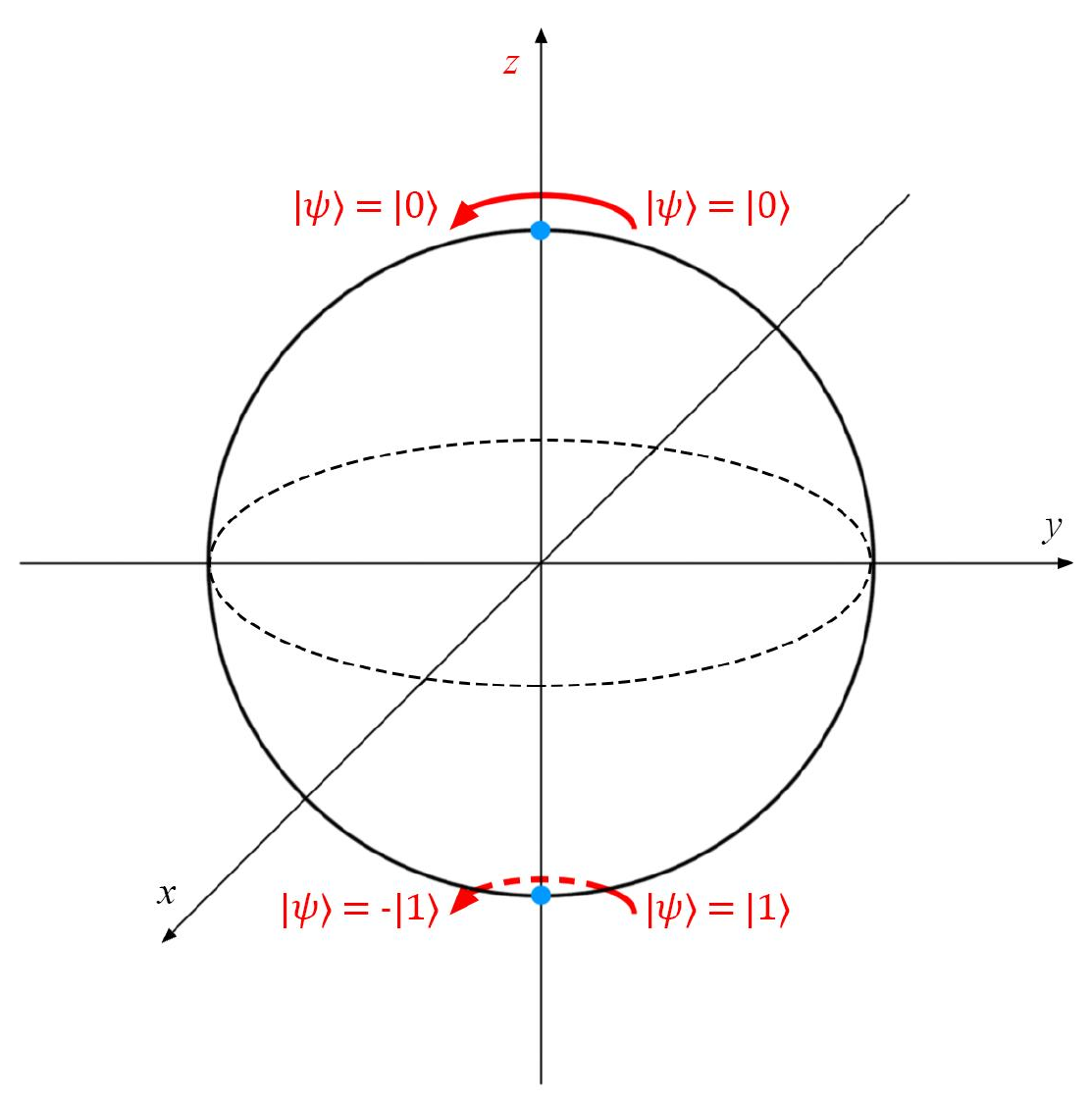} \label{fig:PauliZ}
	} \caption{\textbf{(a)} Effects of the Pauli $X$ gate represented on the Bloch sphere. \textbf{(b)} Effects of the Pauli $Y$ gate represented on the Bloch sphere. \textbf{(c)} Effects of the Pauli $Z$ gate represented on the Bloch sphere.} \label{cp3-fig:Pauli-bloch}
\end{figure*}

\item \textbf{Pauli $Y$ Matrix/Gate}: The Pauli $Y$ matrix is defined as $$ Y \equiv \sigma_2 \equiv \sigma_y \equiv \begin{pmatrix}
0 & -i  \\
i & 0
\end{pmatrix} \ . $$ On the Bloch sphere, the Pauli $Y$ gate acts by rotating a quantum state about the $y$-axis by $180^\circ$ (See Figure \ref{fig:PauliY}). This means that it swaps the amplitudes of the base states and introduces a phase shift of $\pi$ between them. Mathematically, the action of the Pauli $Y$ gate on any quantum state $\ket{\psi} = \alpha\ket{0} + \beta\ket{1}$ can be written as

\begin{align*} 
\begin{split} 
        \ket{\psi'} &= Y\ket{\psi} =  \begin{pmatrix}
0 & -i  \\
i & 0
\end{pmatrix} \begin{pmatrix}
\alpha   \\
\beta 
\end{pmatrix} = \begin{pmatrix}
-i\beta   \\
i\alpha 
\end{pmatrix} \\ &= -i(\beta\ket{0} - \alpha\ket{1}) = e^{-i\frac{\pi}{2}}(\beta\ket{0} + e^{i\pi}\alpha\ket{1}) \ .
\end{split}
\end{align*}

\item \textbf{Pauli $Z$ Matrix/Gate}: The Pauli $Z$ matrix is defined as $$ Z \equiv \sigma_3 \equiv \sigma_z \equiv \begin{pmatrix}
1 & 0  \\
0 & -1
\end{pmatrix} \ . $$ The Pauli $Z$ gate, also known as the phase-flip gate, rotates a quantum state about the $z$-axis by $180^\circ$ (see Figure \ref{fig:PauliZ}). This can be understood as the gate leaving the basis state $\ket{0}$ unchanged and changing the sign of the $\ket{1}$ basis state, or as the introduction of a phase shift of $\pi$ (a phase flip) between the basis states. More explicitly, the action of the Pauli $Z$ gate on any quantum state $\ket{\psi} = \alpha\ket{0} + \beta\ket{1}$ can be written as $$ \ket{\psi'} = Z\ket{\psi} =  \begin{pmatrix}
1 & 0  \\
0 & -1
\end{pmatrix} \begin{pmatrix}
\alpha   \\
\beta 
\end{pmatrix} = \alpha\ket{0} - \beta\ket{1} \ .$$

\end{itemize}

Based on these definitions, it is clear that the three Pauli matrices are related. We can express this relationship mathematically as $Y = iXZ$. This means that the effects of the Pauli $Y$ gate are those of applying a Pauli $X$ gate followed by a Pauli $Z$ gate, hence why the Pauli $Y$ gate is sometimes referred to as the bit-and-phase flip gate. We can write the products between Pauli matrices in a more general manner as

\begin{equation}\label{eq:pauli-rel}
\sigma_a\sigma_b=I\delta_{a,b}+i\sum_{c=1}^{3}\epsilon_{abc}\sigma_c,
\end{equation}
where $\delta$ represents the Kronecker delta and
\[ \epsilon_{abc} = \left\{ \begin{array}{rl}
         1 & \mbox{if $(a,b,c)=\{(1,2,3),(3,1,2),(2,3,1)\}$},\\
        -1 & \mbox{ if $(a,b,c)=\{(3,2,1),(1,3,2),(2,1,3)\}$},\\
         0 & \mbox{if $a=b$ or $b=c$ or $a=c$.}\end{array} \right. \] 
         
By carefully inspecting the product relationships shown in \eref{eq:pauli-rel} we find that the Pauli matrices also exhibit the algebraic property of anticommuting with themselves. This can be written as, 

$$ \{\sigma_i,\sigma_j\} = \sigma_i\sigma_j + \sigma_j\sigma_i = 0 \ ,  $$

where $\{\cdot\}$ denotes the anticommutator\footnote{The anticommutator of two operators $A \in \mathbb{C}^{N\times N}$ and $B \in \mathbb{C}^{N\times N}$ is defined as $ \{A, B\} = AB + BA$, where if $\{A, B\} = 0$ the operators are said to anticommute.}, $i \neq j$, and $i,j \in \{x,y,z\}$. 

\subsubsection*{Hadamard Gate}

The Hadamard gate is a single qubit gate defined as $$ H \equiv \frac{1}{\sqrt{2}} \begin{pmatrix}
1 & 1  \\
1 & -1
\end{pmatrix} \ . $$ It acts by mapping the pure basis states of the computational basis onto superpositions of these states as shown below

\begin{align*} 
\begin{split} 
H\ket{0} &= \frac{1}{\sqrt{2}} \begin{pmatrix}
1 & 1  \\
1 & -1
\end{pmatrix} \begin{pmatrix}
1  \\
0
\end{pmatrix} = \frac{1}{\sqrt{2}}\begin{pmatrix} 1  \\ 1 \end{pmatrix}  \\ &= \frac{1}{\sqrt{2}}\bigg[\begin{pmatrix} 1  \\ 0 \end{pmatrix} + \begin{pmatrix} 0  \\ 1 \end{pmatrix}\bigg] = \frac{1}{\sqrt{2}}(\ket{0} + \ket{1}) = \ket{+}\ ,
\end{split}
\end{align*}

\begin{align*} 
\begin{split} 
H\ket{1} &= \frac{1}{\sqrt{2}} \begin{pmatrix}
1 & 1  \\
1 & -1
\end{pmatrix} \begin{pmatrix}
0  \\
1
\end{pmatrix} = \frac{1}{\sqrt{2}}\begin{pmatrix} 1  \\ -1 \end{pmatrix}  \\ &= \frac{1}{\sqrt{2}}\bigg[\begin{pmatrix} 1  \\ 0 \end{pmatrix} + \begin{pmatrix} 0  \\ -1 \end{pmatrix}\bigg] = \frac{1}{\sqrt{2}}(\ket{0} - \ket{1}) = \ket{-} \ .
\end{split}
\end{align*}

The states $\ket{+}$ and $\ket{-}$ are known as the Hadamard basis states. On the Bloch sphere, the operation of the Hadamard gate can be visualized as a $180^\circ$ rotation around the $x$-axis followed by a $90^\circ$ rotation around the $y$-axis. The effects of applying the Hadamard gate on the base state $\ket{0}$ are shown on the Bloch sphere in Figure \ref{fig:Had}.

\subsubsection*{Phase shift gate}

\begin{figure*}[!h]
	\centering
	\subfloat[]{%
		\includegraphics[width=.415\textwidth]{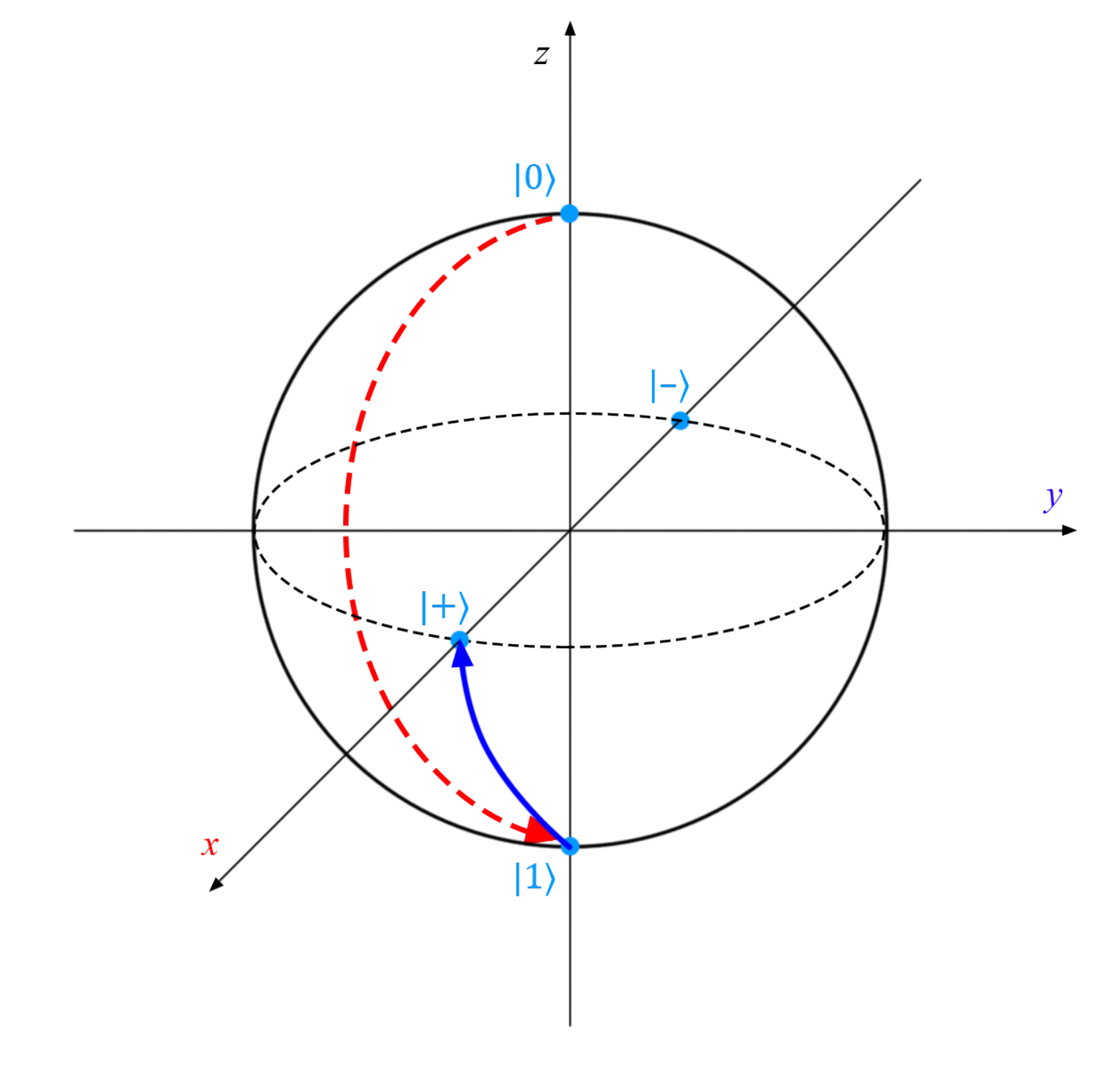}\label{fig:Had}
	} 
	\subfloat[]{%
		\includegraphics[width=.425\textwidth]{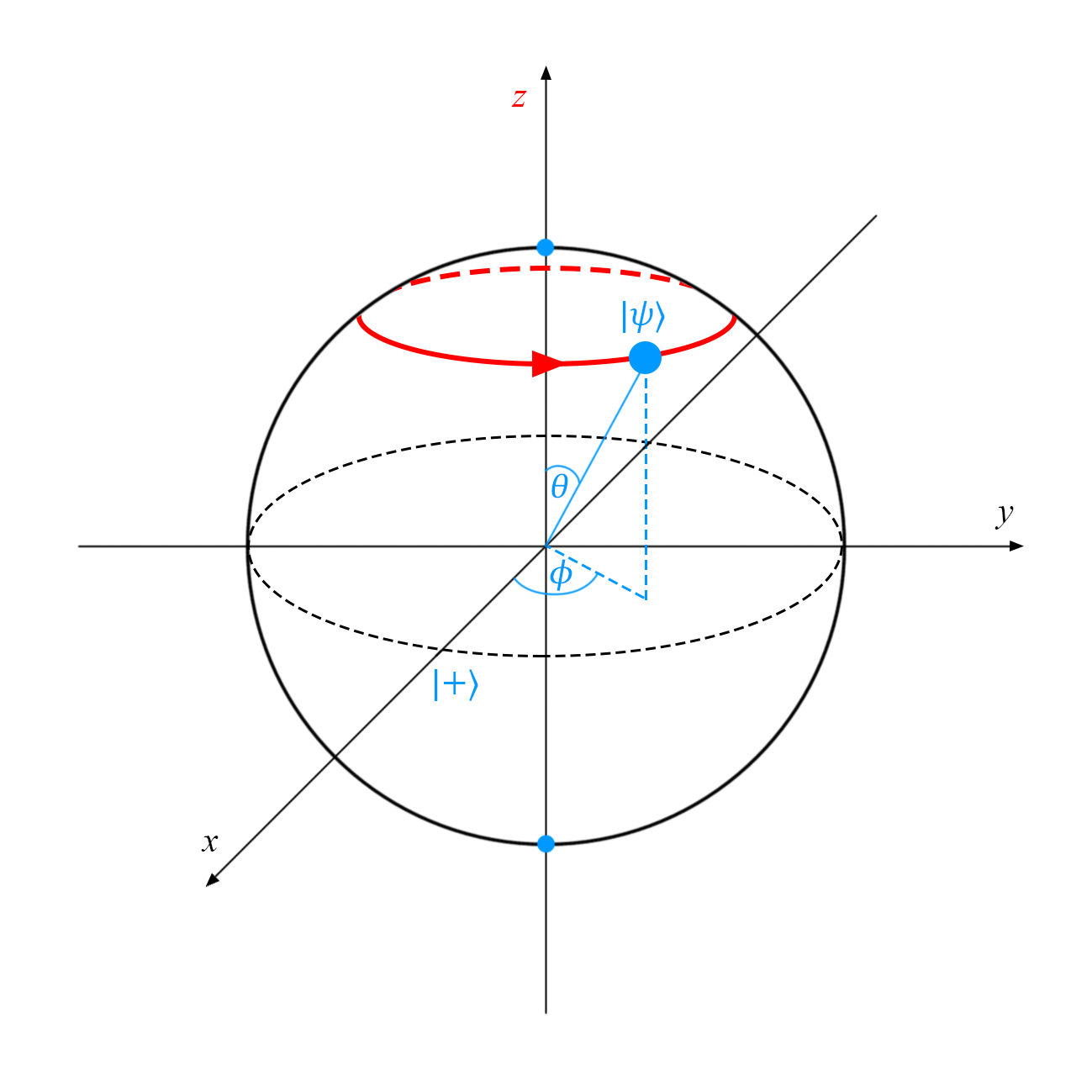} \label{fig:PhaseShift}
} \caption{\textbf{(a)} Representation on the Bloch sphere of $H\ket{0} = \ket{+}$. Recall that the operation of the Hadamard gate can be visualized on the Bloch sphere as a $180^\circ$ rotation around the $x$-axis (red arrow) followed by a $90^\circ$ rotation around the $y$-axis (blue arrow). \textbf{(b)} Phase shift operation shown on the Bloch sphere.} \label{cp3-fig:other-unitaries}
\end{figure*}

The Phase shift gate is a single qubit gate defined as $$ R_\phi \equiv \begin{pmatrix}
1 & 0 \\
0 & e^{i\phi}
\end{pmatrix} \ . $$ It acts by changing the phase of the basis state $\ket{1}$ while leaving the basis state $\ket{0}$ intact, i.e, it rotates the quantum state $\ket{\psi}$ around the $z$-axis by an amount determined by $\phi$ (see Figure \ref{fig:PhaseShift}). For instance, if we set $\phi = \frac{\pi}{2}$, we will shift the quantum state $90^\circ$ around the $z$-axis. The most common phase shift gates are the $S$ and $T$ gates, which are shown below and are obtained by setting $\phi = \frac{\pi}{2}$ and $\phi = \frac{\pi}{4}$, respectively. Note that the Pauli $Z$ gate is a phase shift gate with $\phi = \pi$.

$$ S \equiv \begin{pmatrix}
1 & 0  \\
0 & i
\end{pmatrix} \ , \ T \equiv \begin{pmatrix}
1 & 0  \\
0 & e^{i\frac{\pi}{4}}
\end{pmatrix} \ . $$

\subsubsection*{Multiqubit Gates}

Multiqubit gates are quantum gates that act on two or more qubits (up to this point we have only seen single-qubit gates). Below we present two examples.

\begin{itemize}
    \item \textbf{Swap gate}: The swap gate is defined as $$ \text{SWAP} = \begin{pmatrix}
1 & 0 & 0 & 0  \\
0 & 0 & 1 & 0  \\
0 & 1 & 0 & 0  \\
0 & 0 & 0 & 1 
\end{pmatrix} \ . $$ As befits its name, the swap gate re-orders the input qubits, i.e,  $\text{SWAP}\ket{0}\ket{1} = \ket{1}\ket{0}$. 
    \item \textbf{Controlled NOT (CNOT) gate}: The CNOT gate is a quantum analogue of the classical XOR gate. As with any controlled gate, the CNOT gate has an input control qubit and an input target qubit. If the control qubit is in the base state $\ket{1}$, the gate will perform a NOT operation (a Pauli X) on the target qubit. Otherwise, the target qubit is left unchanged. The CNOT gate is defined by the matrix $$ \text{CNOT} = \begin{pmatrix}
1 & 0 & 0 & 0  \\
0 & 1 & 0 & 0  \\
0 & 0 & 0 & 1  \\
0 & 0 & 1 & 0 
\end{pmatrix} \ . $$

More explicitly, the CNOT gate acts on two qubit quantum states as follows

\begin{align*} 
\begin{split} 
\mathrm{CNOT}\ket{00}\rightarrow\ket{00}, \ &\mathrm{CNOT}\ket{01}\rightarrow\ket{01}, \\
\mathrm{CNOT}\ket{10}\rightarrow\ket{11}, \ &\mathrm{CNOT}\ket{11}\rightarrow\ket{10},
\end{split}
\end{align*}
 where the first qubit is the control qubit and the second qubit is the target qubit.

\end{itemize}

\subsubsection{Diagrammatic notation of quantum systems}

\begin{figure*}[!htp]
	\centering
	\subfloat[]{%
		\includegraphics[width=.19\textwidth]{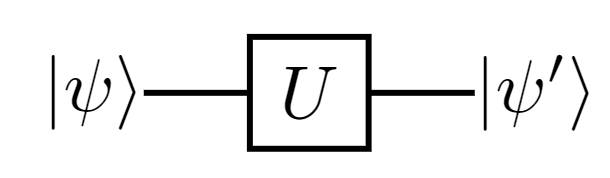}\label{fig:single-qubit}
	} 
	\subfloat[]{%
		\includegraphics[width=.19\textwidth]{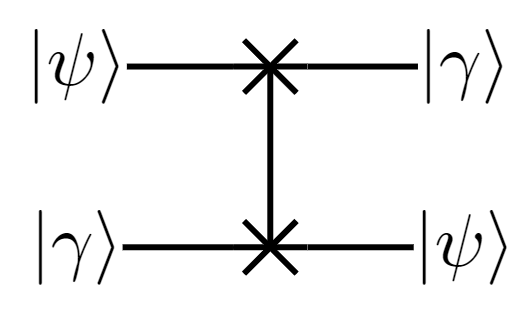} \label{fig:swap}
} \subfloat[]{%
		\includegraphics[width=.19\textwidth]{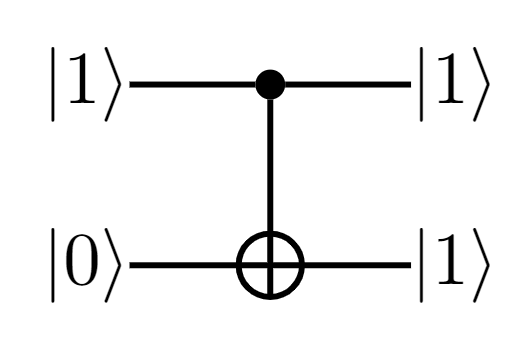} \label{fig:CNOT}
} \caption{\textbf{(a)} Schematic of the single-qubit unitary operator $U$ acting on the input state $\ket{\psi}$. \textbf{(b)} Schematic of the SWAP gate in quantum circuits. \textbf{(c)} Schematic of the CNOT gate in quantum circuits.}  \label{cp3-diagrams}
\end{figure*}

Quantum information theory allows us to model a quantum computation as a sequence of quantum gates. We do this by drawing graphical depictions of quantum circuits based on the Penrose graphical notation \cite{penrose}, which was originally conceived to visually represent tensors in physics. These diagrams must be read from left-to-right and are composed of wires, drawn as straight lines, and quantum gates. These wires do not necessarily correspond to physical wires; instead, they may correspond to the passage of time, or
to a physical particle such as a photon moving from one location to another through space. A single-qubit unitary operator $U$ can be represented on a quantum circuit as a box placed over the corresponding wire (which represents the qubit the operator is acting on). Thus, we can represent any of the gates discussed in the previous section by simply substituting $U$ by the appropriate symbol, i.e, $X$ for the Pauli $X$ gate or $H$ for the Hadamard gate. This is shown in Figure \ref{fig:single-qubit}. 

Similarly, we can represent multiqubit gates by joining the wires that model the evolution of the qubits in our quantum system. The quantum circuit representation of the SWAP gate and the CNOT gate is shown in figures \ref{fig:swap} and \ref{fig:CNOT}. This diagrammatic notation of quantum systems will come in handy later on in the dissertation, especially when representing QEC circuits. 

\subsubsection{$N$-qubit generalization \& The Pauli Group} \label{sec:N-qubits}

So far in the dissertation we have only concerned ourselves with single qubits. Despite being useful when first presenting the framework of quantum information, quantum computers require a much larger number of qubits to run quantum algorithms. This requires that we generalize the framework that we have seen up to this point to quantum systems composed of $N$-qubits. 

\subsubsection*{$N$-qubit systems}

Previously, we employed the first postulate of quantum mechanics to establish that a qubit is an element of the two dimensional complex Hilbert space. Now, by applying the fourth postulate of quantum mechanics, we can define an $N$-qubit composite system as an element of the $2^N$-dimensional Hilbert space, $\mathcal{H}_2^{\otimes N}$, which can be conveniently described as the tensor product of each of the individual qubits. Recall that we have actually already applied this concept to describe a $2$-qubit system in section \ref{sec:entanglement}. 

Based on this same principle, we can represent the state of an $N$-qubit system using a vector defined as the tensor product of $N$ independent vectors, where each of these $N$ vectors belongs to a different two-dimensional Hilbert space (each vector represents a single qubit). The basis states of this $2^N$-dimensional Hilbert space will be given by tensor products of the form $\ket{v_1}\otimes\ldots\otimes\ket{v_N}=\ket{v_1\ldots v_N}$, where $v_i=\{0,1\}$ and $i=1,\ldots,N$. Therefore, we can generalize the superposition state of $N$-qubits as 
$$\alpha_0\ket{00\ldots0} + \alpha_2\ket{00\ldots1}+\ldots+\alpha_{2^N-1}\ket{11\ldots1} \ ,
$$ where $\alpha_i \in \mathbb{C}$ and $\sum_{i = 0}^{2^N-1} \mathopen|\alpha_i\mathclose|^2 = 1$.

\subsubsection*{The Pauli Group}

We will later see that the key to the functioning of most QEC strategies lies in the ingenious application of group theory. This requires that we introduce the concept of the Pauli group.

Let $\Pi=\{I,X,Y,Z\}$ denote the set of single-qubit Pauli matrices defined in section \ref{sec:unitaries}. Based on the product relationships between these matrices (see expression \eref{eq:pauli-rel}) and including the phase factors $\{\pm1, \pm i\}$, we define the set $\widetilde{\Pi}$ as

\begin{align} \label{eq:one-qubit-pauli}
\begin{split} 
        \widetilde{\Pi} &= \{\Theta_1 I, \Theta_2 X, \Theta_3 Y, \Theta_4 Z\} \\
        &= \{\pm I, \pm iI, \pm X, \pm iX, \pm Y, \pm iY, \pm Z, \pm iZ\} \ ,
\end{split}
\end{align}

where $\Theta_l = \{\pm 1, \pm i\}$ and $l=1,2,3,4$. This set, together with the product defined in \eref{eq:pauli-rel}), which we will denote by $\times$ to ensure notation clarity, compose a non-abelian\footnote{An abelian group, also called a commutative group, is a group in which the result of applying the group operation to two group elements does not depend on the order in which they are written. That is, the group operation is commutative.} group known as the single-qubit Pauli group. We will write this group as $\mathcal{G}_1 = (\Tilde{\Pi}, \times)$. 

The single-qubit Pauli group can be extended to $N$-qubit systems by means of the tensor product. To that end, let $\Pi^{\otimes N}=\{I,X,Y,Z\}^{\otimes N}$ denote the set of $N$-fold tensor products of the single qubit Pauli operators
and let $\widetilde{\Pi}^{\otimes N}$ be the set defined as $\widetilde{\Pi}^{\otimes N}=\cup_{k=0}^3 i^k\Pi^{\otimes N}$. Note that $\widetilde{\Pi}^{\otimes N}$ is the set of $N$-fold tensor products of the single qubit Pauli operators and the possible overall factors $\{\pm1, \pm i\}$. That is,

\begin{equation} \label{eq:set-Pauli-N}
    \widetilde{\Pi}^{\otimes N} = \{\Theta_1I, \Theta_2X, \Theta_3Y, \Theta_4Z\}^{\otimes N} \ ,
\end{equation} where $\Theta_l = \{\pm 1, \pm i\}$ and $l=1,2,3,4$. 

Notice that the cardinality of the sets $\Pi^{\otimes N}$ and $\widetilde{\Pi}^{\otimes N}$ is $\mathopen|\Pi^{\otimes N}\mathclose|=2^{2N}$ and $\mathopen|\widetilde{\Pi}^{\otimes N}\mathclose|=2^{2N+2}$, respectively. Let us now define the operation $\cdot$ as

\begin{equation}\label{eq:product-gn}
\forall\;\mathrm{A},\mathrm{B}\in \widetilde{\Pi}^{\otimes N},\;\; \mathrm{A}\cdot \mathrm{B}=A_1\times B_1\otimes B_2\times A_2\otimes\ldots \otimes A_N\times B_N,
\end{equation}

where the products of single qubit operators are given by the product defined in \eref{eq:pauli-rel}. This operation together with the set $\widetilde{\Pi}^{\otimes N}$ compose what is known as the $N$-fold or $N$-qubit Pauli group, which we will denote by $\mathcal{G}_N = (\widetilde{\Pi}^{\otimes N}, \cdot)$. Note that any two operators in $\mathcal{G}_N$ either commute or anticommute, hence why this group is not abelian.

\subsubsection{Error discretization}

Previously we discussed how, as a consequence of the second postulate of quantum mechanics, it is possible for any operator to act on a qubit so long as it is unitary. This means that there is a continuum of errors that may affect a qubit, i.e, qubits may suffer an almost negligible error such as a phase shift of $\frac{\pi}{263}$ or they could be impacted by an apparently catastrophic error that may remove the qubit entirely \cite{NielsenChuang}. Extraordinarily, it can be shown that the entire continuum of possible errors can be corrected by correcting only a discrete subset of those errors. This is known as error discretization and it is the central tenet of quantum error correction (should the entire continuum of possible errors have to be considered, it is obvious that error correction would not be feasible).

The discretization of errors in the quantum paradigm is a direct result of the \textit{theorem for the necessary and sufficient conditions for error correction}. In homage to its discoverers, this theorem is often referred to as the Knill-Laflamme theorem \cite{Knil1}, although the conditions for error correction where also formulated by Bennett at a similar time \cite{Bennett1}. The theorem plays a pivotal role in QEC because it defines the conditions that an error correcting code must fulfil in order to protect against decoherence. 

\begin{theorem} (\textrm{Knill-Laflamme Theorem for the necessary and sufficient conditions of quantum error correction}) Let $\mathrm{C}$ be a quantum error correcting code defined as a subspace of $\mathcal{H}_2^{\otimes N}$ and let $\mathcal{E} \subset \mathbb{C}^{2^N \times 2^N}$ denote a set of errors. Then $\mathrm{C}$ will be able to correct all the errors in $\mathcal{E}$ if and only if 

$$ \bra{\psi_i}E_a^{\dagger}E_b\ket{\psi_j} = c_{ab}\delta_{ij} \ ,$$

where $\ket{\psi_i}, \ket{\psi_j}$ are a basis for the subspace defined by $\mathrm{C}$, i.e, they run over all possible basis codewords, $E_a, E_b \in \mathcal{E}$, $c_{ab} \in \mathbb{C}$ is independent of $i$ and $j$, and $\delta$ denotes the Kronecker delta. 

\end{theorem}

Let us discuss how the theorem actually provides the necessary and sufficient conditions for quantum error correction. In order for a code to correct two different errors, $E_a$ and $E_b$ for instance, it must be capable of distinguishing the action of $E_a$ on a basis codeword $\ket{\psi_i}$ from the action of $E_b$ on a different basis codeword $\ket{\psi_j}$. 
The only possible way that this can be done is if $E_a\ket{\psi_i}$ is orthogonal to $E_b\ket{\psi_j}$, which we can write as $\bra{\psi_i}E_a^{\dagger}E_b\ket{\psi_j} = 0$ and $i\neq j$. However, if a code is to function as a quantum error correcting code, we must also ensure that when making measurements to diagnose an error we learn nothing about the encoded information. This is due to the third postulate of quantum mechanics, which tells us that we cannot directly measure quantum systems, else the superposition states will collapse. We learn about errors that our quantum information may have suffered by computing $\bra{\psi_i}E_a^{\dagger}E_b\ket{\psi_i}$, and since we must not learn anything about the state of the code itself, then these measurements must be the same for all basis codewords: $\bra{\psi_i}E_a^{\dagger}E_b\ket{\psi_i} = \bra{\psi_j}E_a^{\dagger}E_b\ket{\psi_j}$. If we now reconsider the single equation given in the Knill-Laflamme theorem, we can see that it encompasses both of the conditions we have just explained. 

The discretization of errors is perhaps the most critical result of the Knill-Laflamme Theorem, because it allows us to correct all possible errors by only considering a subset of them. It is generally formulated as a corollary of the theorem as follows.

\begin{corollary}
(\textrm{Discretization of errors}) For any quantum error correcting code $\mathrm{C}$ defined as a subspace of $\mathcal{H}_2^{\otimes N}$, if $\mathrm{C}$ is capable of correcting all the errors in a set $\mathcal{E}\subset \mathbb{C}^{2^N\times2^N}$, then $\mathrm{C}$ is capable of correcting all the errors in the linear span of $\mathcal{E}$.  
\end{corollary}

To truly understand the implications of this corolary, let us begin by defining a basis for $\mathbb{C}^{2\times 2}$. For this purpose we need to look no further than the set $\Pi$ of the single qubit Pauli matrices themselves. Showing that the Pauli matrices are a basis for $\mathbb{C}^{2\times 2}$ is straightforward: Firstly, the Pauli matrices cannot be written as linear combinations of themselves, which means that they are linearly independent. Secondly, because $\mathbb{C}^{2\times 2}$ is a complex vector space of dimension 4, any element of this space can be written as the linear combination of $4$ linearly independent matrices that also belong to $\mathbb{C}^{2\times 2}$. Thus, the Pauli matrices are a basis for $\mathbb{C}^{2\times 2}$. 

Having established this, we will now show that the set ${\Pi}^{\otimes N}$ of $N$-fold tensor products of single qubit Pauli matrices is a basis for $\mathbb{C}^{2^N\times 2^N}$. Thanks to the properties of the tensor product, we know that given a vector subspace $V$ with basis $\ket{i}$ and a vector subspace $W$ with basis $\ket{j}$, the tensor product of the basis' $\ket{i}\otimes\ket{j}$ is a basis for the tensor product space $V\otimes W$. Then, knowing that the set $\Pi$ is a basis for $\mathbb{C}^{2\times 2}$ and that $\mathbb{C}^{2^N\times 2^N} = (\mathbb{C}^{2\times 2})^{\otimes N}$, the set $\Pi^{\otimes N}$ is a basis for $\mathbb{C}^{2^N\times 2^N}$.

Now, by focusing on the wording of the error discretization corollary: ``if $\mathrm{C}$ is capable of correcting all the errors in a set $\mathcal{E}\subset \mathbb{C}^{2^N\times2^N}$, then $\mathrm{C}$ is capable of correcting all the errors in the linear span of $\mathcal{E}$'', it becomes clear why this result is so significant. If we can design codes with the capacity to correct errors in the set $\Pi^{\otimes N}$, because this set is a basis for $\mathbb{C}^{2^N\times 2^N}$ ($\Pi^{\otimes N}$ spans $\mathbb{C}^{2^N\times 2^N}$), our codes will also be capable of correcting any error in $\mathbb{C}^{2^N\times2^N}$. Essentially, this means that for the purposes of error correction, instead of having to work with the entire space $\mathbb{C}^{2^N\times2^N}$, it is enough to consider only the set $\Pi^{\otimes N}$. 

\subsubsection{Quantum Channels} \label{sec:channels}

 Earlier we mentioned that decoherence can be understood as the undesired entanglement of a quantum state with its environment. This corrupts quantum superposition states and makes it impossible for quantum computers to function without reverting the effects of this corruption. Although they are caused by various different physical mechanisms, we know from the postulates of quantum mechanics that all decoherence-related processes must be unitary (if the quantum state and the environment are considered as a closed system \cite{NielsenChuang, introQIC}) and that their effects must be diagnosed without direct measurement of the state of our data qubits (else their superposition would be disturbed). Additionally, decoherence is often assumed to affect each qubit differently, i.e, the noise process that affects each qubit is considered to be independent \cite{scheme-decoherence}. 
 
 To truly comprehend the manner in which decoherence affects quantum information, it is useful to study single-qubit examples. First, let us consider how this phenomenon affects the basis states of the computational basis. The basis sates $\ket{0}$ and $\ket{1}$ can be said to decohere as

\begin{align} \label{eq:decoherence-basic}
\begin{split} 
        \ket{\psi}_E\ket{0} \rightarrow \ket{a_1}_E\ket{0} + \ket{a_2}_E\ket{1}, \\
        \ket{\psi}_E\ket{1} \rightarrow \ket{a_3}_E\ket{0} + \ket{a_4}_E\ket{1},
\end{split}
\end{align}

where $\ket{\psi}_E$ is the state of the environment before any interaction takes place \cite{josurev} and $\{\ket{a_i}_E\}_{i=1}^4$ are states of the environment. Essentially, the expressions given in \eref{eq:decoherence-basic} are a mathematical representation of the undesired entanglement of the basis states with the environment. Based on this description, we can say that a qubit in state $\ket{\psi'} = \alpha\ket{0} + \beta\ket{1}$ will decohere as

\begin{equation} \label{eq:decoherence-one-qubit}
    \ket{\psi}_E\ket{\psi'} \rightarrow \alpha(\ket{a_1}_E\ket{0} + \ket{a_2}_E\ket{1}) + \beta(\ket{a_3}_E\ket{0} + \ket{a_4}_E\ket{1}).
\end{equation}

Despite being a useful example, the description given in \eref{eq:decoherence-one-qubit} is an impractical model for more than one qubit \cite{scheme-decoherence}. Instead, we can use quantum channels, as they provide better and more practical ways of modelling decoherence. To be more precise, quantum channels are mathematical abstractions that describe the effects of decoherence by mapping input quantum states onto ``noisy'' output quantum states. Among them, the amplitude damping channel, which describes the energy loss suffered by a quantum system as a consequence of its interaction with the environment, and the phase damping channel, which characterizes the loss of quantum information without energy loss, are the most popular. By combining both of these models we obtain the combined phase-and-amplitude damping channel, which provides the most accurate possible model for decoherence. Unfortunately, simulation of these channel models requires an amount of resources that increases exponentially for every additional qubit, which makes it impossible to efficiently simulate quantum channels directly on classical machines for sufficiently large qubit counts. Thankfully, the following theorem provides a solution to this predicament.

\begin{theorem} (\textrm{Gottesman-Knill Theorem}) Consider a quantum computation performed using only the following elements: state preparations in the computational basis, Hadamard gates, phase gates, controlled-NOT gates, Pauli gates, and measurements in the computational basis. Such a computation can be efficiently simulated on a classical computer.

\end{theorem}

The Gottesman-Knill theorem \cite{NielsenChuang, GotKnill} tells us that certain quantum computations that involve complex and highly entangled states (such as keeping track of how multi-qubit quantum states decohere) may actually be tractable on classical computers. Although the theorem does not apply to all possible quantum computations, it resolves the exponential resource dilemma by providing a way to model decoherence without the need for a functioning quantum computer\footnote{It is worth nothing that the problem is somewhat paradoxical in nature: we actually need the very same machine we are trying to make work, a quantum computer, to completely track the decoherence of qubits.}. Additionally, the theorem also holds for all quantum circuits that can be described using the stabilizer formalism, which, as will be shown in the following chapter, encompasses many QEC code families and allows us to simulate QEC codes efficiently on classical computers \cite{josurev}. 

It must be mentioned that the classically tractable model for decoherence that can be built thanks to the Gottesman-Knill theorem is actually an approximation of the aforementioned quantum channels (amplitude damping, phase damping, and combined amplitude and phase damping). Although not as precise a model as the combined amplitude and phase damping channel, by means of quantum information theory techniques such as ``twirling'', this decoherence model has been shown to be a valid approximation. For the sake of brevity, herein we will only discuss the Pauli channel model for decoherence, which is the name given to the decoherence model that classical machines can simulate. For a thorough and in-depth discussion on the topic of quantum channels and their approximations, the reader should refer to \cite{josurev}. 

\subsubsection*{The Pauli channel} 

The Pauli channel is a classically tractable quantum channel model that represents the decoherence effects suffered by quantum information. The effect of the Pauli channel $\xi_P$ upon an arbitrary single-qubit quantum state with density\footnote{It is significantly easier to describe quantum channels using the density matrix representation of quantum mechanics (the state vector notation can become too complicated in some instances). The reader is referred to \cite{NielsenChuang, josurev} for a more complete discourse on this topic.} matrix $\rho$ can be written as

\begin{equation} \label{eq:pauli-channel}
 \xi_P(\rho) = (1-p_x-p_y-p_z)\rho + p_xX\rho X + p_yY\rho Y + p_zZ\rho Z ,
 \end{equation}

where $\{p_x, p_y, p_z\}$ are the probabilities that the state $\rho$ suffers each respective operator. More explicitly, a qubit that traverses the Pauli channel experiences a bit-flip ($X$ operator) with probability $p_x$, a phase-flip ($Z$ operator) with probability $p_z$ or a combination of both (a $Y$ operator) with probability $p_y$. In summary, the action of the Pauli channel on a quantum state $\ket{\psi}$ can be understood as a mapping onto a linear combination of the original state ($I\ket{\psi}$), the phase flipped state ($Z\ket{\psi}$), the bit flipped state ($X\ket{\psi}$), and the bit-and-phase flipped state ($Y\ket{\psi}$), where the sum is weighted by the probabilities $\{p_x, p_y, p_z\}$. Note that the reason behind the Pauli channel being a manageable decoherence model for classical machines is that it can be completely described by means of the Pauli matrices (see expression \eref{eq:pauli-channel}). In a similar manner to the generalization shown in section \ref{sec:N-qubits}, the Pauli channel will act on a multi-qubit quantum state by applying an error operator $E \in \mathcal{G}_N$, where $E$ can be understood as the tensor product of single-qubit Pauli matrices.

Based on the nature of the probabilities $\{p_x, p_y, p_z\}$ of each single-qubit  Pauli operator, different versions of the general Pauli channel can be derived. The most popular quantum channel model in the QEC literature is the independent depolarizing channel model \cite{jgf1, jgf2, patrick, QTC, josu1, josu2, degen3, nonCSS1, nonCSS2}, which is a specific instance of the Pauli channel in which the individual operator probabilities are all equal, i.e., $p_x = p_z = p_y = \frac{p}{3}$. This channel model is completely characterized by the depolarizing probability $p$. When quantum states of $N$ qubits are considered, the errors that take place belong to the $N$-fold Pauli group $\mathcal{G}_N$ and they will act independently on each qubit, causing an $X$, $Z$, or $Y$ error with probability $p/3$ and leaving the qubit unchanged with probability $(1-p)$. Later on in this dissertation we will present different channel models based on the properties of the individual probabilities $\{p_x, p_y, p_z\}$.

\newpage

\section{Classical Error Correction}\label{cp3:class} 

The field of classical error correction was first conceived following the publishing of Claude E. Shannon's seminal work \cite{Shannon}. Within it, Shannon proved that reliable transmission over a communication channel is not possible if the information transfer rate exceeds a quantity known as the channel capacity. Furthermore, this work also showed that in order to transmit information at rates close to the channel capacity, information protection strategies kown as channel codes are necessary. This gave birth to a new scientific field dedicated to the study and optimization of channel codes, which would later become known as the discipline of Classical Error Correction. In this dissertation we focus on a specific family of classical error correcting codes called linear block codes. Thus, this section begins with an introduction to this family of error correcting codes. Then, we introduce the linear block code family of Low Density Parity Check (LDPC) codes in section \ref{sec:LDPC}. This section also includes basic communication and graph theory notions like factor graphs an iterative decoding. For the sake of simplicity, in this introduction we deal only with binary codes and all the arithmetic will be mod $2$. The reader is referred to \cite{nonbinblock1, nonbinblock2, nonbinblock3, nonbinblock4} for discussions on non-binary linear block codes.

\subsection{Linear Block Codes} \label{sec:linear-code}

Suppose we wish to transmit a message made up of $k$ information bits. Then, a linear block code is defined as a linear mapping between a \textit{k}-bit message vector $[\mathbf{u}]_{1\times k}$ and an \textit{N}-bit codeword vector $[\mathbf{x}]_{1\times N}$. Linear block codes exhibit the linearity property because linear combinations of codewords also belong to the code (they are codewords). In other words, a binary linear block code constitutes a linear mapping from the space $\mathbb{F}_2^{k}$ to the space $\mathbb{F}_2^{N}$ which maps the information word $[\mathbf{u}]_{1\times k}$ to the codeword $[\mathbf{x}]_{1\times N}$ by computing the matrix product $[\mathbf{x}]_{1\times N}=[\mathbf{u}]_{1\times k}[\mathbf{G}]_{k\times N}$. The matrix $[\mathbf{G}]_{k\times N}$ is a size $k\times N$ matrix known as the generator matrix whose columns form a basis for the $k$-dimensional coding subspace of $\mathbb{F}_2^{N}$, and represent basis codewords. A linear block code can also be represented using its Parity Check Matrix (PCM) \textbf{H}, which defines a basis for the nullspace of the code, i.e, the product $\textbf{G}\textbf{H}^\top$ will always be $\mathbf{0}$, where $\mathbf{0}$ denotes the all zero matrix of size $k\times (N-k)$. Thus, $[\mathbf{H}]_{(N-k)\times N}$ is a size $(N-k)\times k$ matrix whose rows represent the linear constraints to which the codewords of the code are subjected and that ensures \textbf{c$\textbf{H}^\top$\ } = \textbf{0} whenever $[\mathbf{c}]_{1\times N}$ is a codeword. In other words, the PCM provides a simple way of checking if any length $N$ binary vector belongs to the code. For context, the generator and parity check matrix pair of the [7,4,3] Hamming code \cite{ham14} is shown below,

\begin{equation} \label{eq:ham}
\mathbf{G}_{\text{ham}} = 
\begin{pmatrix}
1 & 0 & 0 & 0 & 1 & 1 & 1 \\
0 & 1 & 0 & 0 & 1 & 1 & 0 \\
0 & 0 & 1 & 0 & 1 & 0 & 1 \\
0 & 0 & 0 & 1 & 0 & 1 & 1
\end{pmatrix}, \
\mathbf{H}_{\text{ham}} = \begin{pmatrix}
1 & 1 & 1 & 0 & 1 & 0 & 0 \\
1 & 1 & 0 & 1 & 0 & 1 & 1 \\
1 & 0 & 1 & 1 & 0 & 0 & 1
\end{pmatrix}.
\end{equation}

Linear block codes are conventionally represented using the [$N, k , d$] notation, where $N$ represents the block length of the code (the size of a codeword), $k$ represents the number of encoded bits, and $d$ represents a concept known as the minimum distance of the code. The rate of the code can be computed from these parameters as $R = \frac{k}{N}$. The distance of a code provides a measure of the error-correcting properties of a code. To better understand what it represents, it is useful to introduce the concept of the Hamming distance. The Hamming distance between two vectors is the minimum number of bits that must be flipped to convert one vector to the other. The distance between two binary vectors $\mathbf{a}$ and $\mathbf{b}$ is equal to the
weight (the number of $1$s in the vector) of $\mathbf{a}+\mathbf{b}$. This means that, for a code to correct $t$ single-bit errors, it must have distance at least $2t + 1$ between any two codewords. A $t$
bit error will take a codeword exactly a $t$ distance away from its original value, so when the distance between codewords is at least 2$t$+1, we can distinguish errors on different codewords and correct them to the proper codewords.

Let us now recall our initial example of wanting to transmit $k$ information bits. Once the information message \textbf{u} has been encoded into the codeword \textbf{x} using the generator matrix, the next step is to transmit the codeword through a communication channel. Typically this results in noise being added to the codeword and a noisy vector $[\textbf{r}]_{1\times N}=[\textbf{c}]_{1\times N}\oplus[\textbf{e}]_{1\times N}$ being received, where $[\textbf{e}]_{1\times N}$ is a length $N$ vector that represents the action of the particular channel in question. The next step would be to perform decoding, which refers to the process of attempting to revert the action of the channel on the transmitted codeword and recovering the original information message \textbf{u}. There is a wide variety of different decoding strategies for the families of linear block codes, but the most basic and easy to understand is known as syndrome detection. Essentially, when the noisy vector \textbf{r} is received, the decoder will compute a length $N-k$ binary vector $[\textbf{z}]_{1\times (N-k)}$, known as the error syndrome, based on the product $[\textbf{r}]_{1\times N}[\textbf{H}^{\top}]_{N\times(N-k)} = [\textbf{c}\oplus\textbf{e}]_{1\times N}[\textbf{H}^\top]_{N\times(N-k)} = [\textbf{c}]_{1\times N}[\textbf{H}^\top]_{N\times(N-k)} \oplus [\textbf{e}]_{1\times N}\textbf[\textbf{H}^\top]_{N\times(N-k)} = [\textbf{e}]_{1\times N}[\textbf{H}^\top]_{N\times(N-k)} = \textbf{z}_{1\times (N-k)}$. If the syndrome is equal to zero, it is likely that the channel has not corrupted the codeword and that $\mathbf{x}$ has been received. If not, then the decoder can use the syndrome in combination with a decoding algorithm to find the noise vector \textbf{e} that was added to \textbf{c} and attempt to revert its impact by re-adding \textbf{e} to \textbf{r}.

\subsection{Low Density Parity Check codes} \label{sec:LDPC}

Sparse codes, generally referred to as Low Density Parity Check codes, are a class of linear block codes whose parity check matrices, as befits their name, are low density\footnote{When a matrix is said to be sparse or low density, it will have a significantly larger number of zero entries than non-zero entries.} (sparse) matrices. LDPC codes were discovered by Gallager in 1961 \cite{ldpc1, ldpc2}, but quickly fell from grace given that they could not be efficiently decoded. A couple of decades later, following the proposal of BP decoding \cite{BP} and a drastic improvement in communications technology, LDPC codes returned to the limelight as one of the best families of capacity achieving error correcting codes \cite{ldpc3, ldpc4, ldpc5, ldpc6}. 

The most common way of describing LDPC codes is by means of factor graph \cite{spa} and Tanner\footnote{Tanner graphs have been shown to be particular instances of factor graphs \cite{Wiberg}.} graph \cite{Tanner1, Tanner2} representations of their PCMs. Factor graphs are generic, edge-connected graphs that are practical to represent mathematical expressions because of the straightforward manner in which they portray the dependencies and factorizations of these expressions. Moreover, LDPC codes can be efficiently decoded by running the SPA algorithm over a factor graph representation of their PCM. For these reasons, in what follows (which is an adaptation of \cite{imagra}), factor graphs and the SPA algorithm are discussed in detail.

\subsubsection{Factor Graphs} \label{sec:SPA}

\begin{figure}[!h]
	\centering
	\includegraphics[width=0.3\columnwidth]{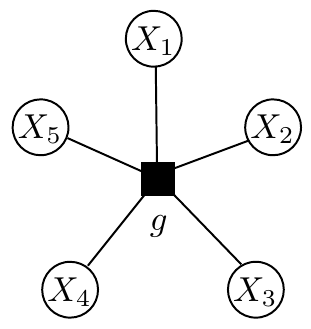}
	\caption{Factor graph representation of the global multi-variable function $g(x_1, x_2, x_3, x_4, x_5)$.}
	\label{fac-graph1}
\end{figure}

Algorithms that must deal with complicated multi-variable global functions often exploit the manner in which these functions factor into products of local functions, each of which depends on a subset of the variables. Such a factorization can be portrayed by means of a bipartite graph that is commonly known as a factor graph. For instance, let $g(x_1, x_2, x_3, x_4, x_5)$ be a multi-variable global function acting on the set of variables $\{X_1,X_2,X_3,X_4,X_5\}$. This function is represented in Fig. \ref{fac-graph1}, where the \textit{function} and \textit{variable nodes} are represented as squares and circles, respectively. Function nodes and variable nodes are connected with an edge if and only if the corresponding variable is an argument of the function. Note that a factor graph is always bipartite, i.e., edges are only allowed between vertices of different types.

Factor graphs are most useful to represent the way in which global functions can be factorized as a product of several local functions. Let us assume that the function $g(x_1,x_2,x_3,x_4,x_5)$ can be written as
\begin{equation*}	
g(x_1, x_2, x_3, x_4, x_5) = f_A(x_1) \cdot f_B(x_2) \cdot f_C(x_1,x_2,x_3) \cdot f_D(x_3,x_4) \cdot f_E(x_3,x_5).
\end{equation*}
\begin{figure}[!h]
	\centering
	\includegraphics[width=0.4\columnwidth]{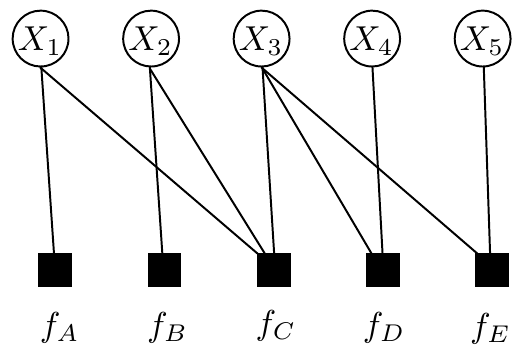}
	\caption{Factorized factor graph of $g(\cdot)$}
	\label{fac-graph2}
\end{figure}

In this case, the factor graph can be expanded to represent the local functions $f_A,f_B,f_C,f_D,f_E$. This expansion is depicted in Fig. \ref{fac-graph2}, where the variable nodes comprise the top layer of the graph and the function nodes make up the bottom layer of the graph. Although this bipartite graph is unique, it can be redrawn differently. For convenience, let us rearrange it as in Fig. \ref{fac-graph3}.  

Aside from the simplicity with which they represent multi-variable functions, the primary appeal of factor graphs comes from the fact that the SPA algorithm \cite{spa} can be run over them. It is for this reason that factor graphs are so often used in conjunction with LDPC codes, as the SPA algorithm can be used to decode LDPC codes by using the factor graph representation of their PCMs. In fact, many other decoding algorithms, such as BP or Viterbi decoding \cite{BP, Viterbi}, have been shown to be specific instances of the SPA \cite{spa}.

The SPA is a message passing algorithm that exploits the way in which factor graphs represent the factorization of multi-variable global functions into simpler local functions to compute target functions of interest known as marginals. The computational efficiency of the algorithm comes precisely from the way in which factor graphs represent global functions. The algorithm can be summarized as a set of rules and procedures that govern the way in wich message passing takes place over a factor graph. Based on the specific structure of a factor graph, two different scenarios can be encountered:

\begin{figure}[!h]
	\centering
	\includegraphics[width=0.6\columnwidth]{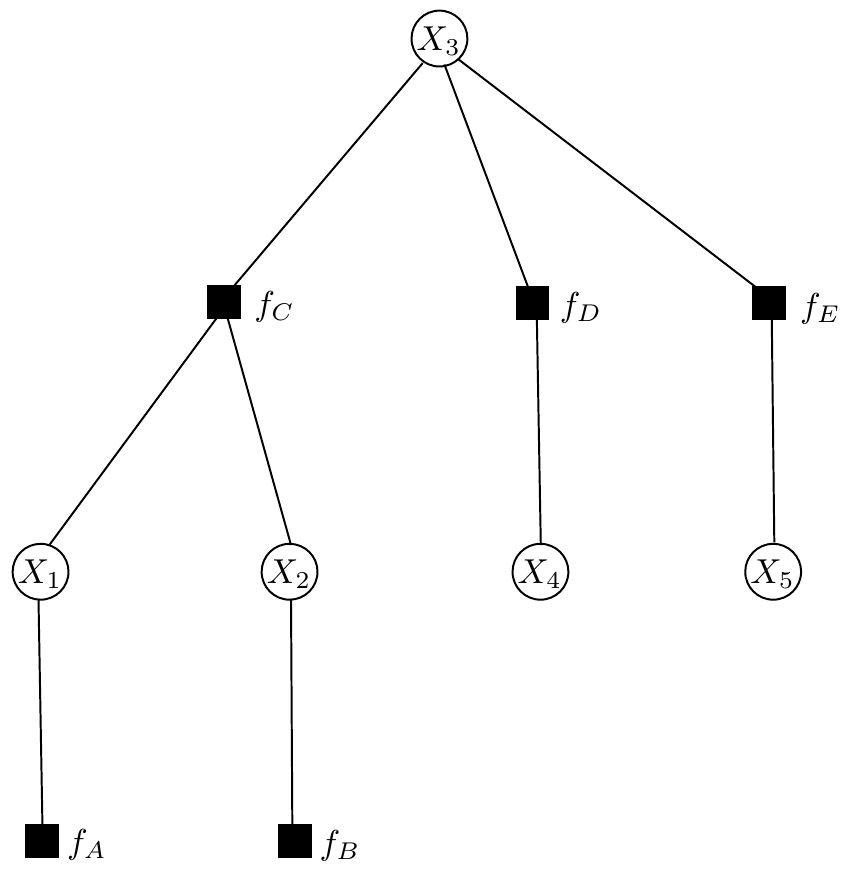}
	\caption{Rearrangement of the graph shown in Figure \ref{fac-graph2}.}
	\label{fac-graph3}
\end{figure}

\subsubsection*{The Sum-Product Algorithm over cycle-free Factor Graphs}

Recall the rearranged factor graph shown in Fig. \ref{fac-graph3}  corresponding to the global function $g(\cdot)$. This global function can be factorized as 
\begin{equation*}	
	g(x_1, x_2, x_3, x_4, x_5) = f_A(x_1) \cdot f_B(x_2) \cdot f_C(x_1,x_2,x_3) \cdot f_D(x_3,x_4) \cdot f_E(x_3,x_5).
\end{equation*}
Let us now assume that we are interested in calculating the marginal function $\hat{g}_1(x_1)$, given by
\begin{equation}\label{eq1_marg}
	\hat{g}_1(x_1) = \sum_{x_2,x_3,x_4,x_5}g(x_1,x_2,x_3,x_4,x_5) = \sum_{\sim x_1}g(x_1,x_2,x_3,x_4,x_5).
\end{equation}

If certain criteria are met, then the SPA can be used to compute $\hat{g}_1(x_1)$ based on the exchange of messages between the nodes of the factor graph. The messages exchanged by the factor graph nodes are described as follows:
\begin{itemize}
	\item We denote by $\psi(X_i)$ the set of function nodes $f_j$ connected to variable node $X_i$. Similarly, $\psi(X_i)\setminus f_k$ denotes the set of function nodes $f_j$ connected to the variable node $X_i$ excluding the function node $f_k$. For example, $\psi(X_1)=\{f_A, f_C\}$ and $\psi(X_1)\setminus f_A=\{f_C\}$  in Fig. \ref{fac-graph3}. 
	\item We denote by $\mu_{X_i\rightarrow f_j}(x_i)$ and $\mu_{f_j\rightarrow X_i}(x_i)$ those messages transmitted from the variable node $X_i$ to the function node $f_j$ and from $f_j$ to $X_i$, respectively. Note that these messages are only transmitted if $f_j \in \psi(X_i)$ and $X_i \in \psi(f_j)$.
\end{itemize}

These messages are propagated through all edges of the graph based on the operational rules of the algorithm and can be reused to compute different marginal functions. The operational rules define how each message should be computed and are summarized as:

\begin{description}
	\item[Messages transmitted from variable nodes to function nodes:] The message sent from variable node $X_i$ to function node $f_j$ is computed as
	\begin{equation}\label{cp2_FG_1}
		\mu_{X_i\rightarrow f_j}(x_i) = \prod_{f_k\in\psi(X_i)\setminus f_j} \mu_{f_k\rightarrow X_i}(x_i).
	\end{equation}
	\item[Messages transmitted from function nodes to variable nodes:] The message sent from function node $f_j$ to variable node $X_i$ is computed as
	\begin{equation}\label{cp2_FG_2}
		\mu_{f_j\rightarrow X_i}(x_i) = \sum_{X_k\in\psi(f_j)\setminus X_i} f_j(x_k) \cdot \mu_{X_k\rightarrow f_j}(x_k).
	\end{equation}
	\item[Computation of the marginal functions:] The marginal function $\hat{f}_i(x_i)$ is computed as
	\begin{equation}\label{cp2_FG_3}
		\hat{f}_i(x_i) = \prod_{f_j\in\psi(X_i)} \mu_{f_j\rightarrow X_i}(x_i).
	\end{equation}
\end{description}

\begin{figure*}[!h]
	\centering
	\subfloat[ \label{cp3_fig:bervssnr1a}]{%
		\includegraphics[width=.52\textwidth]{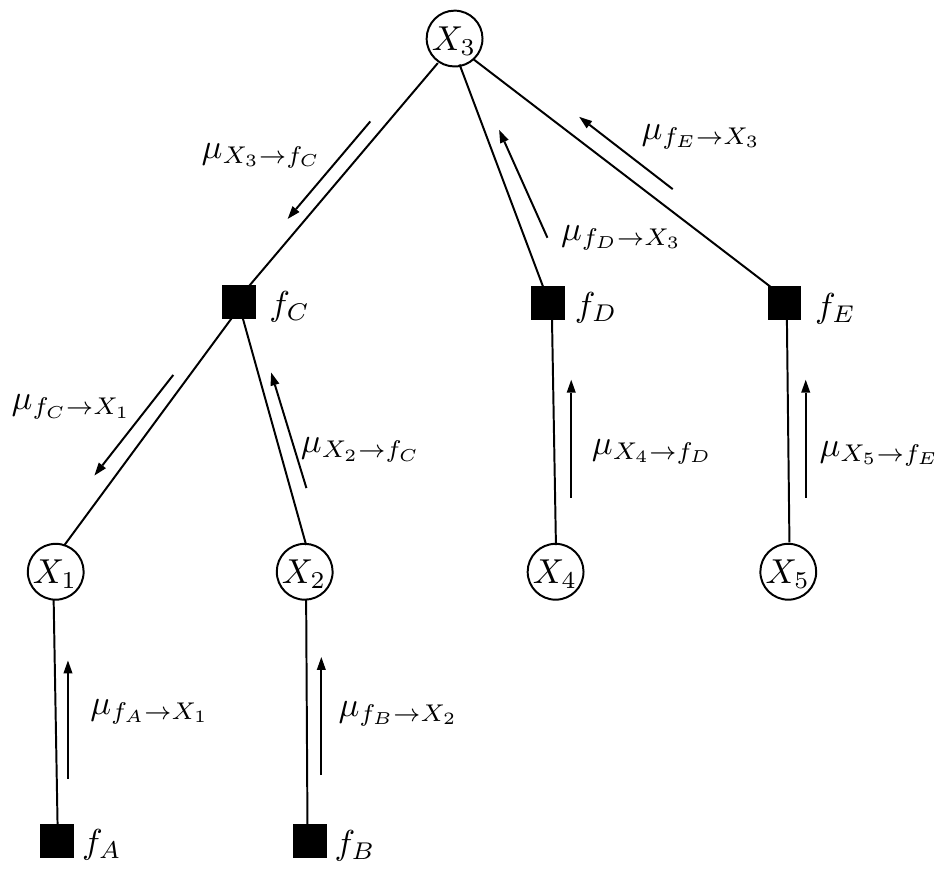}\label{cp2_fig:margf1a}
	} \hfill
	\subfloat[ \label{cp3_fig:bervssnr1b}]{%
		\includegraphics[width=.44\textwidth]{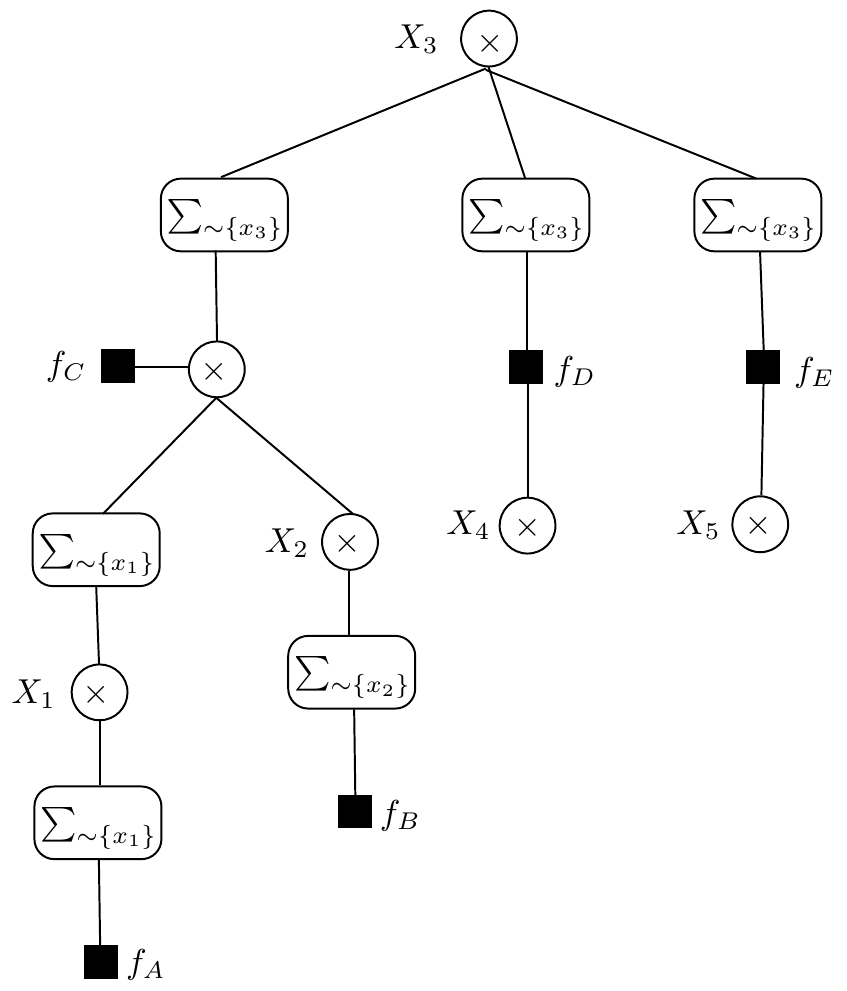} \label{cp2_fig:margf1b}
	}\caption{\textbf{(a)} Messages involved in the marginalization of $\hat{g}_1(x_1)$. \textbf{(b)} Update rules of the marginalization of $\hat{g}_1(x_1)$.} \label{cp2_fig:margf1}
\end{figure*}

Figure \ref{cp2_fig:margf1} portrays the message exchange process of the SPA when computing the marginal function $\hat{g}_1(x_1)$. This can also be seen by breaking up the expression given in (\ref{eq1_marg}) into its compounding factors as
\begin{align} 
	\resizebox{1\textwidth}{!}{$\hat{g}_1(x_1) =\sum\limits_{x_2}\sum\limits_{x_3}\sum\limits_{x_4}\sum\limits_{x_5} f_A(x_1)\cdot f_B(x_2) \cdot f_C(x_1,x_2,x_3) \cdot f_D(x_3,x_4) \cdot f_E(x_3,x_5)$} \notag \\
	\resizebox{1\textwidth}{!}{$=\underbrace{f_A(x_1)}_{\mu_{f_A\rightarrow X_1}(x_1)}\cdot\underbrace{ \sum_{x_2}\sum_{x_3} f_C(x_1,x_2,x_3)\underbrace{\cdot \underbrace{f_B(x_2)}_{\mu_{f_B\rightarrow X_2}(x_2)}}_{\mu_{X_2\rightarrow f_C}(x_2)}\cdot \underbrace{\underbrace{ \sum_{x_4} f_D(x_3,x_4) \cdot \underbrace{1}_{\mu_{X_4\rightarrow f_D}(x_4)}}_{\mu_{f_D\rightarrow X_3}(x_3)} \cdot \underbrace{\sum_{x_5} f_E(x_3,x_5) \cdot \underbrace{1}_{\mu_{X_5\rightarrow f_E}(x_5)}}_{\mu_{f_E\rightarrow X_3}(x_3)}}_{\mu_{X_3\rightarrow f_C}(x_3)}}_{\mu_{f_C\rightarrow X_1}(x_1)},$}	\label{cp2_margf1opt}	
\end{align}

which allows us to derive that the marginalization of $X_1$ is actually given by,
\begin{equation}
	\hat{g}_1(x_1)=\mu_{f_A\rightarrow X_1}(x_1)\cdot \mu_{f_C\rightarrow X_1}(x_1).
\end{equation}

\begin{figure}[!ht]
	\centering
	\includegraphics[width=0.5\columnwidth,angle=-90]{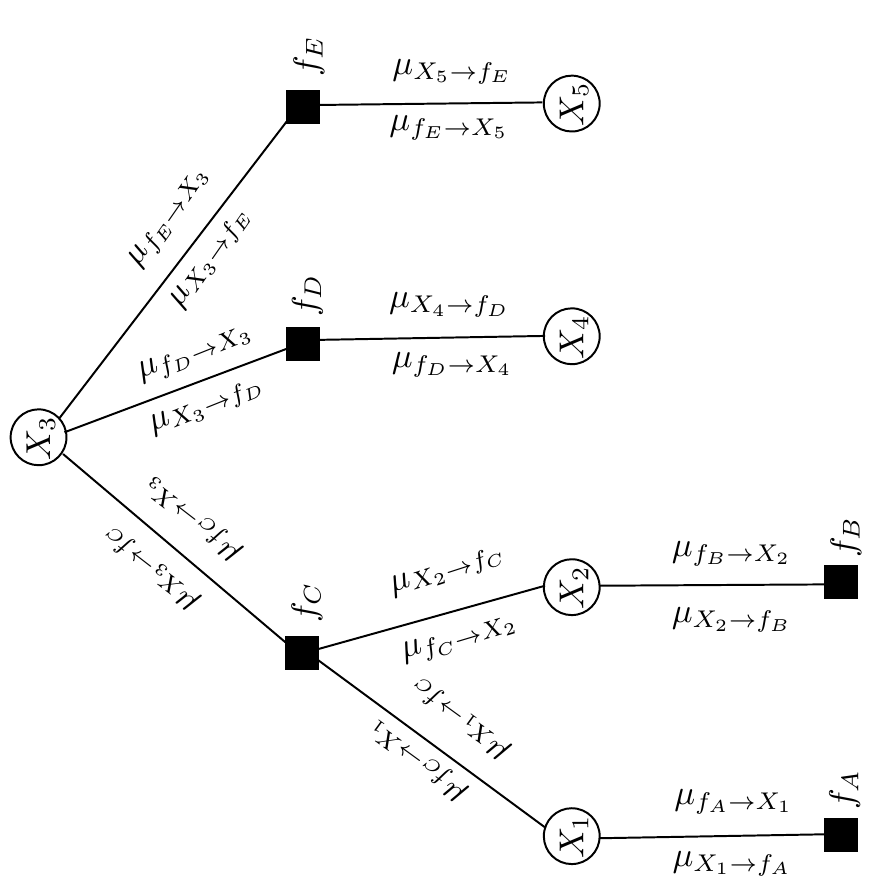}
	\caption{All necessary messages required to compute the marginal functions $\hat{g}_1(x_1), \hat{g}_2(x_2)$, $\hat{g}_3(x_3)$, $\hat{g}_4(x_4)$ and $\hat{g}_5(x_5)$. Messages that propagate upwards/downwards are placed on the right/left side of the edges.}
	\label{cp2_SPA_graph}
\end{figure}

 The other marginals $\hat{g}_2(x_2)$, $\hat{g}_3(x_3)$, $\hat{g}_4(x_4)$ and $\hat{g}_5(x_5)$ are computed in a similar fashion, each particular marginal requiring the computation of a different set of messages. All the necessary messages required to compute these marginals are shown in Fig. \ref{cp2_SPA_graph}.
 
\begin{figure}[!ht]
	\centering
	\includegraphics[width=0.6\columnwidth, height=1.25in]{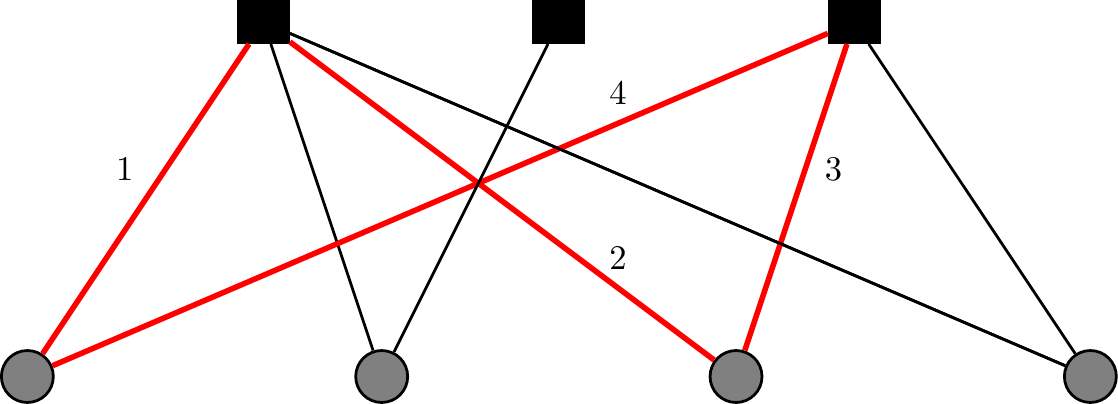}
	\caption{Example of a cycle in a factor graph. The edges that conform the cycle are numbered from left to right.}
	\label{cycle}
\end{figure}

 At this point, it is important to note that execution of the SPA for the computation of these marginals requires a finite number of steps, i.e, the message exchange procedure halts on its own. This occurs because the factor graph representation of the global function $g(\cdot)$ (See Figures \ref{fac-graph1}, \ref{fac-graph2}, and \ref{fac-graph3}) has no cycles. A cycle or loop can be understood as a closed path in the factor graph that begins and ends at the same variable node and that involves the passing of a single message over each of the edges that comprise the path. Cycles can be described by their length, which refers to the number of edges that conform the cycle itself. An example of a length-$4$ cycle is shown in Figure \ref{cycle}. Execution of the SPA over a factor graph with cycles produces an “iterative” algorithm with no natural termination in which messages are  passed multiple times on a given edge. This results in the marginalizations computed by the algorithm not being exact function summaries. However, in many of the SPAs practical applications, like the decoding of LDPC codes or turbo codes, execution over graphs with cycles is involved. Fortunately, despite its inexact marginalization of global functions in the presence of cycles, extensive simulation results have shown that SPA-based decoding of very long codes can achieve astonishing performance (within a small fraction of a decibel of the Shannon capacity on a Gaussian channel) \cite{ldpc3, factor-cite1,factor-cite2}.
 
\subsubsection*{The Sum-Product algorithm over factor graphs with cycles}

Most of the factor graphs that appear in the context of error correction contain cycles. This makes it impossible to compute an exact marginalization of the global function, as the algorithm can no longer terminate in a finite number of steps \cite{cp2_mackay}. However, this does not mean the SPA cannot be executed over a factor graph with cycles, instead, it requires that the iterative unending version of the SPA that manifests over loopy graphs be forcibly halted. This intervention or lack of a natural termination is the reason why the SPA is a sub-optimal marginalization algorithm in the presence of cycles.

When running over a loopy factor graph, operation of the SPA can be summarised in the following steps:
\begin{enumerate}
	\item All factor graph messages are initialized.
	\item Messages are updated according to a specific schedule: (\ref{cp2_FG_1}) and (\ref{cp2_FG_2}). This schedule may vary from step to step. 
	\item After each step (or a predefined number of steps) the marginal functions are computed: (\ref{cp2_FG_3}).
	\item The algorithm output is derived from the current value of the marginal function.
	\item Based on the output, a decision is made: if the output satisfies a set of conditions then the algorithm is stopped and if not, it continues to run through the previous steps until those conditions are met or until a specific number of iterations is reached.
\end{enumerate}

Decoding of LDPC codes can be understood as a particular example of running the SPA algorithm over a factor graph with cycles. 

\subsubsection*{Factor graph representation of a PCM}

The factor graph representation of a linear block code is nothing more than a visual representation of the specific code constraints that the parity check matrix of the code defines. In fact, SPA-decoding of an LDPC code can be summarized succinctly as using the algorithm to check if a received message belongs to the code, which is appropriately represented by a factor graph. We know from the previous discussion that factor graphs are bipartite graphs that contain two distinct types of nodes; variable nodes and function nodes. Deriving such a graph for a linear block code is achieved by relating variable nodes to columns of the PCM and function nodes to the rows of the PCM; there will be a variable node for every column of the PCM and there will be a function node for every row of the PCM, respectively. The graph is completed by connecting variable nodes and function nodes with a directionless edge\footnote{Nodes of the same type are never connected.} for every nonzero entry in the parity check matrix. As is done in the literature and in conventional error correction jargon, we will refer to the function nodes of the factor graph of linear block code as parity check nodes\footnote{The rows of the PCM represent the equations that define the code. These equations enforce a parity on the code, hence why the function nodes of the factor graph corresponding to a linear code can be seen as performing parity checks.}. 

\begin{figure}[htp]
\centering
  \includegraphics[width=0.8\columnwidth,  height=1.5in]{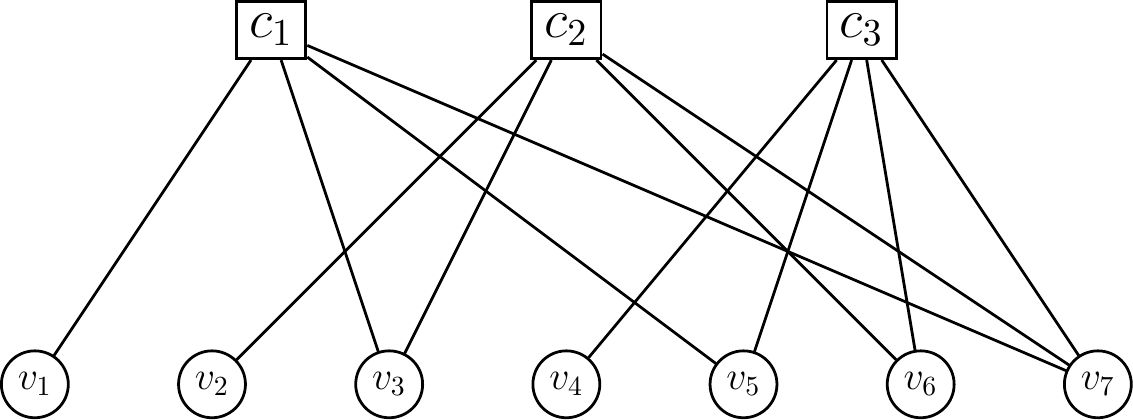}
\caption{Factor graph representation of the  [7,4,3] Hamming code, which corresponds to the parity check matrix shown in equation (\ref{eq:hamPCM}).}
\label{fig:factorham}
\end{figure}

Figure \ref{fig:factorham} portrays the factor graph representation of a column-permuted version of the PCM of the [7,4,3] classical Hamming code shown in (\ref{eq:hamPCM}). Note that the property of linearity ensures that performing linear operations on the PCM of a linear block code does not alter the code itself, hence the PCM of (\ref{eq:ham}) and the one shown below are essentially analogous. 

\begin{equation} \label{eq:hamPCM}
\mathbf{H}_{\text{ham}} = 
\begin{pmatrix}
1 & 0 & 1 & 0 & 1 & 0 & 1 \\
0 & 1 & 1 & 0 & 0 & 1 & 1 \\
0 & 0 & 0 & 1 & 1 & 1 & 1
\end{pmatrix}
\end{equation}

The matrix shown in (\ref{eq:hamPCM}), given its reduced size, is not a low density matrix and so the [7,4,3] Hamming code is not actually an LDPC code. However, because LDPC codes are typically much larger and sparser, utilizing the Hamming code for this example serves the intended purpose. It must also be mentioned that, because the Hamming code is binary, the PCM-to-factor graph mapping is one-to-one \cite{qldpc15, nonbin1, nonbin2}. This is not always the case for non-binary codes. 

\subsubsection{Syndrome-based decoding of LDPC Codes} \label{sec:classical-decoding}

Consider a classical linear block code with generator matrix $[\mathbf{G}]_{k\times N}$ and parity check matrix $[\mathbf{H}]_{(N-k)\times N}$ such that $\mathbf{G}\mathbf{H}^\top = \mathbf{H}\mathbf{G}^\top = \mathbf{0}$. Assume we encode the information word $[\mathbf{u}]_{1\times k}$ as $[\mathbf{u}]_{1\times k}[\mathbf{G}]_{k\times N} = [\mathbf{x}]_{1\times N}$ and transmit the codeword $[\mathbf{x}]_{1\times N}$ through a classical Binary Symmetric Channel (BSC) completely specified by the flip probability $p$ (the channel flips each input bit with probability $p$). Upon reception of the noisy channel output sequence $[\mathbf{r}]_{1\times N} = [\mathbf{x}]_{1\times N} \oplus [\mathbf{e}]_{1\times N}$, the receiver computes the error syndrome, $[\mathbf{z}]_{1 \times (N-k)}$, as $\mathbf{z} = \mathbf{x} \mathbf{H}^\top= (\mathbf{c} \oplus \mathbf{e}) \mathbf{H}^\top= \mathbf{e} \mathbf{H}^\top$. The error syndrome can be understood as a Boolean representation of which code constraints are fulfilled by the received noisy sequence and which ones are not\footnote{Recall that the constraints of a classical code are represented by a set of parity check equations, each one associated to a different row of the parity check matrix of the code.}. Having determined the syndrome, the decoder will use it to produce an estimate $\hat{\mathbf{e}}$ of the error pattern induced by the channel. Following this, the inverse of the estimated error pattern, $-\hat{\mathbf{e}}=\hat{\mathbf{e}}$ (recall that arithmetic is mod$2$), is added to the received noisy vector $\mathbf{r}$. If the error estimate and the channel error match, $\hat{\mathbf{e}}=\mathbf{e}$, the codeword $\mathbf{r}$ will be recovered, and if not, a decoding mistake will have taken place.

Based on this entire process, it is clear that the purpose of the decoding process is to produce an estimate of the most likely error pattern $\hat{\mathbf{e}}$ given the syndrome $\mathbf{z}$ so that the probability of a word\footnote{Error patterns or sequences are sometimes referred to as words.} error $\mathrm{P}(\hat{\mathbf{e}} \neq \mathbf{e})$, commonly known as the Word Error Rate (WER), is minimized. This is formally referred to as syndrome-based Maximum Likelihood (ML) decoding, which we can express mathematically as 

\begin{equation} \label{eq:ML0}
\hat{\mathbf{e}}^{\text{ML}} = \argmax_\mathbf{e} \mathrm{P}(\mathbf{e}|\mathbf{z}).
\end{equation}

By applying Bayes' rule, the expression shown in (\ref{eq:ML0}) can be expanded into

\begin{equation} \label{eq:ML2}
\hat{\mathbf{e}} = \argmax_\mathbf{e} \mathrm{P}(\mathbf{e}|\mathbf{z}) = \argmax_\mathbf{e} \frac{\mathrm{P}(\mathbf{z}|\mathbf{e})\mathrm{P}(\mathbf{e})}{\mathrm{P}(\mathbf{z})}
\end{equation}

Notice that the computation of expressions (\ref{eq:ML0}) and (\ref{eq:ML2}) requires the brute force calculation of the probability of all the possible errors given the obtained syndrome $\mathbf{z}$. In the most simple of scenarios (codes of reduced block length), it is possible to implement the ML decoding rule using a pre-computed Look-Up Table (LUT) that associates error syndromes with error patterns \cite{LUT1}. However, since the set of possible errors grows exponentially as a function of the code block length, brute force and LUT computations quickly become intractable. In fact, decoding a generic code based on equation (\ref{eq:ML0}) has been shown to be an NP problem\footnote{It is actually an NP-complete problem, which is the particular subclass that contains the hardest NP problems.} \cite{degen4, MLhard}.

For this reason, decoding is generally approached via a less computationally demanding approximation of the ML decoding rule known as Symbol-Wise Maximum Likelihood (SWML) decoding. The SWML decoding rule can be written as 


\begin{align} \label{eq:bwML}
\begin{split} 
\hat{\mathrm{e}}_{j}^{\text{SW}} = \argmax_{\mathrm{e}} P(\mathrm{e}_j = \mathrm{e}|\mathbf{z}) \\ = \argmax_{\mathrm{e}_j \in \{0,1\}} \sum_{\mathrm{e}_1,\ldots,\mathrm{e}_{j-1},\mathrm{e}_{j+1},\ldots,\mathrm{e}_N} P(\mathrm{e}_1, \ldots, \mathrm{e}_N|\mathbf{z}),
\end{split}
\end{align}
where we have used regular lower case romans to denote the components of an error $\mathbf{e} \in \mathbb{F}_2^{N}$, i.e., $\mathbf{e} = [\mathrm{e}_1,\ldots,\mathrm{e}_N]$. The reason why SWML (\ref{eq:bwML}) is less computationally complex than ML (\ref{eq:ML0}) comes from the fact that SWML maximizes the individual marginal probabilities instead of the global maximum, i.e, it minimizes the symbol-wise error probability $\mathrm{P}(\mathrm{\hat{e}}_k \neq \mathrm{e}_k$) instead of the WER. For general channels, this difference between both decoding rules can result in SWML yielding different results from those that would be obtained using conventional ML decoding, since the marginal optimum $\hat{\mathbf{e}}
^{\text{SW}} = [\hat{\mathrm{e}}_1
^\text{SW}, \ldots,  \hat{\mathrm{e}}_N^\text{SW}$] need not be equal to the  optimal sequence $\hat{\mathbf{e}}^\text{ML}$. However, for memoryless channels (as are all the channels considered in this dissertation), both criteria agree and decoding based on either the ML or SWML rules will yield the same outcome \cite{class-IT}. 

The SWML decoding rule is generally implemented by running the SPA algorithm over the factor graph representation of the code. Although this procedure does not directly find a solution for (\ref{eq:ML0}) (this only occurs for memoryless channels and non-cyclic factor graphs), the SPA algorithm provides an efficient manner to perform SWML decoding of LDPC codes. Specifically, for the case of memoryless channels, and assuming that the graphical representation of the code does not contain cycles, the SPA obtains the solution for (\ref{eq:ML0}). Refer to chapter \ref{app:spa} in the Appendix for a detailed example on how SWML decoding of LDPC codes is performed by means of the SPA. 

\FloatBarrier

\clearemptydoublepage
\chapter{Degeneracy and its impact on Decoding} \label{chapter4}

\epigraph{\textit{``Not only is the universe stranger than we think, it is stranger than we can think''}}{\textbf{Werner Heisenberg}.}

\noindent\hrulefill

Oftentimes, there is enough similarity between the classical and quantum paradigms to warrant the application of classical methods to the quantum framework. This allows us to interpret, design, and implement quantum error correction strategies based on tried-and-true classical methods and enables us to make headway in our journey towards error-protected qubits. However, there is a particular mechanism by which quantum error correcting codes can exhibit properties that have never before been seen in classical error correction. In the literature, this phenomenon is known as \textit{degeneracy}, and codes that display this particular trait are known as degenerate codes. Quantum codes are said to be degenerate when different error sequences have the same effect on their codewords. This stands in stark contrast to classical codes who will experience different effects (codewords will be corrupted differently) when subjected to different error sequences. 

The manifestation of the degeneracy phenomenon has important implications. The fact that multiple errors can have the same effect on a degenerate code implies that, for this type of code, we will be capable of correcting these errors based on the same recovery operator. Essentially, this means that degeneracy allows quantum codes to ``\textit{pack more information}'' than classical codes \cite{NielsenChuang}, which should ultimately have a positive impact on performance. However, because many QEC strategies make use of classical methods that cannot exploit degeneracy (they were designed for a paradigm in which this phenomenon is absent), this is not always the case. In fact, there are specific instances in which the use of classical error correction stratagems not only neglects the benefits of degeneracy, but also negatively impacts the performance of degenerate quantum codes. Aside from performance related issues, the existence of this strictly-quantum mechanism also disallows the use of classical proof techniques to derive the theoretical bounds on quantum error-correction. Thus, at present, degeneracy presents an interesting conundrum: it should technically improve the performance of quantum codes but it cannot be exploited using classical methods, and it also invalidates the use of classical methodologies to derive the theoretical error correction limits of degenerate quantum codes.  

Beyond these observations and although significant progress has been made recently, especially with regard to improving the performance of quantum codes, the true impact of degeneracy on QEC codes is not yet completely understood. Part of this is due to the difficulty of accurately presenting the idea of degenerate codes, as is reflected by the different and varying notation that can be found in the literature. For this reason, in this chapter we attempt to derive an accurate and easy to follow interpretation of degenerate codes. To do so, we apply group theory to the framework of QEC codes and we use it to describe and study the degeneracy phenomenon in a simple and straightforward manner. Furthermore, this group theoretic approach enables us to discuss the increased nuance of the quantum decoding problem when compared to the classical problem and allows us to show how the presence of degeneracy can be both a blessing and a curse depending on the context.

The chapter begins by presenting the well-known stabilizer formalism in the context of our group theoretic approach. Prior to doing so, we introduce important QEC concepts such as the effective Pauli group or the symplectic representation. Alghough some readers may find this content to be better suited to the previous chapter, given that the interpretation and notation that is employed herein differs from that used in the literature, we believe that including these notions at the beginning of this chapter eases the overall reading experience. Once stabilizer codes have been introduced, we present the quantum decoding problem and discuss its intricate nature. We close the chapter by providing a three-qubit example in which the effects of degeneracy and its impact on the decoding process can be illustrated in a more practical manner. 

\section{Stabilizer codes}

Stabilizer codes, also known as additive quantum codes, are an important class of quantum codes whose construction is analogous to classical linear codes. In fact, the stabilizer theory of quantum error correction allows us to import any binary or quaternary classical code for use as a quantum code, so long as it fulfils a specific condition. This is momentous, as it implies that, if classical codes that satisfy this particular constraint can be found, we may design quantum codes from pre-existing classical codes. Prior to introducing the theory of stabilizer codes, it is worthwhile to extend some of the concepts related to the group structure of Pauli operators. This discussion will come in handy later on, as we will see that stabilizer codes are intricately related to the Pauli group (see section \ref{sec:N-qubits}).

\subsection{The effective Pauli Group} \label{sec:effPauli}

Earlier in section \ref{sec:N-qubits} we defined the $N$-qubit Pauli group as $\mathcal{G}_N = (\widetilde{\Pi}^\otimes, \cdot)$, where the set $\widetilde{\Pi}^\otimes$ is the set of $N$-fold tensor products of single qubit Pauli operators together with the overall factors $\{\pm 1, \pm i\}$ and $\cdot$ is the group operation defined in \eref{eq:product-gn}. We also showed in section \ref{sec:qubit} that quantum states that differ only by an overall phase factor are physically indistinguishable, i.e, that this phase factor has no observable consequence \cite{catalytic}. This means that it makes physical sense to neglect the phase factors $\{\pm1, \pm i\}$ included in $\mathcal{G}_N$. 

Thus, we can define the \textit{effective $N$-fold Pauli Group}, $\overline{\mathcal{G}}_N=(\Pi^{\otimes N}, \star)$, where $\Pi^{\otimes N}$ is the set of $N$-fold tensor products of single qubit Pauli operators (without the overall factors) and $\star$ behaves as in \eref{eq:product-gn} but with the operation between single Pauli operator products defined not as in \eref{eq:pauli-rel} but as
\begin{equation} \label{eq:eff-gn-prod}
     \sigma_a\star \sigma_b=I\delta_{a,b}+\sum_{c=1}^{3}|\epsilon_{abc}|\sigma_c,
\end{equation}
where $\epsilon_{abc}$ is the same as in \eref{eq:pauli-rel}.

Observe that, in contrast to $\mathcal{G}_N$, $\overline{\mathcal{G}}_N$ is an abelian group. This means that the commutation relations exhibited by the elements of $\overline{\mathcal{G}}_N$ under the operation $\cdot$ \eref{eq:product-gn} are lost under the $\star$ operation defined in \eref{eq:eff-gn-prod}.

The relationship between the $N$-qubit Pauli group and its effective counterpart goes beyond a difference in commutation relations. In what follows, we present a set of propositions that further characterize the effective Pauli group and establish an important isomorphism between subsets of the general Pauli group (where the operators in a subset differ only by a phase factor) and the elements of the effective Pauli group. 

\begin{Pro}\label{p4}:
Let $\mathcal{P}\subset \mathcal{G}_N$ be the abelian subgroup $\mathcal{P}=(\{\pm I, \pm i I\}^{\otimes N},\cdot)$. Then, for all $\mathrm{A}\in \mathcal{G}_N$, there is a unique
operator $\mathbf{A}\in \mathcal{G}_N$ such that $\mathbf{A}\in \Pi^{\otimes N}\subset \widetilde{\Pi}^{\otimes N}$ and
\begin{equation}\label{eq:p3}
\mathbf{A}=\mathrm{P}\cdot \mathrm{A}=\mathrm{A}\cdot \mathrm{P}, \;\;\mbox{for some $\mathrm{P}\in \mathcal{P}$,}
\end{equation}
where for the sake of clarity, capital romans are used to denote operators that belong to $\mathcal{G}_N$, and capital boldface is used for those particular operators in $\mathcal{G}_N$ that belong to the subset $\Pi^{\otimes N}$. Note that $\mathrm{A}=\mathbf{A}$ iff $\mathrm{A}\in \Pi^{\otimes N}\subset \widetilde{\Pi}^{\otimes N}$.
\end{Pro}

Proof: We prove it by contradiction. Assume that for $\mathrm{A}\in \mathcal{G}_N$ there are two different operators $\mathbf{A},\mathbf{B}\in \Pi^{\otimes N}$ in $\mathcal{G}_N$ such that
\[\mathbf{A}=\mathrm{P}_1\cdot \mathrm{A}\;\mbox{and $\mathbf{B}=\mathrm{P}_2\cdot \mathrm{A}$ for some $\mathrm{P}_1$ and $\mathrm{P}_2$ in $\mathcal{P}$.}\]
Therefore,
\[\mathbf{A}\cdot \mathbf{B}=\mathrm{P}_1\cdot \mathrm{P}_2 \cdot \mathrm{A}^2  =\pm \mathrm{P}_1\cdot \mathrm{P}_2,   \]
where we have applied that  $\mathrm{A}^2=\mathrm{A} \cdot \mathrm{A}= \mathrm{A} \cdot \mathrm{A}^\dagger = \pm \ I^{\otimes N}$ (The third step, where we have applied $\mathrm{A} = \mathrm{A}^\dagger$, holds because $\mathrm{A}\in\mathcal{G}_N$). Since $\pm\mathrm{P}_1\cdot \mathrm{P}_2 \in \mathcal{P}$, then necessarily $\mathbf{A}=\mathbf{B}$.

$\Box$

Throughout the rest of this chapter and dissertation, an arbitrary operator $\mathrm{D}\in \mathcal{G}_N$ and its unique corresponding operator in $\mathcal{G}_N$ that belongs to $\Pi^{\otimes N}$ will be denoted by the same letter, i.e., $\mathrm{D}$ and $\mathbf{D}$. Furthermore, we employ the symbolic notation $\mathrm{D}\equiv \mathbf{D}$ to represent the fact that for all physical purposes or from a quantum operator perspective, $\mathrm{D}$ and $\mathbf{D}$ will be equal up to a phase. That is,

\begin{equation}\label{eq:car3}
\mathrm{D}\equiv \mathbf{D}\;\Leftrightarrow \; \mathrm{D}=i^k \mathbf{D},\; \;\mbox{for some $k \in \{0,1,2,3\}$}.
\end{equation}

From the definition of the effective $N$-fold Pauli group $\overline{\mathcal{G}}_N=({\Pi}^{\otimes N},\star$), the following proposition easily follows.

\begin{Pro}\label{p5}:
 Given any $\mathrm{A}, \mathrm{B} \in \mathcal{G}_N$ and $\mathbf{A},\mathbf{B} \in \overline{\mathcal{G}}_N$, such that $\mathrm{A}\equiv \mathbf{A}$ and $\mathrm{B}\equiv \mathbf{B}$, then, $\mathrm{A}\cdot \mathrm{B} \equiv \mathbf{A}\star \mathbf{B}$.
\end{Pro}
Proof:
From \eref{eq:car3}, $\mathrm{A}\cdot \mathrm{B}=(i^{k_1}\mathbf{A})\cdot(i^{k_2}\mathbf{B})= i^{k_1+k_2}(\mathbf{A}\cdot \mathbf{B})$ for some $k_1,k_2 \in \{0,1,2,3\}$. Based on the group operations, $\cdot$ \eref{eq:product-gn} and $\star$ \eref{eq:eff-gn-prod}, we know that $\mathbf{A}\cdot \mathbf{B}$ can only differ from  $\mathbf{A}\star \mathbf{B}$ by a phase factor, i.e., $\mathbf{A}\cdot \mathbf{B}= i^{k_3} \mathbf{A}\star \mathbf{B}$, for some $k_3 \in \{0,1,2,3\}$. Then,
 $\mathrm{A}\cdot \mathrm{B}= i^{k_1+k_2}(\mathbf{A}\cdot \mathbf{B})=i^{k_1+k_2+k_3} \mathbf{A}\star \mathbf{B}= i^k\mathbf{A}\star \mathbf{B}$, where $k=(k_1+k_2+k_3)\text{mod}4$.
Therefore, $\mathrm{A}\cdot \mathrm{B} \equiv \mathbf{A}\star \mathbf{B}$.

$\Box$

Let us now define the equivalence relation $\sim_{\mathcal{P}}$ on $\mathcal{G}_N$. For all  $\mathrm{A},\mathrm{B}\in \mathcal{G}_N$, $\mathrm{A}$ will be equivalent to $\mathrm{B}$, i.e., \[\forall\, \mathrm{A},\mathrm{B}\in \mathcal{G}_N,\;\;\mathrm{A}\sim_{\mathcal{P}}\mathrm{B}\]
if and only if $\mathrm{B}=\mathrm{P}\cdot \mathrm{A}=\mathrm{A}\cdot \mathrm{P}$ for some $\mathrm{P}\in\mathcal{P}$, where $\mathcal{P}$ is given in \eref{eq:p3}.

Based on this equivalence relation, the equivalence class or coset of $\mathcal{P}$ in $\mathcal{G}_N$ that contains operator $\mathrm{A} \in \mathcal{G}_N$ is defined as
\[\mathcal{P} \mathrm{A}=\{\mathrm{P}\cdot \mathrm{A}: \forall\,\mathrm{P}\in \mathcal{P}\}.\]
We say that $\mathrm{A}$ is the representative of coset $\mathcal{P} \mathrm{A}$.
If we now look back to Proposition \ref{p4} and expression \eref{eq:p3}, it is straightforward to derive the following proposition.
\begin{Pro}\label{p6}
Given an arbitrary coset $\mathcal{P}\mathrm{A}$ of $\mathcal{P}$ in $\mathcal{G}_N$, then $\mathbf{A}\equiv \mathrm{A}$ is the unique operator in $\mathcal{G}_N$ such that $\mathbf{A}\in \mathcal{P}\mathrm{A}\cap \Pi^{\otimes N}$, i.e., it belongs to the coset and to the subset $\Pi^{\otimes N}$.
\end{Pro}

Based on this last proposition, we can take all the elements of $ \Pi^{\otimes N}$ as the representatives of all the cosets of $\mathcal{P}$ in $\mathcal{G}_N$. The total number of cosets, which we denote as $|\mathcal{P}:\mathcal{G}_N|$, will thus be given by the cardinality of $\Pi^{\otimes N}$, i.e., $2^{2N}$. Proposition \ref{p6} also allows us to partition the underlying set of the $N$-fold Pauli group $\mathcal{G}_N$, $\widetilde{\Pi}^{\otimes N}$, as
\[\widetilde{\Pi}^{\otimes N}=\bigcup_{\mathbf{A}\in \Pi^{\otimes N}} \mathcal{P} \mathbf{A},\] where $\mathbf{A}$ runs through all the elements of $\Pi^{\otimes N}$.

Let us now consider the set of all cosets of $\mathcal{P}$ in $\mathcal{G}_N$, that is, $Q=\{\mathcal{P}\mathbf{A}\subset \widetilde{\Pi}^{\otimes N}: \forall \,\mathbf{A}\in \Pi^{\otimes N}\}$. The quotient group $\mathcal{G}_N/\mathcal{P}$ is defined as $\mathcal{G}_N/\mathcal{P}=(Q,\bullet)$ where the group operation $\bullet$ is defined as
\begin{equation}\label{quo-prod}
(\mathcal{P}\mathrm{A})\bullet(\mathcal{P}\mathrm{B})= \mathcal{P}[\mathrm{A}\cdot \mathrm{B}],\end{equation}
where $\mathcal{P}[\mathrm{A}\cdot \mathrm{B}]$ denotes the coset that contains the operator $\mathrm{A}\cdot \mathrm{B}$. From Proposition \ref{p5}, we can use  $\mathbf{A}\star \mathbf{B} \in \Pi^{\otimes N}$ as the representative of coset  $\mathcal{P}[\mathrm{A}\cdot \mathrm{B}]$. That is to say,
\begin{equation}\label{eq:car4}
(\mathcal{P}\mathrm{A})\bullet(\mathcal{P}\mathrm{B})=\mathcal{P}[\mathbf{A}\star \mathbf{B}].
\end{equation}
Having introduced the quotient group $\mathcal{G}_N/\mathcal{P}$, we can now define an isomorphism between $\mathcal{G}_N/\mathcal{P}$ and $\mathcal{\overline{G}}_N$.

\begin{Def}\label{D3}
Let $\alpha$ be the (one to one) mapping:
\[\alpha: Q\rightarrow \Pi^{\otimes N},\]
defined as $\alpha(\mathcal{P}\cdot \mathrm{A})=\mathbf{A}$. In other words, $\alpha$ maps a coset of $\mathcal{P}$ in $\mathcal{G}_N$ to its unique representative in $\Pi^{\otimes N}$.
\end{Def}

\begin{Pro}
The mapping $\alpha$ is an isomorphism between $\mathcal{G}_N/\mathcal{P}$ and $\overline{\mathcal{G}}_N$.
\end{Pro}

Proof:
First, by Proposition \ref{p6}, $\alpha$ is a bijective (one to one) mapping. Second, we must show that for all $\mathrm{A,B}\in \mathcal{G}_N$
\[\alpha(\mathcal{P} \mathrm{A}\bullet\mathcal{P}\mathrm{B})=\alpha(\mathcal{P} \mathrm{A})\star \alpha(\mathcal{P} \mathrm{B}).\]
From \eref{eq:car4}, $(\mathcal{P}\mathrm{A})\bullet(\mathcal{P}\mathrm{B})=\mathcal{P}[\mathbf{A}\star \mathbf{B}]$. Thus,
\[\alpha(\mathcal{P}\mathrm{A}\bullet\mathcal{P}\mathrm{B})=\alpha(\mathcal{P}[\mathbf{A}\star \mathbf{B}])=\mathbf{A}\star \mathbf{B}=\alpha(\mathcal{P}\mathrm{A})\star \alpha(\mathcal{P} \mathrm{B}),\]
as we wanted to prove.

$\Box$

The isomorphism $\alpha$ establishes that considering the effective Pauli group $\overline{\mathcal{G}}_N=(\Pi^{\otimes N},\star)$ is analogous to working with equivalence classes of the Pauli group. This distinction is sometimes neglected in the literature, which may result in confusion when the notation is abused by referencing the Pauli group $\mathcal{G}_N$ instead of the effective Pauli group $\overline{\mathcal{G}}_N$. Because the global phase of a quantum state cannot be measured, quantum error correction will exclusively deal with elements of the effective Pauli group, meaning that quantum codes are capable of correcting qubits only up to a phase. Once again, given that the global phase has no observable consequence, being unable to consider the global phase has no impact on the error correcting capabilities of quantum codes. Instead, this result serves to simplify the framework of QEC codes as, strictly for the purposes of error correction, we only need to consider elements of the effective Pauli group. 

Unfortunately, because the group operation $\star$ is not able to convey the commutation properties that exist among Pauli operators under the $\cdot$ product, the effective Pauli group is not sufficient on its own for appropriate QEC design. This will become clearer when we present the concept of quantum syndromes later on in this chapter. For now, assume that it is not possible to perform error correction based only on $\overline{\mathcal{G}}_N$ without recovering the commutation properties that exist in $\mathcal{G}_N$.

\subsection{The Symplectic Representation}

This issue of lost commutation relations can be overcome by adopting the symplectic representation \cite{NielsenChuang,EAQECC} of the Pauli operators in $\overline{\mathcal{G}}_N$. More specifically, by means of the symplectic mapping $\beta:\Pi^{\otimes N} \rightarrow \mathbb{F}_2^{2N}$, which is an isomorphism between the group $\overline{\mathcal{G}}_N=(\Pi^{\otimes N},\star$) and the group ($\mathbb{F}_2^{2N}, \oplus$) of $2N$ binary-tuples under the mod 2 sum operation, we are able to recover the commutation properties that are lost when working with the operators of $\overline{\mathcal{G}}_N$. For clarity, throughout the remainder of this chapter, lower case boldface romans without a subscript will be used to denote $2N$ binary-tuples that belong to $\mathbb{F}_2^{2N}$, and lower case boldface romans with a subscript will be used to denote $N$ binary-tuples that belong to $\mathbb{F}_2^{N}$.


The symplectic mapping $\beta$ is defined as

\[\beta(\mathbf{A})=\mathbf{a}=(\mathbf{a}_x|\mathbf{a}_z),\;\mathbf{a}_x,\mathbf{a}_z\in \mathbb{F}_2^N,\]
where the values of the entries of $\mathbf{a}_x$ and $\mathbf{a}_z$ at position $i=1,\ldots, N$, are directly dependent on the single qubit Pauli operator, $[\mathbf{A}]_i$, located at the $i$-th position in the tensor product that makes up $\mathbf{A}$. More specifically,

\begin{equation}\label{eq:Paulimapping}
\begin{split}
\beta([\mathbf{A}]_i=I) = ([\mathbf{a}_x]_i=0\;|\;[\mathbf{a}_z]_i=0) \\
\beta([\mathbf{A}]_i=X) = ([\mathbf{a}_x]_i=1\;|\;[\mathbf{a}_z]_i=0) \\
\beta([\mathbf{A}]_i=Z) = ([\mathbf{a}_x]_i=0\;|\;[\mathbf{a}_z]_i=1) \\
\beta([\mathbf{A}]_i=Y) = ([\mathbf{a}_x]_i=1\;|\;[\mathbf{a}_z]_i=1).
\end{split}
\end{equation}

\begin{Pro}\label{car1}
The symplectic map $\beta$ is an isomorphism $\overline{\mathcal{G}}_N\simeq (\mathbb{F}_2^{2N}, \oplus)$.
\end{Pro}
Proof: By construction $\beta$ is bijective and for all $\mathbf{A},\mathbf{B}\in \overline{\mathcal{G}}_N$, it can be easily checked that $\beta(\mathbf{A} \star \mathbf{B})=\beta(\mathbf{A})\oplus \beta(\mathbf{B})$. 

$\Box$

Note that the commutation properties of the Pauli operators with regard to the $\cdot$ product are not recovered by just defining the isomorphism $\beta$ (after all, ($\mathbb{F}_2^{2N}, \oplus)$ is an abelian group). For this purpose, we define the symplectic scalar product, $\mathbf{a}\odot \mathbf{b}\in \mathbb{F}_2 \ \forall \ \mathbf{a},\mathbf{b}\in  \mathbb{F}_2^{2N}$ as 
\begin{equation}\label{eq:car34}
\mathbf{a}\odot \mathbf{b}\triangleq (\mathbf{a}_x\circledast \mathbf{b}_z)\oplus(\mathbf{a}_z\circledast \mathbf{b}_x).
\end{equation}
where $\circledast$ is the standard mod 2 inner product defined on  $(\mathbb{F}_2^{N}, \oplus)$ considered as a vector space over the field $\mathbb{F}_2$.

\begin{figure}[!h]
	\centering
  \includegraphics[width=\linewidth,  height=3in]{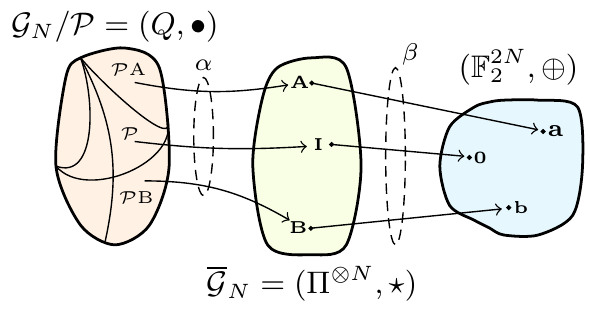}
	\caption{Isomorphism between $\mathcal{G}_N/\mathcal{P}$ and $\mathcal{\overline{G}}_N$ defined by the mapping $\alpha: (Q, \bullet)\longrightarrow (\Pi^{\otimes N},\star)$ and isomorphism between $\mathcal{\overline{G}}_N$ and $\mathbb{F}_2^{2N}$ defined by the symplectic mapping $\beta: (\Pi^{\otimes N}, \star) \longrightarrow (\mathbb{F}_2^{2N}, \oplus)$. The operator $\mathbf{I}$ represents the $N$ qubit identity operator, i.e., $\mathbf{I} = I^{\otimes N}$.}
	\label{fig:mapping-simple}
\end{figure}

\begin{Pro}\label{car2}
Any two operators $\mathbf{A},\mathbf{B}$ in $\overline{\mathcal{G}}_N$ will either commute or anticommute (with respect to the group operation $\cdot$ in ${\mathcal{G}}_N$) if and only if
the symplectic scalar product between $\beta(\mathbf{A})=\mathbf{a}$ and $\beta(\mathbf{B})=\mathbf{b}$, i.e., $\mathbf{a}\odot \mathbf{b}$, takes the value 0 or 1, respectively \normalfont{\cite{josurev, catalytic}}.
\end{Pro}
Fig. \ref{fig:mapping-simple} summarizes the isomorphisms $\alpha$ and $\beta$.

\section{Stabilizer-based error correction} \label{sec:stabs}

The fact that quantum states cannot be directly measured has been mentioned numerous times throughout this dissertation. However, we also know that it is impossible to perform error correction without any knowledge regarding how decoherence is acting on our quantum information. Thus, in order for QEC to function, a method capable of extracting information about decoherence-related errors without actually measuring the quantum states themselves is necessary. Fortunately, this issue can be circumvented with the stabilizer formalism, which allows us to gleam information about errors without having to look at the actual quantum information by measuring the so-called \textit{quantum syndrome} \cite{introQIC, NielsenChuang, josurev, catalytic}. Prior to diving into the specifics of stabilizer codes, it is worthwhile to provide a more general view of the stabilizer formalism itself, as its applications extend far beyond the niche of QEC. This is best achieved using the following example, which was originally introduced in \cite{NielsenChuang}.

Consider the EPR state $\ket{\Phi^{+}} = \frac{1}{\sqrt{2}}(\ket{00} + \ket{11}) $ given in \eref{eq:EPRpairs}. Now assume that we apply the operations $X_1X_2$ and $Z_1Z_2$ to $\ket{\Phi^{+}}$, where the notation $X_iZ_j$ describes the action of an $X$ Pauli gate on the $i$-th qubit and a $Z$ Pauli gate on the $j$-th qubit. Notice how these operations leave the state unchanged, i.e, $X_1X_2\ket{\Phi^{+}} = \ket{\Phi^{+}}$ and $Z_1Z_2\ket{\Phi^{+}} = \ket{\Phi^{+}}$. In written form, we say that the state $\ket{\Phi^{+}}$ is stabilized by the operators $X_1X_2$ and $Z_1Z_2$. Although it may not seem advantageous from this small example, being able to represent quantum states by means of the operators that stabilize them is extremely practical. In fact, it becomes even more useful to describe quantum codes, which can be very difficult to represent in state vector notation and are much more compactly described using stabilizers. 

After this short introduction to the stabilizer formalism, we move towards to the description of stabilizer coding principles from the perspective of group theory. Those familiar with these concepts will realize that this interpretation deviates from conventional representations and that the employed notation differs slightly from the one usually found in the literature \cite{introQIC, NielsenChuang, group1, group2, QEClidar}. This is meant to facilitate the comprehension of certain topics in the field of quantum stabilizer codes that are sometimes misunderstood.

\subsection{The Stabilizer Group}

Let $\{\mathbf{s}_1,\mathbf{s}_2, \ldots, \mathbf{s}_{N-k}\}$ be a subset of independent vectors in the vector space $\mathbb{F}_2^{2N}$ over $\mathbb{F}_2$ that satisfies $\mathbf{s}_i\odot \mathbf{s}_j=0$, $\forall\,i\neq j$. Let us now define the subspace $\mathcal{R}$ as the span of the set $\{\mathbf{s}_1,\mathbf{s}_2, \ldots, \mathbf{s}_{N-k}\}$. Thus, $\mathcal{R}$ has cardinality $|\mathcal{R}|=2^{N-k}$. Based on these definitions, we can now use the isomorphism $\beta$ (see proposition \ref{car1}) to define the stabilizer group.

\begin{Def} \label{def:effstabs}
The stabilizer set $\overline{\mathcal{S}}\subset \Pi^{\otimes N}$ is defined as $\mathcal{\overline{S}}=\beta^{-1}(\mathcal{R})$, where $\beta^{-1}$ is the inverse of the symplectic isomorphism in Proposition \ref{car1}.
\end{Def}
\begin{Pro}
The set $\overline{\mathcal{S}}$ together with the $\star $ product, i.e., ($\overline{\mathcal{S}}, \star)$, is an abelian subgroup of $\overline{\mathcal{G}}_N$.
\end{Pro}

Proof: That ($\overline{\mathcal{S}}, \star)$ is a subgroup of $\overline{\mathcal{G}}_N=(\Pi^{\otimes N},\star)$ is straightforward from the fact that $\mathcal{R}$ is a subgroup of $(\mathbb{F}_2^{2N}, \oplus)$ and $\beta$ is an isomorphism $\overline{\mathcal{G}}_N \backsimeq (\mathbb{F}_2^{2N}, \oplus)$. That $\overline{\mathcal{S}}$ is abelian follows from the fact that $\mathcal{\overline{G}}_N$ is abelian.

$\Box$

In what follows, for the purpose of simplicity, we will not distinguish between the group ($\overline{\mathcal{S}}, \star)$ and its underlying set $\overline{\mathcal{S}}$ (the group operation that applies is clear). Similarly, $\mathcal{R}$ will also represent $(\mathcal{R},\oplus)$.

Note that the stabilizer group $\overline{\mathcal{S} }\subset \overline{\mathcal{G}}_N$ can be generated by the set of stabilizer generators $\{\mathbf{S}_v=\beta^{-1}( \mathbf{s}_v)\}_{v=1}^{N-k}$ in $\overline{\mathcal{G}}_N$. That is to say, for any $\mathbf{S}\in \overline{\mathcal{S}}$, there are $b_v\in \{0,1\}$, $v=1,\ldots,N-k$, such that
\[\mathbf{S}=\mathbf{S}^{b_1}_1\star \mathbf{S}^{b_2}_2\star\ldots\star \mathbf{S}^{b_{N-k}}_{N-k}.\]
 Observe how, from Proposition \ref{car2}, all the $\mathbf{S}_v$ operators will commute with respect to the group operation $\cdot$ in ${\mathcal{G}}_N$, since the symplectic products of their binary counterparts in $\mathbb{F}_2^{2N}$ are always zero.
 
\subsubsection{Partition of $\mathbb{F}_{2}^{2N}$ into cosets}
 Next, we derive a partition of the group $(\mathbb{F}_{2}^{2N},\oplus)$ into cosets. We will later relate this partition to the concepts of stabilizer codes.

\begin{Def}\label{D2}
Let $\Gamma_\mathcal{R}\subset \mathbb{F}_2^{2N}$ be the set
\begin{gather*}
\Gamma_\mathcal{R}=\{\mathbf{a}\in \mathbb{F}_2^{2N}:\mathbf{ a}\odot \mathbf{s}=0,\;\forall \mathbf{s}\in\mathcal{ R}\}  \\ =\{\mathbf{a}\in \mathbb{F}_2^{2N}:\mathbf{ a}\odot \mathbf{s}_v=0, \;v=1\ldots N-k\}. \end{gather*}

 \end{Def}

  \begin{Pro}\label{PD66}
 The set $\Gamma_\mathcal{R}$ together with the modulo-2 sum, i.e., ($\Gamma_\mathcal{R},\oplus $), is an abelian subgroup of $(\mathbb{F}_2^{2N}, \oplus)$.

 \end{Pro}

 Proof: The identity element  of $(\mathbb{F}_2^{2N}, \oplus)$, which is $\mathbf{0}$, is in $\Gamma_\mathcal{R}$ since $\mathbf{0\ }\odot \mathbf{\ s}_v=0$ for all $ v\in \{1\ldots N-k\}$.
 On the other hand, one must show that $\mathbf{a } \oplus\mathbf{ b}\in \Gamma_\mathcal{R}$, that is, we must show that ($\mathbf{a \ }\oplus\mathbf{\ b})  \ \odot \ \mathbf{s}_v=0$, for all $ v\in \{1\ldots N-k\}$. Note that if $\mathbf{a},\mathbf{b}\in \Gamma_\mathcal{R}$, then $\mathbf{a}\odot \mathbf{s}_v=0$ and $\mathbf{b \ }\odot \mathbf{\ s}_v=0$ for all $v$, so that,
\begin{gather*}
(\mathbf{a \ }\oplus\mathbf{\ b})  \ \odot \ \mathbf{s}_v=\mathbf{a \ }\odot \mathbf{\ s}_v\oplus\mathbf{\ b\ }\odot \mathbf{\ s }_v \\ =0 \ \oplus \ 0=0 \end{gather*}
Therefore, we conclude that  ($\Gamma_\mathcal{R},\oplus $) is an abelian group.

$\Box$

Once more, in what follows we will not distinguish between the group ($\Gamma_\mathcal{R},\oplus $) and its underlying set $\Gamma_\mathcal{R}$.

\subsubsection{Partition of $\mathbb{F}_2^{2N}$ into cosets of $\Gamma_\mathcal{R}$}
Based on the newly defined subgroup $\Gamma_\mathcal{R}$, we can partition the group $(\mathbb{F}_2^{2N},\oplus)$ into cosets of this set in a similar manner to what was done previously for $\mathcal{G}_N$ with respect to its subgroup $\mathcal{P}$ in section \ref{sec:effPauli}.

\begin{Def} \label{def4}
Define the equivalence relation $\sim_{\Gamma}$ in $\mathbb{F}_2^{2N}$ as
 \[\mathbf{a}\sim_{\Gamma}\mathbf{ b} \; \mbox{iff $\mathbf{b}=\mathbf{c }\oplus \mathbf{a}$ for some $\mathbf{c} \in \Gamma_\mathcal{R}$}.\]
\end{Def}

Then, the equivalence class or coset, $\gamma \mathbf{a}$, of $\Gamma_\mathcal{R}$ in $\mathbb{F}_2^{2N}$ containing $\mathbf{a}$, is the set
\[\gamma \mathbf{a}\triangleq \Gamma_\mathcal{R}\oplus \mathbf{a}=\{\mathbf{c}\oplus \mathbf{a}:\,\forall\,\mathbf{c} \in  \Gamma_\mathcal{R}\}.\]
This means that two elements $\mathbf{a}, \mathbf{b}\in \mathbb{F}_2^{2N}$ will belong to the same coset if and only if $\mathbf{a}\oplus \mathbf{b}\in \Gamma_\mathcal{R}$.

\begin{Pro}\label{pedro1}
The number of cosets of $\Gamma_\mathcal{R}$ in $\mathbb{F}_2^{2N}$ is $|\mathbb{F}_2^{2N}:\Gamma_\mathcal{R}|=2^{N-k}$.
\end{Pro}
Proof: Define the syndrome\footnote{The concept of syndromes in quantum error correction is discussed later on in this chapter. Rigorous explanations on this topic can also be found in \cite{NielsenChuang, josurev}.} vector $\mathbf{w}\in \{0,1\}^{N-k}$ (we abuse notation slightly by using a lower case boldface roman without a subscript to denote a vector in $\mathbb{F}_2^{N-k}$) associated to each $\mathbf{a}\in \mathbb{F}_2^{2N}$ as

\begin{equation}\label{eq:car33}
\mathrm{w}_v=\mathbf{a}\odot \mathbf{s}_v, \; v=1,\ldots, N-k.
\end{equation}

Observe from Definition \ref{D2} that a vector $\mathbf{a}\in \mathbb{F}_2^{2N}$ has syndrome zero, $\mathbf{w}=\mathbf{0}$, where $\mathbf{0} = [0_1, \ldots, 0_{N-k}]$, if and only if $\mathbf{a}\in\gamma \mathbf{0}=\Gamma_{\mathcal{R}}$.
Furthermore, since two vectors $\mathbf{a}$ and $\mathbf{b}$ will belong to the same coset iff $\mathbf{a}\oplus \mathbf{b}\in \Gamma_\mathcal{R}$, all the vectors in a coset have the same syndrome, and two different cosets will have two different syndromes. Consequently, since there are $2^{N-k}$ different syndromes, there will be $2^{N-k}$ different cosets. 

$\Box$

Notice that we have partitioned $\mathbb{F}_2^{2N}$ into cosets of $\Gamma_\mathcal{R}$. We denote the corresponding coset representatives as
$\{\mathbf{t}_1,\ldots, \mathbf{t}_{2^{N-k}}\}\in \mathbb{F}_2^{2N}$. Each $\mathbf{t}_i$ is associated to a different syndrome vector $\mathbf{w}_i$, so that for all $i=1,\ldots,2^{N-k}$,
\begin{equation}\label{eq:car35}
\mathbf{t}_i\odot (\mathbf{s}_1,\ldots,\mathbf{s}_{N-k})=\mathbf{w}_i.
\end{equation}

Thus, the group $(\mathbb{F}_2^{2N},\oplus )$ can be partitioned as
\begin{equation}\label{eq:p100}
\mathbb{F}_2^{2N}=\bigcup_{i=1}^{2^{N-k}}\underbrace{\mathbf{t}_i\oplus\Gamma_\mathcal{R}}_{\gamma \mathbf{t}_i}=\Gamma_\mathcal{R}\bigcup \left[ \bigcup_{i=2}^{2^{N-k}}\mathbf{t}_i\oplus \Gamma_\mathcal{R}\right],
\end{equation}
where, without loss of generality, the representative of coset $\Gamma_\mathcal{R}\oplus\mathbf{t}_1$ has been chosen to be $\mathbf{0}$. We denote this partition of $\mathbb{F}_2^{2N}$ into cosets of $\Gamma_\mathcal{R}$ by $\mathbb{F}_2^{2N}/\Gamma_\mathcal{R}$.

Since the cardinality of $\mathbb{F}_2^{2N}$ is $2^{2N}$ and $|\mathbb{F}_2^{2N}|=|\mathbb{F}_2^{2N}:\Gamma_\mathcal{R}||\Gamma_\mathcal{R}|$, we conclude that $|\Gamma_\mathcal{R}|=2^{N+k}$. Recall that $|\mathbb{F}_2^{2N}:\Gamma_\mathcal{R}| = 2^{N-k}$ (Proposition \ref{pedro1}).

\subsubsection{Partition of $\Gamma_\mathcal{R}$ into cosets of $\mathcal{R}$}

The group $\Gamma_\mathcal{R}$ can itself be partitioned into cosets of its subgroup $\mathcal{R}\subset \Gamma_\mathcal{R}$ by defining the following equivalence relation.

\begin{Def}

Define the equivalence relation $\sim_{\mathcal{R}}$ for all the elements $ \mathbf{a}, \mathbf{b}\in\Gamma_\mathcal{R}$ as
\[ \mathbf{a}\sim_{\mathcal{R}} \mathbf{b}\;\; \mbox{iff $\mathbf{b}=\mathbf{a}\oplus \mathbf{s}$, for some $\mathbf{s} \in \mathcal{R}$}.\]
\end{Def}
This coset of $\mathcal{R}$ in $\Gamma_\mathcal{R}$ containing $\mathbf{a}$ induced by this relation, $\mathcal{R}\mathbf{a}$, is the subset of $\Gamma_\mathcal{R}$

\[\mathcal{R}\mathbf{a}\triangleq \mathbf{a}\oplus \mathcal{R}=\{\mathbf{a}\oplus\mathbf{s}_v: \forall\,\mathbf{s}_v\in \mathcal{R}\}.\]
As with the previous equivalence relation given in Definition \ref{def4}, two vectors $\mathbf{a}, \mathbf{b}\in \Gamma_\mathcal{R}$ belong to the same coset if and only if $\mathbf{a}\oplus \mathbf{b}\in \mathcal{R}$.
\begin{Pro}
The number of cosets of $\mathcal{R}$ in $\Gamma_\mathcal{R}$ is given by $|\Gamma_\mathcal{R}:\mathcal{R}|=2^{2k}$.
\end{Pro}

Proof:
Knowing that $|\Gamma_\mathcal{R}|=|\Gamma_\mathcal{R}:\mathcal{R}||\mathcal{R}|$, with $|\mathcal{R}|=2^{N-k}$, and $|\Gamma_\mathcal{R}|=2^{N+k}$, we can conclude that the number of cosets of $\mathcal{R}$ in $\Gamma_\mathcal{R}$ is $|\Gamma_\mathcal{R}:\mathcal{R}|=2^{2k}$.

$\Box$

If we denote the set of representatives as $\{\mathbf{l}_j\in \Gamma_\mathcal{R} \}_{j=1}^{2^{2k}}$, one obtains the following $\Gamma_\mathcal{R}/\mathcal{R}$ partition:

\begin{equation}\label{eq:p77}
\Gamma_\mathcal{R}=\bigcup_{j=1}^{2^{2k}}\underbrace{\mathbf{l}_j\oplus \mathcal{R}}_{\mathcal{R}\mathbf{l}_j}= \mathcal{R}\bigcup \left[ \bigcup_{j=2}^{2^{2k}}\mathbf{l}_j\oplus \mathcal{R}\right],
\end{equation}

\noindent where, once again, we have chosen  $\mathbf{l}_1=\mathbf{0}$ for convenience. Note that all these coset representatives satisfy

\begin{equation}\label{eq:car36}
\mathbf{l}_j\odot (\mathbf{s}_1,\ldots,\mathbf{s}_{N-k})={0},
\end{equation}
where $j = 1,\ldots,2^{2k}$.


\subsubsection{Partition of $\mathbb{F}_2^{2N}$ into cosets of $\mathcal{R}$}

By combining the partitions $\mathbb{F}_2^{2N}/\Gamma_\mathcal{R}$ in \eref{eq:p100} and $\Gamma_\mathcal{R}/\mathcal{R}$ in \eref{eq:p77}, the partition $\mathbb{F}_2^{2N}/\Gamma_\mathcal{R}/\mathcal{R}$ can be obtained. That is, $\mathbb{F}_2^{2N}$ can be partitioned as the union of the cosets of $\mathcal{R}$ in $\mathbb{F}_2^{2N}$ with coset representatives $\{\mathbf{t}_i\oplus \mathbf{l}_j\}$:

\begin{align}  \label{p99}
\begin{split}
\mathbb{F}_2^{2N} &=\bigcup_{i=1}^{2^{N-k}}\bigcup_{j=1}^{2^{2k}}(\mathbf{t}_i\oplus \mathbf{l}_j)\oplus \mathcal{R}.
\end{split}
\end{align}


\subsection{Partition of $\mathcal{\overline{G}}_N$ into cosets}

At this point, we are now in a position to use the partition $\mathbb{F}_2^{2N}/\Gamma_\mathcal{R}/\mathcal{R}$ and the inverse of the isomorphism $\beta$ to establish a partition over $\mathcal{\overline{G}}_N$. We begin by defining the effective centralizer group of a stabilizer $\mathcal{\overline{S}}$ in $\mathcal{\overline{G}}_N$.

\begin{Def} \label{centra}
The effective centralizer ${\overline{\mathcal{Z}}(\overline{\mathcal{S}})}\subset \Pi^{\otimes N}$ of stabilizer $\mathcal{\overline{S}}$ is the set obtained by applying the inverse of the isomorphism $\beta$ to the set $\Gamma_\mathcal{R}$ (refer to Definition \ref{D2}). That is to say, ${\overline{\mathcal{Z}}(\overline{\mathcal{S}})}=\beta^{-1}(\Gamma_\mathcal{R})$, where
\[\beta^{-1}(\Gamma_\mathcal{R})=\{\beta^{-1}(\mathbf{a})\in \overline{\mathcal{G}}_N:\forall\,\mathbf{a}\in \mathbb{F}_2^{2N},\,\mathbf{a}\odot \mathbf{s}_v=0, \]
\[ v = 1, \ldots, N-k \}. \]
\end{Def}

Therefore, ${\overline{\mathcal{Z}}(\overline{\mathcal{S}})}$ is the set of all operators in $\overline{\mathcal{G}}_N$ that commute with all the stabilizer generators $\mathbf{S}_v\in \overline{\mathcal{S}}\subset \overline{\mathcal{G}}_N$ with regard to the group operation $\cdot$ defined in $\mathcal{G}_N$. Thus,
\[{\overline{\mathcal{Z}}(\overline{\mathcal{S}})}=\{\mathrm{A}\in \mathcal{G}_N: \mathrm{A}\cdot \mathrm{S}_v=\mathrm{S}_v\cdot \mathrm{A},\;v=1,\ldots, N-k,\;v=1,\ldots, N-k\} \]
\[ =\{\mathbf{A}\in \overline{\mathcal{G}}_N: \underbrace{\beta(\mathbf{A})}_{\mathbf{a}}\odot \underbrace{\beta(\mathbf{S}_v)}_{\mathbf{s}_v} = 0,\;v=1,\ldots, N-k\}.\]

\begin{Pro}
The effective centralizer ${\overline{\mathcal{Z}}(\overline{\mathcal{S}})}$ is a subgroup of $\overline{\mathcal{G}}_N=(\Pi^{\otimes N},\star)$.
\end{Pro}

Proof: This is straightforward from the fact that $\beta$ is an isomorphism $\mathcal{\overline{G}}_N \backsimeq (\mathbb{F}_2^{2N},\oplus)$ and $\Gamma_\mathcal{R}$ is a subgroup of $\mathbb{F}_2^{2N}$ (Proposition \ref{PD66}).

$\Box$

\textbf{Remark 1}:
In the literature of stabilizer codes, the stabilizer is defined as a subgroup of $(\mathcal{G}_N,\cdot)$ instead of $(\overline{\mathcal{G}}_N,\star$), and it is denoted by $\mathcal{S}$. The relationship between $\mathcal{S}$ and $\mathcal{\overline{S}}$ as defined in Definition \ref{def:effstabs} is
\[\mathcal{S}\triangleq\{ \mathrm{A}\in \mathcal{G}_N: \mathrm{A}\equiv \mathbf{A},\,\forall \mathbf{A}\in \overline{\mathcal{S}}\}.\]

Based on $\mathcal{S}$, the corresponding centralizer in $\mathcal{G}_N$ is
\[\mathcal{Z}(\mathcal{S})\triangleq\{\mathrm{M}\in \mathcal{G}_N: \mathrm{M}\cdot \mathrm{S}=\mathrm{S}\cdot \mathrm{M},\,\forall\,\mathrm{S}\in \mathcal{S}\}.\]\

\textbf{Remark 2}: In the context of stabilizer codes, the concept of the normalizer set of $\mathcal{S}$, $\mathcal{N(S)}$, is often referenced. The normalizer of a stabilizer is defined as
\[\mathcal{N}(\mathcal{S})=\{\mathrm{M}\in \mathcal{G}_N: \mathrm{M}\cdot \mathrm{S}\cdot \mathrm{M}^\dag\in \mathcal{S},\,\forall \,\mathrm{S}\in \mathcal{S}\}.\]

As it turns out, due to the product properties of Pauli operators, the normalizer and centralizer of a stabilizer group are actually the same set, $\mathcal{N}(\mathcal{S})=\mathcal{Z}(\mathcal{S})$ \cite{NielsenChuang, QEClidar}. This can be seen by multiplying the conditional statement for the centralizer on the right by $\mathrm{M}^\dagger \in \mathcal{G}_N$ as 

\[
   \mathrm{M} \cdot \mathrm{S} \cdot \mathrm{M}^\dagger = \mathrm{S} \cdot \mathrm{M} \cdot \mathrm{M}^\dagger \rightarrow
   \mathrm{M} \cdot \mathrm{S} \cdot \mathrm{M}^\dagger = \mathrm{S} \rightarrow \mathrm{M}\cdot \mathrm{S}\cdot \mathrm{M}^\dag\in \mathcal{S}.
\]

where $\mathrm{M}^\dagger \cdot \mathrm{M} = I^{\otimes N}$ holds because $\mathrm{M} \in \mathcal{G}_N$. Consequently, the effective normalizer over $\mathcal{\overline{G}}_N$ will satisfy $\overline{\mathcal{N}}(\mathcal{\overline{S}})=\overline{\mathcal{Z}}(\overline{\mathcal{S}})$.

Having shown that it makes good physical sense to neglect the global phase (see section \ref{sec:qubit}), and given that the commutation properties of Pauli operators with regard to the $\cdot$ product are recovered based on the isomorphism $\beta$, in order to preserve notation integrity, moving forward we will work with the concepts defined over the group $\mathcal{\overline{G}}_N=(\Pi^{\otimes N}, \star)$.

\subsubsection{Partition $\overline{\mathcal{G}}_N/ \overline{\mathcal{Z}}(\overline{\mathcal{S}})$ }

By applying the inverse isomorphism $\beta^{-1}$ to the partition $\mathbb{F}_2^{2N}/\Gamma_\mathcal{R}$ in \eref{eq:p100}, we obtain

\begin{align}  \label{eq:pauli-centra}
\begin{split}
\overline{\mathcal{G}}_N &=\bigcup_{i=1}^{2^{N-k}}\mathbf{T}_i\star \overline{\mathcal{Z}}(\overline{\mathcal{S}})\\&=\overline{\mathcal{Z}}(\overline{\mathcal{S}})\bigcup \left[\bigcup_{i=2}^{2^{N-k}}
\mathbf{T}_i\star \overline{\mathcal{Z}}(\overline{\mathcal{S}})\right],
\end{split}
\end{align}

\noindent where $\mathbf{T}_i=\beta^{-1}(\mathbf{t}_i)$. Note from expression \eref{eq:car35} that, except for $\mathbf{T}_1 = \beta(\mathbf{t}_1) = \beta(\mathbf{0})=I^{\otimes N}$, which is associated to the zero syndrome vector, each $\mathbf{T}_i$ is related to a unique non-zero syndrome vector $\mathbf{w}_i$. That is, $\forall i = 2,\ldots,2^{N-k}$,
\begin{equation}\label{eq:car37}
\beta(\mathbf{T}_i)\odot (\beta(\mathbf{S}_1),\ldots,\beta(\mathbf{S}_{N-k})) = \mathbf{w}_i\neq \mathbf{0}.
\end{equation}

\subsubsection{Partition $\overline{\mathcal{Z}}(\overline{\mathcal{S}})/\overline{\mathcal{S}}$ } \label{sec:logicalgroup}
Applying this same inverse isomorphism $\beta^{-1}$ to the partition \eref{eq:p77} provides us with the corresponding partition of the effective centralizer,

\begin{equation} \label{eq:centra-cosets}
\overline{\mathcal{Z}}(\overline{\mathcal{S}})=\bigcup_{j=1}^{2^{2k}} \mathbf{L}_j\star \overline{\mathcal{S}}=\overline{\mathcal{S}}\bigcup \left[
\bigcup_{j=2}^{2^{2k}}\mathbf{L}_j\star \overline{\mathcal{S}}\right],
\end{equation}
where $\mathbf{
L}_j=\beta^{-1}(\mathbf{l}_j)$ and  $\overline{\mathcal{S}}=\beta^{-1}(\mathcal{R})$.

\begin{figure}[!h]
	\centering
  \includegraphics[width=\linewidth,  height=2.75in]{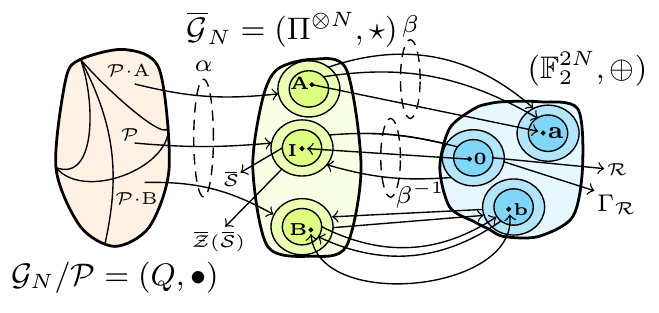}
	\caption{Partitions of the effective Pauli group $\overline{\mathcal{G}}_N=(\Pi^{\otimes N},\star)$ and the group of length $2N$ binary tuples $(\mathbb{F}_2^{2N},\oplus)$ and how they are related by the isomorphism $\beta$ and its inverse $\beta^{-1}$. Operator $\mathbf{I}$ represents the $N$ qubit identity operator, i.e., $\mathbf{I} = I^{\otimes N}$.}
	\label{fig:isos_final}
\end{figure}

Finally, by combining these last two partitions, we obtain the partition $\overline{\mathcal{G}}_N/ \overline{\mathcal{Z}}(\overline{\mathcal{S}})/\overline{\mathcal{S}} $. In other words, we derive the partition of the entire effective Pauli group into cosets of the stabilizer $\mathcal{\overline{S}}$,

\begin{align} \label{eq:pauli-stab-3}
\begin{split}
\overline{\mathcal{G}}_N &=\bigcup_{i=1}^{2^{N-k}}\bigcup_{j=1}^{2^{2k}}(\mathbf{T}_i\star \mathbf{L}_j)\star \mathcal{\overline{S}} .
\end{split}
\end{align}

Figure \ref{fig:isos_final} serves to graphically portray the relationships between the concepts that have been discussed up to this point. 

From our earlier definition of the representatives $\{\mathbf{l}_j\}_{j=1}^{2^{2k}}$, we know that, except for $\mathbf{l}_1 = \mathbf{0}$ which is the representative of the set $\mathcal{R}$ itself, they belong to the set $\Gamma_\mathcal{R}$ but not to the set $\mathcal{R}$. This means that these same representatives over $\mathcal{\overline{G}}_N$, $\{\mathbf{L}_j\}_{j=2}^{2^{2k}}$, belong to the set $\overline{\mathcal{Z}}(\overline{\mathcal{S}})-\overline{\mathcal{S}}$ and satisfy
\begin{equation}\label{eq:car38}
\beta(\mathbf{L}_j)\odot (\beta(\mathbf{S}_1),\ldots,\beta(\mathbf{S}_{N-k})) = \mathbf{0}.
\end{equation}
In other words, the representatives $\{\mathbf{L}_j\}_{j=2}^{2^{2k}}$ commute with all the stabilizer elements but are not within the stabilizer.

\subsection{Construction of stabilizer codes} 

Recall that we defined the stabilizer group, $\overline{\mathcal{S}}$, as an abelian subgroup of $\mathcal{\overline{G}}_N$ that is obtained by applying the inverse isomorphism $\beta^{-1}$ to the subspace $\mathcal{R} \in \mathbb{F}_{2}^{2N}$. The stabilizer $\mathcal{\overline{S}}$ has $2^{N-k}$ distinct elements, and is generated by $N-k$ independent generators $\mathbf{S}_i$, where $i=1 \dots N-k$. A stabilizer code can be described by a minimal set of these $N-k$ independent generators (the rest of the generators are linear combinations of the elements of the minimal set), as this provides a compact representation of the code. Based on this, we are now in the position to define a stabilizer code. 

\begin{Def} \label{stab-codes}
A stabilizer code $\mathcal{C}({\overline{\mathcal{S}}})$ encodes $k$ logical qubits into $N$ physical qubits where its codespace, which is defined by the stabilizer $\mathcal{\overline{S}}$, can be written as
\[\mathcal{C}(\mathcal{\overline{S}}) = \{|\psi\rangle \in \mathcal{H}_2^{\otimes N}:\mathbf{S}_i|\psi\rangle=|\psi\rangle,i=1 \dots N-k\},\] where $\mathbf{S}_i|\psi\rangle\in \mathcal{H}_2^{\otimes N}$ is the evolution of state $|\psi\rangle$ under stabilizer generator $\mathbf{S}_i$.
\end{Def}

Note that $\mathcal{C}(\mathcal{\overline{S}})$ is the subspace of $\mathcal{H}_2^{\otimes N}$ formed by the simultaneous $+1$-eigenspaces of all the operators in the stabilizer group $\mathcal{\overline{S}}$. Here $\mathcal{H}_2^{\otimes N}$ denotes the complex Hilbert space of dimension $2^{N}$ that comprises the state space of $N$-qubit systems.

\subsubsection{Pure Errors, Logical Operators \& Encoded Pauli operators} \label{sec:PE&LO}

In the literature, stabilizer codes are generally defined over $\mathcal{G}_N$ (by using $\mathrm{S}_i$ in definition \ref{stab-codes}). Furthermore, the operators $\mathrm{L} \in \mathcal{G}_N$ that commute with the stabilizers elements (they map the codespace to itself) are called \textit{logical operators} and those operators $\mathrm{T} \in \mathcal{G}_N$ that commute with the logical operators and anticommute with at least one stabilizer element are known as \textit{pure errors} \cite{logical,softVit,neural,Hard,softPoul,rewiring}. In our group theoretical interpretation, it is easy to see that the coset representatives $\{\mathbf{T}_i\}_{i=1}^{2^{N-k}}$ and $\{\mathbf{L}_j\}_{j=1}^{2^{2k}}$, of the cosets of $\overline{\mathcal{Z}}(\overline{\mathcal{S}})$ in $\overline{\mathcal{G}}_N$ \eref{eq:pauli-centra} and of the cosets of $\overline{\mathcal{S}}$ in $\mathcal{\overline{Z}(\overline{S})}$ \eref{eq:centra-cosets}, are the equivalent operators over $\mathcal{\overline{G}}_N$ of these pure errors and logical operators, respectively. This means that we can write $\{\mathrm{T}\in \mathcal{G}_N: \mathrm{T}\equiv \mathbf{T}\}$ and $\{\mathrm{L}\in \mathcal{G}_N: \mathrm{L}\equiv \mathbf{L}\}$. 

Another important concept in the literature is that of the logical group $\mathcal{L} \in \mathcal{G}_N$ defined as $\mathcal{L}  = \mathcal{N(S)/S}$ \cite{degen3}. Once again, we have an equivalent concept of this logical group over $\mathcal{\overline{G}}_N$ in our group theoretical framework: the partition of the effective centralizer into cosets of the stabilizer (see section \ref{sec:logicalgroup}). The logical group is generated by a set of operators known as encoded Pauli operators which are denoted by $\overline{Z}_q, \overline{X}_l \in \mathcal{G}_N$. Thus, we can write $\mathcal{L} = \{\overline{Z}_q,\overline{X}_l\}$ where $\{q,l\}\in\{1,\ldots,k\}$. This means that all the logical operators of a stabilizer code $\{\mathrm{L} \in \mathcal{G}_N\}$, can be obtained as linear combinations of the encoded Pauli operators. The encoded Pauli operators of an $N$-qubit stabilizer code are defined as those operators in $\mathcal{G}_N$ that commute with the elements of the stabilizer group and whose action on an encoded state can be understood as an $X$ or $Z$ operation on each of the logical qubits encoded by the stabilizer code. In this manner, $\overline{Z}_q$ represents an operator in $\mathcal{G}_N$ whose action is analogous to performing a $Z$ operation (phase flip) on the $q$-th logical qubit. Thus $\overline{Z}_q \in \mathcal{G}_N$ maps to $Z_q \in \mathcal{G}_k$, where $Z_q$ denotes the action of a $Z$ operator on the $q$-th logical qubit and the action of $I$ operators on the remaining $k-1$ logical qubits (identity operators are omitted). In consequence, each stabilizer code has $2k$ encoded Pauli operators. As previously, we define the equivalent encoded Pauli operators over the effective $N$-fold Pauli group as $\overline{Z} \equiv \overline{\mathbf{Z}}$ and $\overline{X} \equiv \overline{\mathbf{X}}$, where $\overline{\mathbf{Z}}, \overline{\mathbf{X}} \in \mathcal{\overline{G}}_N$.

Given the equivalence between the operators of $\mathcal{G}_N$ and $\mathcal{\overline{G}}_N$ (recall that they are equivalent in all relevant physical manners) and the commutation property preserving effect of the symplectic map, it makes sense in the context of QEC to consider that the stabilizer coding framework operates over the effective Pauli group. Thus, we adopt the nomenclature employed in the literature (which refers to concepts in $\mathcal{G}_N$) to reference the equivalent operators over $\mathcal{\overline{G}}_N$. For instance, we will refer to the coset representatives $\mathbf{T}_i$ and $\mathbf{L}_j$ as pure errors and logical operators, respectively. This terminology (the reasoning behind the names), can be better understood upon closer inspection of expressions \eref{eq:car37} and \eref{eq:car38}.

From \eref{eq:car37}, we know that each $\mathbf{T}_i$ (except $\mathbf{T}_1$) is related to a unique non-zero syndrome vector $\mathbf{w}_i$. This means that each $\mathbf{T}_i$ has a unique commutation relation (with respect to the group operation $\cdot$ over $\mathcal{G}_N$) with the stabilizer generators $\{\mathbf{S}_v\}_{v=1}^{N-k}$: it anticommutes with at least one generator and commutes with the remaining ones. The term ``pure'' makes reference to how the singular commutation properties of each representative $\mathbf{T}_i$ are reflected solely by the corresponding syndrome $\mathbf{w}_i$.

From \eref{eq:car38}, we know that all the representatives $\mathbf{L}_j$ commute with all the stabilizer elements, but they do not belong to the stabilizer. This means that whenever an operator $\mathbf{L}_j$ acts on a quantum codeword, it will shift it from one stabilizer coset to another while leaving it within the same effective centralizer coset (it maps the codespace to itself). This means that the effects of each logical operator on a codeword can be interpreted as the operation of different single qubit Pauli operators on each of its logical qubits. It is for this reason that these operators are named ``logical" operators. More specifically, in a scenario in which $k$ logical qubits have been encoded into $N$ physical qubits using a stabilizer code, logical operators are $N$-qubit operators whose action on the code results in an $X,Y,Z$ or $I$ operation being applied to each of the logical qubits. Since there are $k$ logical qubits and 4 operators can be applied to each qubit, there are a total of $4^k=2^{2k}$ logical operators. 

The similarity between logical operators and encoded Pauli operators is worth noting. Because encoded Pauli operators define a generating set for the logical group, it is clear that they are also logical operators. Similarly, logical operators can be understood as linear combinations of encoded Pauli operators. Thus, one could simply refer to the encoded Pauli operators as the logical group generators. Although it may appear to complicate matters presently, the convention of referring to these concepts with different terms will be shown to serve an important purpose in the next chapter. For now, the primary takeaway is the fact that a stabilizer code has $2k$ encoded Pauli operators but $2^{2k}$ logical operators (just like how there are $N-k$ stabilizer generators but $2^{N-k}$ stabilizer elements).

\begin{figure}[!h]
	\centering
  \includegraphics[width=0.8\linewidth,  height=3.5in]{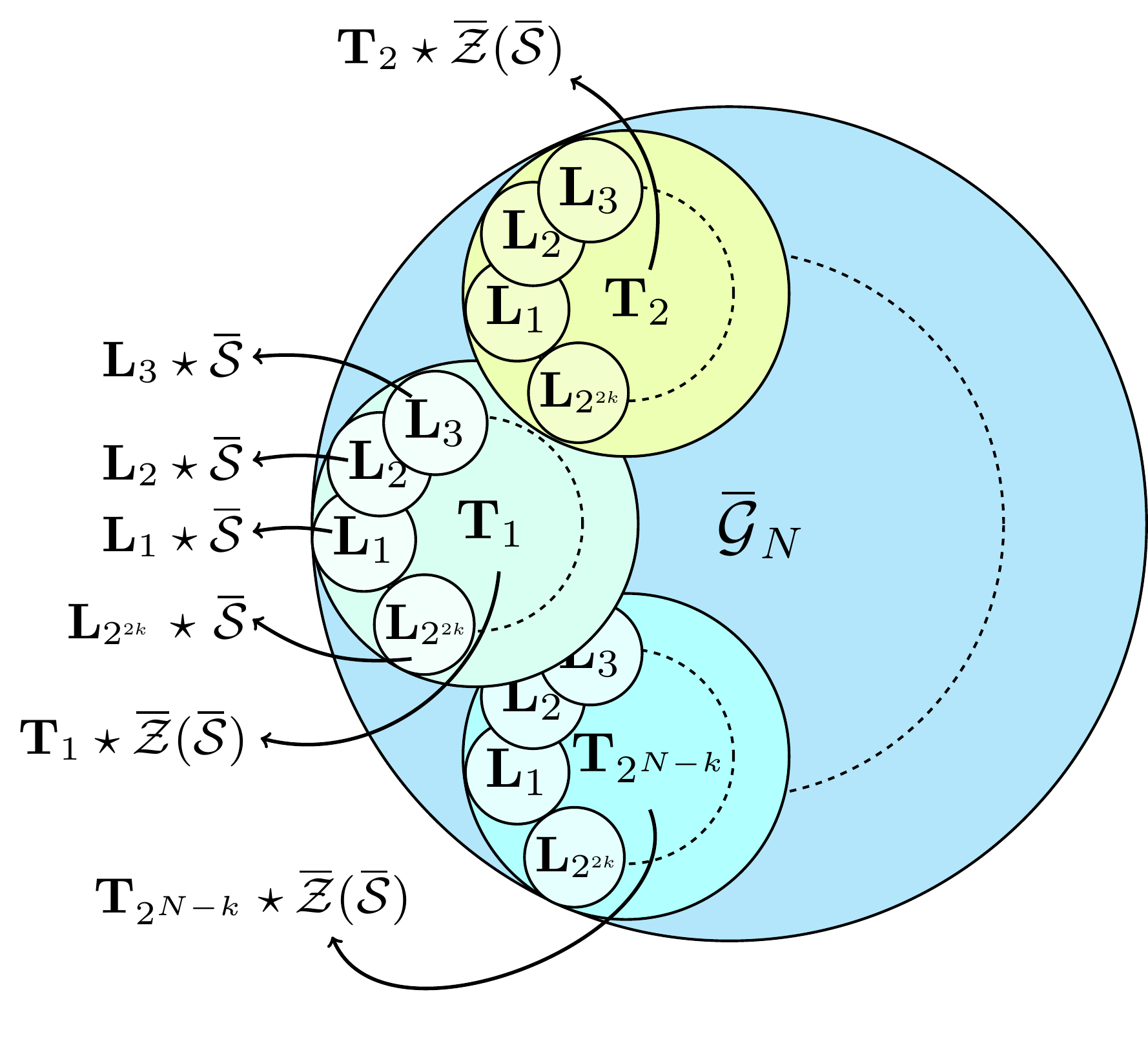}
\caption{Graphical representation of the partition of $\mathcal{\overbar{G}}_N$ into cosets of the effective centralizer, $\mathcal{\overbar{G}}_N=\bigcup_{i=1}^{2^{N-k}} \mathbf{T}_i\star \mathcal{\overbar{Z}(\overbar{S})}$, and of the partition of the effective centralizer into cosets of the stabilizer group, $\mathcal{\overbar{Z}(\overbar{S})}=\bigcup_{j=1}^{2^{2k}} \mathbf{L}_j\star \mathcal{\overbar{S}}$. Each effective centralizer coset is assigned a pure error $\mathbf{T}_i$ as its representative, and each stabilizer coset is assigned a logical operator $\mathbf{L}_j$ as its representative.}
    \label{fig:division}
\end{figure}

Figure \ref{fig:division} depicts the partition of the effective $N$-fold Pauli group into the cosets indexed by logical operators and pure errors. Intuitively, this figure reflects how any operator $\mathbf{A} \in \overline{\mathcal{G}}_N$ can be decomposed into three distinct terms, each one associated to a pure error, a logical operator, and a stabilizer element. We will later show how this decomposition of the operators in $\overline{\mathcal{G}}_N$ can be used to better approach and understand the task of decoding quantum stabilizer codes.

\subsubsection{Error detection using stabilizer codes} 

Let us assume we wish to transmit the information word $|\psi\rangle\in\mathcal{H}_2^{\otimes k}$ through some quantum channel $\xi$. First, we use a stabilizer code to encode the information word into the quantum codeword $|\overline{\psi}\rangle\in \mathcal{C}(\mathcal{\overline{S}})\subset\mathcal{H}_2^{\otimes N}$. When the codeword $|\overline{\psi}\rangle$ is sent through the quantum channel $\xi$, it is exposed to the decoherence-emulating harmful effects of said channel, and a corrupted quantum state $\ket{\overline{\psi}}_{\xi}$ is obtained at the channel output. To recover $\ket{\overline{\psi}}$ from $\ket{\overline{\psi}}_{\xi}$, which can then be used to obtain $|\psi\rangle$, the appropriate recovery operation must be applied. This requires some knowledge regarding the error induced by the channel, which must inevitably be derived from the corrupted output. However, we know that direct measurement of a quantum state needs to be avoided\footnote{Recall that measurement of a quantum state forces its superposition state to collapse, which in this case would result in the loss of the information regarding the initial state $|\psi\rangle$.}. Instead, a methodology that avoids the measurement of quantum states while still gleaning sufficient information from the corrupted channel output to implement recovery operations is necessary. Fortunately, as we mentioned previously in section \ref{sec:stabs}, this can be achieved by means of measuring the quantum syndrome, a strategy reminiscent of error syndrome measurements in classical coding scenarios \cite{josurev, NielsenChuang, newcite, classicaltoquantum}. Let us discuss it.

Assume now that the quantum channel $\xi$ is the generic Pauli channel $\xi_P$ discussed in section \ref{sec:channels}. Any error induced by this channel will be an operator $\mathrm{E} \in \mathcal{G}_N$ that represents an $N$-fold tensor product of single qubit error operators $\mathrm{E}_u$, where each $\mathrm{E}_u$ belongs to the single qubit Pauli group and $u=1,...,N$. This error acts on the encoded quantum state as $\ket{\overline{\psi}}_{\xi} = \mathrm{E}\ket{\overline{\psi}}$. As has been discussed previously, the global phase has no observable consequence, hence, using the notation introduced in \eref{eq:car3}, any $\mathrm{E}\in \mathcal{G}_N=(\widetilde{\Pi}^{\otimes N}, \cdot)$ will be taken as the operator $\mathbf{E}\equiv \mathrm{E}$ belonging to the effective Pauli group (i.e., $\mathbf{E}\in \overline{\mathcal{G}}_N=(\Pi^{\otimes N}, \star)$). Recall that although the commutation properties (in terms of the group operation $\cdot$) of the operators in $\overline{\mathcal{G}}_N$ are lost under the group operation $\star$, they can be recovered by applying the symplectic isomorphic mapping $\beta: \overline{\mathcal{G}}_N\rightarrow \mathbb{F}_2^{2N}$, together with the symplectic product $\odot$ in $\mathbb{F}_2^{2N}$ (refer to Proposition \ref{car2}). Consequently, any error operator $\mathbf{E}\in\overline{\mathcal{G}}_N \subset \mathcal{G}_N$, will commute, in terms of the group operation in $\mathcal{G}_N$, with all the stabilizer generators $\mathbf{S}_v \in \mathcal{\overline{S}}\subset \overline{\mathcal{G}}_N \subset \mathcal{G}_N$, $v = 1,\ldots,N-k$, iff
\[\beta(\mathbf{E})\odot \beta(\mathbf{S}_v)=0.\]


On the other hand, if there is some index $v'\in \{1,\ldots, N-k\}$ where the above product takes the value 1, then $\mathbf{E}$ and $\mathbf{S}_j$ will anticommute, i.e., $\beta(\mathbf{E})\odot \beta(\mathbf{S}_v')=-\beta(\mathbf{S}_j)\odot \beta(\mathbf{E})$.
Thus, the commutation properties of any $\mathbf{E}\in\overline{\mathcal{G}}_N$ with regard to the stabilizer generators $\{\mathbf{S}_v\}_{v=1}^{N-k}$, will be completely characterized by the syndrome vector $\mathbf{w}\in \mathbb{F}_2^{N-k}$ defined in \eref{eq:car33}. That is to say,

\begin{equation}\label{eq:synd}
\mathbf{w}=\beta(\mathbf{E})\odot (\beta(\mathbf{S}_1),\ldots,\beta(\mathbf{S}_{N-k}))=\mathbf{e}\odot (\mathbf{s}_1,\ldots,\mathbf{s}_{N-k}).
\end{equation}

We can also write this as 

\begin{equation}\label{eq:synd-1}
\mathrm{E}\cdot\mathrm{S}_v = (-1)^{\mathrm{w}_v}\mathrm{S}_v\cdot\mathrm{E}.
\end{equation}

\begin{figure}[!h]
	\centering
  \includegraphics[width=0.7\linewidth,  height=1.5in]{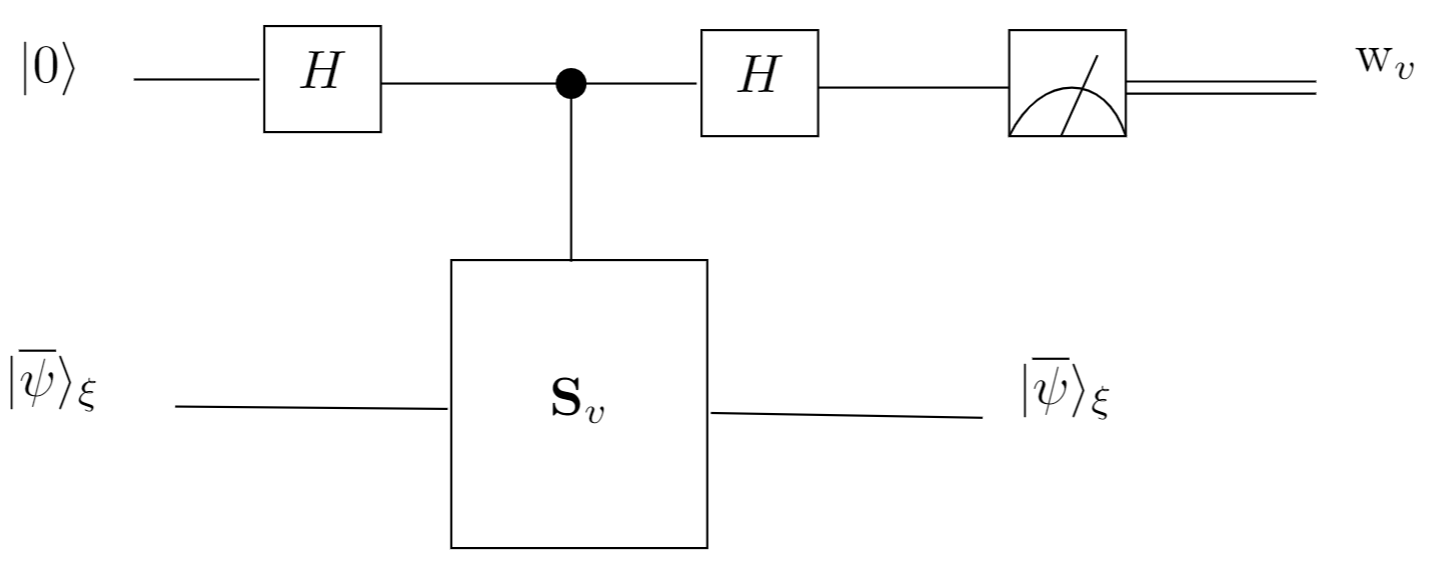}
\caption{Quantum circuit that measures the syndrome associated to each of the stabilizer generators. The H block represents the Hadamard gate defined in section \ref{sec:unitaries} and the $\mathbf{S}_v$ block represents a controlled unitary gate where the stabilizer generator $\mathbf{S}_v$ is the unitary that is applied.}
    \label{fig:synd-circuit}
\end{figure}

Based on these expressions, and considering that $\ket{\overline{\psi}}_{\xi_P} = \mathrm{E}\ket{\overline{\psi}}$, we can write

\begin{align} \label{eq:pauli-stab-2}
\begin{split}
\mathrm{S}_v\ket{\overline{\psi}}_{\xi_P} = \mathrm{S}_v\cdot\mathrm{E}\ket{\overline{\psi}} = (-1)^{\mathrm{w}_v}\mathrm{E}\cdot\mathrm{S}_v\ket{\overline{\psi}} \\
= (-1)^{\mathrm{w}_v}\mathrm{E}\ket{\overline{\psi}} = (-1)^{\mathrm{w}_v}\ket{\overline{\psi}}_{\xi_P},
\end{split}
\end{align}

where we recover the notation over $\mathcal{G}_N$ to preserve the commutation relations. From \eref{eq:pauli-stab-2}, we can tell that $\ket{\overline{\psi}}_{\xi_P}$ is an eigenstate of each stabilizer generator associated to the $\pm 1$ eigenvalues. We can also tell from this expression that the components of the syndrome $\mathbf{w}$ will determine the value of each particular eigenvalue. Therefore, we know that it is possible to determine the commutation relations between the channel error and the stabilizer generators, i.e, measure the syndrome $\mathbf{w}$, by measuring the eigenvalues of the stabilizer generators of the code \cite{EAQECC}. This can be achieved by using a circuit like the one shown in Figure \ref{fig:synd-circuit}.

\subsubsection{Quantum Parity Check Matrix of a Stabilizer Code} \label{sec:QPCM}

In section \ref{sec:linear-code} we defined the PCM of a linear block code as the matrix whose rows represent the linear constraints (the parity check equations) that make up the code. We can adapt this notion to stabilizer codes by building a matrix whose rows are the stabilizer generators. We define the Quantum Parity Check Matrix (QPCM) of a stabilizer code as the $N-k\times 2N$ matrix in $\mathbb{F}_2^{N-k\times 2N}$

\[ \mathrm{\mathbf{H}}_\mathcal{\overline{S}}=(\mathbf{\mathrm{H}}_x|\mathbf{\mathrm{H}}_z)  \triangleq \left( \begin{array}{c}
         \mathbf({\mathbf{s}}_{1,x}|{\mathbf{s}}_{1,z}) \\
         \vdots\\
         \mathbf({\mathbf{s}}_{N-k,x}|{\mathbf{s}}_{N-k,z})\end{array} \right).\]

Using this QPCM representation, we can rewrite the expression in \eref{eq:synd} as
$\mathbf{w}=\mathbf{e} \odot  \mathbf{\mathbf{H}}_\mathcal{\overbar{S}}=\mathbf{\mathbf{H}}_\mathcal{\overbar{S}}\odot \mathbf{e}$. Recall that expression \eref{eq:car34} relates the symplectic product $\odot$ with the mod 2 operations of $(\mathbb{F}_2^{N},\oplus)$. Therefore, based on this relationship we can write

\begin{equation} \label{synd_q}
\mathbf{w}=\mathrm{\mathbf{H}}_{\mathcal{\overbar{S}}} \odot \mathbf{e}=\mathbf{e}_z \mathbf{\mathrm{H}}^\top_x\oplus \mathbf{e}_x \mathbf{\mathrm{H}}^\top_z.
\end{equation}

Observe also that by construction $\beta(\mathbf{S}_i)\odot \beta(\mathbf{S}_j)=0$, $i,j=1,\ldots N-k$ (i.e., all the stabilizer generators $\{\mathbf{S}_v\}_{v=1}^{N-k}$ commute with respect to the group operation $\cdot$ in $\mathcal{G}_N$). Therefore, matrix  $\mathrm{\mathbf{H}}_\mathcal{\overbar{S}}=(\mathbf{\mathrm{H}}_x|\mathbf{\mathrm{H}}_z)$ should verify the following constraint, which is commonly known as the \textit{symplectic criterion},

\begin{equation} \label{eq:symplec}
     \mathrm{H}_x  \mathrm{H}_z^\top \oplus \mathrm{H}_z  \mathrm{H}_x^\top = 0.
\end{equation}

Constraint \eref{eq:symplec} is significant, because it specifies which existing classical codes can be used to design stabilizer codes. For instance, given two parity check matrices of any two classical codes of the same length and dimension, $\mathrm{H}_1$ and $\mathrm{H}_2$, the parity check matrix obtained as $\mathbf{H}=(\mathrm{H}_1|\mathrm{H}_2)$ will only define a stabilizer code if it satisfies (\ref{eq:symplec}), i.e., $\mathrm{H}_1 \mathrm{H}_2^\top \oplus \mathrm{H}_2 \mathrm{H}_1^\top = 0$.

\subsection{Decoding Quantum Stabilizer codes}

From what we have seen thus far, we know that direct measurement of quantum information states is avoided by measuring a binary vector known as the quantum syndrome. Additionally, we know that the quantum syndrome can be written using QPCM notation as shown in \eref{synd_q} and that it can be obtained by measuring the eigenvalues of the generators of the stabilizer code. Physically, this task is performed by a circuit like the one shown in Figure \ref{fig:synd-circuit}. In what follows we discuss how stabilizer codes can be decoded based on the quantum syndrome, and we draw parallels to the classical decoding problem presented in section \ref{sec:classical-decoding}. It is worth mentioning that his has become a prominent research topic in the field of QEC \cite{QSC, bicycle, qldpc15, degen2, degen3, degen4, NielsenChuang, QEClidar, newcite}. 

For the sake of simplicity, we will assume operation in a framework that employs a stabilizer code to encode $k$ logical qubits into $N$ physical qubits and that transmits the encoded states over a generic Pauli channel. Furthermore, we will restrict the discussion to the sum-product based decoding of stabilizer codes, which (mostly) concerns QLDPC and QTC codes \cite{turbo, ldpc1, ldpc2, ldpc3}. Certain families of stabilizer codes, such as topological codes \cite{toricphd1, toricphd2, sabo}, can be decoded differently and so this discussion may not apply to them.

Based only on our knowledge of the decoding process of classical linear block codes (see section \ref{sec:classical-decoding}), we would approach the decoding of stabilizer codes by using a factor graph to represent \eref{synd_q} and then running the SPA algorithm to find the most likely error pattern (its symplectic representation to be more precise) associated to the measured syndrome $\mathbf{w}$. This can be written as 
\begin{equation} \label{eq:ML1-deg}
\hat{\mathbf{e}} = \argmax_\mathbf{e} P(\mathbf{e}|\mathbf{w}).
\end{equation}

However, in contrast to what happens in the classical domain, the solution given by \eref{eq:ML1-deg} does not necessarily lead to optimal decoding performance for a stabilizer code. Let us discuss this counter-intuitive result.

\subsubsection{Optimal decoding of  quantum stabilizer codes}

We showed earlier (see section \ref{sec:PE&LO}) that any error operator $\mathbf{E} \in \mathcal{\overbar{G}}_N$, can be decomposed into three different terms: a pure error, $\mathbf{T}$, which represents the effective centralizer coset that contains $\mathbf{E}$, a logical operator, $\mathbf{L}$, which represents the stabilizer coset that contains $\mathbf{E}$, and a stabilizer component, $\mathbf{S}$, which represents the specific operator in the stabilizer associated with $\mathbf{E}$. That is, any arbitrary $N$ qubit effective Pauli error operator $\mathbf{E} \in \mathcal{\overbar{G}}_N$ can be decomposed as

\begin{equation} \label{eq:decomp}
\mathbf{E} = \mathbf{T}\star\mathbf{L}\star\mathbf{S},
\end{equation}
where $\mathbf{T} \in \{\mathbf{T}_i\}_{i= 1}^{2^{N-k}}$, $\mathbf{L} \in \{\mathbf{L}_j\}_{j= 1}^{2^{2k}}$, and $\mathbf{S} \in \mathcal{\overbar{S}}$, denote the pure error, the logical operator, and the stabilizer element involved in the decomposition of the error operator $\mathbf{E}$, respectively.

We know from \eref{eq:car37} that the pure errors $\{\mathbf{T}_i = \beta^{-1}(\mathbf{t}_i)\}_{i=2}^{2^{N-k}}$ will anticommute, in terms of the group operation in $\mathcal{G}_N$, with at least one stabilizer generator and commute with the rest of them. This means that each $\mathbf{T}_i$ will have a unique combination of commutation relations with the stabilizer generators. However, we know from \eref{eq:car38} that the logical operators $\{\mathbf{L}_j = \beta^{-1}(\mathbf{t}_j)\}_{j=1}^{2^{2k}}$ will commute (with regard to the group operation over $\mathcal{G}_N$) with the stabilizer generators, and we also know that all the stabilizer generators commute with each other. This means that, among the operators $\mathbf{T}$, $\mathbf{L}$, and $\mathbf{S}$, only the pure error component can possibly anti-commute (in terms of the group operation in $\mathcal{G}_N$) with the stabilizer generators. Therefore, while syndrome measurements serve to reveal the pure error component of the error, the logical and stabilizer components remain unknown. 

This limitation in quantum syndrome decoding is similar to what happens when syndrome decoding is applied in the classical domain. In the classical scenario, the syndrome only identifies the coset of the code and not the specific error pattern. In order to optimize performance, classical syndrome decoders generally choose the coset leader (the most likely error pattern in the coset) as the error operator that has taken place. However, the quantum version of this problem is significantly more nuanced, as choices from various different cosets can be made: the identified centralizer coset is composed of $2^{2k}$ cosets of the stabilizer, each one of these represented by a logical operator $\mathbf{L}_j, \;j= 1\ldots2^{2k}$. Fortunately, the stabilizer component of the decomposition, $\mathbf{S}$, is irrelevant in the error correction procedure and can be ignored in the decoding process\footnote{The reason for this is that any error operator contained in the same stabilizer coset as $\mathbf{S}$ will have the same effect on a codeword. This is discussed further in the next section. }.

Assuming that the quantum decoder finds the appropriate pure error component from the set \{$\mathbf{T}_i\}_{i= 1}^{2^{N-k}}$ using the syndrome, the next step is to find the correct logical component, $\mathbf{L} \in \{\mathbf{L}_j\}_{j= 1}^{2^{2k}}$, of the error  $\mathbf{E}$. Therefore, optimal decoding for quantum stabilizer codes can be understood as the task of locating the most likely logical operator in  \{$\mathbf{L}_j\}_{j= 1}^{2^{2k}}$ involved in the decomposition of the error operator, $\mathbf{E}$, given the measured quantum syndrome $\mathbf{w}$. That is,
\begin{equation} \label{eq:DQMLD}
\hat{\mathbf{L}} = 
\argmax_{\mathbf{L} \in \{\mathbf{L}_j\}_{j= 1}^{2^{2k}}} P(\mathbf{L}|\mathbf{w}).
\end{equation}

This decoding task can be seen as obtaining the coset $\mathbf{L}_j \star \mathcal{\overbar{S}}$ that has the highest probability amidst those that have the same pure error component, $\mathbf{T}$. This means that the decoder has to locate the stabilizer coset indexed by the representative $\mathbf{L}$ that has the highest probability among all the stabilizer cosets that belong to the effective centralizer coset with representative $\mathbf{T}$, i.e., it must find the most likely coset $\mathbf{T}\star\mathbf{L}\star\mathcal{\overbar{S}}$. Notice how the complexity increases in comparison to the classical syndrome decoding problem (\ref{eq:ML1-deg}): In the classical domain, the most likely error pattern associated with a given syndrome is obvious (for a BSC, the error pattern with minimum weight in the coset), while in quantum decoding the solution to  (\ref{eq:DQMLD}) is not obvious\footnote{Notice that this solution is independent of the error operator introduced by the channel and only depends on the channel parameters and the code structure. Thus, it could be calculated for each one of the pure errors prior to the decoding process.}. Therefore, the results obtained by a decoder following the classical decoding rule in (\ref{eq:ML1-deg}) may differ from those obtained following the optimal quantum decoding rule given in (\ref{eq:DQMLD}). If a classical decoder is employed to decode a stabilizer code, the estimated error will be the Pauli operator of minimum weight among those belonging to the effective centralizer coset $\mathbf{T} \star \mathcal{\overbar{Z}(\overbar{S})}$ associated to the syndrome. This is shown in \cite{Hard}, where by weight we refer to the same concept as that of the Hamming weight (see section \ref{sec:linear-code}); the number of non-identity elements of the error pattern. In \cite{Hard}, the strategy of decoding quantum stabilizer codes based on the classical rule given in (\ref{eq:ML1-deg}) is called non-degenerate decoding or Quantum Maximum Likelihood Decoding (QMLD), while the optimal rule given in (\ref{eq:DQMLD}) is called degenerate decoding or Degenerate QMLD (DQMLD). Essentially, a decoder based on the QMLD strategy is 
``blind'' to the partition of the coset $\mathbf{T} \star \mathcal{\overbar{Z}(\overbar{S})}$ into cosets of the stabilizer. In contrast, the optimal quantum decoder is capable of finding the stabilizer coset $(\mathbf{T} \star \mathbf{L}) \star \mathcal{\overbar{S}}$ which has the highest probability and will simply choose any operator contained in it. To do so, it must compute the probability of each stabilizer coset by adding the probabilities of all the operators that make up the coset, which is significantly harder than choosing the minimum-weight operator \cite{degen3, degen4, Hard}.

\subsubsection{Optimal vs SPA decoding of quantum stabilizer codes}

\begin{figure}[!htp]
	\centering
  \includegraphics[width=\linewidth,  height=6.5in]{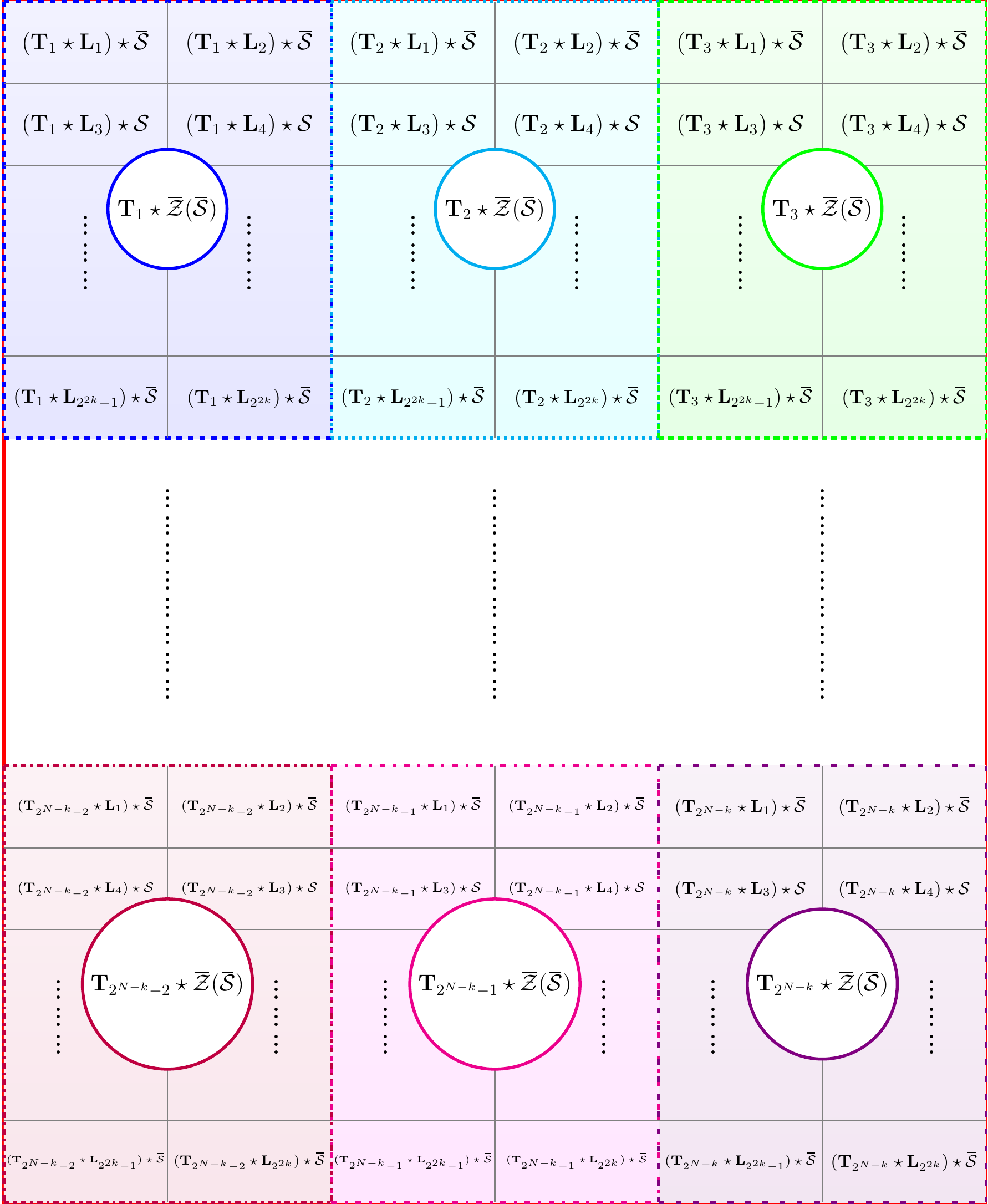}
\caption{Cosets of $\mathcal{\overbar{S}}$ and $\mathcal{\overbar{Z}(\overbar{S})}$ in $\mathcal{\overbar{G}}_N$. The first coset partition is represented by the differently colored shaded rectangles and given in \eqref{pauli-centra}. The second coset partition is represented by the rectangles found within each of the effective centralizer cosets and given in \eref{eq:centra-cosets}. }
    \label{fig:cosets}
\end{figure}

In order to discuss the differences between decoding based on (\ref{eq:ML1-deg}) and (\ref{eq:DQMLD}), we redefine the coset partition of \cite{degen1}. This is shown graphically in Figure \ref{fig:cosets}. To comprehend this figure, recall the division of the effective $N$-fold Pauli group shown previously in Figure \ref{fig:division}, where $\mathcal{\overbar{G}}_N$ is partitioned into effective centralizer cosets $\mathbf{T}_i \star \mathcal{\overbar{Z}(\overbar{S})}$, which themselves are partitioned into stabilizer cosets $\mathbf{L}_j \star \mathcal{\overbar{S}}$, where $i = 1,\ldots,2^{N-k}$ and $j = 1,\ldots,2^{2k}$. In Figure \ref{fig:cosets}, $\mathcal{\overbar{G}}_N$ is represented by the outermost red bordered rectangle that contains the smaller shaded rectangles. These shaded rectangles portray the coset partition that divides $\mathcal{\overbar{G}}_N$ into $2^{N-k}$ cosets of the effective centralizer. Each of these cosets is assigned a different representative from the set $\{ \mathbf{T}_j\}_{j=1}^{2^{N-k}}$, where $\mathbf{T}_1 = \beta^{-1}(\mathbf{t}_1) = \beta^{-1}(\mathbf{0}) = I^{\otimes N}$. In addition, each of the shaded rectangles is itself divided into smaller rectangles. This subdivision represents the coset partition of each effective centralizer coset into cosets of the stabilizer. These stabilizer cosets are each indexed by a different logical operator $\{ \mathbf{L}_j\}_{j=1}^{2^{2k}}$, where $\mathbf{L}_1 = \beta^{-1}(\mathbf{l}_1) = \beta^{-1}(\mathbf{0}) = I^{\otimes N}$. Within each stabilizer coset, effective Pauli operators that differ in terms of their stabilizer component can be found. However, knowing that all stabilizer elements have the same effect on a codeword, errors within each stabilizer coset need not be distinguished (this is studied in the next section).

Based on this graphical representation, the generic classical decoder defined by the rule in (\ref{eq:ML1-deg}) will obtain the most likely error operator (i.e., the error operator with the lowest weight) for the received syndrome. Thus, it will identify the correct coset $\mathbf{T} \star \mathcal{\overbar{Z}(\overbar{S})}$ and the corresponding pure error $\mathbf{T}$. However, as explained before, Figure \ref{fig:cosets} illustrates that this solution does not necessarily correspond to the one obtained by a quantum decoder that follows \eref{eq:DQMLD}. To solve \eref{eq:DQMLD}, the logical operator component of the error, $\mathbf{L}$, must be appropriately estimated. This can be understood graphically in Figure \ref{fig:cosets} as finding the appropriate stabilizer coset within the effective centralizer coset corresponding to the received syndrome. The classical decoder of (\ref{eq:ML1-deg}) treats the entire effective centralizer coset as a whole, without considering its partition into stabilizer cosets. Graphically, this entails being limited to choosing the error with the highest symbol-wise probability (the error with the lowest weight) within the effective centralizer coset, which in some cases will not belong to the most likely coset. This issue can be resolved by using a decoder that operates based on the rule shown in \eref{eq:DQMLD}, which is capable of considering the partition of the effective centralizer cosets into cosets of the stabilizer and finding the most likely stabilizer coset. Alas, no decoder for QLDPC codes that efficiently implements the Degenerate QMLD decoding rule of \eref{eq:DQMLD} \cite{degen4, Hard} exists at the time of writing. 

It is important to mention that the differences between \eref{eq:ML1-deg} and \eref{eq:DQMLD} are only relevant when the elements of the stabilizer group have low weight, i.e., when they have a small amount of non-identity operators \cite{QEClidar}. Because low weight stabilizer elements are quite common in QLDPC codes, developing implementations of the optimal decoder is germane to the field. However, understanding why the differences between  (\ref{eq:ML1-deg}) and  (\ref{eq:DQMLD}) are negligible when there are no low weight stabilizer elements also merits discussion.

Assume that we have a stabilizer code whose stabilizer elements have sufficiently large weight. Then, there will be an error operator in each effective centralizer coset (colored shaded rectangles in Figure \ref{fig:cosets}), specifically one belonging to the stabilizer coset indexed by $\mathbf{L}_1 = I^{\otimes N}$, with a much smaller weight than the rest of the error operators contained in the coset. This will occur regardless of the partition of the effective centralizer coset into stabilizer cosets \cite{QEClidar}. This low weight error will dominate the probability of the entire effective centralizer coset, and will force the probability of the stabilizer coset $\mathbf{T}_i\star\mathbf{L}_1\star\mathcal{\overbar{S}}$ to be higher than that of any other stabilizer coset. Therefore, decoding based on the QMLD classical rule (\ref{eq:ML1-deg}) will lead to the lowest weight error operator $\mathbf{\hat{E}} \in \mathbf{T}_i\star\mathbf{L}_1\star\mathcal{\overbar{S}}$, while decoding based on (\ref{eq:DQMLD}) will produce a solution from the coset $\mathbf{T}_i\star\mathbf{L}_1\star\mathcal{\overbar{S}}$. As is shown in the next section, due to degeneracy any element of the coset $\mathbf{T}_i\star\mathbf{L}_1\star\mathcal{\overbar{S}}$ will perform the same recovery operation, leading to the same result. Thus, in this case the performance of a decoder based on the classical rule (QMLD) will be identical to that of optimal quantum decoding (DQMLD). On the other hand, when the code has low weight stabilizer elements, there are many low weight error operators that may contribute to the probability of the different stabilizer cosets in the effective centralizer coset. Thus, the most likely error operator obtained by the classical rule \eref{eq:ML1-deg} may not belong to the most likely stabilizer coset, which is the solution obtained by the optimal quantum decoder \eref{eq:DQMLD}, and this would mean that the performance of a decoder based on the classical rule would be worse than that of optimal quantum decoding. We illustrate this with an example in section \ref{sec:example}.

\section{Degeneracy and why it arises}\label{sec:degen}

In earlier sections of this chapter, we mentioned that due to degeneracy the stabilizer component of the error can be ignored in the decoding process, i.e, errors that differ only by a stabilizer element need not be distinguished by the decoder. Let us study why this happens. 

For this purpose we introduce a similar example to the one given in \cite{degen3}. Consider a scenario where the quantum state $|\psi\rangle$ has been encoded using a stabilizer code into the quantum codeword $\ket{\overbar{{\psi}}}$ and transmitted through a quantum channel $\xi$. At the channel output, the noisy quantum state $\ket{\overbar{{\psi}}}_{\xi} = \mathbf{E}'\ket{\overbar{{\psi}}}$ is obtained\footnote{Recall that the notation $\mathbf{E}'\ket{\overbar{{\psi}}} \in \mathcal{H}_2^{\otimes N}$ denotes the evolution of state $\ket{\overbar{{\psi}}}$ under the action of the error operator $\mathbf{E}'$.}, where $\mathbf{E}'=\mathbf{E} \star \mathbf{S}_v$, i.e., the error operator $\mathbf{E}'$ that has corrupted the quantum codeword can be expressed as the operation of another error operator $\mathbf{E}$ and a stabilizer element $\mathbf{S}_i$. Since by definition\footnote{All elements of the stabilizer commute amongst themselves and have the same effect on any given quantum codeword \cite{NielsenChuang,QEClidar}, i.e, they leave the state unchanged (see section \ref{sec:stabs}).} of a stabilizer code $\mathbf{E}'\ket{\overbar{{\psi}}} = \mathbf{E}\star\mathbf{S}_v\ket{\overbar{{\psi}}} = \mathbf{E}\ket{\overbar{{\psi}}}$, the measurable effects of error operator $\mathbf{E}'$ on the codeword will be the same as those of $\mathbf{E}$. In consequence, if we were to decompose $\mathbf{E}'$ and $\mathbf{E}$ based on \eref{eq:decomp}, we would see that they would differ only in terms of their stabilizer component $\mathbf{S}_v$, i.e., their logical and pure error components would be identical. 

From a graphical perspective, this scenario can be understood as $\mathbf{E}$ and $\mathbf{E}'$ belonging to the same stabilizer coset (gray bordered rectangles within colored shaded rectangles) in Figure \ref{fig:cosets}. Based on this depiction, in order to accurately distinguish both errors, a decoder capable of differentiating operators contained within the same stabilizer coset would be required. Although the task of finding the correct stabilizer component of an error remains out of the reach of current quantum decoders, when focusing on the recovery of the transmitted quantum codeword, devising a decoder that can solve this task is unnecessary: the same reason that leads to the inability of current decoders to estimate the stabilizer component of the errors also makes the endeavor itself pointless. To see this, consider a recovery operator $\mathbf{R}$ that nullifies the effects of $\mathbf{E}$ on the transmitted quantum codeword, i.e., $ \mathbf{R}\star\mathbf{E}\ket{\overbar{{\psi}}}$ = $\ket{\overbar{{\psi}}}$. Then, this same recovery operator reverts the impact of $\mathbf{E}'$, since 
\begin{equation}\label{eq:eq-jgf}
\mathbf{R}\star\mathbf{E}'\ket{\overbar{{\psi}}} = \mathbf{R}\star\mathbf{E}\star\mathbf{S}_i\ket{\overbar{{\psi}}} = \mathbf{R}\star\mathbf{E}\ket{\overbar{{\psi}}} = \ket{\overbar{{\psi}}}.
\end{equation}

Essentially, this shows that errors that differ only in the stabilizer component (errors that belong to the same stabilizer coset) can all be corrected using the same recovery operator. Hence, the decoder does not need to distinguish these errors to begin with. Similarly, all operators belonging to a specific stabilizer coset would recover the same quantum state, and any of these operators would be able to correct any error belonging to the coset. As a result, degeneracy should theoretically improve the performance of quantum codes, as it enables many errors to be corrected based on the same recovery operation. In particular instances, the positive effects that degeneracy can have on performance have already been observed \cite{degen1, degen-improv1, degen-improv2}. 

\subsection{Degeneracy in sparse quantum codes} \label{symdeg}

Previously we mentioned that the performance improvements provided by the optimal quantum decoder of \eref{eq:DQMLD} over the classical decoder of \eref{eq:ML1-deg} may become more notable when a large amount of stabilizer elements have low weight. When this happens, it is possible for many low weight operators to be spread out across the stabilizer cosets of the code. In such circumstances, properly summing the probabilities over these cosets is paramount to perform optimal quantum decoding following \eref{eq:DQMLD}. Such is the case for many quantum LDPC codes, whose sparse nature typically results in them having a large amount of low weight stabilizer generators \cite{degen3, degen4, Hard, reviewPat, QEClidar, TPS}. The number of low weight stabilizer operators is higher for sparser codes, and given that the sparsity of these codes is their primary appeal\footnote{Typically, sparser codes require less quantum gates. This is beneficial because quantum gates are faulty elements that can introduce additional errors in the recovery process.}, QLDPC codes will generally have a large number of low weight stabilizer elements. 

Should we be equipped with an optimal QLDPC decoder, we would be able to exploit the high degree of degeneracy of sparse quantum codes to improve performance, as typical (low weight) errors are likely to correspond to degenerate error operators \cite{degen3}. In fact, it is likely that such a decoding strategy would result in a significant leap forward in terms of the performance of QLDPC codes. Unfortunately, implementation of a DQMLD decoder for sparse quantum codes remains an open research problem, so most QLDPC decoding schemes are based on the classical SPA decoder of \eref{eq:ML1-deg}. Given that in some cases the presence of low-weight stabilizer elements may result in the most likely error pattern not being contained in the most likely coset, decoding QLDPC codes based on the QMLD classical rule \eref{eq:ML1-deg}, rather than the DQMLD rule \eref{eq:DQMLD}, can worsen their performance. This idea can be better understood by introducing the concept of end-to-end errors.

\subsection{End-to-end errors} \label{sec:end-to-end}
As was shown in \eref{eq:decomp}, any error operator can be understood as the combination of a pure error, a logical operator and a stabilizer element, i.e,  $\mathbf{E} = \mathbf{T} \star \mathbf{L} \star \mathbf{S}$. After decoding is performed, the estimated error operator, $\hat{\mathbf{E}} = \hat{\mathbf{T}} \star \hat{\mathbf{L}} \star \hat{\mathbf{S}}$, may be the same as the error operator introduced by the Pauli channel or it may be different. In the former case, the transmitted codeword will be recovered perfectly. In the latter however, i.e, when $\mathbf{\hat{E}} \neq \mathbf{E}$, we can have different situations:
\begin{enumerate}
    \item \textbf{End-to-end degenerate errors:} These events take place when the estimated error pattern and the channel error both belong to the same stabilizer coset but they do not match, i.e., $\mathbf{T}_i = \mathbf{\hat{T}}_i$ and $\mathbf{L}_j = \mathbf{\hat{L}}_j$ but $\hat{\mathbf{E}} \neq \mathbf{E}$. However, because stabilizer elements have no effect on the transmitted quantum codeword, this difference is irrelevant and the encoded state will be recovered perfectly. In Figure \ref{fig:division}, this situation occurs when $\mathbf{E}$ and  $\hat{\mathbf{E}}$  belong to the same stabilizer coset $(\mathbf{T}_i \star \mathbf{L}_j) \star \mathcal{\overbar{S}}$, where $i = 1,\ldots,2^{N-k}$ and $j = 1,\ldots,2^{2k}$.

    \item \textbf{End-to-end identical syndrome errors:} These events take place when the estimated error sequence and the channel error both belong to the same centralizer coset but each of them belongs to a different stabilizer coset, i.e., $\mathbf{T}_i = \mathbf{\hat{T}}_i$ and $\mathbf{L}_j \neq \mathbf{\hat{L}}_j$ (the channel logical operator and estimated logical operator do not match). This means that $\hat{\mathbf{E}}$ and $\mathbf{E}$ exhibit identical commutation relations with the stabilizer generators, and so they have the same syndrome $\mathbf{w}$. However, because they differ in their logical operator components, they will each act on the transmitted codeword in a distinct non-trivial manner. In consequence, the recovered quantum codeword will be different from the transmitted one. In Figure \ref{fig:division}, this situation occurs when $\hat{\mathbf{E}}$ and  ${\mathbf{E}}$ belong to the effective centralizer coset $\mathbf{T}_i \star \mathcal{\overbar{Z}(\overbar{S})}$ but not to the same stabilizer coset. The impact of these errors, which are impossible to avoid in noisy channels, may be alleviated by employing the as of yet nonexistent optimal quantum decoder given in \eref{eq:DQMLD} rather than the classical decoder in \eref{eq:ML1-deg}. The classical counterpart of this situation would be the case in which the error introduced by the classical channel belongs to the codespace and so the transmitted codeword is mapped onto another codeword, which results in the information being corrupted in an undetectable manner. It should be noted that some literature refers to these end-to-end identical syndrome error events as logical errors.

     \item \textbf{End-to-end errors with different syndromes:} These events occur when the error sequence estimated by the decoder and the real error sequence belong to different centralizer cosets, i.e., $\mathbf{T}_i\neq\mathbf{\hat{T}}_i$. Thus, they take place because the syndrome of  $\hat{\mathbf{E}}$ is different from the syndrome of $\mathbf{E}$. Equipped with a perfect decoder, such scenarios would not exist, as the syndrome of the estimated error pattern $\hat{\mathbf{E}}$ would always match the measured syndrome associated to the channel error $\mathbf{E}$. However, because the decoding algorithm for sparse quantum codes is not ideal for the task, these errors can take place with varying probability \cite{degen3, reviewPat, mod-BP}. 
     
     \end{enumerate}
     
     The existence of end-to-end errors with different syndromes merits further discussion. As mentioned previously in section \ref{sec:SPA}, the SPA decoder assumes that the individual marginal probabilities it computes are exact. However, this only holds whenever the factor graph is a tree, which, given the well-documented notoriety of LDPC and QLDPC codes for having short cycles, results in this requirement almost never being met \cite{qldpc15, cycle14, cycle15, cycle16}. In the classical paradigm, as a result of the negative impact of short-cycles on SPA-based decoding methods, a large body of research exists on how to minimize the presence of short-cycles in LDPC factor graphs \cite{cycle1, cycle2, cycle3, cycle4, cycle5}. This work becomes even more relevant for QEC, since the presence of short cycles of length-$4$ is essentially unavoidable in the construction of QLDPC codes. This occurs because of the commutativity constraint of stabilizer codes (recall that all the elements of the stabilizer must commute), which results in the QPCM of the code having an even number of row overlaps and in the appearance of length-$4$ cycles in the corresponding factor graph \cite{qldpc15}. The negative effects of these cycles can be mitigated by using specific QLDPC construction strategies like the bicycle codes of \cite{bicycle}, LDGM based CSS and non-CSS codes \cite{jgf1, jgf2, patrick}, or the quasi-cyclic constructions of \cite{hag1, hag2, hag3, hag4}. Recently, some results have shown that there is some merit in preserving a number of these cycles \cite{short-cycle-improv}, as they can help with spreading information throughout the factor graph during the decoding process. This last result is further reaffirmed by the work of \cite{degen3}, where the methods employed to correct the so-called \textit{symmetric degeneracy errors} benefit from the presence of short cycles.
     
     Symmetric degeneracy errors \cite{qldpc15, degen3, TPS,  efb} are a particular type of end-to-end error with different syndrome that are caused by specific flaws in the structure of the factor graphs of sparse quantum codes (short cycles for instance). The term was coined in the work of Poulin et al. \cite{degen3}, which introduces a $2$-qubit example where the SPA decoder produces a ``symmetric" estimate of the channel error sequence, even though neither the true channel error nor the operators in its equivalence class exhibit this symmetry (hence a decoding mistake is made). The reader is referred to \cite{TPS, TPS-thes} for a rigourous discussion on the symmetric degeneracy error phenomenon.

  All in all, in order for QLDPC codes to realize their full potential, means of minimizing the frequency with which end-to-end errors with identical syndromes and end-to-end errors with different syndromes occur are necessary. To do so, we must first have the capacity to distinguish between different types of end-to-end errors, which, as will be shown in the next chapter, is a far from trivial task. 
  
  \subsection{A useful example} \label{sec:example}
  
We close this chapter by providing a practical example to illustrate and facilitate the understanding of many of the concepts that have been presented thus far. The exercise is designed with the goal of helping the reader comprehend the phenomenon of degeneracy and the coset partitions of the effective $N$-fold Pauli group.

\begin{figure}[!htp]
	\centering
  \includegraphics[width=0.7\linewidth,  height=1.25in]{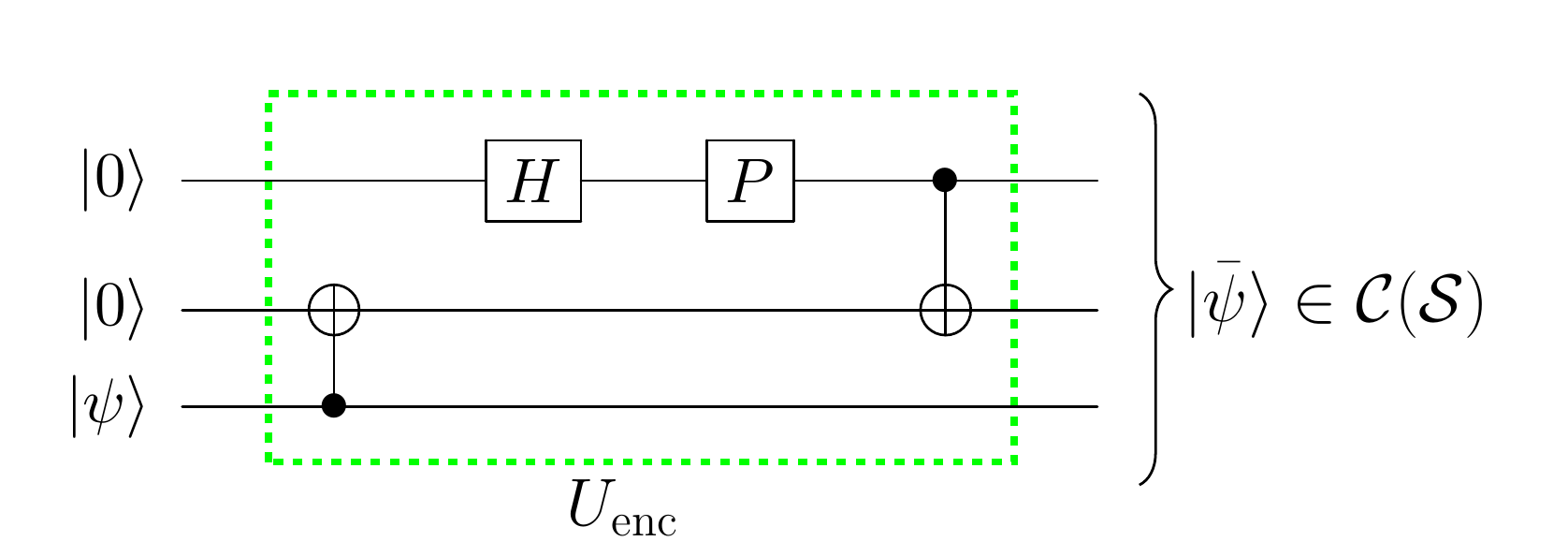}
\caption{Encoding circuit for the stabilizer code with the generators shown in (\ref{example-gens}). $U_\text{enc}$ represents the unitary that performs the encoding operation, which, with the aid of the ancilla qubits $\ket{0}^{\otimes 2}$ takes the logical qubit $\ket{\psi}\in\mathcal{H}_2^{\otimes 1}$ to the codespace $\ket{\overbar{\psi}}\in\mathcal{C}(\mathcal{\overbar{S}})\subset\mathcal{H}_2^{\otimes 3}$.}
   \label{fig:encoding}
\end{figure}

For the purpose of simplicity, we will be working with a very reduced number of qubits. Consider a scenario in which we wish to encode a single logical qubit into three physical qubits ($k=1$ and $N=3$). To build this system, we define a 3-qubit stabilizer code with generators 

\begin{equation}\label{example-gens}
\begin{aligned}
&\mathbf{S}_1 = XYZ \\
&\mathbf{S}_2 = YXI, \\
\end{aligned}
\end{equation}

\noindent whose encoding circuit, which requires the use of two ancilla qubits\footnote{In the context of encoding operations, ancilla qubits are the extra qubits that are used to encode logical qubits into physical qubits.}, is shown in Figure \ref{fig:encoding}. The Phase (P), Hadamard (H), and CNOT gates that make up this encoding circuit are defined in Chapter \ref{chapter3} (see section \ref{sec:unitaries}). 

\begin{figure*}[!htp]
\centering
\includegraphics[width=\columnwidth,  height=5.75in]{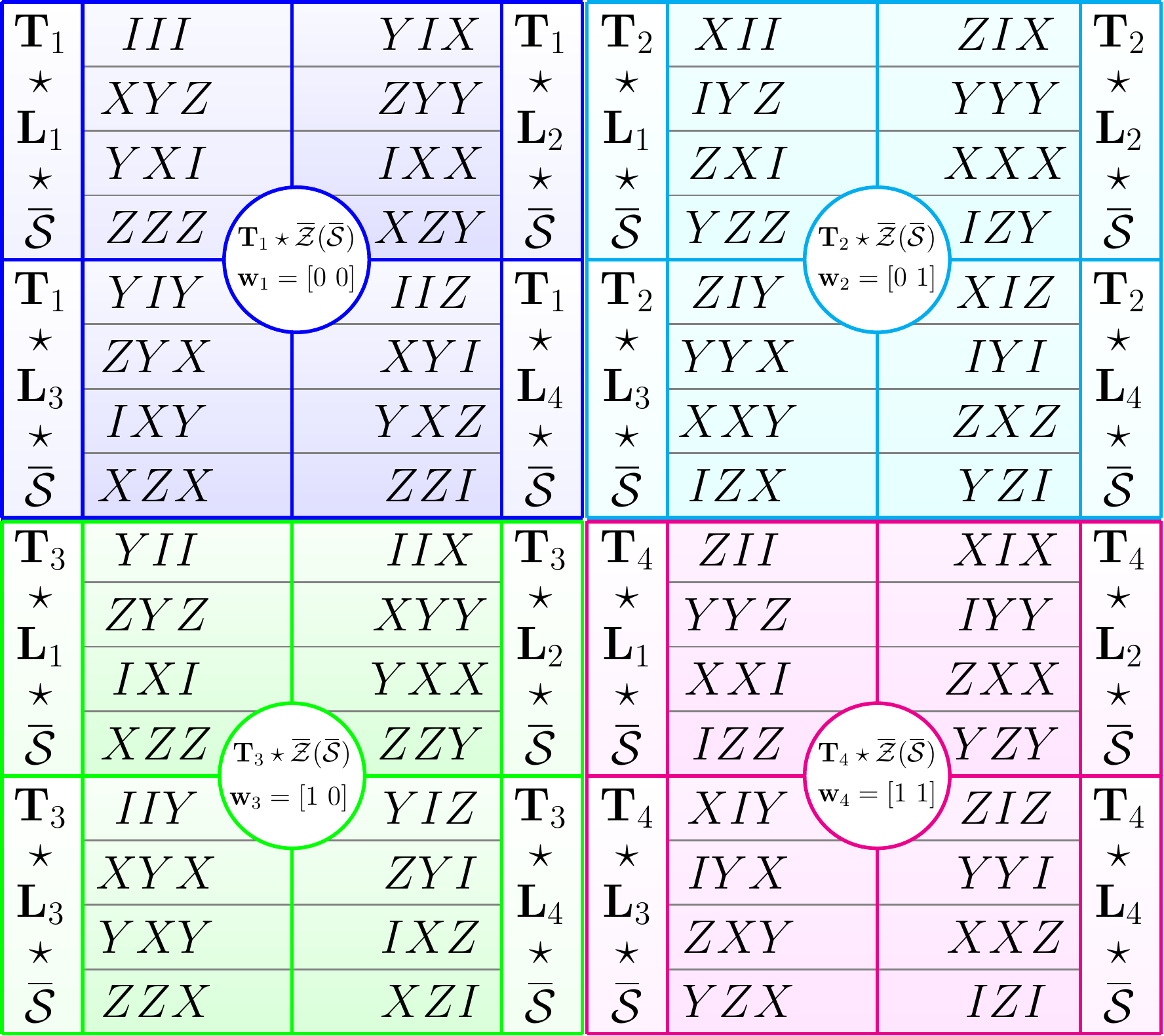}
    \caption{Partition of $\mathcal{\overbar{G}}_3$ into cosets when the $3$-qubit stabilizer code with generators $\{XYZ, YXI\}$ is used. The first coset partition divides $\mathcal{\overbar{G}}_3$ into $2^{N-k} = 4$ cosets of the effective centralizer, which are distinguished by different colours, indexed by their corresponding pure error representative $\mathbf{T}_i$ and denoted as  $\mathbf{T}_i\star\mathcal{\overbar{Z}(\overbar{S})}$, where $i = 1 \ldots 4$. Each of these cosets is also associated to its corresponding syndrome $\mathbf{w}_i$. The second coset partition is represented by the subdivision of each coset $\mathbf{T}_i\star\mathcal{\overbar{Z}(\overbar{S})}$ into the $2^{2k} = 4$ cosets associated to their corresponding logical operator representatives $\mathbf{L}_{j}$, where $j = 1 \ldots 4$. This partition divides every effective centralizer coset into cosets of the stabilizer. Within each stabilizer coset, the degenerate errors associated to each logical operator can be found.}
    \label{fig:cosets_example}
\end{figure*}

If the encoded quantum state, codeword $\ket{\overbar{\psi}}$, is transmitted through an arbitrary quantum channel $\xi$, a corrupted quantum state $\ket{\overbar{\psi}}_\xi$ will be obtained at the channel output. The error syndrome associated to this corrupted quantum state will be an $N-k$ binary vector that represents the commutation status of the stabilizer generators and the Pauli error introduced by the channel\footnote{Recall that, up to an overall phase, assuming that errors induced by a Pauli channel belong to $\mathcal{\overbar{G}}_N$ is identical to assuming that they belong to $\mathcal{G}_N$.}\cite{josurev}. Figure \ref{fig:cosets_example} shows the coset partition of $\mathcal{\overbar{G}}_N$ into cosets of the effective centralizer of the code $\mathcal{\overbar{Z}(\overbar{S})}$. Since for this example we have $k=1$ and $N=3$, the effective $3$-qubit Pauli space $\mathcal{\overbar{G}}_3$ will be divided into $2^{(3-1)}$ cosets of $\mathcal{\overbar{Z}(\overbar{S})}$. In Figure \ref{fig:cosets_example}, the effective centralizer cosets are depicted in different colours. Moreover, each coset is headlined by its representative $\mathbf{T}_i$ as well as being associated to its respective syndrome $\mathbf{w}_i$, where $i = 1,2,3,4$. 

Let us now consider the centralizer coset $\mathbf{T}_1 \star \mathcal{\overbar{Z}(\overbar{S})}$. Since $\mathbf{T}_1 = \beta^{-1}(\mathbf{t}_1)$ and $\mathbf{t}_1 = \mathbf{0}$, it is easy to see that the coset $\mathbf{T}_1 \star \mathcal{\overbar{Z}(\overbar{S})}$ is actually the effective centralizer itself, since $\mathbf{T}_1 =\beta^{-1}(\mathbf{0}) = {I}^{\otimes 3} = III$. The effective centralizer of a stabilizer code is comprised of $2^{2k}$ stabilizer cosets. In our example, given that $k=1$, each effective centralizer coset is divided into $4$ stabilizer cosets. This is represented in Figure \ref{fig:cosets_example} by the subdivision of the distinctly coloured rectangles (which represent the cosets of $\mathcal{\overbar{Z}(\overbar{S})}$) into even smaller rectangles (which represent the stabilizer cosets). Each of these rectangles is indexed by the appropriate representative $\mathbf{L}_j$, where $j = 1,2,3,4$. These representatives embody non-trivial operations on the logical qubit. More explicitly, $\mathbf{L}_1$ represents the action of the $I$ Pauli gate on the logical qubit, $\mathbf{L}_2$ represents the action of the $X$ Pauli gate on the logical qubit, i.e., $U_\text{enc}(\ket{0}^{\otimes 2}X\ket{\psi})=\mathbf{L}_2 \bar{\ket{\psi}}$, $\mathbf{L}_3$ represents the action of the $Y$ Pauli gate on the logical qubit, and $\mathbf{L}_4$ represents the action of the $Z$ Pauli gate on the logical qubit. We can also relate these logical operators to the concept of encoded Pauli operators. For this purpose, recall that the $2k$ encoded Pauli operators form a generating set for the logical group, i.e, all the operators $\{\mathbf{L}_j\}_{j=1}^{4}$ can be generated from $\overline{\mathbf{X}}_1$ and $\overline{\mathbf{Z}}_1$. Knowing that the Pauli matrices are related as $Y = iXZ$ (see section \ref{sec:unitaries}), then it is clear that in this example\footnote{In more complex scenarios the encoded Pauli operators of a stabilizer code are found by computing the standard form of the QPCM of the code. This is discussed in the next chapter.}, $\overline{\mathbf{X}}_1 = \mathbf{L}_2$ and $\overline{\mathbf{Z}}_1 = \mathbf{L}_4$.

All the error operators of a specific stabilizer coset $\mathbf{T}_i \star \mathbf{L}_j \star \mathcal{\overbar{S}}$ (the degenerate Pauli operators associated to the logical operator $\mathbf{L}_j$ and pure error $\mathbf{T}_i$) have the same effect on the transmitted codeword and they will be reversible via the same recovery operator. Notice that the coset associated to $\mathbf{L}_1$ within $\mathcal{\overbar{Z}(\overbar{S})}$ is actually the stabilizer of the code. As shown in Figure \ref{fig:cosets_example}, the stabilizer is made up of the $3$-qubit identity operator $III$, the generators $XYZ$ and $YXI$, and the product of the generators $ZZZ$. 

Table \ref{tab:info} shows the logical operators $\{\mathbf{T}_1 \star \mathbf{L}_j = \mathbf{L}_j\}_{j=1}^{4}$ that serve as the coset representatives of the stabilizer cosets found within the first centralizer coset, $\mathbf{T}_1 \star \mathcal{\overbar{Z}(\overbar{S})} = \mathcal{\overbar{Z}(\overbar{S})}$. It also includes the pure errors $\{\mathbf{T}_i\}_{i=1}^{4}$ that serve as the coset representatives of the cosets of $\mathcal{\overbar{Z}(\overbar{S})}$. Based on the information provided by this table, along with the stabilizer generators, we can generate the entire partition $\mathcal{\overbar{G}}_3/\mathcal{\overbar{Z}(\overbar{S})}/\mathcal{\overbar{S}}$. First, we obtain the set $\{\mathbf{S}_v\}_{v=0}^{3}$ of elements that make up $\mathcal{\overbar{S}}$ from the stabilizer generators $\mathbf{S}_1 = XYZ$ and $\mathbf{S}_2 = YXI$, where $\mathbf{S}_0$ is the trivial generator $III$. Following this, we can obtain the set of operators that define the effective centralizer $\mathcal{\overbar{Z}(\overbar{S})}$ by computing the products $\mathbf{L}_j\star\mathbf{S}_v$, where $j=1,\ldots 4$. Lastly, we can obtain the rest of the $3$-qubit effective Pauli space $\mathcal{\overbar{G}}_3$ by computing the products $(\mathbf{T}_i  \star  \mathbf{L}_j)  \star  \mathbf{S}_v$, where $i=1,\ldots,4$, $j=1,\ldots,4$.

\begin{table}[h!]
\centering
    \caption{Coset representatives $\{\mathbf{T}_i\}_{i=1}^{4}$ and $\{\mathbf{L}_j\}_{j=1}^{4}$.}
    \label{tab:info}
    \setlength{\tabcolsep}{6pt}
    \begin{tabular}{|p{30pt}|p{30pt}|p{30pt}|p{30pt}|}
        \hline
{$\mathbf{L}_1$}
                    &  $III$ & $\mathbf{T}_1$ &$III$ \\ [0.1cm]
        \hline
{$\mathbf{L}_2$}
                    &  $YIX$ & $\mathbf{T}_2$ &$XII$\\[0.1cm]
 \hline
$\mathbf{L}_3$
                    &  $YIY$ & $\mathbf{T}_3$ &$YII$\\[0.1cm]
        \hline
        $\mathbf{L}_4$
                    &  $IIZ$ & $\mathbf{T}_4$ &$ZII$\\[0.1cm]
        \hline
    \end{tabular}
\end{table}

Table \ref{tab:info} also serves to provide practical examples of the decomposition given in \eref{eq:decomp} for operators in $\overbar{\mathcal{G}}_N$. For instance, consider the operator $\mathbf{A}_1 = XYX$ which is the second element of the coset $(\mathbf{T}_3 \star \mathbf{L}_3) \star \mathcal{\overbar{S}}$ (leftmost bottom green rectangle in Figure \ref{fig:cosets_example}). Based on the aforementioned decomposition, we should be able to write this operator as the $\star$ product of its pure, logical and stabilizer components. This can be easily obtained as 
$$\mathbf{A}_1 = XYX = YII \star YIY \star XYZ\ , $$ where $\mathbf{T}(\mathbf{A}_1) = YII$, $\mathbf{L}(\mathbf{A}_1) = YIY$, and $\mathbf{S}(\mathbf{A}_1) = XYZ$, denote the pure error, logical and stabilizer components, respectively.

\begin{figure*}[!htp]
	\centering
	\subfloat[ \label{irre}]{%
		\includegraphics[width=.5\textwidth, height = 2.25in]{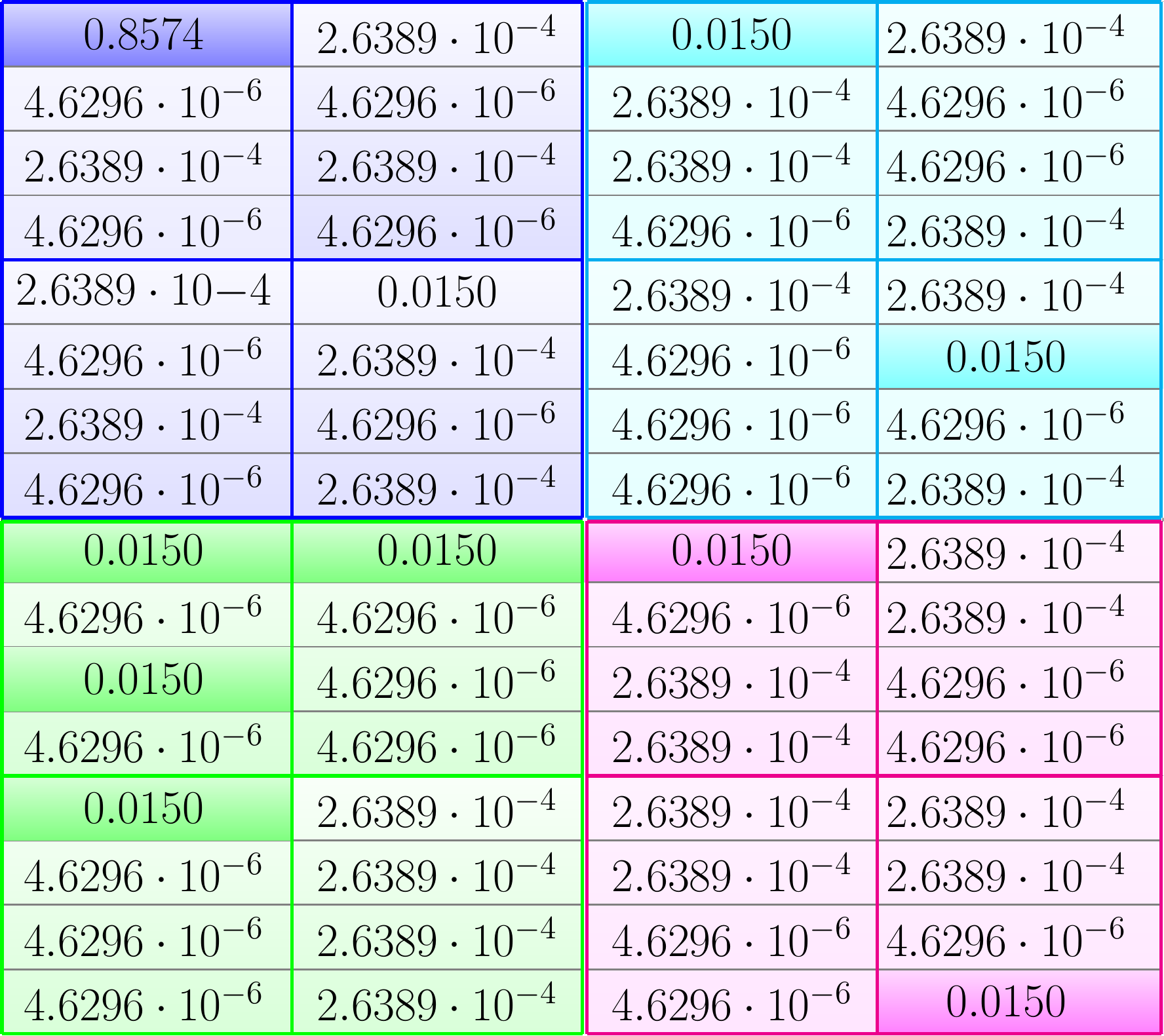}\label{fig:probabilities_example}
	} \hfill
	\subfloat[ \label{irre2}]{%
		\includegraphics[width=.46\textwidth, height = 2.25in]{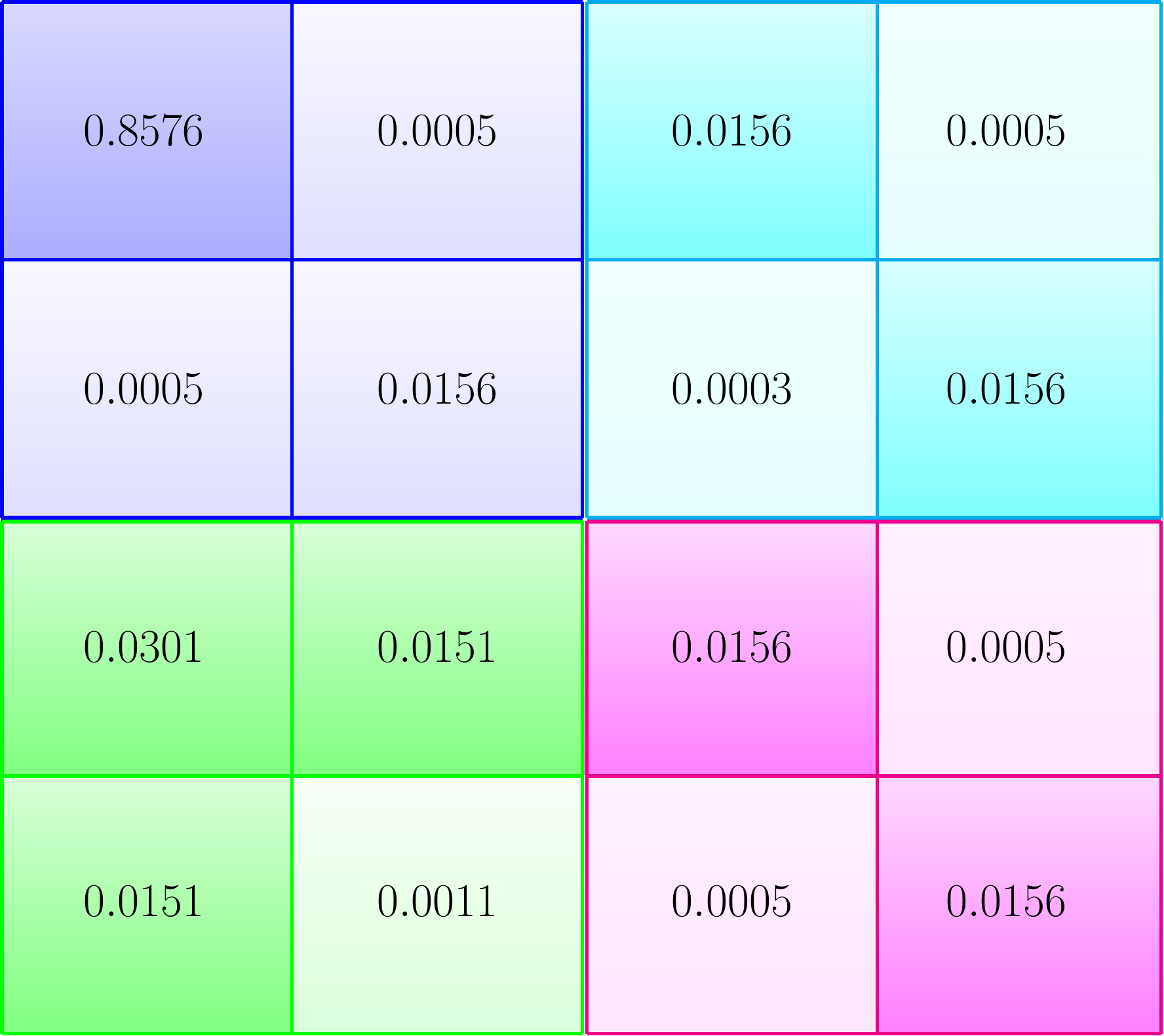} \label{fig:probabilities_cosets_example}
	}\caption{\textbf{(a)} Probability of the Pauli operators shown in Figure \ref{fig:cosets_example} when a depolarizing channel with parameter $p = 0.05$ is considered. The darker shaded rectangles denote the operators from each centralizer coset that could be obtained by the classical SPA decoder of \eref{eq:ML1-deg}. \textbf{(b)} Probability of each stabilizer coset considered in the partition $\mathcal{\overbar{G}}_3$ shown in Figure \ref{fig:cosets_example} when a depolarizing channel with parameter $p = 0.05$ is considered. The darker shaded rectangles denote the stabilizer cosets from which the classical SPA decoder of (\ref{eq:ML1-deg}) can produce solutions. Notice that for syndrome $\mathbf{w}_1$, the result produced by the classical SPA decoder belongs to the stabilizer coset that has the highest probability, and thus classical decoding produces the same result as optimal quantum (degenerate) decoding.} \label{fig:probabilities_1}
\end{figure*}

Most importantly, Figure \ref{fig:cosets_example} can be used to illustrate the effects of degeneracy and the impact of employing suboptimal methods to decode QLDPC codes. Let us assume that we use our $3$-qubit stabilizer code to transmit quantum codeword $\ket{\overbar{\psi}}$ over a depolarizing channel with depolarizing probability $p = 0.05$, and that we obtain the noisy quantum state $\ket{\overbar{\psi}}_{\xi_P}$ at the channel output. Based on the probability distribution of the depolarizing channel shown in section \ref{sec:channels}, we can compute the probability of every possible $3$-qubit error operator shown in Figure \ref{fig:cosets_example} that may have acted on the transmitted quantum codeword. These probabilities are shown in Figure \ref{fig:probabilities_example}, which also shows the operators that the traditional SPA decoder of \eref{eq:ML1-deg} would estimate out of each of the effective centralizer cosets. For instance, if the syndrome extracted from  $\ket{\overbar{\psi}}_{\xi_P}$ is $\mathbf{w}_1 = [0 \ 0]$, decoding based on \eref{eq:ML1-deg} would yield $\mathbf{\hat{E}} = III$, which is the most likely error operator in the coset $\mathbf{T}_1 \star \mathcal{\overbar{Z}(\overbar{S})}$. In this case, having a decoder with the ability to consider the partition into stabilizer cosets would not yield any gain, because the probability of the coset indexed by $\mathbf{T}_1 \star \mathbf{L}_1$ is vastly greater than the probability of the remaining stabilizer cosets. This is shown in Figure \ref{fig:probabilities_cosets_example}, where the probabilities of the stabilizer cosets (the sum over each of the stabilizer equivalence classes) are depicted. Thus, for this scenario, a decoder based on the classical SPA algorithm and a degenerate decoder would attain the same results.

Let us now consider the unlikely\footnote{The likelihood of such an outcome will increase as the depolarizing probability of the channel increases.} outcome that performing measurements on $\ket{\overbar{\psi}}_{\xi_P}$ yields the syndrome $\mathbf{w}_4 = [1 \ 1]$. Successful decoding based on the traditional SPA decoder would yield the lowest weight operator from the coset $\mathbf{T}_4 \star \mathcal{\overbar{Z}(\overbar{S})}$ which is either the error operator $ZII$ belonging to coset $\mathbf{T}_4\star\mathbf{L}_1\star\mathcal{\overbar{S}}$ or the error operator $IZI$ belonging to coset $\mathbf{T}_4\star\mathbf{L}_4\star\mathcal{\overbar{S}}$. Notice, as shown in Figure \ref{fig:probabilities_cosets_example}, that the probability of the coset $\mathbf{T}_4\star\mathbf{L}_1\star\mathcal{\overbar{S}}$ is identical to that of the coset $\mathbf{T}_4\star\mathbf{L}_4\star\mathcal{\overbar{S}}$. As in the previous case, a degenerate decoder capable of considering the partition of the effective centralizer cosets into stabilizer cosets would not outperform the traditional SPA decoder in this scenario. Note that, for both of the previous examples, this happens because the error sequence with highest probability is in the stabilizer coset with highest probability. As shown in \eref{eq:eq-jgf}, regardless of whether the estimated error sequence and the channel error match, the traditional SPA decoder will correctly solve the decoding problem if the estimated error sequence is in the same stabilizer coset as the channel error.

\begin{figure*}[!htp]
	\centering
	\subfloat[ \label{irre3}]{%
		\includegraphics[width=.5\textwidth, height = 2.25in]{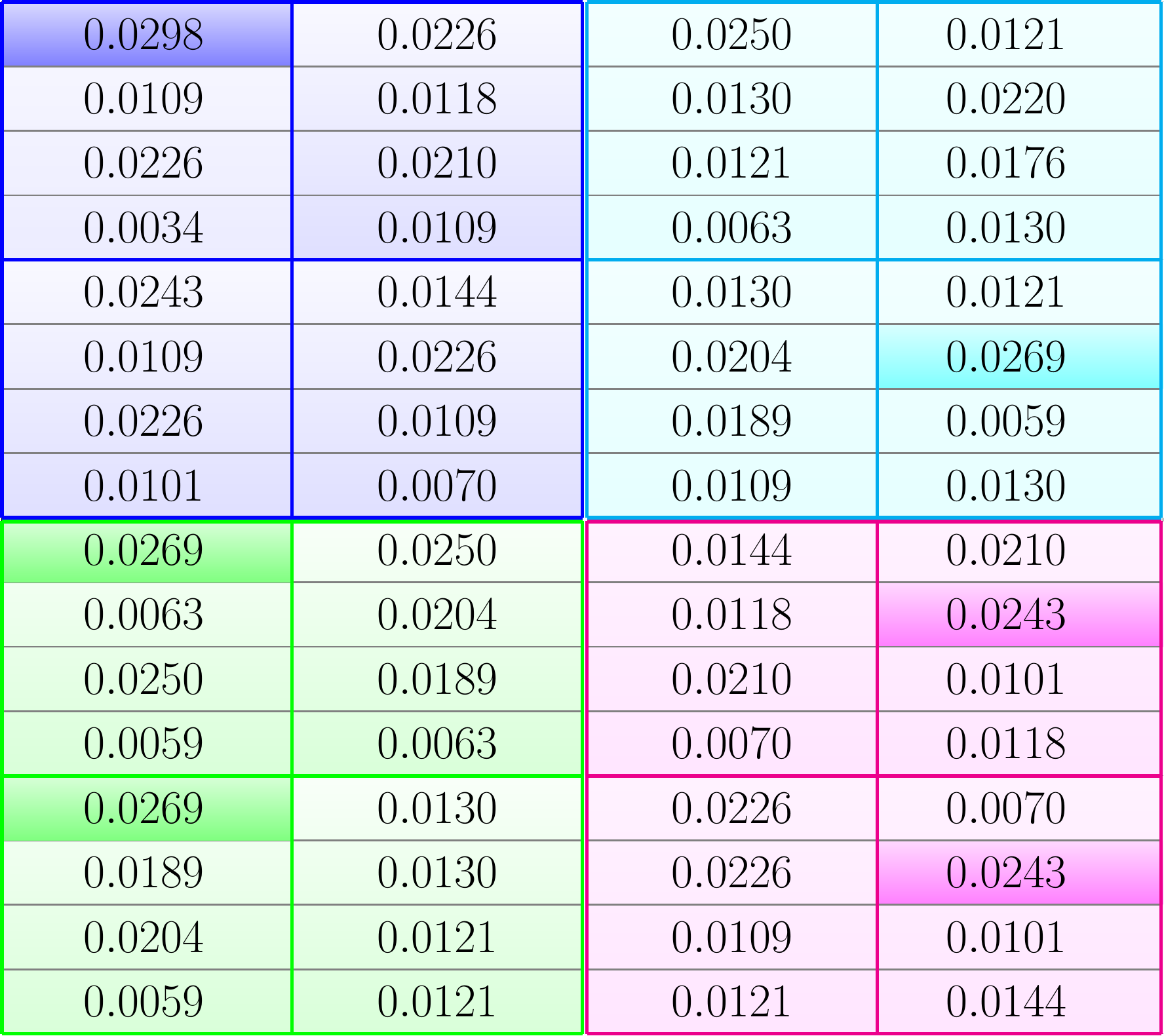}\label{fig:probabilities_2_A}
	} \hfill
	\subfloat[ \label{irre4}]{%
		\includegraphics[width=.46\textwidth, height = 2.25in]{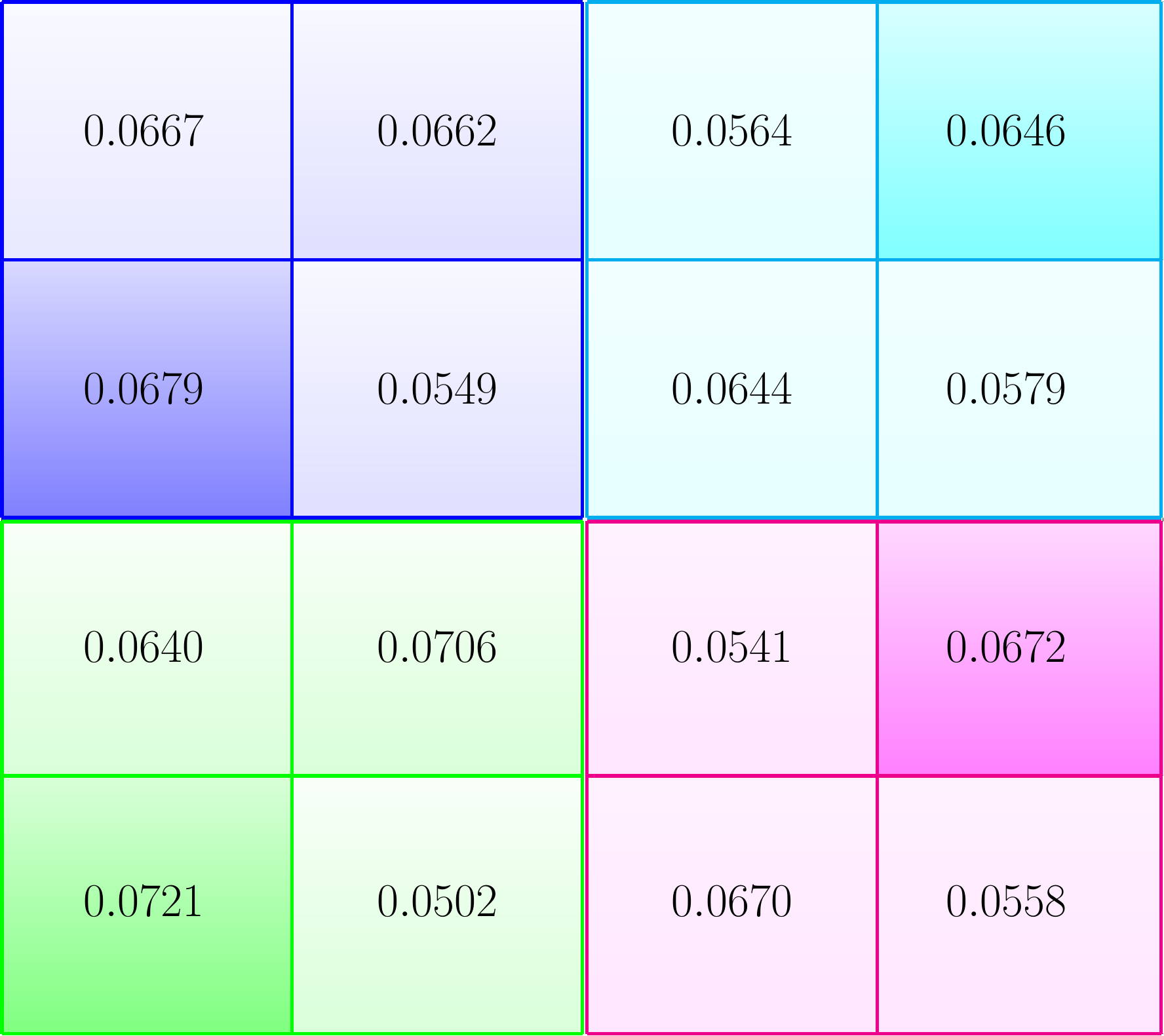} \label{fig:probabilities_2_B}
	}\caption{\textbf{(a)} (a) Probability of the Pauli operators that are shown in Figure \ref{fig:cosets_example} when a Pauli channel with parameters $p_x = 0.26$, $p_y=0.28$, and $p_z = 0.15$ is considered. The darker shaded rectangles denote the error operators from each centralizer coset that could be obtained by the classical SPA decoder of \eref{eq:ML1-deg}. \textbf{(b)} Probability of each stabilizer coset considered in the partition $\mathcal{\overbar{G}}_3$ shown in Figure \ref{fig:cosets_example} when a Pauli channel with parameters $p_x = 0.26$, $p_y=0.28$, and $p_z = 0.15$ is considered. The darker shaded rectangles denote the stabilizer cosets with the highest probability within each effective centralizer coset. Notice that for syndrome $\mathbf{w}_1$, the result produced by the classical SPA decoder does not belong to the stabilizer coset that has the highest probability, and thus classical decoding underperforms when compared to optimal quantum (degenerate) decoding.} \label{fig:probabilities_2}
\end{figure*}

To provide an example in which the degenerate decoder clearly outperforms the traditional SPA decoder, let us now consider transmission of encoded quantum states over a Pauli channel with $p_x = 0.26$, $p_y=0.28$, and $p_z = 0.15$. Figures \ref{fig:probabilities_2_A} and \ref{fig:probabilities_2_B} portray the individual probabilities of each of the operators shown in Figure \ref{fig:cosets_example} and the probabilities of each stabilizer coset, respectively, after transmitting over the Pauli channel with $p_x = 0.26$, $p_y=0.28$, and $p_z = 0.15$. In this case, there are multiple scenarios in which the optimal (degenerate) QLDPC decoder would outperform the traditional SPA decoder. Let us discuss some of them.

Assume, as previously, that we have transmitted a quantum codeword over this channel and that an error $\mathbf{E}_2 \in \mathbf{T}_1 \star \mathcal{\overbar{Z}(\overbar{S})}$ has been introduced by the Pauli channel, which results in the syndrome measurement yielding $\mathbf{w}_1 = [0\ 0]$. A traditional SPA decoder would be capable of determining that $\mathbf{T}(\mathbf{E}_2) = \mathbf{T}_1 = III$, and would choose the error operator contained in the effective centralizer coset $\mathbf{T}_1 \star \mathcal{\overbar{Z}(\overbar{S})}$ that has the highest probability of occurring, i.e., $\hat{\mathbf{E}}_2^{\text{SPA}} = III$ (see Figure \ref{fig:probabilities_2_A}). However, Figure \ref{fig:probabilities_2_B} shows that the probability of the stabilizer coset $(\mathbf{T}_1 \star \mathbf{L}_1) \star \mathcal{\overbar{S}}$, which contains the most likely operator, $\hat{\mathbf{E}}_2^{\text{SPA}} = III$, is lower than the probability of the stabilizer coset $(\mathbf{T}_1 \star \mathbf{L}_3) \star \mathcal{\overbar{S}}$. This means that producing an estimate $\hat{\mathbf{E}}_2 \in (\mathbf{T}_1 \star \mathbf{L}_3) \star \mathcal{\overbar{S}}$ will result in better performance than choosing the most likely error operator in $\mathbf{T}_1 \star \mathcal{\overbar{Z}(\overbar{S})}$. Such an estimate can only be produced by the decoder given in \eref{eq:DQMLD}, which is capable of considering the partition of the effective centralizer into cosets of the stabilizer. Therefore, in this instance, the degenerate decoder would outperform the traditional SPA decoder. This would also occur if an error  $\mathbf{E}_3 \in \mathbf{T}_2 \star \mathcal{\overbar{Z}(\overbar{S})}$ were to take place. By comparing Figures \ref{fig:probabilities_2_A} and \ref{fig:probabilities_2_B}, we can see how, in this scenario, the most likely error operator does not belong to the most likely stabilizer coset.

In the case that $\mathbf{E} \in \mathbf{T}_3 \star \mathcal{\overbar{Z}(\overbar{S})}$ or $\mathbf{E} \in \mathbf{T}_4 \star \mathcal{\overbar{Z}(\overbar{S})}$, there are multiple error operators that have the highest probability to occur, but only one of them belongs to the most likely stabilizer coset. This implies that, although performance of the traditional SPA decoder would be better than in the two previous cases, it would sometimes produce suboptimal estimates that would not belong to the most likely stabilizer coset. Therefore, in instances such as this one, the optimal degenerate decoder would also outperform traditional SPA decoding methods. 

Lastly, it is important to mention that when a traditional SPA-based decoder is used to estimate operators from $\mathcal{\overbar{G}}_N$ and it fails to find the correct logical component of the error, this goes unnoticed. Recall that although the effect of logical operators on the codeword is non-trivial, their actions preserve the codespace. Based on this fact, it is easy to see how these undetected decoding errors can quickly become a significant nuisance when using stabilizer codes with a large number of stabilizer elements. As mentioned earlier, this is a common occurrence when operating with QLDPC codes that utilize a large number of qubits, as they may sometimes have many low weight operators spread out across their stabilizer cosets. In such scenarios, hefty performance prices may potentially be paid in consequence of employing a suboptimal decoder based on traditional SPA decoding.

\section{Chapter Summary}

The quantum phenomenon known as degeneracy should theoretically improve the performance of quantum codes. Unfortunately, degeneracy can end up sabotaging the performance of sparse quantum codes because its existence is neglected in the decoding process. In this chapter we have provided a broad overview of the role this phenomenon plays in the realm of quantum error correction and, more specifically, in the field of QLDPC codes. We began by introducing a group theoretical explanation of the most relevant concepts in the field of quantum error correction. Following this, we examined the differences between the classical decoding problem and its quantum counterpart. Despite the intricate similarities between classical and quantum decoders, the coset partition of the $N$-fold Pauli space in the context of stabilizer codes reveals the higher complexity of the quantum decoding task. Based on this discussion, we were later able to show why applying the classical decoding algorithm for sparse codes to the quantum problem is suboptimal. Then, we provided a detailed explanation regarding the origin of degeneracy and how it may have detrimental effects on traditional SPA-based decoders. We finished with an example that serves to facilitate the comprehension of the topics discussed throughout the chapter.

\clearemptydoublepage
\chapter{Detecting Degeneracy and Improved Decoding Strategies}

\epigraph{\textit{``Our need will be the real creator''}}{\textbf{Plato}.}

\noindent\hrulefill

In the previous chapter we analyzed the phenomenon of degeneracy through the lens of group theory. We introduced the coset structure of stabilizer codes and showed how degeneracy is responsible for the existence of error equivalence classes (stabilizer cosets) whose individual elements act identically on the transmitted information (encoded quantum states). Following this, we presented how, when using decoding based on the QMLD rule, different decoding scenarios can be encountered. We referred to these scenarios as end-to-end errors, and showed how they represent particular instances that must be distinguished in order to both assess and improve the performance of stabilizer codes. For example, it is necessary to differentiate between end-to-end degenerate errors and end-to-end identical syndrome errors, since the former type of end-to-end error has no impact on encoded quantum states but the latter one does (it acts non-trivially on the codespace). This is especially relevant when studying sparse quantum codes, whose degenerate nature will more than likely increase the frequency with which degenerate errors occur \cite{QSC, degen3, degen4, logical, softVit, neural, Hard, softPoul, reviewPat}. 

In this chapter, we employ the group theoretic framework presented in Chapter \ref{chapter4} to derive a strategy capable of accurately assessing the impact of degeneracy on the performance of sparse quantum codes. We also explain another method that has been used in \cite{osd1, osd2, refined, exploiting} to do so. Then, we use our methodology to show how sparse quantum codes should always be assessed using a metric known as the logical error rate. We back our claims up by using our strategy to analyze the frequency with which different types of errors occur when using sparse quantum codes. Finally, we provide insight on how the design and decoding of these codes can be improved, and we discuss the most relevant advancements that have appeared in the literature \cite{bicycle, qldpc15, degen3, mod-BP, efb,  osd1, osd2, refined, jgf3}.

\section{Performance metrics}

Despite its theoretical performance benefits, limited research exists on how to quantify the true impact that degeneracy has on QLDPC codes. This has resulted in the performance of sparse quantum codes being assessed differently throughout the literature; while some research considers the effects of degeneracy by computing a metric known as the \textit{logical error rate} \cite{degen3, mod-BP, osd1, osd2, refined, exploiting, Til}, other works employ the classical strategy of computing the \textit{physical error rate} \cite{qldpc15, jgf1, jgf2, patrick, efb, bab11}, a metric which provides an upper bound on the performance of these codes. Let us analyze the differences between the two.

\subsection{Physical and Logical Error rate}

The manifestation of the degeneracy phenomenon in the realm of quantum error correction makes it impossible to accurately predict the performance of QEC codes with methods that disregard its presence. The more degenerate the code family that is being studied, the larger the discrepancy between results computed based on degeneracy-ignoring metrics and the true performance of these codes will be. Thus, in the paradigm of sparse quantum codes appropriate performance assessment is of paramount importance. As was mentioned in the previous chapter, this requires a method that can distinguish between the different types of end-to-end errors. For reference, the characteristics of these end-to-end errors are summarized in table \ref{errors} (see section \ref{sec:end-to-end} for a more complete discussion on end-to-end errors). Fortunately, a metric capable of telling different end-to-end errors apart already exists. It is known as the logical error rate and it has been widely employed to assess the performance of various other families of QEC codes such as QTCs and quantum topological codes \cite{softVit, EAQTC, toricphd1, toricphd2, sabo}. Generally, the decoders used in QTC or topological schemes are capable of producing estimates of the stabilizer coset representatives (the logical operators) of the channel errors, hence these decoders ``inherently'' compute the logical error rate. In contrast, the classical decoding algorithms \cite{BP, spa} used to decode QLDPC codes by solving the QMLD rule \eref{eq:ML1-deg} are unable to produce estimates of stabilizer coset representatives, and so their logical error rate must be computed differently. 

\begin{table*}[htp!]

    \centering\caption{Characteristics of the different types of end-to-end errors that can arise when using stabilizer codes. The hat notation is used to represent estimations made by the decoder, i.e., $\hat{\mathbf{w}}$, $\hat{\mathbf{E}}$, $\mathbf{\hat{T}}_i$, and $\mathbf{\hat{L}}_j$ represent the estimated syndrome, estimated error sequence, centralizer coset representative of the estimated error sequence and stabilizer coset representative of the estimated error sequence, respectively. }
    \vspace{0.15mm}\begin{tabular}{cccc}
    \toprule 
    Type of &Defining  &Outcome &Outcome\\ error &Characteristics &(Physical) &(Logical)\\
    \midrule
    End-to-end &$\hat{\mathbf{w}} \neq \mathbf{w}$\\
    error with &$\mathbf{\hat{T}}_i\neq \mathbf{T}_i$ &Failure &Failure \\
     different syndrome  &$\hat{\mathbf{E}}\neq \mathbf{E}$ \\
     \midrule
    End-to-end &$\hat{\mathbf{w}} = \mathbf{w}$\\
    identical &$\mathbf{\hat{T}}_i = \mathbf{T}_i$ and $\mathbf{\hat{L}}_j\neq \mathbf{L}_j$  &Failure &Failure\\
     syndrome error &$\hat{\mathbf{E}}\neq \mathbf{E}$ \\
     \midrule
    End-to-end  &$\hat{\mathbf{w}} = \mathbf{w}$\\
    degenerate &$\mathbf{\hat{T}}_i = \mathbf{T}_i$ and $\mathbf{\hat{L}}_j = \mathbf{L}_j$  &Failure &Success\\
    error  &$\hat{\mathbf{E}}\neq \mathbf{E}$ \\
    \bottomrule
    \label{errors}
    \end{tabular}
    \end{table*}

With our knowledge regarding the coset structure of sparse quantum codes, intuition would point towards calculating the logical error rate by finding and comparing the stabilizer cosets of the estimated error sequences and the stabilizer cosets of the channel errors. Unfortunately, the task of computing stabilizer cosets has been shown to be computationally hard \cite{degen4, Hard, efb}, which is the reason why the performance of some sparse quantum codes \cite{jgf1, jgf2, qldpc15, patrick, bab11} and some improved decoding strategies \cite{efb} has been assessed based on the \textit{physical error rate}. This metric is computed by comparing the error sequence estimated by the decoder, $\hat{\mathbf{E}} \in \mathcal{\overbar{G}}_N$, to the channel error, $\mathbf{E} \in \mathcal{\overbar{G}}_N$. Essentially, if the estimation matches the channel error, the decoder has been successful, and if not, a decoding failure has occurred. Note, however, that because this metric ignores the degenerate nature of stabilizer codes, the physical error rate overestimates the number of decoding failures and actually represents an upper bound on the performance of stabilizer codes.

Despite the use of the physical error rate in some works, other literature has successfully computed the logical error rate of specific sparse quantum codes \cite{degen3,  mod-BP, osd1, osd2, refined, exploiting, Til}. These works succeed in computing the logical error rate because they do not approach the issue from the perspective of stabilizer cosets. Instead, in most of these works \cite{degen3, osd1, osd2, refined, exploiting}, they use Gaussian elimination to obtain the parity check matrix of the code in what is known as \textit{standard form} \cite{QSC, Benny, Wilde}, and then use it to extract a basis for the encoded Pauli operators of the corresponding codes. Then, they employ this basis to differentiate between different types of end-to-end errors. Additionally, other research \cite{mod-BP} employs a different, albeit much more computationally demanding, method to compute the logical error rate. These strategies and concepts are explained in what follows.

\subsubsection*{Undetected classical errors}

Before studying how the logical error rate can be computed, it should be mentioned that the concept of \textit{undetected} errors is not exclusive to the quantum paradigm. In fact, even though degeneracy does not exist in the classical coding realm, \textit{undetected} or \textit{logical} errors in classical LDPC codes and classical turbo codes have previously been studied. This idea was introduced by the early work of MacKay et al. \cite{ldpc3}, where a classical undetected error is defined as a decoding estimate that is not equal to the original error sequence and that is produced when the decoder exits before the maximum number of decoding iterations (it produces a valid syndrome). More explicitly, a classical undetected error occurs when the estimate of the channel error is not equal to the real error but the estimated syndrome and the measured syndrome match. In their analysis of classical LDPC codes, MacKay et al. showed that all of the decoding mistakes they encountered were detected errors (classical undetected errors were only observed in turbo codes). This was also shown in \cite{refined} for a slightly different decoding algorithm. Similar outcomes were observed in the quantum paradigm for the failed recoveries of the modified decoding strategies of \cite{degen3, efb}. The techniques of \cite{degen3, efb} serve to improve standard SPA decoding of quantum codes by post-processing the initial error estimates and producing new estimates of the channel error. If these new estimates do not revert the channel error, then they are referred to as failed recoveries or failed error corrections. In \cite{degen3}, all of the failed error corrections were shown to be end-to-end errors with different syndromes, whereas in \cite{efb} a small percentage of failed estimates were shown to be end-to-end identical syndrome errors and end-to-end degenerate errors\footnote{The authors of this work do not distinguish between end-to-end identical syndrome errors and end-to-end degenerate errors.}.

Aside from these failed recovery analyses, the literature is limited when it comes to assessing the percentage of decoding failures that is caused by each type of end-to-end error. It is reasonable to believe that QLDPC codes, given their large number of degenerate operators \cite{degen3, degen4, Hard, reviewPat, QEClidar, TPS}, will experience a large number of end-to-end degenerate errors. We confirm this intuition in the final section of this chapter, where we show how up to $30\%$ of the end-to-end errors that take place when using the QLDPC codes of \cite{jgf1, jgf2, patrick} are degenerate. 

\subsection{Discriminating between different types of end-to-end errors}

It is clear from the defining characteristics of each specific type of end-to-end error (see table \ref{errors}) that end-to-end errors with different syndromes are the easiest type of error to identify. In fact, doing so is trivial, as all that is required is a comparison of the syndrome estimate, $\mathbf{\hat{w}} \in  \mathbb{F}_2^{N-k}$, and the measured syndrome, $\mathbf{w} \in \mathbb{F}_2^{N-k}$, where $k$ and $N$ represent the number of logical qubits and physical qubits (block length) of the quantum code, respectively. Similarly, knowing that either an end-to-end identical syndrome error or an end-to-end degenerate error has occurred is simple. This can be done by comparing the estimate of the error sequence $\mathbf{\hat{E}}$ to the channel error $\mathbf{E}$ whenever $\mathbf{\hat{w}} = \mathbf{w}$, i.e., if $\mathbf{\hat{E}} \neq \mathbf{E}$ either an end-to-end degenerate error or an identical syndrome error has occurred, and if $\mathbf{\hat{E}} = \mathbf{E}$, no error has taken place. The issue arises when trying to distinguish between these two families of end-to-end errors. Notice that the comparison $\mathbf{\hat{E}} \neq \mathbf{E}$ does not reveal if the error estimate belongs to the same stabilizer coset as the channel error. Hence, we have no way of discriminating between end-to-end degenerate errors and end-to-end identical syndrome errors.

The first strategy that comes to mind to resolve this problem is to compute the stabilizer of the code in question, calculate the $\star$ product of the stabilizer with the channel error $\mathbf{E}$ to extract the specific stabilizer coset of the channel error, and then check if $\mathbf{\hat{E}}$ belongs to this coset. This works because the coset representative choice for the coset $(\mathbf{T}_i\star\mathbf{L}_j)\star\mathcal{\overline{S}}$ is irrelevant (any operator belonging to the coset serves as a valid representative). Thus, computing $\mathbf{E} \star \mathcal{\overline{S}}$ will yield the stabilizer coset of the channel error, i.e.,  $\mathbf{E} \star \mathcal{\overline{S}} = (\mathbf{T}_i\star\mathbf{L}_j)\star\mathcal{\overline{S}}$. Therefore, whenever $\mathbf{\hat{w}} = \mathbf{w}$ and $\mathbf{\hat{E}} \neq \mathbf{E}$, we will know that an end-to-end degenerate error has occurred if the estimated error sequence $\mathbf{\hat{E}}$ is in the coset $\mathbf{E} \star \mathcal{\overline{S}}$. If this does not occur, then an end-to-end identical syndrome error will have taken place. 

Unfortunately, since extracting the stabilizer of a quantum code becomes increasingly complex as its block length increases, this strategy will only be applicable to short quantum codes. We know from Chapter \ref{chapter4} that the number of elements in the stabilizer of a quantum code with block length $N$ and rate $R_Q$ is given by $2^{N-k}=2^{N(1-R_Q)}$, which grows exponentially with $N$ and can rapidly become intractable on a classical machine as this parameter increases. In light of this, it is apparent that more practical methods to differentiate between end-to-end degenerate errors and end-to-end identical syndrome errors are necessary. Both the strategy we propose herein and that employed in \cite{degen3, refined, osd1, osd2, exploiting}, which are explained in the next two sections, resolve this issue.

\section{An algebraic perspective on end-to-end degenerate errors}

The problem of differentiating between end-to-end identical syndrome errors and end-to-end degenerate errors can also be formulated as a set of linear equations. We know from section \ref{sec:QPCM} that the QPCM of a stabilizer code that encodes $k$ logical qubits into $N$ physical qubits (a rate $R_Q = \frac{k}{N}$ code with block length $N$) can be written as
\begin{equation}\label{eq:pcm}
\mathbf{H}_{\mathcal{\overline{S}}} = \begin{pmatrix}
\mathbf{h}_1 \\
\mathbf{h}_2 \\
\vdots \\
\mathbf{h}_{N-k}
\end{pmatrix},
\end{equation}
where $\mathbf{h}_v = \mathbf{s}_v$ denotes the symplectic representation of the generators $\{\mathbf{S}_v\}_{v=1}^{N-k} \in \mathcal{\overline{G}}_N$ that define the stabilizer group. Each of the elements of $\mathcal{\overline{S}}$ is a linear combination of the $N-k$ generators, hence, if $\mathbf{S}$ is an element of the stabilizer and $\mathbf{s}$ is the symplectic representation of this stabilizer element, then
\begin{equation}\label{eq:stabsarelinearcombs}
\mathbf{s} = \left(\sum_{v=1}^{N-k} \mathrm{a}_v \mathbf{h}_v\right)\mod{2},
\end{equation}
where $\mathbf{a} = (\mathrm{a}_1,\ldots,\mathrm{a}_{N-k})$ is a unique binary vector.

Whenever a channel error $\mathbf{E}$ takes place, the decoder will compute an estimate of this error and produce an estimate of the syndrome associated to it. As discussed previously, this syndrome only determines which specific effective centralizer coset the channel error belongs to. In other words, the syndrome diagnoses the effective centralizer coset representative $\mathbf{T}_i$ of the channel error. Assuming that an end-to-end error with different syndrome does not occur, the estimated syndrome will match the measured syndrome, hence the centralizer coset representative of the estimated error sequence and the centralizer coset representative of the channel error will also be the same, i.e., $ \mathbf{\hat{T}}_i = \mathbf{T}_i$. Thus, if we compute the $\star$ operation of the channel error $\mathbf{E}$ and the estimated error $\mathbf{\hat{E}}$, which can also be understood as the mod2 sum of their symplectic representations over $\mathbb{F}_2^{2N}$: $\beta(\mathbf{E})\oplus \beta(\mathbf{\hat{E}}) = \mathbf{e}\oplus\mathbf{\hat{e}}$, where $\beta$ denotes the symplectic map, the sequence $\mathbf{E}$ will be shifted to the effective centralizer $\mathcal{\overline{Z}}(\mathcal{\overline{S}})$. Note that we can also write this using the symplectic map as $\mathbf{E}\star\mathbf{\hat{E}} \in \mathcal{\overline{Z}}(\mathcal{\overline{S}}) \rightarrow \beta(\mathbf{E})\oplus \beta(\mathbf{\hat{E}}) \in \beta(\mathcal{\overline{Z}}(\mathcal{\overline{S}})) = \mathbf{e}\oplus\mathbf{\hat{e}}\in\Gamma_R$. Based on this, the issue of determining whether an end-to-end error is degenerate can be resolved by answering the following question:
\begin{equation}\label{eq:question}
\exists \ \mathbf{a} : \mathbf{e}\oplus\mathbf{\hat{e}} = \left(\sum_{i=1}^{N-k}\mathrm{a}_i \mathbf{h}_i\right)\mod{2} \ \text{ ?}
\end{equation}

Essentially, if a set of coefficients $\mathbf{a}$ exists such that the above equation holds, i.e., if $\mathbf{e}\oplus\mathbf{\hat{e}} \in \mathbb{F}_2^{2N}$ is a linear combination of the symplectic representation of the stabilizer generators, then $\mathbf{E}\star\mathbf{\hat{E}}$ belongs to the stabilizer and an end-to-end degenerate error will have occurred. If such a set of coefficients does not exist, then and end-to-end identical syndrome error has taken place.

The expression shown in \eref{eq:question} defines a linear system of Diophantine equations over the binary field. An answer to this question can be found by writing the augmented matrix $[\mathbf{H}_{\mathcal{\overline{S}}}^\top | (\mathbf{e}\oplus\mathbf{\hat{e}})^\top]$ in its row-echelon form. This means that, based on this procedure, it is possible to determine the type of end-to-end errors that occur and subsequently compute the logical error rate. This is done in \cite{mod-BP}. However, although the procedure is conceptually simple, rewriting the augmented matrix in such a manner becomes increasingly computationally complex as matrices grow in size. Unfortunately, for sparse quantum codes to be good, the block length must be large, which implies that the PCMs of these codes will also be large\footnote{The PCMs of QEC codes are of size $N-k\times 2N$. A common size in the literature of QLDPC codes is $10000$ qubits, thus the PCM associated to such a code would be of size $(10000-k)\times 20000$.}. Furthermore, the row-echelon form of the matrix $[\mathbf{H}_{\mathcal{\overline{S}}}^\top | (\mathbf{e}\oplus\mathbf{\hat{e}})^\top]$ must be computed during every simulation iteration (whenever the estimated syndrome and the measured syndrome match) to determine what type of end-to-end error has taken place, which may significantly increase simulation time. For these reasons, calculating the logical error rate based on this procedure can become a cumbersome and lengthy endeavour. Hence, the task at hand is to find a more practical and less computationally demanding way to determine if the congruence equation system given in \eref{eq:question} has a solution, as this suffices to determine if the end-to-end error is degenerate (we do not actually need to solve the system itself).

\subsection{Classical coding inspired strategy} \label{sec:method}

It is possible to find an answer to \eref{eq:question} by casting the problem in the framework of classical linear block codes. In classical coding theory, the encoding matrix or generator matrix $\mathbf{G}_c$ of a binary linear block code and its corresponding parity check matrix $\mathbf{H}_c$ fulfil $(\mathbf{G}_c\mathbf{H}_c^\top)\mod2 = (\mathbf{H}_c\mathbf{G}_c^\top)\mod2 = 0$. This means that the parity check matrix defines a basis for the null space of the generator matrix and viceversa. In the classical paradigm, having a basis for the null space of a code enables us to determine whether the decoding outcome $\mathbf{x}$ belongs to the code by simply computing its product with the parity check matrix of the code, i.e., if $(\mathbf{H}_c\mathbf{x}^\top)\mod 2 = 0$ then $\mathbf{x}$ is a codeword. Essentially, whenever $(\mathbf{H}_c\mathbf{x}^\top)\mod 2 = 0$, the decoding outcome is a linear combination of the rows of the generator matrix $\mathbf{G}_c$ and it belongs to the code, and whenever $(\mathbf{H}_c\mathbf{x}^\top)\mod 2 \neq 0$, $\mathbf{x}$ does not belong to the code. 

Notice that, based on this formulation, the quandary posed in \eref{eq:question} is reminiscent of the classical decoding scenario. The main difference is that instead of determining if the decoding outcome belongs to the code, we must discover if the sum of the symplectic representations of the channel error and the estimated error belong to the stabilizer. This parallelism between the classical and quantum problems allows us to apply the classical resolution strategy to the quantum paradigm with only a slight caveat: answering \eref{eq:question} requires an inverse approach to the classical method. Since the generators of the stabilizer code are given by the rows of the parity check matrix $\mathbf{H}_{\mathcal{\overline{S}}}$, the corresponding kernel generator matrix\footnote{We refer to the matrix $\mathbf{G}_{\mathcal{\overline{S}}}$ as the kernel generator matrix to avoid the term stabilizer generator matrix, as this latter term implies that the the matrix can be used for encoding purposes (which may not be true in the present case).} $\mathbf{G}_{\mathcal{\overline{S}}}$ (instead of the parity check matrix like in the classical paradigm) must be used to discover if $\mathbf{e}\oplus\mathbf{\hat{e}}$ can be written as a linear combination of the stabilizer generators. The matrix $\mathbf{G}_{\mathcal{\overline{S}}}$ defines a basis for the nullspace of the stabilizer code, hence it will suffice to compute $[\mathbf{G}^\top_{\mathcal{\overline{S}}}(\mathbf{e}\oplus\mathbf{\hat{e}})^\top]\mod 2$ to find the answer to \eref{eq:question}. If $[\mathbf{G}^\top_{\mathcal{\overline{S}}}(\mathbf{e}\oplus\mathbf{\hat{e}})^\top]\mod 2 = 0 \rightarrow \mathbf{E}\star\mathbf{\hat{E}} \in \mathcal{\overline{S}}$ which means that an end-to-end degenerate error has occurred, and if $[\mathbf{G}^\top_{\mathcal{\overline{S}}}(\mathbf{e}\oplus\mathbf{\hat{e}})^\top]\mod2 \neq 0 \rightarrow \mathbf{E}\star\mathbf{\hat{E}} \not\in\mathcal{\overline{S}}$, meaning that an end-to-end identical syndrome error will have taken place.

This strategy provides us with a simple and computationally efficient method to determine the type of end-to-end error that has taken place. The only requirement is obtaining the matrix $\mathbf{G}_{\mathcal{\overline{S}}}$, which can be computed once (by finding a basis for the nullspace of its parity check matrix $\mathbf{H}_{\mathcal{\overline{S}}}$) and can then be stored offline for any stabilizer code. In this manner, we have designed a simple method to solve \eref{eq:question} that does not require the computation of the stabilizer and so avoids the complexity issues that this entails.

\subsection{ Detecting end-to-end degenerate errors using encoded Pauli operators}

 There is another manner of distinguishing between end-to-end identical syndrome errors and end-to-end degenerate errors. It involves obtaining the encoded Pauli operators (see section \ref{sec:PE&LO}) of a code following a method derived by Gottesman in his seminal work \cite{QSC}, and then using these operators to determine whether the error estimate produced by the decoder is in the stabilizer coset of the channel error. This strategy was first applied to QLDPC codes in \cite{degen3}, and has since been used in \cite{osd1, osd2, refined, exploiting}. 
 
 Based on the definition of the encoded Pauli operators given in the previous chapter, we know that an encoded Pauli operator $\overline{\mathbf{Z}}_q$ commutes with all the elements of $\mathcal{\overline{S}}$ as well as with all other encoded Pauli operators except for the operator $\overline{\mathbf{X}}_l$ when $q=l$. This means that, if the encoded Pauli operators of a stabilizer code are known, we can determine if an operator $\mathbf{A} \in \mathcal{\overline{Z}(\overline{S})} \subset \mathcal{\overline{G}}_N$ belongs to $\mathcal{\overline{S}}$ by checking the commutation relations of $\mathbf{A}$ with the encoded Pauli operators. If $\mathbf{A}$ commutes with all the encoded Pauli operators it is within the stabilizer and if it does not (it anti commutes with one encoded Pauli operator) it is not in the stabilizer. Against this backdrop, it is easy to see how this strategy can be applied to solve the issue of discriminating between end-to-end identical syndrome errors and end-to-end degenerate errors. After successful decoding (the decoder produces a matching estimate of the syndrome), we know that $\mathbf{E}\star\mathbf{\hat{E}} \in\mathcal{\overline{Z}(\overline{S})}$. Now, we determine if $\mathbf{E}\star\mathbf{\hat{E}}$ is in $\mathcal{\overline{S}}$ by checking its commutation status with $\{\overline{\mathbf{Z}}_q\}_{q=1}^k$ and $\{\overline{\mathbf{X}}_l\}_{l=1}^k$. If $\mathbf{E}\star\mathbf{\hat{E}}$ commutes with all the encoded Pauli operators, $\mathbf{E}\star\mathbf{\hat{E}} \in \mathcal{\overline{S}}$ and an end-to-end degenerate error has occurred. If not, an end-to-end identical syndrome error has occurred. It is based on these comparisons that the logical error rate was successfully computed in \cite{osd1, osd2, refined, exploiting}.
  
  Naturally, to be able to apply this method, one must first have knowledge of the encoded Pauli operators of the code. This is similar to the classical coding-based strategy we proposed earlier, which requires the computation of the kernel generator matrix $\mathbf{G}_{\mathcal{\overline{S}}}$. The encoded Pauli operators of a stabilizer code can be found based on the concept of the \textit{standard form} (see Chapter 4 of \cite{QSC}). In this work, Gottesman showed how, by applying row operations (i.e. Gaussian elimination) together with the necessary qubit permutations (i.e. column permutations) on the parity check matrix of a stabilizer code, one can obtain a special row reduced echelon form of the parity check matrix: the standard form. Once the standard form is known, the encoded Pauli operators $\{\overline{\mathbf{Z}}_q\}_{q=1}^k$ and $\{\overline{\mathbf{X}}_l\}_{l=1}^k$ can be directly obtained from it using matrix algebra \cite{QSC, Benny}. 
  As with the kernel generator matrix $\mathbf{G}_{\mathcal{\overline{S}}}$, the encoded Pauli operators of a specific code need only be computed once and can then be stored offline.  
  
\section{Frequency of each type of end-to-end error} \label{sec:firstresults}

We now employ our method to study the frequency with which the different types of end-to-end error occur when using QLDPC codes. For this purpose, we simulate the CSS QLDGM codes of \cite{jgf1, jgf2, patrick} with different rates and block lengths over the depolarizing channel. These codes are presented in detail in the next chapter. If desired, the reader can skip ahead for a rigorous discussion on their design. However, this is not necessary to understand the remainder of this chapter. For reference, the particular characteristics of the simulated codes are detailed in table \ref{tab:codes}.

\begin{figure*}[!htp]
	\centering
	\subfloat[ \label{irre_12}]{%
		\includegraphics[width=.48\textwidth, height = 2.25in]{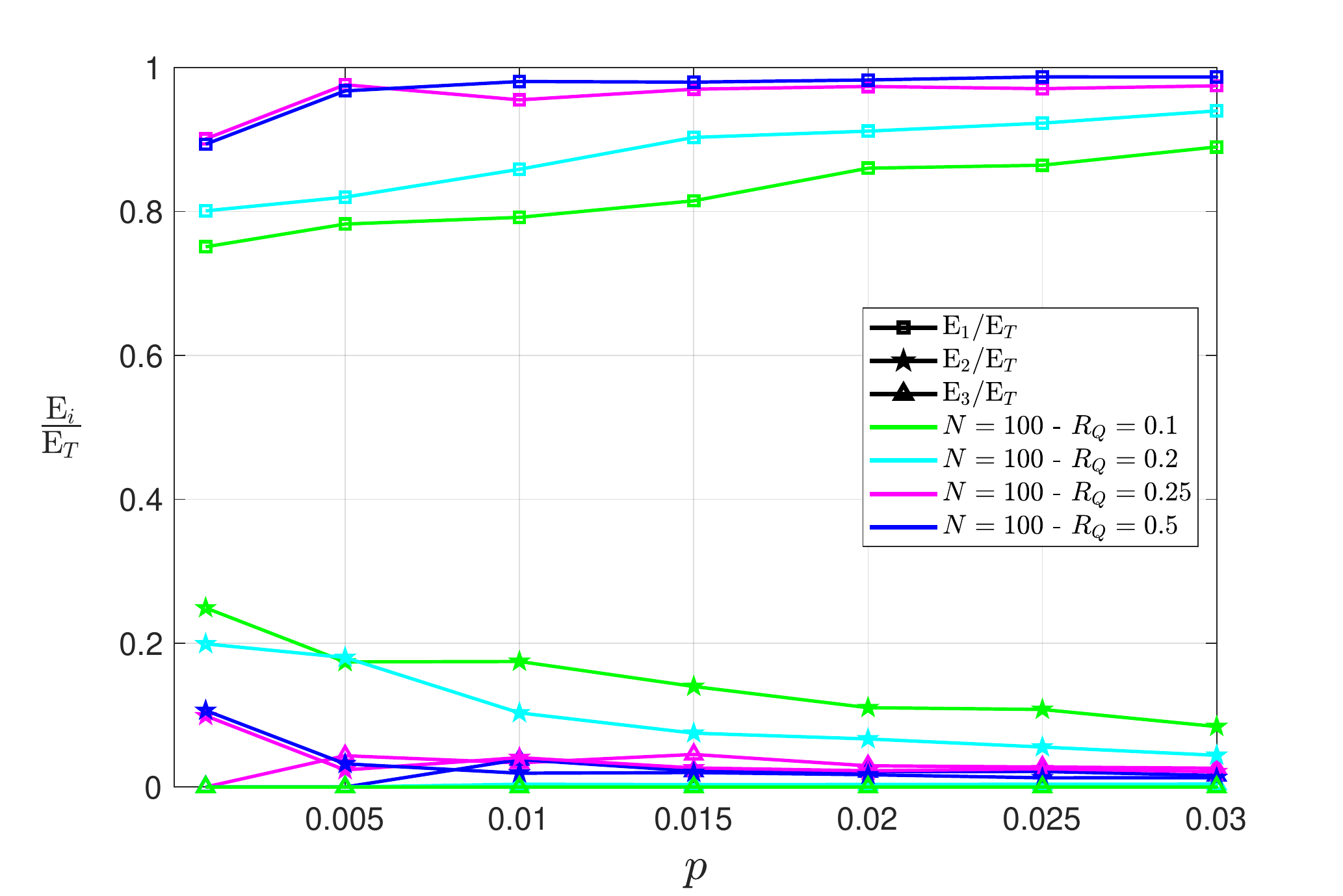}\label{fig:results_A}
	} \hfill
	\subfloat[ \label{irrelevant_af}]{%
		\includegraphics[width=.48\textwidth, height = 2.25in]{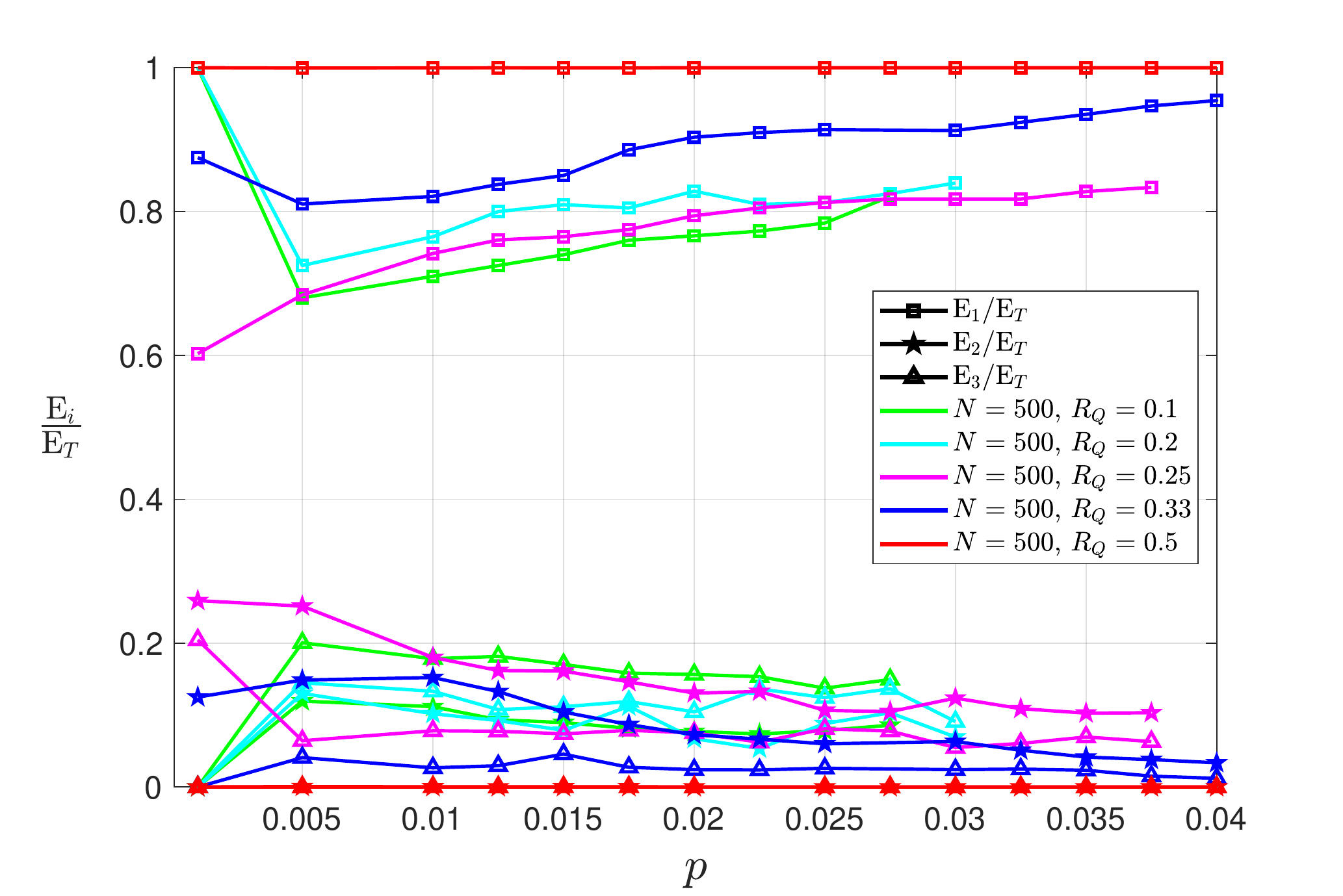} \label{fig:results_B}
	}\hfill
	\subfloat[ \label{irre234}]{%
		\includegraphics[width=.5\textwidth, height = 2.25in]{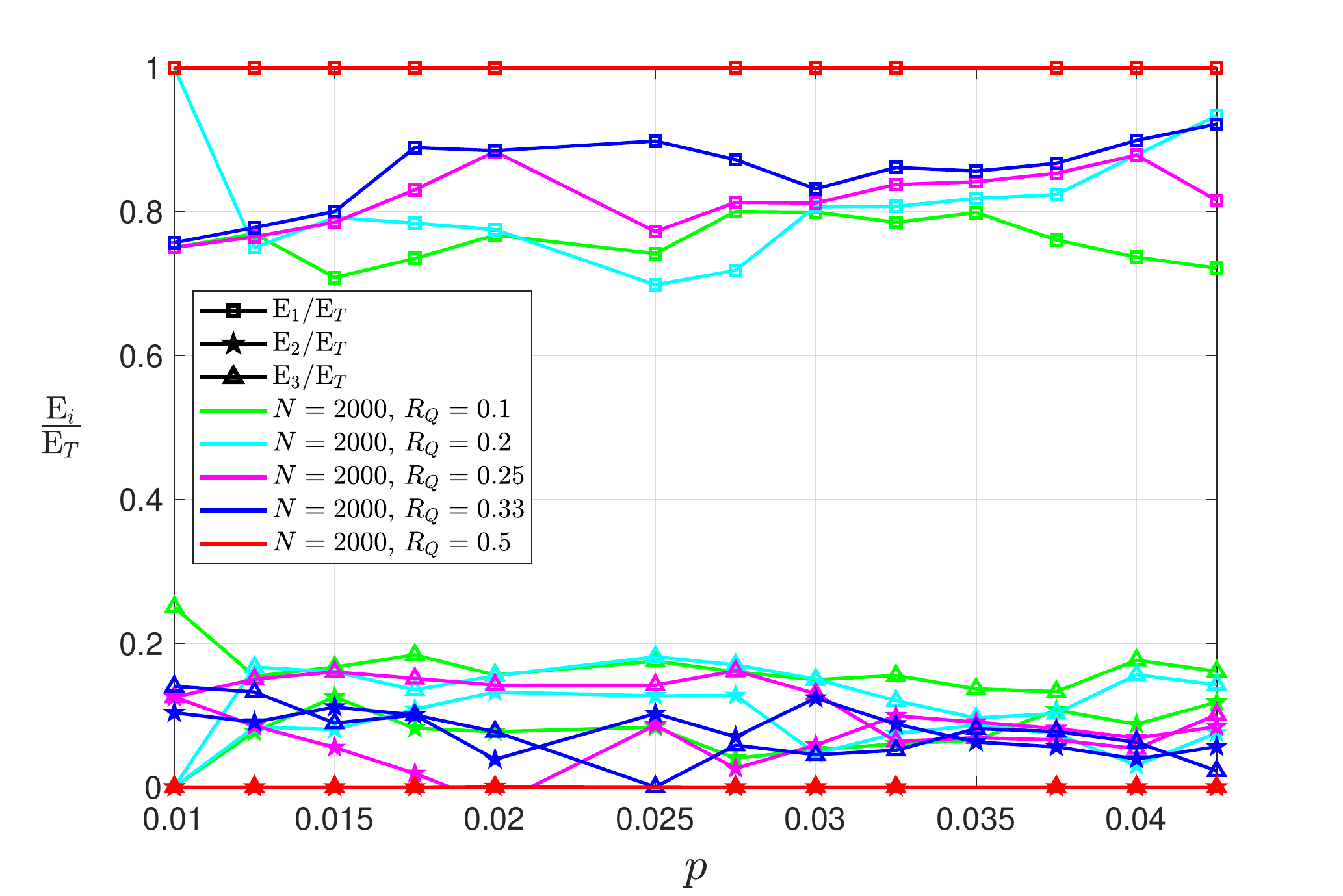} \label{fig:results_C}
	} 
	\caption{Ratios of end-to-end different syndrome errors, identical syndrome errors, and degenerate errors for codes of various rates and block lengths: \textbf{(a)} $N=100$ \textbf{(b)} $N=500$ \textbf{(c)} $N=2000$. .} \label{fig:results-codes}
\end{figure*}

The results of our simulations are shown in Figure \ref{fig:results-codes}, where each subfigure groups the results by block length, i.e, each of the subfigures portrays the results for all the codes with the same value of $N$. The graphs plot the ratio of a specific type of end-to-end error against the depolarizing probability. The aforementioned ratio is computed as $\frac{\mathrm{E}_i}{\mathrm{E}_T}$, where $\mathrm{E}_i$ denotes the total number of end-to-end errors of a specific type ($i = 1,2,3$.): 
\begin{itemize}
    \item $\mathrm{E}_1 \rightarrow$ end-to-end errors with different syndromes,
    \item $\mathrm{E}_2 \rightarrow$ end-to-end errors with identical syndromes,
    \item $\mathrm{E}_3 \rightarrow$ end-to-end degenerate errors,
    
\end{itemize} and $\mathrm{E}_T$ represents the total number of end-to-end errors. To ensure that the simulation results are precise, the ratios have been computed after a total of $1000$ decoding mistakes have been made (following the Monte Carlo simulation rule of thumb provided in \cite{MonteCarlo}), i.e., $\mathrm{E}_T = 1000$. For a complete discussion on how these simulations are conducted the reader is referred to Appendix \ref{app:sims}.

\begin{table}[h!]    
\centering
    \caption{\normalsize{Parameter values and configurations of simulated codes. }}
    \begin{tabular}{ccccc}
    \toprule 
    $N$ &$R_Q$ &Classical LDGM &$[m, p, x, y]$\\
    \midrule
    $100$ &$0.1$ &P($3,3$) &[$45, 24, 6, 3$] \\
    $100$ &$0.2$ &P($3,3$) &[$40, 18, 6, 3$] \\
    $100$ &$0.25$ &P($3,3$) &[$38, 15, 6, 3$] \\
    $100$ &$0.5$ &P($3,3$) &[$25, 6, 7.57, 3$] \\\midrule
    $500$ &$0.1$ &P($5,5$) &[$225, 170, 11, 3$] \\
    $500$ &$0.2$ &P($5,5$) &[$200, 144, 11, 3$] \\
    $500$ &$0.25$ &P($5,5$) &[$188, 130, 11, 3$] \\
    $500$ &$0.33$ &P($5,5$) &[$163, 102, 11, 3$] \\
    $500$ &$0.5$ &P($5,5$) &[$125, 60, 11, 3$] \\\midrule
    $2000$ &$0.1$ &P($9,9$) &[$900, 691, 11, 3$] \\
    $2000$ &$0.2$ &P($9,9$) &[$800, 581, 11, 3$] \\
    $2000$ &$0.25$ &P($9,9$) &[$750, 526, 11, 3$] \\
    $2000$ &$0.33$ &P($9,9$) &[$670, 438, 11, 3$] \\
    $2000$ &$0.5$ &P($9,9$) &[$500, 251, 11, 3$] \\
    \bottomrule
    \end{tabular}
    \label{tab:codes}
    \end{table}

The outcomes portrayed in Figure \ref{fig:results-codes} confirm our initial intuition that sparse quantum codes are degenerate. It is easy to see that for all of the simulated block lengths and rates (except for $R_Q = 0.5$), the percentage of end-to-end errors that are not of the $\mathrm{E}_1$ type is not negligible, i.e., $\frac{\mathrm{E}_2}{\mathrm{E}_T} + \frac{\mathrm{E}_3}{\mathrm{E}_T} \neq 0$. Furthermore, these results speak towards the higher precision of the logical error rate compared to the physical error rate when assessing the performance of these codes. For instance, at a noise level of $p = 0.005$, $\frac{\mathrm{E}_3}{\mathrm{E}_T} = 0.198$ for the $N=500$ $R_Q = 0.1$ code. This means that $19.8\%$ of the end-to-end errors are degenerate and should not be counted as decoding failures. Thus, in this scenario, the physical error rate overestimates the number of decoding failures and does not provide an accurate representation of the performance of the code. In fact, regardless of the noise level of the channel, the rate of the code (except for $R_Q = 0.5$), and the block length of the code, end-to-end degenerate errors take place, and so the physical error rate will always provide an inaccurate representation of the performance of these sparse quantum codes. Therefore, as is stated in \cite{exploiting}, it is clear that performance results assessed based on the physical error rate ($\mathbf{\hat{E}} = \mathbf{E}$ as the decoding success criterion) \cite{qldpc15, jgf1, jgf2, patrick, efb, bab11} are inaccurate (they report an upper bound).

Furthermore, the results shown in Figure \ref{fig:results-codes} also reveal how the frequency with which each type of end-to-end error takes place varies as a function of different parameters:

\begin{itemize}
    \item End-to-end errors with different syndromes represent a large percentage of the total number of end-to-end errors when the rate of the code is high. This percentage decreases as the rate of the codes goes from $R_Q = 0.5$ to $R_Q = 0.1$ (see ratio $\frac{\mathrm{E}_1}{\mathrm{E}_T}$ in Figure \ref{fig:results-codes}). This trend becomes further exacerbated as the block length of the simulated codes increases, i.e., for low rate large block length codes the ratio $\frac{\mathrm{E}_1}{\mathrm{E}_T}$ will be significantly smaller than for low rate short codes. 
    \item End-to-end identical syndrome errors represent the smallest percentage of the total number of end-to-end errors in most of the simulated cases. This is reflected by the fact that $\frac{\mathrm{E}_2}{\mathrm{E}_T} < \frac{\mathrm{E}_1}{\mathrm{E}_T}, \frac{\mathrm{E}_3}{\mathrm{E}_T}$ in all of our simulation outcomes. 
    \item As the noise level of the channel grows, the ratio $\frac{\mathrm{E}_1}{\mathrm{E}_T}$ becomes larger and the ratio $\frac{\mathrm{E}_2}{\mathrm{E}_T}$ diminishes. The ratio of end-to-end degenerate errors $\frac{\mathrm{E}_3}{\mathrm{E}_T}$ stays relatively constant. 

\end{itemize}

The relationships between these parameters and different types of end-to-end errors serve to draw important conclusions. For instance, the large values of $\frac{\mathrm{E}_1}{\mathrm{E}_T}$ in many of the simulated instances can be understood as a sign that performance gains may be attained by improving the decoding algorithm. This means that applying modified decoding strategies, such as those of \cite{mod-BP, efb, osd1, osd2, refined}, will aid in reducing the presence of end-to-end errors with different syndromes and improve performance. A matter that should also be considered is how often these strategies produce failed error corrections in the form of end-to-end identical syndrome errors. For the methodology of \cite{efb}, such events were shown to be rare, hence we expect these strategies to be a good approach to improve the performance of QLDPC codes. We discuss modified decoding strategies for QLDPC codes in the next section of this chapter. 

In terms of other end-to-end error ratios, the large values of $\frac{\mathrm{E}_3}{\mathrm{E}_T}$ when compared to $\frac{\mathrm{E}_2}{\mathrm{E}_T}$, especially at higher block lengths, show that end-to-end identical syndrome errors are the least frequent of all the end-to-end error types. Despite the relatively small percentage that end-to-end identical syndrome errors represent, it is possible that their relevance will grow when the amount of end-to-end errors with different syndromes is reduced (using modified decoding strategies) or when the degenerate content of the code is increased (through design). At this point, it may be that further improvements in performance will only be possible by designing an optimal degenerate decoder with the capability to correct end-to-end identical syndrome errors. Finally, given that the method proposed in \ref{sec:method} is valid to detect end-to-end degenerate errors, it could be interesting in future work to employ this methodology to specifically design codes to be degenerate. This might lead to code constructions whose probability of suffering end-to-end degenerate errors is maximized, which would allow the positive effects of degeneracy (improved error correction capabilities without a decoding complexity increase) to be completely exploited for quantum error correction purposes.

\section{Improved decoding strategies for QLDPC codes} \label{sec:heuristics}

We close this chapter by providing a succinct summary of the most relevant improved decoding strategies that have been proposed in the literature. A more thorough look into each of these methods is provided in Appendix \ref{app:modified}.

In the previous section we discussed how $\frac{\mathrm{E}_1}{\mathrm{E}_T}$, the ratio of end-to-end errors with different syndromes, accounts for a large portion of the decoding mistakes of the simulated QLDGM codes. Fortunately, although they cause decoding mistakes, these errors are always detected (the estimated syndrome is not equal to the measured syndrome). This means that the decoder will know when these errors occur, an outcome that can be used to apply certain modifications to the original QMLD strategy to try to revert the errors. Eliminating the presence of these end-to-end errors may lead to drastic performance improvements, and although it does not address the issue of constructing a degenerate decoder (end-to-end identical syndrome errors will still occur), it represents an important frontier for QLDPC code design. 

The earliest mention of improved decoding strategies for QLDPC codes dates back to the work of Mackay et al. \cite{bicycle}, who proposed a strategy to adapt the conventional SPA decoder to the depolarizing channel. Later on, this technique was applied by Lou et al in \cite{jgf2}, where a modified decoder was built for the depolarizing channel and shown to surpass traditional BP decoders. Following this, a set of heuristic methods that successfully improved the ability of the classical qubit-wise BP decoder to handle detected harmful errors was built in \cite{degen3}. In this work, the authors also discussed the concept of symmetric degeneracy errors and proposed a host of techniques (freezing, collision, and random perturbation) capable of improving the performance of QLDPC codes under BP decoding. Simultaneously, an enhanced-feedback decoder specifically designed for the depolarizing channel was derived in \cite{efb}. Later on, in \cite{qldpc15}, a modified decoder for dual-containing CSS QLDPC codes which also led to performance improvements was introduced. Recently, in  \cite{mod-BP, osd1, osd2, refined, exploiting}, a number of new and improved decoding strategies have been proposed. A timeline of the most significant contributions to the field of modifed QLDPC decoding is provided in table 1.

Except for the decoders proposed in \cite{jgf2, qldpc15, refined}, all of these heuristic methods operate according to the same working principle: prior to applying any changes to the decoder, the original qubit-wise BP decoder must fail. Generally, these modified decoding processes begin by running a standard unaltered version of the qubit-wise BP decoder. If the standard decoder finds an error estimate whose corresponding syndrome is equal to the real one, i.e., $\hat{\mathbf{w}} = \mathbf{w}$, then the procedure halts and assumes that the error estimate is correct. However, if the standard decoder fails to find a matching estimate of the syndrome after the allotted number of iterations, changes are made to the decoder (according to specific techniques) and decoding is reattempted in order to fix this decoding impasse.

These decoding alterations combine a range of strategies to revert the impact of end-to-end errors with different syndromes. Some techniques, like those of \cite{bicycle, jgf2, degen3, mod-BP, efb}, utilize the incorrect error sequence and the syndrome estimates of the original decoder to compute new a priori probabilities that, when used by the decoder, may result in a matching syndrome estimate. Others, like the augmented decoder of \cite{mod-BP} and the supernode decoder of \cite{qldpc15}, rely on changing the structure of the original factor graph. Also in \cite{mod-BP}, the authors apply the two proposed approaches in unison, so that when one fails the other is applied and viceversa, a technique called the combined decoder. Finally, the most recent strategies of \cite{osd1, osd2, refined, exploiting} focus more on post-processing and direct modifications to the decoding algorithm.

\begin{table}[h!] \label{tab:time23}
\caption{Most significant modified decoding strategies for QLDPC codes.}
\centering
\hspace{-0.1cm}
\begin{minipage}[t]{0.8\linewidth}
\color{gray}
\rule{\linewidth}{1pt}
\ytl{2004}{Initial Proposal for a Modified SPA-based decoding Strategy \cite{bicycle}}
\ytl{2005}{Correlation Exploiting Decoder \cite{jgf2}}
\ytl{2008}{Freezing, Collision \& Random Perturbation Decoders  \cite{degen3}}
\ytl{2012}{Enhanced Feedback Decoder \cite{efb}}
\ytl{2015}{Supernode Decoder \cite{qldpc15}}
\ytl{2019}{Adjusted \& Augmented Decoders \cite{mod-BP}}
\ytl{2019/2020}{Ordered Statistics Decoder \cite{osd1,osd2}}
\ytl{2020}{Refined Belief Propagation Decoding \cite{refined}}
\ytl{2021}{Degeneracy Exploiting Decoder \cite{exploiting}}
\bigskip
\rule{\linewidth}{1pt}%
\end{minipage}%
\end{table}
\newpage

Regardless of the specific modified decoding method that is employed, the impact of its modifications spreads quickly during the new decoding phase, mostly due to the presence of short cycles in QLDPC codes. This means that, while short cycles hinder the traditional BP decoder, they benefit these modified decoding strategies \cite{degen3, mod-BP, efb}. The downside of these heuristic schemes comes in the form of an increased decoding complexity. Because most of these techniques rely on the failure of a standard decoder and a re-execution of the modified process anew, the chosen method sometimes requires numerous decoding attempts before it finds the correct syndrome estimate. With regard to performance, the degree to which it can be improved varies depending on the chosen strategy; the techniques of \cite{degen3, mod-BP} can yield performance improvements of about 30\% in terms of the word error rate over traditional decoding, while the enhanced feedback decoding technique of \cite{efb} has been shown to be approximately 10 times better than the traditional SPA decoder, even though it only requires a 25-35\% increase in the number of decoding iterations. In \cite{osd1, osd2} the Ordered Statistics Decoder (OSD) is shown to outperform all of the aforementioned modified decoding techniques.

In any case, aside from a potential increase in complexity, these heuristic methods are relatively simple tools that can be easily included within conventional qubit-wise BP decoders to improve the performance of QLDPC codes under BP/SPA decoding.

\section{Chapter summary}

In this chapter we have presented a method to detect degenerate errors in sparse quantum codes in a computationally efficient manner. Based on this method, we have shown how the physical error rate provides an inaccurate representation of the performance of sparse quantum codes and we have shown that logical error rate should be used instead. The discrepancy between the logical error rate and the physical error rate is especially relevant to the field of sparse quantum codes because of their degenerate nature. This is reflected by the results we have obtained for a specific family of QLDPC codes, whose performance can be up to $20\%$ better than would be expected based on previous results in the literature. In addition, these simulation outcomes serve to show how performance may be improved by constructing degenerate quantum codes, and they also speak toward the positive impact that modified decoding strategies can have on the performance of sparse quantum codes. The last portion of this chapter surveys the literature on these modified decoding strategies, most of which are thoroughly discussed in the Appendix.
\clearemptydoublepage
\chapter{Non-CSS QLDPC codes} \label{chapter6}

\epigraph{\textit{``The most important step a man can take. It's not the first one, is it? It's the next one. Always the next step...''}}{\textbf{Brandon Sanderson}.}

\noindent\hrulefill

Up to this point in the dissertation our focus has been (for the most part) theoretical. Aside from studying the phenomenon of degeneracy, what we have seen thus far has served to lay out the necessary groundwork to understand how QEC works and why it is important. This context should be useful to understand the remaining chapters of the dissertation, where our attention will shift to the design and performance of a specific family of QLDPC codes. Before getting started with Chapter $6$, in the following discussion we provide some insight into why we chose to work with sparse quantum codes.

\subsection*{Why build QEC codes using LDPC codes?}

In the realm of classical communications, turbo codes \cite{turbo, turbo2} and LDPC codes \cite{ldpc1, ldpc2, ldpc3, LDPCshan1, LDPCshan2, LDPCshan3} are known to exhibit capacity approaching performance at a reasonable decoding computational complexity. In particular, turbo codes offer great flexibility when it comes to choosing their block length and rate, whereas the sparsity of LDPC codes makes them easy to decode. Given their appealing characteristics, the scientific community has searched diligently for the quantum equivalents of classical turbo and LDPC codes. Quantum codes based on turbo codes first appeared in \cite{QTC, EAQTC}, and have since been modified and improved \cite{josu1, josu2, EXITQTC,EAQIRCC,QSBC}. In turn, QLDPC codes were first analyzed in \cite{bicycle}, were their sparse nature was shown to be especially advantageous in the quantum paradigm because it ensures that a smaller number of quantum interactions per qubit is necessary during the error correction procedure. Because quantum gates are faulty, having less quantum interactions per qubit avoids additional quantum operations, which minimizes errors and facilitates fault-tolerant computing \cite{introQIC, fault11, fault22}. Thus, QLDPC codes can be said to be particularly well suited for quantum error correction.

Out of the existing types of LDPC codes, LDGM codes \cite{classicalLDGM} provide a seamless manner for code design in the quantum domain. LDGM codes are a specific subset of LDPC codes whose generator matrices are also sparse, and thus their encoding complexity is similar to that of turbo codes, and much lesser than for standard LDPC codes. Given that LDGM codes form a special subclass of the LDPC code family, they can be decoded in the same manner and with the same complexity as any other LDPC code. LDGM codes have been extensively studied \cite{jgf3, classicalLDGM, jgf4} and used in classical communications \cite{EXITRCMLDGM,BWTRCMLDGM}. In \cite{ldpc3}, regular LDGM codes were studied and shown to be asymptotically bad, displaying error floors that do not decrease with the block length. In \cite{jgf3} and \cite{jgf4}, a concatenated LDGM scheme was shown to achieve irregular LDPC code-like performance at a very low encoding/decoding complexity. 

Based on what was shown in Chapter 4, we know that quantum codes can be built from classical codes by casting them in the framework of stabilizer codes \cite{QSC}. In \cite{qldpc15}, the authors document the design of QLDPC codes from their classical counterparts based on the stabilizer formalism by detailing numerous construction and decoding techniques along with their flaws and merits. Among the discussed methods, the construction of QLDPC codes based on LDGM codes is shown to yield performance and code construction improvements, albeit at an increase in decoding complexity. This method was originally proposed in \cite{jgf1} and \cite{jgf2}, where CSS quantum codes based on regular LDGM classical codes were shown to surpass the best quantum coding schemes of the time. Later, performance was significantly improved in \cite{jgf3} and \cite{jgf4} by utilizing a parallel concatenation of two regular LDGM codes. 

As with most QLDPC designs, QLDGM-based quantum code implementations are based on CSS constructions. CSS codes, simultaneously proposed by Calderbank, Shor, and Steane in \cite{CSS1}, \cite{CSS2}, are a particular subset of the stabilizer code family. They provide a straightforward method to design quantum codes via existing classical codes. However, due to the specific nature of CSS constructions, their performance is limited by an unsurpassable bound, referred to as the \textit{CSS lower bound} \cite{CSSbound}. This inspires the search for non-CSS constructions, as they should theoretically be able to outperform CSS codes if designed optimally. Non-CSS LDPC-based codes were proposed in \cite{nonCSS1} and \cite{nonCSS2}. However, despite showing promise, they fail to outperform existing CSS QLDPC codes for comparable block lengths.

In this chapter, we introduce a non-CSS scheme based on LDGM codes and compare its performance to existing CSS QLDGM codes. We begin by explaining how CSS codes are constructed and how they can be modified to create non-CSS quantum codes. Following this, we provide insight regarding how our construction is optimized, and we show how the performance of our non-CSS codes is similar to that of the CSS schemes they are derived from, despite the fact that their quantum rate is higher. When their rate is the same, the non-CSS scheme outperforms the original CSS design over the depolarizing channel. Finally, we compare the non-CSS structures proposed here with other existing QLDPC codes in the literature, illustrating that our method surpasses such error correction schemes.

The chapter is structured as follows. We commence with a brief presentation of important preliminary topics, such as CSS codes or LDGM codes, that did not appear in Chapter \ref{chapter3}. We proceed by presenting CSS LDGM based codes in section \ref{sec:CSS} and our non-CSS LDGM-based strategy in section \ref{sec:nonCSS}. In section \ref{sec:ldgmres}, we compare the performance of the proposed scheme to existing CSS quantum codes. 

\section{Preliminaries}

In this section we provide an overview of some preliminary topics in the field of QEC that were not presented in Chapter \ref{chapter3}. 

\subsection{Pauli channels}\label{depchan}

We saw in Chapter \ref{chapter3} that the most popular quantum channel model is the generic Pauli channel $\xi_p$. To recap, the effect of the Pauli channel $\xi_p$ upon an arbitrary quantum state described by its density matrix $\rho$ can be written as $$ \xi_p(\rho) = (1-p_x-p_y-p_z)\rho + p_xX\rho X + p_yY\rho Y + p_zZ\rho Z .$$

This expression can be interpreted as the quantum state experiencing a bit-flip ($X$ Pauli operator) with probability $p_x$, a phase-flip ($Z$ Pauli operator) with probability $p_z$, or a combination of both ($Y$ Pauli operator) with probability $p_y$. 
 
\subsubsection{Depolarizing channel}

We also mentioned previously that most of the QEC literature considers the independent depolarizing channel model. This model is a specific instance of the Pauli channel in which the depolarizing probabilities are all equal, i.e, $p_x = p_z = p_y=p$, and the channel is characterized by the depolarizing probability $p$. When quantum states of $N$ qubits are considered, the errors that take place belong to the $N$-fold Pauli\footnote{Recall that the global phase has no observable consequence, so we can consider the channel errors to be elements of the effective $N$-fold Pauli group.} group $\mathcal{G}_N$. These $N$-qubit error operators are made up of single qubit Pauli operators that act independently on each qubit causing an $X$, $Z$, or $Y$ error with probability $p/3$ and leaving it unchanged with probability $(1-p)$.  

\subsubsection{i.i.d. $X/Z$ Channel}

A simpler quantum channel model, known as the \textit{standard flipping channel} or \textit{i.i.d. X/Z channel}, was introduced in \cite{bicycle}, where $Z$ and $X$ errors are modelled as independent events identically distributed (i.i.d.) according to the flip probability $f_m$. This quantum channel model is analogous to two independent Binary Symmetric Channels (BSCs) with marginal bit flip probability $f_m = 2p/3$, where the separate BSCs can be seen as $Z$ and $X$ error channels, respectively. Given that $Y$ errors occur when both a phase and a bit-flip happen to the same qubit, the simplified notion of the i.i.d. X/Z channel ignores any correlation that exists between $X$ and $Z$ errors in the depolarizing channel. 

\subsection{The Hashing bound}

In the introduction to section \ref{cp3:class} we mentioned how the classical channel capacity establishes the maximum information transfer rate at which reliable communications can take place over a classical channel. Unsurprisingly, we can also derive the quantum counterpart of this concept and use it to study the theoretical limits of quantum channels. 

The capacity of a quantum channel is defined as the highest possible achievable rate at which quantum information can be asymptotically transmitted in an error-less manner. Unfortunately, a closed formula for the capacity of the Pauli channel is not known \cite{josurev, Wilde1, josuarxiv}, which means that a closed formula for the capacity of the depolarizing channel (the most prominent channel model for decoherence in the literature) is also unknown. Instead, we work with a quantity known as the Hashing bound, which defines a lower bound on the capacity of the depolarizing channel and is computed as $C_{\text{Hash}}(p) = 1 - H_2(p) -p\text{log}_23$, where $H_2(p)$ is the binary entropy function and $p$ represents the depolarizing probability \cite{Bennett1}. This means that for a given value of $p$, $C_{\text{Hash}}(p)$ represents a lower bound on the highest possible coding rate at which asymptotically error-free quantum communication is possible. In theory, it is possible for quantum codes to surpass the Hashing bound due to the phenomenon of degeneracy (which is also the reason why a closed formula for Pauli channels is difficult to obtain) \cite{degen2}. Alternatively, for a specific quantum rate $R_Q$, where we have $R_Q = C_{\text{Hash}}(p^*)$, $p^*$ represents a bound on the channel's depolarizing probability \cite{newcite}. As in the classical domain, we can refer to $p^*$ as the \textit{noise limit}. 

\subsubsection{Distance to the Hashing bound}

Ideally, quantum codes that are properly designed should ensure error-free communications close to the noise limit $p^*$. Thus, we can efficiently characterize the quality of quantum codes of a specific quantum rate $R_Q$ built for the depolarizing channel by assessing how far away they are from the Hashing bound. The distance from the Hashing bound can be computed based on the expression 

\begin{equation}
    \delta = 10\text{log}_{10}\bigg(\frac{p^*}{p}\bigg) \,,
    \label{eq:hash_distance}
\end{equation}
where we use $\delta$ to represent the distance to the Hashing bound in decibels (dB), $p^*$ is the noise limit of the depolarizing channel for a specific quantum coding rate $R_Q$, and $p$ is the highest depolarizing probability at which the code in question can operate in an error-free manner. 

\subsection{CSS codes}

Based on our previous discussion on stabilizer codes (see section \ref{sec:QPCM}), we know that two binary classical LDPC codes can only be used to construct a quantum stabilizer code if they satisfy the symplectic criterion \eref{eq:symplec}. With this knowledge, the first design strategy one could devise to construct quantum stabilizer codes would be to randomly select pairs of classical LDPC codes and combine them into a QPCM. However, finding two LDPC codes of reasonable block size that satisfy \eref{eq:symplec} is highly unlikely. Calderbank-Shor-Steane (CSS) codes \cite{CSS1}, \cite{CSS2}, provide a more efficient design strategy than random selection of classical codes. The quantum parity check matrix of these codes is written as

\begin{equation} \label{eq:CSS}
\mathbf{H}_Q = (\mathrm{H}_x|\mathrm{H}_z) =
\begin{pmatrix} \mathrm{H}_x^{'} &0 \\
    0 &\mathrm{H}_z^{'} \end{pmatrix} , 
\end{equation}

where $\mathrm{H}_x = \begin{pmatrix} \mathrm{H}_x^{'} \\ 0 \end{pmatrix}$ and $\mathrm{H}_z = \begin{pmatrix} 0 \\ \mathrm{H}_z^{'} \end{pmatrix}$.

In this construction, $\mathrm{H}_x'$ and $\mathrm{H}_z'$ are the parity check matrices of two binary classical LDPC codes $C_1$ and $C_2$, respectively, where each matrix is used to correct either bit-flips or phase-flips. The classical codes are chosen so that $C_2^\perp \subseteq C_1$, where $C_2^\perp$ is the dual of the classical LDPC code $C_2$. This design constraint,  generally referred to as the \textit{CSS condition}, reduces \eref{eq:symplec} to $(\mathrm{H}_x^{'}\mathrm{H}_z^{'\top})=0$.

\subsection{Systematic classical LDGM codes} \label{sec:classLDGM}

Systematic LDGM codes are useful, both in classical and quantum environments, because of the particular structure of their generator and parity check matrices. Let $C$ be a binary systematic LDGM code. Then, its generator matrix $\tilde{\mathrm{G}}$ and its parity check matrix $\tilde{\mathrm{H}}$ can be written as 

\begin{align} \label{eq:systematicLDGM}
\begin{gathered}
\tilde{\mathrm{G}} = (\mathrm{I} \ \mathrm{P}) \\
\tilde{\mathrm{H}} = (\mathrm{P}^\top \ \mathrm{I}), 
\end{gathered}
\end{align}
where $\mathrm{I}$ denotes the identity matrix, and $\mathrm{P}$ is a sparse matrix. 

Because LDGM codes belong to the family of linear block codes, these matrices will satisfy $\tilde{\mathrm{G}}\tilde{\mathrm{H}}^\top = \tilde{\mathrm{H}}\tilde{\mathrm{G}}^\top = 0 $. Those systematic LDGM codes whose variable and parity check nodes have degrees\footnote{The degree of the variable nodes is the number of nonzero entries per column of the PCM. The degree of the parity check nodes is given by the number of nonzero entries per row of the PCM. An LDGM code is said to be regular when all the rows of its PCM have the same number of nonzero entries, $x$, and so do its columns, $y$.} $x$ and $y$, respectively, will be denoted as $(x,y)$ regular LDGM codes. Regular LDGM codes are known to be asymptotically bad \cite{ldpc3}, displaying error floors that do not decrease with the block length. However, in \cite{parallelshannon}, codes built via the parallel concatenation of two regular LDGM codes\footnote{The parallel concatenation of regular LDGM codes is equivalent to an LDGM code with an irregular degree distribution.} were shown to yield significant reduction in these error floors. The parallel concatenation of two regular LDGM codes $C_1$ and $C_2$ with generator matrices $\mathrm{G}_1 = [\mathrm{I}\ \mathrm{P}_1]$ and $\mathrm{G}_2 = [\mathrm{I}\ \mathrm{P}_2]$, where $\mathrm{P}_1$ and $\mathrm{P}_2$ have degree distributions $(y_1, y_1)$ and $(y_2, z_2)$, is the irregular LDGM code with generator matrix $\mathrm{G} = [\mathrm{I}\ \mathrm{P}_1 \mathrm{P}_2]$. Generally, this concatenation is accomplished by using a high rate code $C_2$ that is able to reduce the error floor of $C_1$, while also causing negligible degradation of the original convergence threshold. 

With regard to decoding, because LDGM codes are a specific subset of LDPC codes, they are decoded in exactly the same manner as generic LDPC codes. In Chapter \ref{chapter3} we showed that classical LDPC syndrome-based decoding is performed by solving the equation $\mathbf{z}=\mathbf{H}\mathbf{e}$, where $\mathbf{z}$ represents the received classical syndrome, $\mathbf{H}$ is the PCM of the code, and $\mathbf{e}$ is the error pattern that has corrupted our information (a rigorous description on the details of this procedure is provided in Appendix \ref{app:spa}). It should be mentioned that it is also possible to decode systematic LDGM codes by solving equation $\mathbf{c}=\mathrm{P}\mathbf{u}$, where $\mathbf{c}$ is the vector of parity bits generated at the encoder, $\mathrm{P}$ is the constituent sparse matrix of the LDGM generator matrix (see equation \eref{eq:systematicLDGM}), and $\mathbf{u}$ is the information message we want to obtain. This means that the decoding algorithm for LDGM codes can also be implemented by applying the SPA \cite{BP, spa} over the graph associated to the equation $\mathbf{c}=\mathrm{P}\mathbf{u}$.

\section{CSS LDGM-based codes} \label{sec:CSS}

Based on what we have seen thus far, our first intuition to derive the QPCM of a QLDGM CSS code would be to select any classical LDGM code with parity check and generator matrices $\tilde{\mathrm{H}}$ and $\tilde{\mathrm{G}}$, and set $\mathrm{H}'_z = \tilde{\mathrm{H}}$ and $\mathrm{H}'_x = \tilde{\mathrm{G}}$ in \eref{eq:CSS}, since the property $\tilde{\mathrm{G}}\tilde{\mathrm{H}}^\top = \tilde{\mathrm{H}}\tilde{\mathrm{G}}^\top = 0$ would ensure the fulfilment of the symplectic criterion. However, this results in a QPCM $\mathbf{H}_Q$ of size $N\times2N$, which cannot be used for encoding purposes. This is easy to see based on the following discussion.

Consider a stabilizer code that has $(N-k)$ stabilizer generators and whose QPCM is of of size $(N-k)\times2N$. Such a quantum code encodes $k$ logical qubits into $N$ physical qubits, which implies that the code has a quantum rate $R_Q = \frac{k}{N}$. This means that the quantum rate of a code with QPCM $\mathbf{H}_Q$ of size $N\times2N$ is $R_Q=0$. Therefore, to build a valid quantum code from classical LDGM codes, we must reduce the number of rows in $\mathbf{H}_Q$ while ensuring that the \textit{CSS condition} is fulfilled. In \cite{jgf1}, the authors successfully achieve this via linear row operations. They do so by applying the following theorem. 

\begin{theorem} \label{theo:css} Given the generator and parity check matrices of a systematic LDGM code \eref{eq:systematicLDGM}, define $\mathrm{H}_{m_1\times N} = [\mathrm{M}_1]_{m_1\times n_1}[\tilde{\mathrm{H}}]_{n_1\times N}$ and $\mathrm{G}_{m_2\times N} = [\mathrm{M}_2]_{m_2\times n_2}[\tilde{\mathrm{G}}]_{n_2\times N}$, where $n_1 + n_2 = N$ and $\mathrm{M}_1$ and $\mathrm{M}_2$ are low-density full-rank binary matrices whose number of rows satisfy $m_1 < n_1$ and $m_2 < n_2$, respectively. Then, the quantum PCM shown in \eref{eq:CSSLDGMPCM}, obtained by setting $\mathrm{H}'_x = \mathrm{H}$ and $\mathrm{H}'_z = \mathrm{G}$ in \eref{eq:CSS}, is the quantum PCM of an LDGM-based CSS code with rate $R_Q = \frac{N-m_1-m_2}{N}$.

\begin{equation} \label{eq:CSSLDGMPCM}
\mathbf{H}_Q = (\mathrm{H}_x|\mathrm{H}_z) =
\begin{pmatrix} \mathrm{H} &0 \\
    0 &\mathrm{G} \end{pmatrix}  = \begin{pmatrix} \mathrm{M}_1\tilde{\mathrm{H}} &0 \\ 0 &\mathrm{M}_2\tilde{\mathrm{G}} \end{pmatrix}.
\end{equation}
\end{theorem}

We showed in Chapter \ref{chapter4} that sparse quantum codes are decoded by running the SPA algorithm over the factor graph associated to the expression $$\mathbf{w}=\mathrm{\mathbf{H}}_{\mathcal{\overbar{S}}} \odot \mathbf{e}=\mathbf{e}_x \mathrm{H}^\top_z\oplus \mathbf{e}_z \mathrm{H}^\top_x,$$ where $\mathbf{H}_{\mathcal{\overbar{S}}}$ denotes the QPCM of the stabilizer code in question.

Thus, CSS LDGM codes will be decoded by applying the SPA over the factor graph defined by $\mathbf{w}=\mathrm{\mathbf{H}}_{Q} \odot \mathbf{e}$, where $\mathbf{H}_Q$ is given in \eref{eq:CSSLDGMPCM}. We can easily derive this factor graph based on the unique structure of a CSS QPCM and the properties of the symplectic representation, which allows us to split $\mathbf{e} = (\mathbf{e}_x|\mathbf{e}_z)$. Notice how, given the diagonal structure of $\mathbf{H}_Q$, the terms $\mathbf{e}_x \mathrm{H}^\top_z$ and $\mathbf{e}_z \mathrm{H}^\top_x$ have no non-zero overlaps. This means that the syndrome is made up of two separate parts, $\mathbf{w} = [\mathbf{w}_x, \mathbf{w}_z]$, where $\mathbf{w}_x = \mathbf{e}_x \mathrm{H}^\top_z$ and $\mathbf{w}_z = \mathbf{e}_z \mathrm{H}^\top_x$ . Thus, we can actually decode CSS codes separately, i.e, by generating two separate subgraphs associated to $\mathbf{w}_x$ and $\mathbf{w}_z$. In the following, we illustrate this derivation for $\mathbf{w}_x = \mathbf{e}_x \mathrm{H}^\top_z$. The procedure for $\mathbf{w}_z = \mathbf{e}_z \mathrm{H}^\top_x$ is identical but using $\mathrm{G}$ instead of $\mathrm{H}$ in \eref{eq:snodes}. We can write the first half of the syndrome of a CSS code, $\mathbf{w}_x^\top$ (in column form), as

\begin{equation} \label{eq:snodes}
    \mathbf{w}_x^\top = (\mathbf{e}_x\mathrm{H}_z^\top)^\top =  \mathrm{H}_z\mathbf{e}_x^\top = \mathrm{M}_1\tilde{\mathrm{H}}\mathbf{e}_x^\top = \mathrm{M}_1[\mathrm{P}^\top\  \mathrm{I}]\mathbf{e}_x^\top .
\end{equation}
If we now split $\mathbf{e}_x^\top = [\mathbf{e}_{x_1} \ \mathbf{e}_{x_2}]^\top$, we can write 
\begin{align} \label{eq:dnodes}
    \mathrm{d}_x &= [\mathrm{P}^\top \ \mathrm{I}]\mathbf{e}_{x}^\top = [\mathrm{P}^\top \ \mathrm{I}]_{n_1\times N}\begin{pmatrix}
        \mathbf{e}_{x_1} \\ \mathbf{e}_{x_2}
    \end{pmatrix}_{N \times 1}\nonumber \\ &= \mathrm{P}^\top_{n_1\times n_2}[\mathbf{e}_{x_1}^\top]_{n_2 \times 1} + [\mathbf{e}_{x_2}^\top]_{n_1\times 1} . 
\end{align}

We then relate $\mathrm{d}_x$ to $\mathbf{w}_x^\top$ as 

\begin{equation} \label{eq:stodnodes}
    [\mathbf{w}_x^\top]_{m_1\times 1} = \mathrm{M}_{1_{m_1\times n_1}}\mathrm{d}_{x_{n_1\times 1}}.
\end{equation}

\begin{figure}[H]
\centering
\includegraphics[width=\linewidth,height=3.75in]{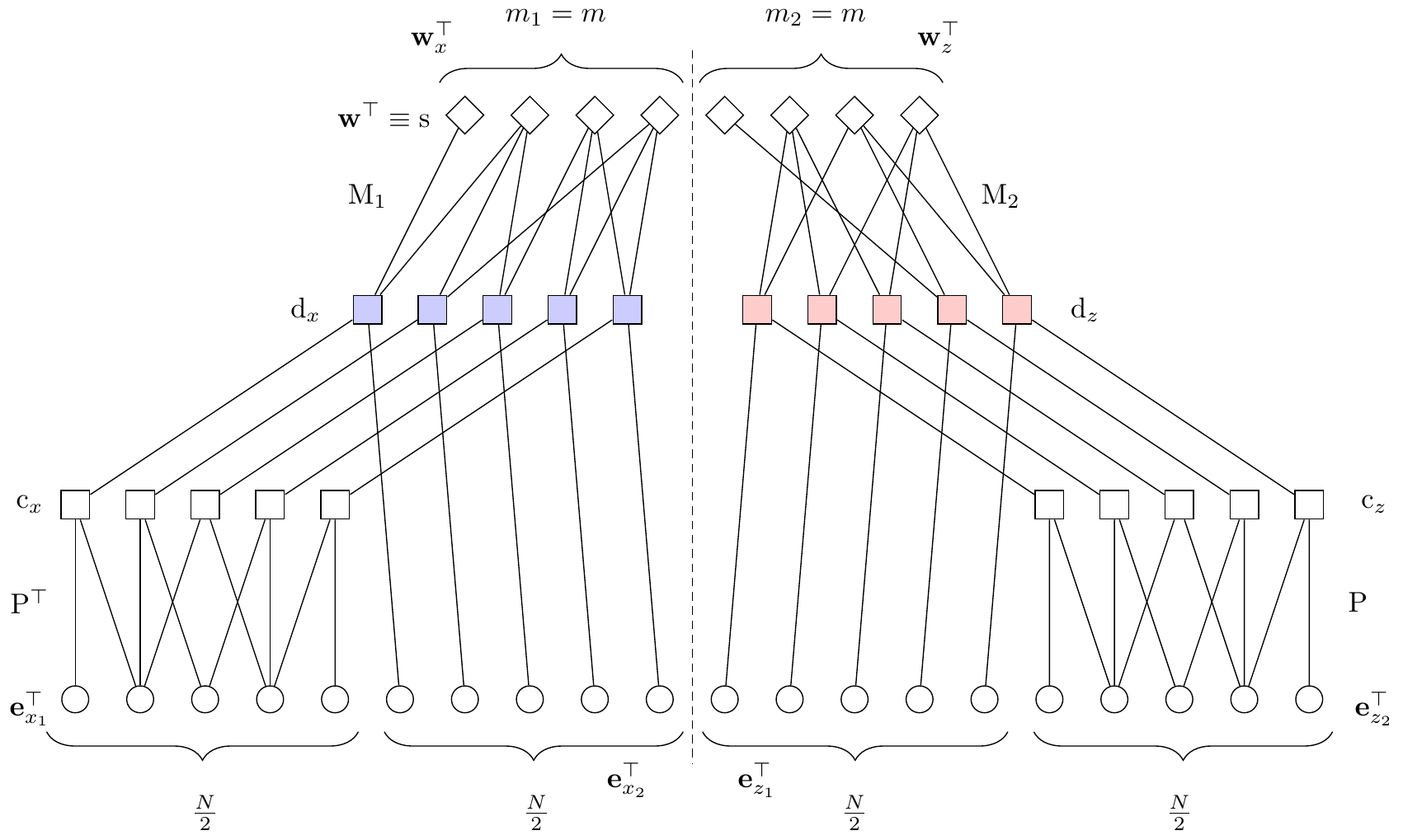}
\caption{\normalsize{Decoding graph for a QLDGM CSS scheme. The dotted line is included to emphasize the separation of the two constituent subgraphs. The top layer nodes (diamonds) represent the syndrome nodes, which we will denote as $\mathrm{s}$ nodes. The leftmost subgraph decodes the $X$ operators while the one on the right decodes the $Z$ operators. We have assumed that $m_1 = m_2$, $n_1 = n_2 = \frac{N}{2}$, and $m = m_1 + m_2$.}}

    \label{fullCSS}
\end{figure}

The factor graph shown in Figure \ref{fullCSS} is obtained based on expressions \eref{eq:dnodes} and \eref{eq:stodnodes}, as well as their equivalents when using $\mathrm{e}_z$ and $\mathrm{G}$ in \eref{eq:snodes}. 

As mentioned previously, and upon closer examination of the QPCM shown in in \eref{eq:CSSLDGMPCM}, it is easy to see that decoding for the $\mathrm{H}$ and $\mathrm{G}$ matrices can be done separately. Notice how this is visible in Figure \ref{fullCSS}, where the leftmost subgraph is associated to the matrix $\mathrm{H}$, which is used to decode $\mathbf{e}_x$, and the rightmost subgraph is associated to the matrix $\mathrm{G}$, which is used to decode $\mathbf{e}_z$. Separate decoding of these matrices is made possible by the nature of CSS constructions, which results in syndrome nodes containing information only of either $X$ or $Z$ operators, hence why we can write $\mathbf{w} = [\mathbf{w}_x, \mathbf{w}_z]$. This is reflected on the factor graph by the fact that a specific $\mathrm{s}$ node connects to either a $\mathrm{d}_x$ or a $\mathrm{d}_z$ node. Note that, for notation purposes, we refer to the syndrome nodes of the factor graph (top level diamonds in Figure \ref{fullCSS}) as $\mathrm{s}$ nodes.

\begin{figure*}[]
\centering
\includegraphics[width=\linewidth,height=3in]{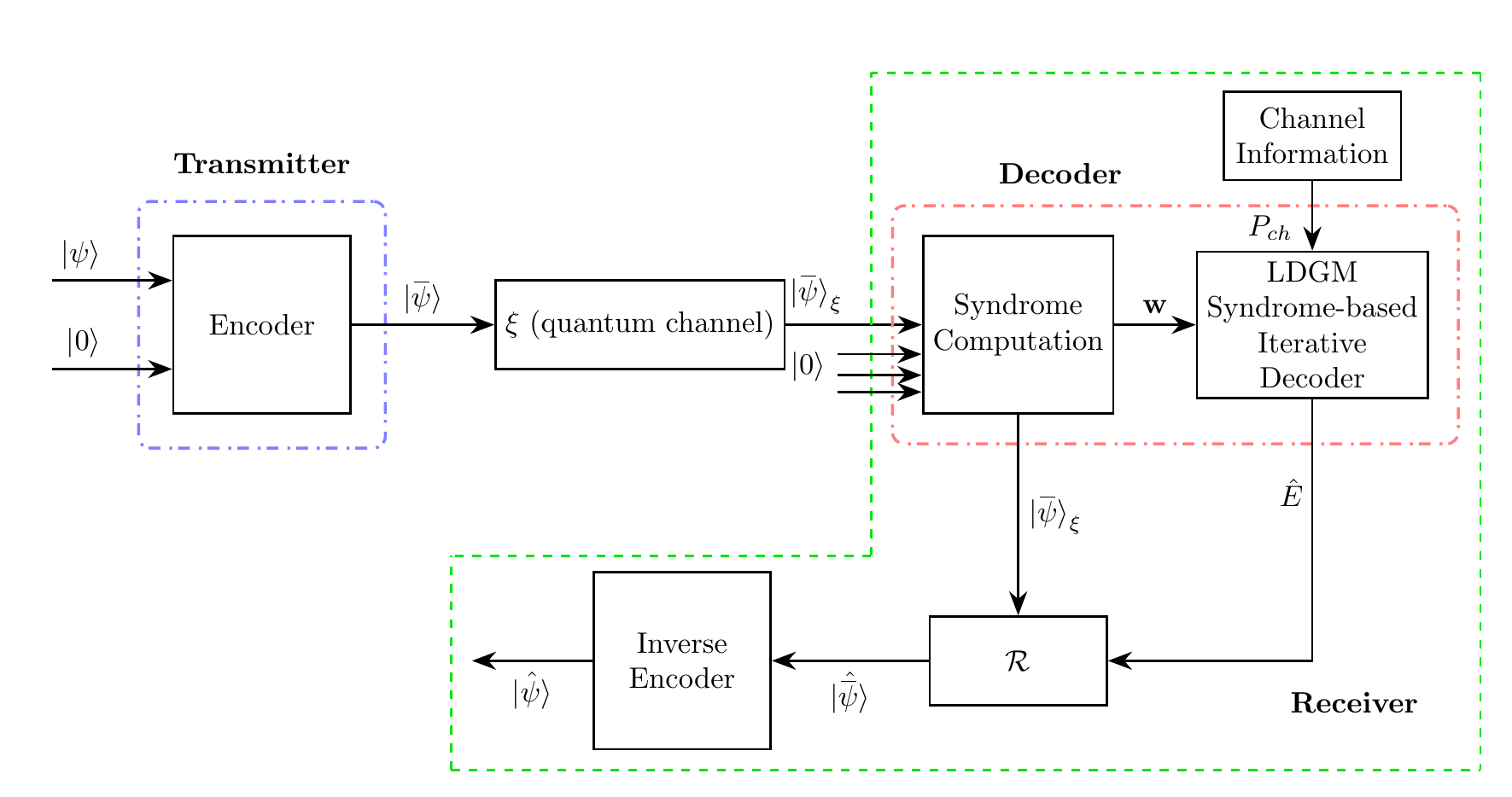}
\caption{\normalsize{Schematic of a quantum communication system using QLDGM codes. At the transmitter, a QLDGM code with QPCM $\mathbf{H}_Q$ maps the quantum state (logical qubits) $\ket{\psi} \in \mathcal{H}_2^{\otimes k}$ onto the codeword (physical qubits) $\ket{\overbar{\psi}} \in \mathcal{H}_2^{\otimes N}$ by using ($N-k$) ancilla qubits (denoted in the figure by $\ket{0})$. At the receiver, the noisy codeword $\ket{\overbar{\psi}}_\xi = E\ket{\overbar{\psi}}$ is received, where $E \in \mathcal{G}_N$ is the error inflicted by the quantum channel. The quantum state $\ket{\overbar{\psi}}_\xi$ is processed by the syndrome computation block at the decoder to compute its error syndrome $\mathbf{w}$. The ancilla qubits at the input of the syndrome computation block are necessary to physically implement the syndrome calculation. Together with the \textit{a priori} channel information $P_{ch}$, $\mathbf{w}$ is provided to the syndrome based LDGM decoder, which yields an estimate of the original error pattern $\hat{E}$. The recovery operator $\mathcal{R}$ uses $\hat{E}$ to correct the noisy codeword. Finally, the inverse encoder yields an estimate of the logical qubits $\hat{\ket{\psi}}$ from the corrected codeword $\hat{\ket{\overbar{\psi}}}$.}}

    \label{comms_system}
\end{figure*}

The matrix multiplications used to perform the linear row operations on $\tilde{\mathrm{H}}$ and $\tilde{\mathrm{G}}$ generate a middle layer, represented by the $\mathrm{c}$ and $\mathrm{d}$ nodes, in both decoding subgraphs of Figure \ref{fullCSS}. This new layer hampers the decoding algorithm, especially during the initial decoding iterations, since \textit{a priori} information regarding the aforementioned middle layer nodes is not be available. This can be seen in Figure \ref{comms_system}, where a generic quantum communication system is shown. The LDGM decoder block of this figure, which runs the SPA over the graph shown in Figure \ref{fullCSS}, has the syndrome $\mathbf{w}$ and the \textit{a priori} probability of the error pattern $P_{ch}$ as its inputs. However, it receives no information pertaining to the $\mathrm{c}$ and $\mathrm{d}$ nodes.

In \cite{jgf1}, the authors circumvent this lack of information by using the so-called \textit{doping} technique of \cite{doping}. This method introduces degree-1 syndrome nodes into the decoding graph. These degree-1 nodes, which we will refer to as $\mathrm{s}_A$ nodes, transmit correct information to the $\mathrm{d}$ nodes they are connected to, ultimately pushing the decoding process in the right direction. They are embodied within the $\mathrm{M}_1$ and $\mathrm{M}_2$ matrices as rows with a single non-zero entry, which corresponds to the edge that connects a given $\mathrm{s}_A$ node to a $\mathrm{d}$ node. The other rows of matrices $\mathrm{M}_1$ and $\mathrm{M}_2$, which correspond to the rest of the $\mathrm{s}$ nodes, have as many non-zero entries as required to guarantee the regularity\footnote{Regularity in this context implies that all the $d$ nodes have the same degree, i.e, that they are all connected to the same number of $\mathrm{s}$ nodes.} of the $\mathrm{d}$ nodes and the necessary number of $\mathrm{s}_A$ nodes. This results in matrices $\mathrm{M}_1$ and $\mathrm{M}_2$ having a special degree distribution which is described by means of the notation $(y; 1, x)$ and the parameter $t$, where $y$ represents the degree of the $\mathrm{d}$ nodes, $t$ is the number of syndrome nodes that are forced to have degree 1 (they become $\mathrm{s}_A$ nodes), and $x$ represents the degree of the remaining syndrome nodes, referred to as $\mathrm{s}_B$ nodes.

Given the particular structure of the $\mathrm{M}_1$ and $\mathrm{M}_2$ matrices and the number of different types of nodes that are present in the factor graph shown in Figure \ref{fullCSS}, the sum-product decoding of these quantum LDGM CSS codes becomes relatively nuanced. The reason for deriving a complex graph like the one shown in Figure \ref{fullCSS} is that decoding over it is different to decoding over the graph associated to the matrix $\mathbf{H}_Q$ that results from the product shown in \eref{eq:CSSLDGMPCM}. This occurs because the matrix products $\mathrm{M}_1\mathrm{\tilde{H}}$ and $\mathrm{M}_2\mathrm{\tilde{G}}$ eliminate some edges and
introduces more cycles, which negatively impacts the performance of the SPA \cite{jgf5}. This is is similar to what happens for serial concatenated LDGM schemes in classical error correction \cite{serialLDGM}. Therefore, although decoding over the graph associated to the set of matrix products shown in Figure \ref{fullCSS} is more complex than doing so over the factor graph associated to the final matrix $\mathbf{H}_Q$, it is worth doing so for the sake of performance. In \cite{jgf4}, a technique known as Discretized Density Evolution (DDE) \cite{ldpc5} is applied to optimize the design of quantum LDGM CSS codes and a complete description of how the decoding process unfolds over the graph shown in Figure \ref{fullCSS} is provided.

\section{Design of non-CSS LDGM-based codes} \label{sec:nonCSS}

We know from the previous section that LDGM-based CSS codes can be decoded over two separate (sub)graphs like the ones shown in Figure \ref{fullCSS}. This is made possible by the specific nature of the quantum PCMs of CSS codes \eref{eq:CSS}, and is visible on a CSS decoding graph by the fact that any given $\mathrm{s}$ node can only be connected to either $\mathrm{d}_x$ or $\mathrm{d}_z$ nodes: a subset of $\mathrm{s}$ nodes is used to decode the $X$ operators (the $X$ component of the symplectic representation of the error sequence $\mathbf{e}_x$) and another subset is used to decode the $Z$ operators (the $Z$ component of the symplectic representation of the error sequence $\mathbf{e}_z$).

The main appeal of non-CSS codes is their ability to exploit redundancy more efficiently than CSS schemes. In our proposed non-CSS construction, we achieve this by allowing edges from a given $\mathrm{s}$ node to go to both $\mathrm{d}_x$ and $\mathrm{d}_z$ nodes. The first method that comes to mind to implement this idea is to randomly distribute the edges of the upper layer of the graph in a manner that ensures that $\mathrm{s}$ nodes are connected to both $\mathrm{d}_x$ and $\mathrm{d}_z$ nodes. However, attempting to decode the $X$ and $Z$ parts over a decoding graph with $\mathrm{s}$ nodes whose edges have been haphazardly assigned to both $\mathrm{d}_x$ and $\mathrm{d}_z$ nodes will cause numerous decoding problems. For instance, not defining a specific distribution for these edges inadvertently causes a reduction in the number of $\mathrm{s}_A$ nodes, and not limiting their total number causes a reduction in the values of the log-likelihood messages exchanged in the decoding process, which severely degrades the decoding performance. Therefore, it is important to optimally design the upper layer of the decoding graph when constructing a non-CSS QLDGM based code. Devising a proper way of distributing the connections among $\mathrm{s}$, $\mathrm{d}_x$, and $\mathrm{d}_z$ nodes in the decoding graph is paramount to construct good non-CSS LDGM-based quantum codes. 

\subsection{Proposed procedure for the construction of non-CSS QLDGM codes} \label{construction}

We begin the non-CSS design process by using a CSS quantum code based on classical LDGM codes \cite{jgf1, jgf2, jgf3, jgf4} as the starting point. For the sake of simplicity and comparison continuity, we maintain the requirements enforced in \cite{jgf3, jgf4}: the matrices used to perform linear row operations are equal to each other $\mathrm{M}_1 = \mathrm{M}_2 = \mathrm{M}$, and the degree distribution of $\mathrm{P}^\top$ and $\mathrm{P}$ is the same. 

The CSS QLDGM code used as a starting point will be associated to two separate decoding subgraphs, one for $\mathrm{H}$ and the other for $\mathrm{G}$. The upper layers of these subgraphs (the number and degree distribution of the $\mathrm{d}$, $\mathrm{s}_A$, and $\mathrm{s}_B$ nodes) will be defined by two identical matrices $\mathrm{M}$ of size $m\times \frac{N}{2}$ described by $(y; 1, x)$ which have $t$ rows with a single non-zero entry.  

We can build our non-CSS scheme using two different strategies. Given that both of them involve very similar procedures, we begin by explaining the simplest construction method. Then, we will present the second proposed design technique.

\subsubsection{Method 1: Syndrome node combination} \label{sec:first}

Non-CSS codes based on the first strategy are constructed as follows:

\begin{enumerate}
    \item First, generate a new matrix, $\mathrm{M}_d$, as \
    
       \begin{equation} \label{eq:nonCSSLDGMPCM}
    \mathrm{M}_d =
\begin{pmatrix} \mathrm{M}_{m\times \frac{N}{2}} &0_{m\times \frac{N}{2}} \\
    0_{m\times \frac{N}{2}} &\mathrm{M}_{m\times \frac{N}{2}} \end{pmatrix}_{2m\times N}.
\end{equation}

\item Select $q$ nodes out of the $2t$ $\mathrm{s}_A$ nodes of matrix $\mathrm{M}_d$\footnote{Note that $\mathrm{M}_d$, as defined in \eref{eq:nonCSSLDGMPCM}, is the matrix representation of the upper layer of the graph in Figure \ref{fullCSS}.}, which we will refer to as $\mathrm{s}_C$ nodes, and add an edge from these nodes to the $\mathrm{d}$ nodes on the side of the decoding graph they are not connected to. We apply a criterion to ensure these new connections are not made randomly: \textit{the edges added to the $q$ selected $s_A$ nodes can only be made to a $\mathrm{d}$ node ($\mathrm{d}_x$ or $\mathrm{d}_z$) that is a $\mathrm{d}_A$ node}. We define $\mathrm{d}_A$ nodes as any $\mathrm{d}$ nodes that are connected to an $\mathrm{s}_A$ node. Of the $q$ $\mathrm{s}_C$ nodes, half of them proceed from $\mathrm{s}_A$ nodes in the CSS subgraph used to decode the $X$ operators, while the other half come from $\mathrm{s}_A$ nodes in the CSS subgraph used to decode the $Z$ operators. Figure \ref{scnode} illustrates how an $\mathrm{s}_C$ node is generated. 

\begin{figure}[!h]
\centering
\includegraphics[width=\linewidth,  height=1.35in]{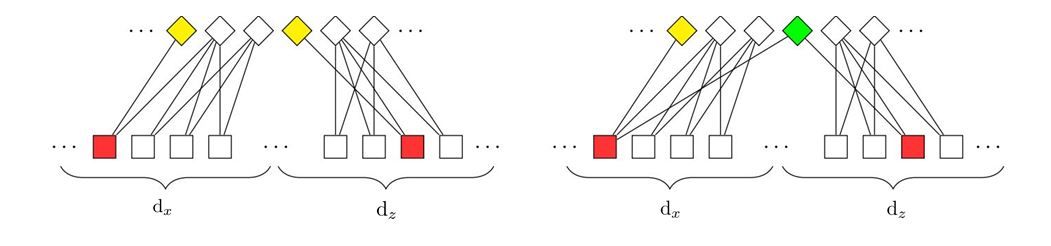}
\caption{\normalsize{Generation of an $\mathrm{s}_C$ node. The upper nodes represent the syndrome nodes while the bottom nodes represent the $\mathrm{d}$ nodes ($\mathrm{d}_x$ and $\mathrm{d}_z$ denote the $\mathrm{d}$ nodes associated to each of the separate CSS decoding subgraphs). The $\mathrm{s}_A$ nodes are represented in yellow, the $\mathrm{d}_A$ nodes are shown in red, and the $\mathrm{s}_C$ node is pictured in green.}}
    \label{scnode}
\end{figure}

The reasoning behind adding edges that traverse the $X$ and $Z$ sides of the graph only to $\mathrm{s}_A$ nodes is based on the following considerations: First, transforming an $\mathrm{s}_A$ node into an $\mathrm{s}_C$ node implies that the new node no longer provides perfect syndrome information, given that it is now connected to two $\mathrm{d}$ nodes. However, the fact that an $\mathrm{s}_C$ node only has two edges implies that its syndrome information, although not transferred exactly, still has high impact when computing messages for associated nodes. At the same time, the edge that traverses from the $\mathrm{s}_C$ node to the other side of the factor graph (either $\mathrm{d}_x$ or $\mathrm{d}_z$) reaches a $\mathrm{d}_A$ node. Considering that messages from $\mathrm{d}_A$ nodes are more likely to be correct (they are connected to an $\mathrm{s}_A$ node), coupled with the fact that $\mathrm{s}_C$ node syndrome information still plays an important role in the messages that the node computes, it is reasonable to assume that $\mathrm{s}_C$ nodes relay accurate information and that they behave in a similar manner to $\mathrm{s}_A$ nodes. Therefore, $\mathrm{s}_C$ nodes provide a way in which reliable messages can be exchanged between both sides of the factor graph, which should have a positive impact on decoding and improve performance.

It is important to note that if we were to add a cross-graph edge to an $\mathrm{s}_B$ node, because of its high degree, the messages received over this new edge would play a limited role in the computations made by the node. By association, the cross-graph messages transmitted over the new edge would also have a very limited effect on the computation of messages exchanged on the other side of the graph and little performance improvement, if any, would be obtained.

Notice that at this stage we have transformed $\mathrm{M}_d$ into a new matrix $\mathrm{M}'_d$, modifying the upper layer of the original CSS decoding graph of Figure \ref{fullCSS} in the following manner:
    
    \begin{itemize}
        \item There are $q$ $\mathrm{s}_C$ nodes that are connected to both sides of the graph.
        \item Some $\mathrm{d}$ nodes are connected to both $\mathrm{s}_A$ and $\mathrm{s}_C$ nodes.
    \end{itemize}
    
These modifications force the $\mathrm{s}$ and $\mathrm{d}$ nodes of the non-CSS decoding graph to have a somewhat irregular edge distribution. Indeed, the ``regularity'' of the $\mathrm{d}$ nodes has been violated in order to connect the separate CSS decoding subgraphs, resulting in $\frac{q}{2}$ $\mathrm{d}_x$ nodes and $\frac{q}{2}$ $\mathrm{d}_z$ nodes having an additional edge. Furthermore, $q$ $\mathrm{s}$ nodes now have two edges, one of them directed towards a $\mathrm{d}_x$ node and the other towards a $\mathrm{d}_z$ node. 

It is intuitive to think that the performance of this novel non-CSS structure should at least be as good, if not better, than that of the CSS scheme utilized as a starting point, provided that the parameter $q$ is chosen properly. If we select $q << m$, the decrease in the number of $\mathrm{s}$ nodes providing perfect information will be small and should have negligible impact in the decoding process\footnote{A total of $q$ $\mathrm{s}_A$ nodes get converted into $\mathrm{s}_C$ nodes, which  do not provide perfect information.}. On the contrary, the degree-2 $\mathrm{s}_C$ nodes allow the exchange of information between both sides of the graph as the iterative decoding process progresses, potentially improving the decoding performance. Therefore, we expect this scheme to present its best performance for a specific range of small values of $q$, with deterioration occurring when $q$ is increased beyond this range.

\end{enumerate}

\subsubsection{Method 2: Syndrome node combination + removal of $\mathrm{s}_A$ nodes}
    
\begin{figure}[!h]
\centering
\includegraphics[width=\linewidth, height=1.35in]{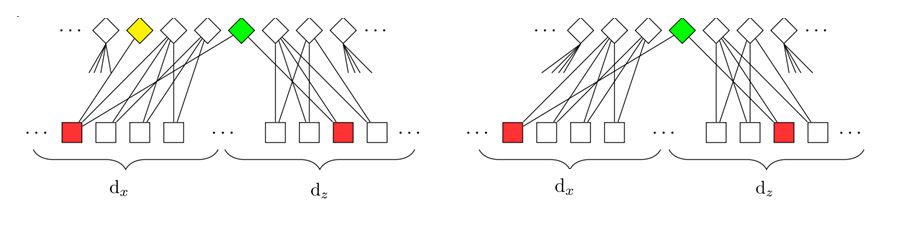}
    \caption{\normalsize{Removal of an $\mathrm{s}_A$ (yellow) node that was previously used to generate an $\mathrm{s}_C$ (green) node.}}
    \label{removal}
\end{figure}

The second design technique removes $q$ syndrome nodes from the decoding graph generated by the first method, specifically the $q$ $\mathrm{s}_A$ nodes connected to a $\mathrm{d}_A$ node that is linked to an $\mathrm{s}_C$ node. The process is shown in Figure \ref{removal}. Notice that with this construction the regularity of the $d$ nodes is maintained and the data rate of the code is increased\footnote{By eliminating syndrome nodes we are decoding the same number of qubits using less syndrome information.}. Moreover, as long as the number of removed $\mathrm{s}_A$ nodes is not too large, the impact on decoding should be minimal: although $\mathrm{s}_C$ nodes do not provide ``perfect'' information, much of the reliability of the messages they transmit to the corresponding $\mathrm{d}$ (previously $\mathrm{d}_A$) nodes will be kept, as they only have two edges. This should mitigate any impact on performance. Because method 2 requires the removal of some $\mathrm{s}_A$ nodes, compared with the CSS code used as a starting point, a larger amount of $\mathrm{s}_A$ nodes may be necessary in the non-CSS structure to ensure acceptable  performance as the value of $q$ grows. This hints towards an increased importance of applying \textit{doping} (increasing the value of $t$) to the constituent matrices $\mathrm{M}$.

Once the specific $s_A$ nodes are removed, we obtain a new matrix $\mathrm{M}'_d$ of size $(2m-q)\times N$, which has the following characteristics:

\begin{itemize}
    \item It has $2(t-q)$ rows with a single nonzero entry. In terms of the decoding graph, this means that there are $2(t-q)$ $\mathrm{s}_A$ nodes.
    \item There are $q$ rows with two nonzero entries. The first entry must be placed in any of the first $\frac{N}{2}$ columns of the matrix while the second one must be placed in any of the last $\frac{N}{2}$ columns. This ensures that $\mathrm{s}_C$ nodes connect both sides of the decoding graph.
    \item The other $2(m-t)$ rows have $x$ nonzero entries. In the decoding graph, these rows correspond to the $\mathrm{s}_B$ nodes, which remain the same as in the CSS structure used as a starting point.
\end{itemize}

\subsubsection{Non-CSS QPCM}

The quantum PCM of the proposed non-CSS code obtained using either of the aforementioned methods is calculated as
\begin{align} \label{eq:FULLCSSLDGMPCM}
\mathbf{H}_{Q_{\text{nonCSS}}} = \mathrm{M}'_d\mathbf{H}_\text{CSS} = \mathrm{M}'_d(\mathrm{H}_x|\mathrm{H}_z)\nonumber \\ =
\mathrm{M}'_d\begin{pmatrix} \tilde{\mathrm{H}} &0 \\ 0 &\tilde{\mathrm{G}} \end{pmatrix} = (\mathrm{H}_x''|\mathrm{H}_z''),
\end{align}
where $\mathbf{H}_{\text{CSS}}$ is defined as in \eref{eq:CSS}, $\tilde{\mathrm{H}}$ and $\tilde{\mathrm{G}}$ are the parity check and generator matrices of a classical LDGM code, and $\mathrm{M}'_d$ is obtained using any of the construction methods.

As shown below, the construction in \eref{eq:FULLCSSLDGMPCM} satisfies the symplectic criterion given in \eref{eq:symplec}. Assume that $\mathrm{M}'_d$ is obtained based on the second construction method presented above. With the goal of simplifying the proof, we write $[\mathrm{M}'_d]_{m_r\times N}$ as the concatenation of two sub-matrices, i.e.,  $ [\mathrm{M}'_d]_{m_r\times N} = [\mathrm{M}'_{\alpha_{m_r\times\frac{N}{2}}} \mathrm{M}'_{\gamma_{m_r\times\frac{N}{2}}} ]$. Substituting this expression into \eref{eq:FULLCSSLDGMPCM}, we obtain

\begin{align*} 
    \mathbf{H}_{Q_{\text{nonCSS}}} &= (\mathrm{H}_x''|\mathrm{H}_z'') \\ &=
 [\mathrm{M}'_d]_{m_r\times N}\mathbf{H}_{\text{CSS}_{N\times 2N}} \\ &=
[\mathrm{M}'_{\alpha_{m_r\times\frac{N}{2}}} \mathrm{M}'_{\gamma_{m_r\times\frac{N}{2}}} ]\begin{pmatrix} \tilde{\mathrm{H}}_{\frac{N}{2}\times N} &0_{\frac{N}{2}\times N} \\ 0_{\frac{N}{2}\times N} &\tilde{\mathrm{G}}_{\frac{N}{2}\times N} \end{pmatrix} \\& 
= \bigg([\mathrm{M}'_{\alpha_{m_r\times\frac{N}{2}}}\tilde{\mathrm{H}}_{\frac{N}{2}\times N}]_{m_r\times N} \ \big| \ [\mathrm{M}'_{\gamma_{m_r\times\frac{N}{2}}}\tilde{\mathrm{G}}_{\frac{N}{2}\times N}]_{m_r\times N} \bigg),\\
\end{align*}
where $m_r=2m-q$ is the total number of rows of $\mathrm{M}'_d$. 

Since for an LDGM code $\tilde{\mathrm{G}}\tilde{\mathrm{H}}^T = \tilde{\mathrm{H}}\tilde{\mathrm{G}}^T = 0$, when checking the symplectic criterion, we obtain
\begin{align*} \label{symplec22}
 \mathrm{H}_x''\mathrm{H}_z''^\top \oplus \mathrm{H}_z''\mathrm{H}_x''^\top = \\&\bigg([\mathrm{M}'_{\alpha_{m_r\times\frac{N}{2}}}\tilde{\mathrm{H}}_{\frac{N}{2}\times N}][\mathrm{\mathrm{M}}'_{\gamma_{m_r\times\frac{N}{2}}}\tilde{\mathrm{\mathrm{G}}}_{\frac{N}{2}\times N}]^{\top} \\ &\oplus [\mathrm{M}'_{\gamma_{m_r\times\frac{N}{2}}}\tilde{\mathrm{G}}_{\frac{N}{2}\times N}][\mathrm{M}'_{\alpha_{m_r\times\frac{N}{2}}}\tilde{\mathrm{H}}_{\frac{N}{2}\times N}]^{\top}\bigg) \\&
= \bigg(\mathrm{M}'_{\alpha_{m_r\times\frac{N}{2}}}\underbrace{\tilde{\mathrm{H}}_{\frac{N}{2}\times N}\tilde{\mathrm{G}}^{\top}_{N\times\frac{N}{2}}}_{ 0_{\frac{N}{2}\times\frac{N}{2}}}\mathrm{M}'^{\top}_{\gamma_{\frac{N}{2}\times m_r}}\\ &\oplus \mathrm{M}'_{\gamma_{m_r\times\frac{N}{2}}}\underbrace{\tilde{\mathrm{G}}_{\frac{N}{2}\times N}\tilde{\mathrm{H}}^{T}_{N\times\frac{N}{2}}}_{0_{\frac{N}{2}\times\frac{N}{2}}}\mathrm{M}'^{\top}_{\alpha_{\frac{N}{2}\times m_r}}\bigg)\\&
= 0_{m_r\times m_r}
\end{align*}
proving that $\mathbf{H}_{Q_{\text{nonCSS}}}$ satisfies the symplectic criterion.

\subsubsection{Mixture of both methods}

Another possibility to design non-CSS codes is to remove only a fraction of the $\mathrm{s}_A$ nodes that are used to generate $\mathrm{s}_C$ nodes in the first construction method. This procedure is identical to the second technique, with the sole exception that instead of removing the entire subset of $q$ $\mathrm{s}_A$ syndrome nodes involved in the generation of the $\mathrm{s}_C$ nodes, only $l < q$ $\mathrm{s}_A$ nodes are removed. In the following section, we focus on codes obtained using the first two methods. Optimizing the performance of codes derived using this third approach may be of interest in future work.

\subsection{Decoding non-CSS QLDGM codes}

\begin{figure}[!h]
\centering
\includegraphics[width=\linewidth, height=1.75in]{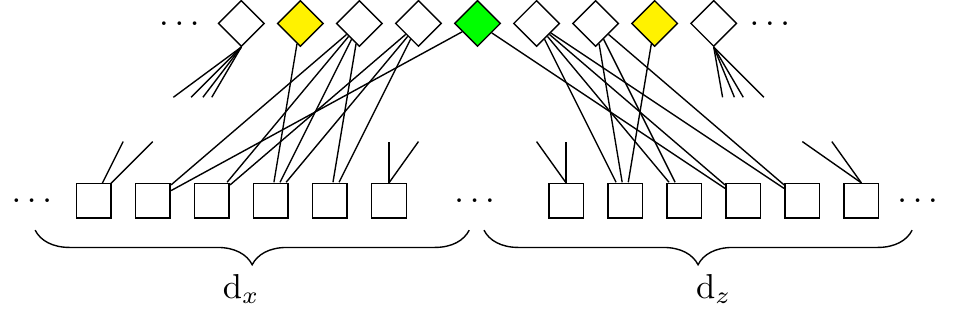}
    \caption{\normalsize{Upper layer of the decoding graph associated to a non-CSS code obtained using method 1. The upper nodes represent the syndrome nodes, while the bottom nodes represent the $\mathrm{d}$ nodes ($\mathrm{d}_x$ and $\mathrm{d}_z$ denote the $\mathrm{d}$ nodes associated to each of the original separated CSS decoding subgraphs). The $\mathrm{s}_A$ nodes are represented in yellow and the $\mathrm{s}_C$ nodes are pictured in green.}}
    \label{feafe}
\end{figure}

Independently of the design method, decoding of the novel non-CSS quantum codes is performed utilizing the sum-product algorithm over the factor graph defined by the product $\mathrm{M}'_d\times \mathbf{H}_\text{CSS}$. Message passing will differ from that performed at a CSS QLDGM decoder, in the sense that both sides of the graph interact during the decoding process (messages are exchanged between nodes on the left and right subsections of the graph). This occurs because of the modified upper layer in the decoding graph of the non-CSS code, as shown in Figure \ref{feafe} for a non-CSS code derived using the first design method.

\subsection{Rate considerations}

Both design methods allow a high degree of flexibility in terms of selecting the rate of the non-CSS quantum code. In fact, different non-CSS codes of the same rate can be obtained depending on the selected design method. Consider an arbitrary non-CSS code of quantum rate $R_Q$ obtained based on the first design method taking as a starting point a CSS code of the same rate. A different non-CSS code of rate $R_Q$ can be obtained by using the second design method, taking as a starting point a CSS code of lower rate. In fact, the second construction technique allows us to build multiple matrices $\mathrm{M}'_d$ of equal size (which will lead to codes of equal quantum rate) by starting from different matrices $\mathrm{M}$ (and thus from CSS codes of different rates) and varying the parameter $q$, which represents number of $\mathrm{s}_A$ nodes that are removed. For instance, we could design a matrix $\mathrm{M}'_{d_1}$ of size $(2m_1-q_1)\times N$ using two matrices $\mathrm{M}_1$ of size $m_1\times \frac{N}{2}$, and construct a second matrix $\mathrm{M}'_{d_2}$ of size $(2m_2-q_2)\times N$ using two matrices $\mathrm{M}_2$ of size $m_2\times \frac{N}{2}$. If $2m_1-q_1 = 2m_2-q_2$, then both codes will have the same rate. Therefore, when the second method is utilized, code optimization depends on the choice of matrix $\mathrm{M}_d$, as well as parameters $m$ and $q$. 

Notice that for a fixed $m \times N$ matrix $\mathrm{M}$, by using the second design method and varying the value of $q$, we obtain different matrices $\mathrm{M}'_d$ of size $(2m - q)\times2N$. This will result in non-CSS codes that encode $N - (2m-q)$ logical qubits into $N$ physical qubits, and thus their rate will be 
\begin{equation}
    R_{Q,\text{non-CSS}} = \frac{N-(2m-q)}{N}.
\end{equation}
Since $q>0$, $R_{Q,\text{non-CSS}}$ is always higher than the rate of the CSS QLDGM code used as a starting point, which is given by
 \begin{equation}
    R_{Q,\text{CSS}} = \frac{N-2m}{N}.
\end{equation}
Notice that if the non-CSS code obtained from method 2 maintains the same performance as the original CSS code, this will be achieved with a higher rate. We introduce the parameter $R_I$, defined as 
\begin{equation} \label{rate_increase}
    R_{I} =  R_{Q,\text{non-CSS}}-R_{Q,\text{CSS}}=\frac{q}{N-2m},
\end{equation}
to quantify the rate increase provided by the non-CSS scheme derived using the second construction method when compared to the original CSS scheme. This rate increase is determined by the value of $q$, which for the second design method represents the number of $\mathrm{s}_C$ nodes in the non-CSS decoding graph, as well as the number of $\mathrm{s}_A$ nodes removed from the decoding graph of the CSS code used as a starting point. The value of $q$ will influence the performance of the resulting non-CSS code: intuitively, large increases in its value should lead to worsened performance, as the doping effect is reduced, but $q>0$ allows for information exchange between the left and the right sides of the decoding graph, which should have a positive effect on performance. The impact of $q$ on the proposed schemes is studied in the following section.

\section{Simulation Results} \label{sec:ldgmres}

In this section we compare the performance  of the proposed non-CSS codes to that of the CSS codes of \cite{jgf3} and \cite{jgf4} when they are used over the i.i.d. $X/Z$ channel and the depolarizing channel. The CSS codes in  \cite{jgf3} and \cite{jgf4} have rate $R_Q \approx \frac{1}{4}$ and block length $N=19014$, encoding $k = 4752$ qubits into $N$ qubits. Matrix $\mathrm{P}$, of size $9507\times9507$, has the same degree distribution as its transpose $\mathrm{P}^\top$, and corresponds to a rate $\frac{1}{2}$ classical LDGM code. Hence, both $\tilde{\mathrm{G}}$ and $\tilde{\mathrm{H}}$ have size $9507 \times 19014$. Matrix $\mathrm{M}$, which is used to transform $\tilde{\mathrm{G}}$ and $\tilde{\mathrm{H}}$ into $\mathrm{G}$ and $\mathrm{H}$, is full-rank, low-density, and has size $7131\times9507$. Results are depicted in terms of either the Qubit Error Rate (QBER) or the Word Error Rate (WER). QBER represents the fraction of qubits that experience an error, while WER is the fraction of transmitted blocks that have at least one qubit error. We use the QBER metric for some of our simulations because it can be estimated with high confidence faster than the WER, which is helpful in shortening the required simulation time of some of our codes. Readers should refer to Appendix \ref{app:sims} for a more complete explanation on how these simulations have been conducted.

First, the codes are simulated over the i.i.d. $X/Z$ channel model of \cite{bicycle}, where $Z$ and $X$ errors are modeled as independent events identically distributed according to the flip probability $f_m$. We begin by using as a starting point the family of CSS codes of the first proposed structure in \cite{jgf3} and \cite{jgf4}, which are individual regular LDGM codes. The simplicity of the channel model and of the code structure allow us to assess, in a rapid and efficient manner, the values of $q$ that optimize the performance of the non-CSS construction. Using these values of $q$, we will repeat the simulations when the CSS codes used as the starting point consist of the parallel concatenation of regular LDGM codes, as described in \cite{jgf3, jgf4}. Both design methods are utilized to obtain the resulting non-CSS codes. Finally, we repeat the same process for the depolarizing channel. 

\subsection{i.i.d. X/Z channel - Non-CSS codes based on individual regular LDGM codes}

For these simulations, the CSS code utilized as a starting point is an individual regular LDGM code. Matrix $\mathrm{P}$ is generated pseudorandomly and corresponds to a regular $(X,X)$ LDGM code. $\mathrm{M}$ is characterized by the parameter values\footnote{The fractional number $8.72$ represents the fact that $72\%$ of the $\mathrm{s}_B$ nodes will have degree 9 while $28\%$ of them will have degree 8.} $\mathrm{M}(3;1,8.72)$ and $t = 4361$. The degrees\footnote{Given that the degree distribution of $\mathrm{P}^\top$ and $\mathrm{P}$ is the same, we refer to them indistinctly throughout this paper.} of $\mathrm{P}$ (and $\mathrm{P}^\top$) are varied between $(9,9)$ and $(13,13)$. In the figures that follow, $f_m$ is the probability of error in each separate $X$ and $Z$ error channels and the analytical error floors of the LDGM codes have been obtained as shown in \cite{jgf5}. 

\subsubsection{Non-CSS codes derived using method 1}

\begin{figure}[h!]
\centering
\includegraphics[width=\linewidth, height=4in]{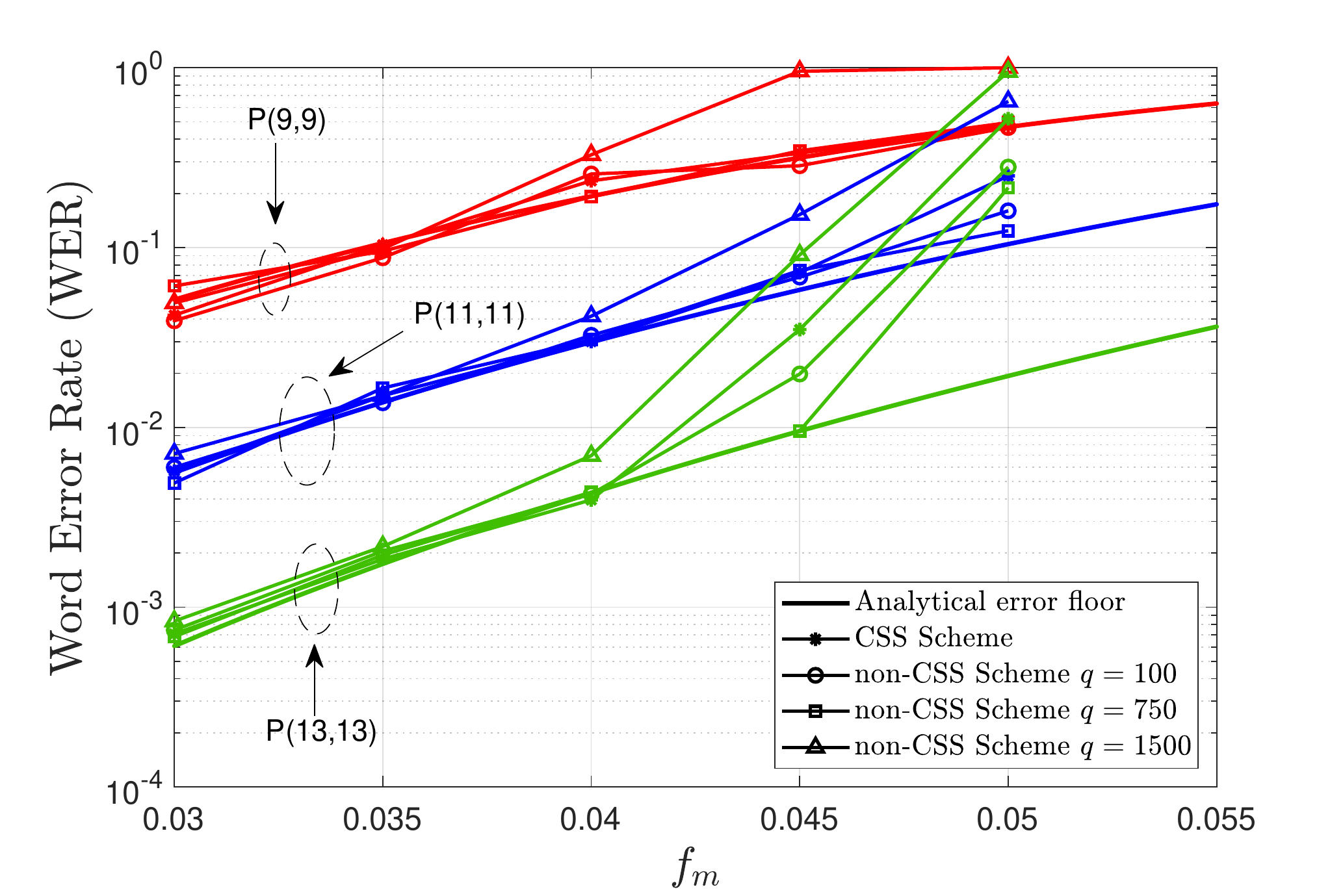}
\caption{\normalsize{Simulated WER for non-CSS QLDGM codes based on individual regular LDGM codes obtained using the first design method when they are applied over the flipping channel. $f_m$ is the probability of error in each separate $X$ and $Z$ error channel.}}
\label{meth1_sims}
\end{figure}

As explained before, non-CSS codes based on this method are obtained by transforming $q$ $\mathrm{s}_A$ nodes of the CSS decoding graph into degree-2 $\mathrm{s}_C$ nodes. All the non-CSS codes obtained in this manner have the same quantum rate ($R_Q = \frac{1}{4}$). Matrix $\mathrm{M}_d$ is built from $\mathrm{M}$ as shown in \eref{eq:nonCSSLDGMPCM}. The rest of the underlying components of the non-CSS configuration are identical to those of the CSS designs\footnote{The only difference between the CSS and non-CSS codes lies in matrix $\mathrm{M}'_d$. The rate $\frac{1}{2}$ classical LDGM code and, therefore, matrices $\mathrm{P}$, $\tilde{\mathrm{G}}$, and $\tilde{\mathrm{H}}$, are identical.}. We test three different configurations, $q= [100, 750, 1500]$, for different degrees of the $\mathrm{P}$ matrices. The simulation results, which are shown in Figure \ref{meth1_sims} reflect how for $q = 100$ and $q = 750$, regardless of the degree of the pseudorandom matrix $\mathrm{P}$, the performance of the proposed non-CSS codes is slightly better than that of the CSS schemes used as a starting point. This performance improvement is most notable for the case of $\mathrm{P}(13,13)$, where, as $f_m$ increases, the non-CSS schemes operate closer than the original CSS code to the analytical error-floor. However, for $q = 1500$, the performance of the non-CSS code is worse than that of the CSS scheme. The best results are observed when the non-CSS code is designed using $q=750$.

\subsubsection{Non-CSS codes derived using method 2}

\begin{figure}[h!]
\centering
\includegraphics[width=\linewidth, height=4in]{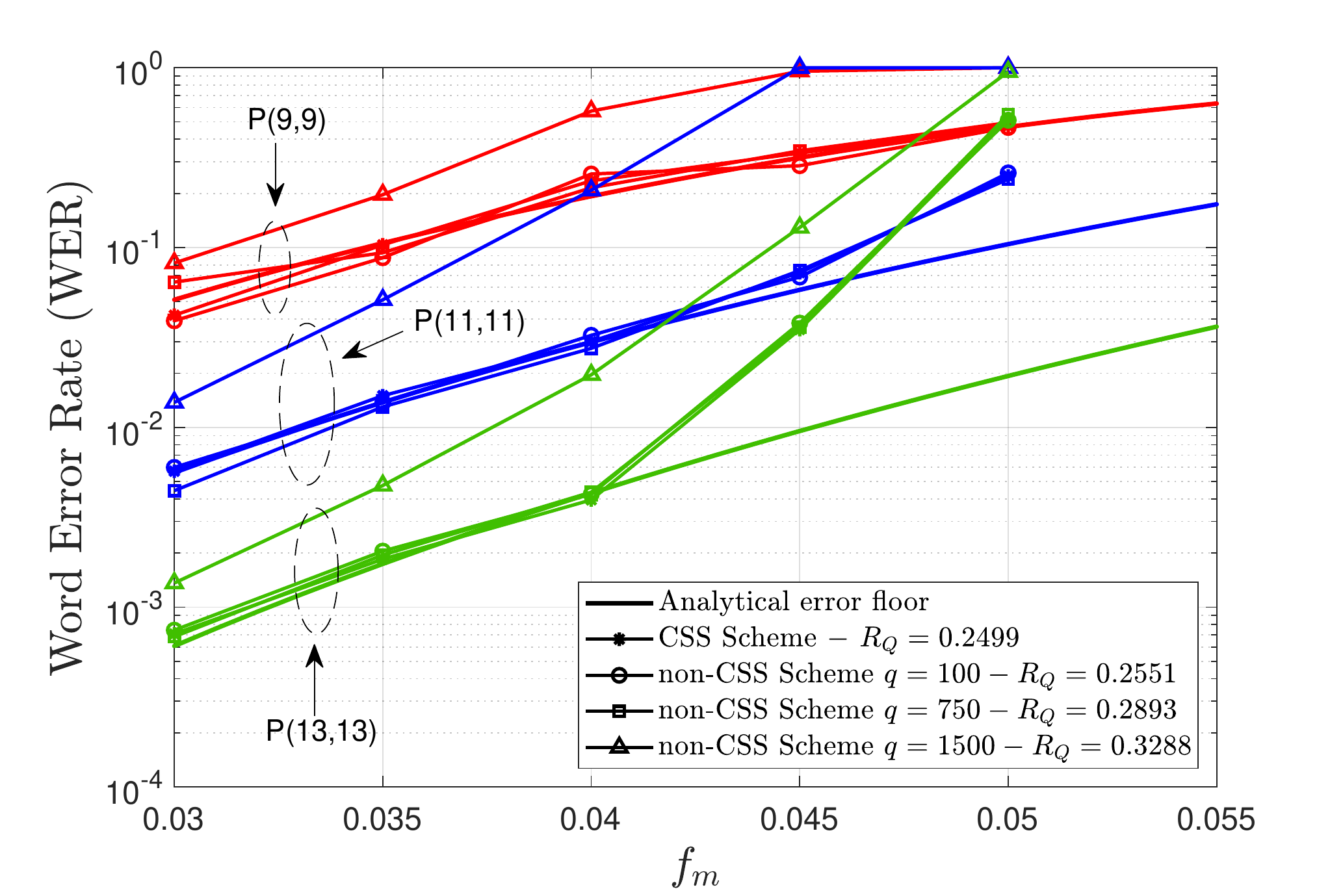}
\caption{\normalsize{Simulated WER for non-CSS QLDGM codes based on individual regular LDGM codes obtained using the second design method when they are applied over the flipping channel. $f_m$ is the probability of error in each separate $X$ and $Z$ error channel.}}

\label{sims}
\end{figure}

Earlier we saw how non-CSS codes derived using method $2$ are obtained by removing $q$ syndrome nodes from the decoding graph generated by the first method, specifically the $q$ $\mathrm{s}_A$ nodes connected to a $\mathrm{d}_A$ node that is linked to an $\mathrm{s}_C$ node. Note that removing syndrome nodes will result in codes with different quantum rates. As previously, we test the values $q = [100, 750, 1500]$ for different degrees of the $\mathrm{P}$ matrices. The results are shown in Figure \ref{sims}.

The results displayed in Figure \ref{sims} are once again consistent regardless of the degree of $\mathrm{P}$. They show how for the two smaller $q$ values, $100$ and $750$, the non-CSS codes yield the same performance as the CSS schemes of \cite{jgf3}. For $q = 1500$, the performance of the non-CSS codes is significantly worse. For instance, at a value of $f_m = 0.04$, the word error rate for the $q=1500$ schemes is around an order of magnitude higher than for the $q \leq 750$ non-CSS schemes. This corroborates our intuition that there is an optimum range of values for $q$, and that utilizing values outside of that range degrades the code performance. Based on the results in Figure \ref{sims}, the optimal value of $q$ should be a small percentage of the total number of syndrome nodes $m$, $q < 0.1\times m$ ($\frac{1500}{14262}\approx 0.1$). 

Table \ref{table} presents the WER performance for the codes considered in Figure \ref{sims} when matrix $\mathrm{P}$ has degrees $(13,13)$ and $f_m=0.03$. The CSS scheme has rate $R_{Q-\text{CSS}} = \frac{19014-2\times7131}{19014} = 0.2499$. Results for the $q=1500$ non-CSS scheme have not been included, since the performance of the original CSS scheme is better. As with the first method, the best result here is obtained for the $q=750$ non-CSS scheme, which achieves similar performance to that of the CSS code with an approximately $15 \%$ higher rate ($R_I \approx 0.158$).

Overall, the results obtained for both design methods illustrate that our proposed non-CSS codes, regardless of the design method, outperform QLDGM CSS codes when individual regular LDGM codes are utilized. The first design method provides a way to obtain non-CSS codes that outperform CSS codes of the same rate. The second method enables us to construct non-CSS codes with error correcting capabilities comparable to those of lower rate CSS schemes. Therefore, to design non-CSS codes of a fixed rate, we could apply the first design method to a CSS code of the same rate, or we could start with a CSS code of lower rate and apply the second method. In this case, we should carefully chose the rate of the original CSS code to obtain the desired rate in the designed non-CSS code. 

 \begin{table}[h!]

    \centering\caption{\normalsize{Comparison between the codes shown in Figure \ref{sims}. The WER data included in the table corresponds to the codes whose $\mathrm{P}$ matrix has degrees $(13,13)$.}}
    \vspace{0.15mm}\begin{tabular}{ccccc}
    \toprule 
    Code Type &$R_Q$ &$q$ &WER @ $f_m=0.03$ \\
    \midrule
    CSS  &0.2499 &- &$7.12\times10^-3$ \\
    non-CSS &0.2551 &100 &$7.26\times10^-3$\\
    non-CSS &0.2893 &750 &$7.13\times10^-3$
    \\\bottomrule
    \label{table}
    \end{tabular}
    \end{table}

\subsection{i.i.d. X/Z Channel - Non-CSS codes based on the parallel concatenation of LDGM codes}

As mentioned in section \ref{sec:classLDGM}, regular LDGM codes used in classical channels present error floors. Fortunately, these error floors can be substantially lowered if we use the parallel concatenation of two regular LDGM codes. In \cite{jgf1}, \cite{jgf2} CSS quantum codes based on the use of single regular LDGM codes were shown to also exhibit error floors. Inspired by the good performance displayed by parallel LDGM codes in the classical domain, a scheme based on the parallel concatenation of LDGM codes was designed and applied to the i.i.d. $X/Z$ channel in \cite{jgf3} and \cite{jgf4}. Similar to the classical scenario, CSS codes built based on the parallel structure display lower error floors and better performance overall.

We now repeat the study carried out in the previous subsection, but using the parallel concatenation of regular LDGM codes as the starting point to derive our proposed non-CSS codes. In \cite{jgf4}, varios parallel\footnote{The notation $\mathrm{P}[(y_1, y_1);(y_2, z_2)]$ indicates the degree distributions of the constituent regular LDGM codes utilized in the parallel concatenation. In \cite{jgf4}, the parameters of the second code were typically chosen as $z_2 = 20y_2$.} LDGM structures $\mathrm{P}[(y_1,y_1);(y_2,z_2)]$ were employed. For our analysis we use the structure with the lowest degrees that appears in \cite{jgf4}: $\mathrm{P}[(8, 8);(3, 60)]$. Although performance is better when the degrees of the second code $(y_2, z_2)$ are larger, codes with smaller values for these degrees require less processing and simulation time. To ease simulation requirements even further, results in this subsection are presented in terms of the QBER. The best configuration for $\mathrm{P}[(y_1,y_1);(y_2,z_2)]$ in \cite{jgf4} will be used later for the simulations over the depolarizing channel.

As we did for individual regular LDGM codes in the previous subsection, we analyze the performance of the non-CSS codes obtained using the two design methods proposed in section \ref{sec:nonCSS}. The CSS code used as a starting point is the same for both methods, and utilizes the same matrix $\mathrm{M}$ as in the scenario for individual regular LDGM codes: $\mathrm{M}_{7131\times9507}(3;1,8.72)$ and $t=4361$. We consider the values of $q=\{100,500,1000,1500\}$. As before, $f_m$ is the probability of error in each separate $X$ and $Z$ error channel and the analytical error floors of the LDGM codes have been obtained as shown in \cite{jgf5}. 

\begin{figure}[h!]
\centering
  \includegraphics[width=\linewidth,height=4in]{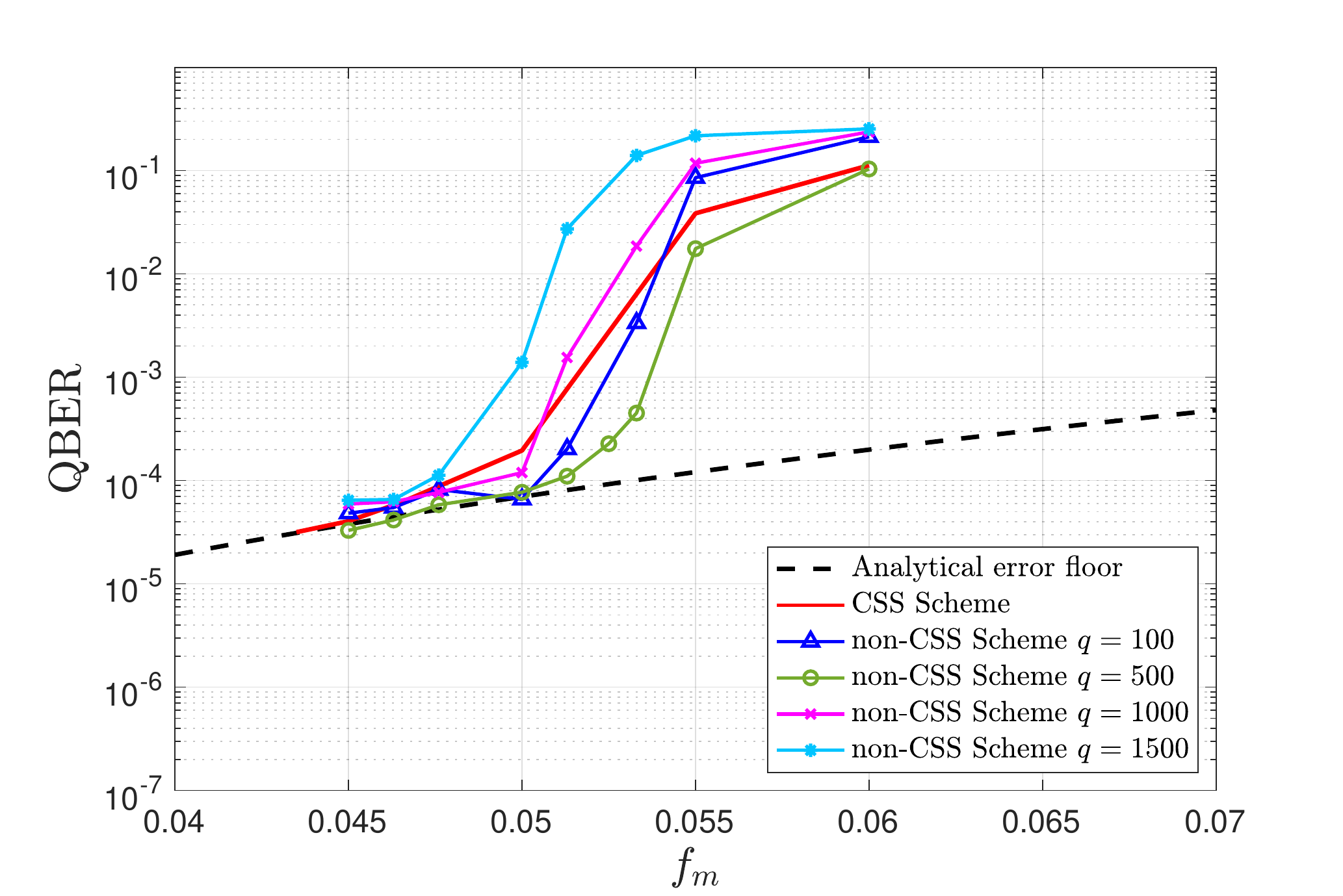}
\caption{Simulated QBER for a $R_Q=\frac{1}{4}$ CSS code and non-CSS codes of $R_Q=\frac{1}{4}$ code derived using the first design methodology. The underlying classical LDGM code is the same for all the codes and has degree distribution $\mathrm{P}[(8, 8);(3, 60)]$. $f_m$ is the probability of error in each separate $X$ and $Z$ error channel.}
\label{parallel_results}
\end{figure}

 Figure \ref{parallel_results} presents the performance of the $R=\frac{1}{4}$ original CSS code and of the $R=\frac{1}{4}$ non-CSS codes obtained by applying the first design method. As shown in the figure, the non-CSS codes derived by selecting $q=100$ and $q=500$ outperform the original CSS structure. For $q > 500$, performance of the non-CSS codes gradually deteriorates, with the result for $q=1500$ being substantially worse than that of the original CSS code.

\begin{figure}[!htp]

  \includegraphics[width=\linewidth,height=4in]{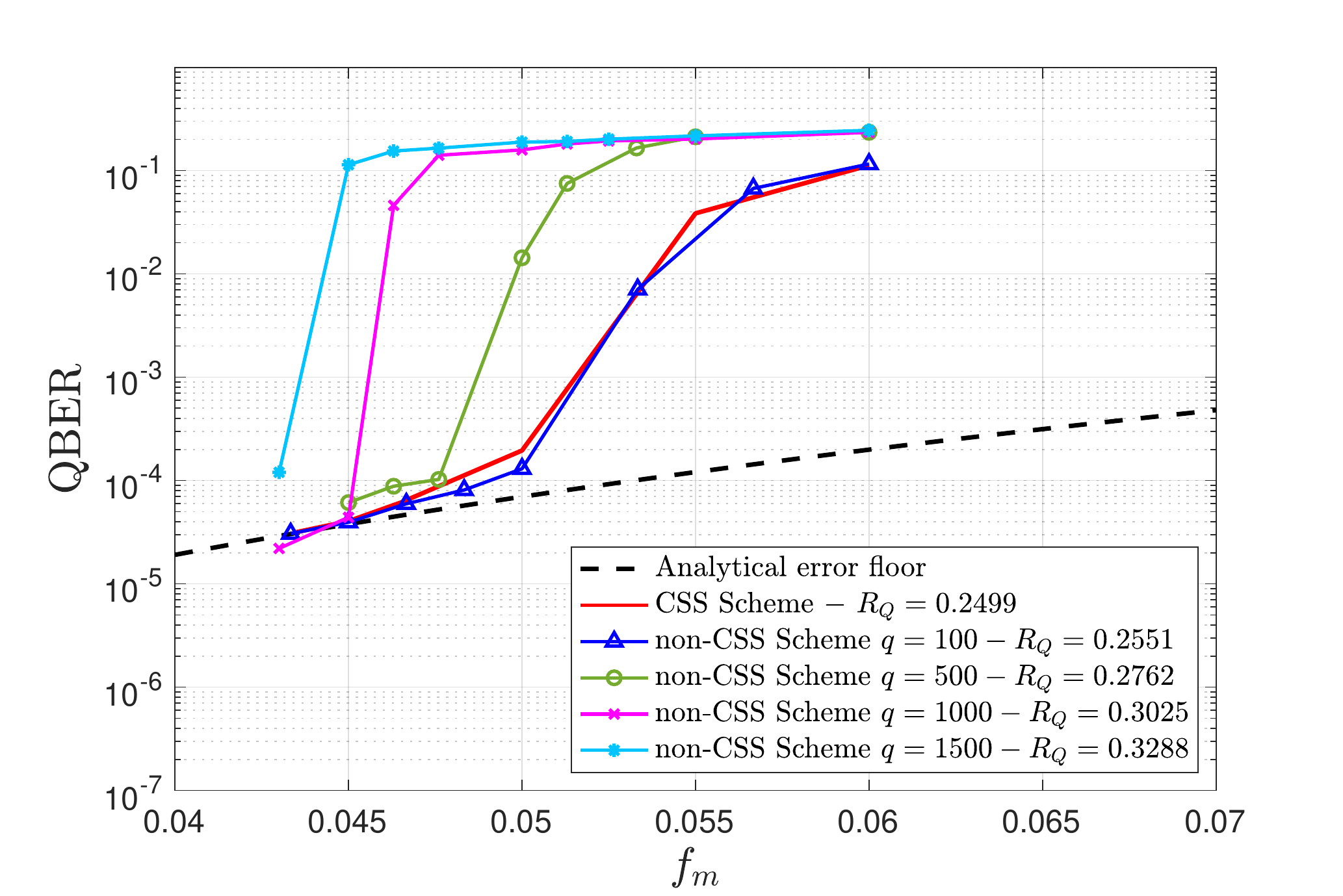}

\caption{Simulated QBER for a $R_Q=\frac{1}{4}$ CSS code and non-CSS codes of different rate derived using the second design methodology. The underlying classical LDGM code is the same for all the codes and has degree distribution $\mathrm{P}[(8, 8);(3, 60)]$. $f_m$ is the probability of error (iid) in each separate $X$ and $Z$ error channel.}
\label{meth2_parallel_results}
\end{figure}

Figure \ref{meth2_parallel_results} shows the results for the non-CSS codes derived using the second proposed design method. The curves shown in this figure portray how the performance of the non-CSS codes is drastically degraded as we increase the value of $q$. This effect is much more noticeable than for codes based on a single regular LDGM code. In fact, the only value for which the non-CSS configuration based on parallel concatenation matches the performance of the original CSS scheme is $q=100$, whereas in Figure \ref{sims} we could see that schemes based on individual regular LDGM codes matched the performance of the original CSS code at least up to $q=750$. In essence, although the $\mathrm{s}_A$ node elimination step explained in section \ref{construction} still yields a small benefit, this is much lower than for the case of single regular LDGM codes. 

  Throughout this chapter we have mentioned that the second non-CSS code design methodology can be used not only to obtain higher rate non-CSS codes with performance similar to that of lower rate CSS codes, but also to generate different non-CSS codes of a fixed rate by varying the CSS codes used as starting points and selecting the appropriate value of $q$. Analyzing the performance of $R_Q=\frac{1}{4}$ non-CSS codes obtained in this manner and comparing the results to those obtained using the first design method will allow us to determine which design technique yields codes with better performance.

\begin{figure}[htp]

  \includegraphics[width=\linewidth,height=4in]{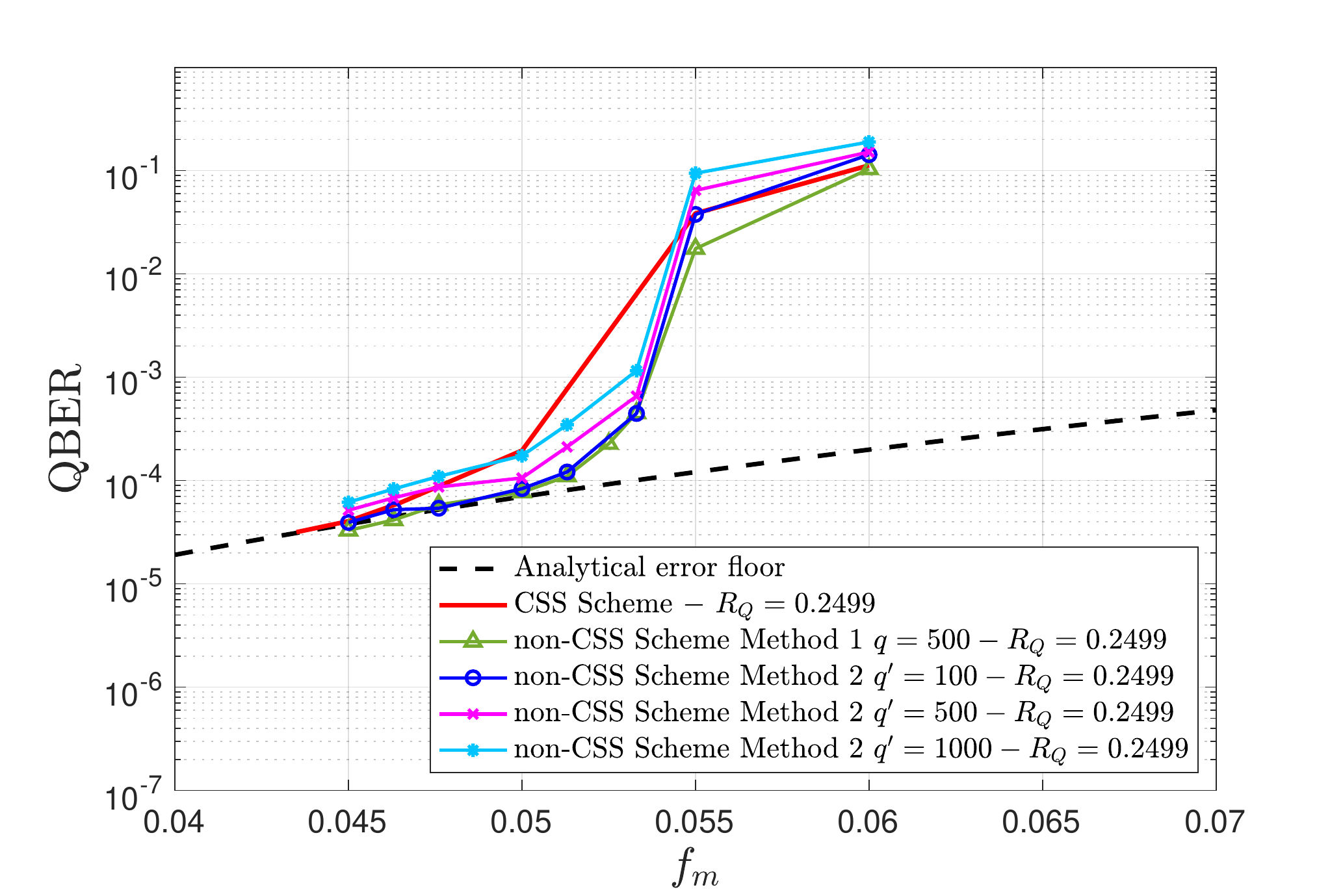}

\caption{\normalsize{Simulated QBER for a $R_Q=\frac{1}{4}$ CSS code and $R_Q=\frac{1}{4}$ non-CSS codes constructed using the two proposed design methods. $q$ denotes how many $\mathrm{s}_A$ nodes have become $\mathrm{s}_C$ nodes in the upper layer of the decoding graph when method 1 is applied. $q'$ denotes how many $\mathrm{s}_A$ nodes have become $\mathrm{s}_C$ nodes and how many $\mathrm{s}_A$ nodes have been removed from the upper layer of the decoding graph when method 2 is used. The underlying classical LDGM code is the same for all the codes and has degree distribution $\mathrm{P}[(8, 8);(3, 60)]$. $f_m$ is the probability of error (iid) in each separate $X$ and $Z$ error channel.}}

\label{newfig}
\end{figure}

 This comparison is shown in Figure \ref{newfig}, where various  $R_Q=\frac{1}{4}$ non-CSS codes obtained using the second design method are compared to the best $R_Q=\frac{1}{4}$ code generated by the first design method ($q=500$ in Figure \ref{parallel_results}) and to the original $R_Q=\frac{1}{4}$ CSS structure. The non-CSS codes generated by method 2 are derived by building matrices $\mathrm{M}'_d$ using matrices $\mathrm{M}$ of different size, which are designed according to the analysis conducted in \cite{jgf4}. We introduce parameter $q'$  to refer to the values of $q$ used to derive these codes and to distinguish them from the code built using the first method. 
 
The $q' = 100$ code uses $\mathrm{M}_{7181\times9507}(3;1,8.56)$ with $t=4361$ (using this $\mathrm{M}$ for a CSS QLDGM scheme would yield a code of rate $R_Q=0.244$). The $q' = 500$ code uses $\mathrm{M}_{7381\times9507}(3;1,8.36)$ with $t=4561$ (using this $\mathrm{M}$ for a CSS QLDGM scheme would yield a code of rate $R_Q=0.2236$). The $q' = 1000$ code uses $\mathrm{M}_{7631\times9507}(3;1,8.27)$ with $t=4761$ (using this $\mathrm{M}$ for a CSS QLDGM scheme would yield a code of rate $R_Q=0.1973$).
 
 Figure \ref{newfig} shows that the non-CSS codes designed using the second method outperform the CSS scheme for all values of $q'$. However, the $q=500$ non-CSS code designed using the first method is still better than any of the aforementioned codes, although the performance of the $q'= 100$ non-CSS code is only slightly worse. 
 
 In this subsection we have discussed results for the i.i.d $X/Z$ Channel, where the $X$ and $Z$ operators are modelled independently according to the same distribution. Our analysis has allowed us to determine the design methodologies and the values of $q$ that lead to the best performance. Making use of this knowledge, we will now consider performance over the depolarizing channel.

 \newpage

\subsection{Depolarizing channel}

\begin{figure}[p!]
\centering

      \includegraphics[width=\linewidth,height=4in]{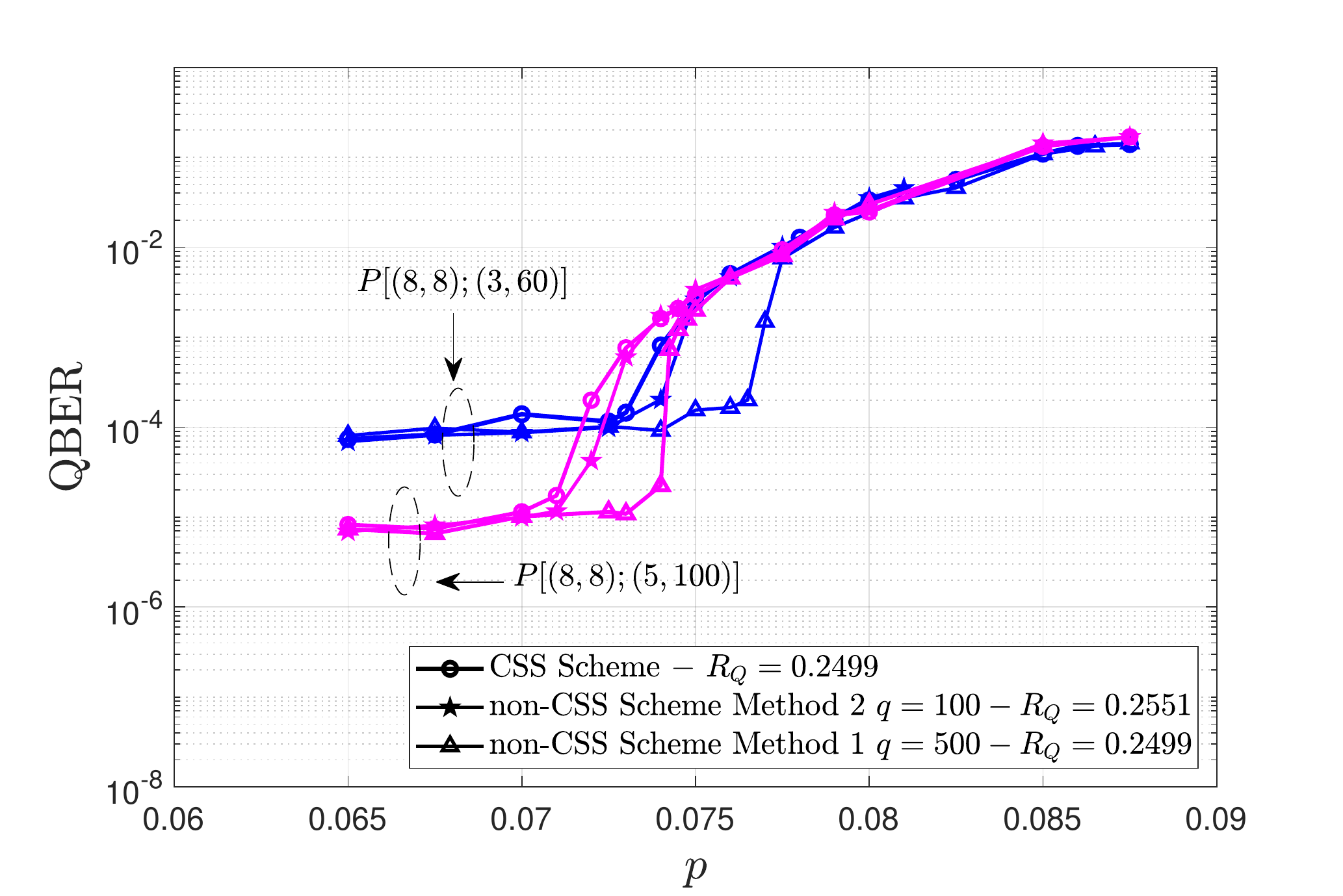}

\caption{\normalsize{Simulated QBER for three quantum codes: An $R_Q=\frac{1}{4}$ CSS QLDGM code, an $R_Q=\frac{1}{4}$ non-CSS QLDGM code obtained using the first proposed method and $q=500$, and an $R_Q=0.255$ non-CSS QLDGM code derived via the second design technique with $q=100$. All three codes have been simulated for two different degree distributions of the underlying parallel concatenated LDGM scheme: $\mathrm{P}[(8, 8);(3, 60)]$ and $\mathrm{P}[(8, 8);(5, 100)]$. $p$ is the depolarizing probability.}}

\label{parallel_dep}

\end{figure}

We now focus on the depolarizing channel introduced in section \ref{depchan}. We consider the best non-CSS codes obtained before for the i.i.d. $X/Z$ channel: The $q=500$ $R_Q=\frac{1}{4}$ non-CSS code obtained using the first design method and the $q=100$ $R_Q=0.255$ non-CSS code obtained using the second method. Figure \ref{parallel_dep} shows the QBER of the aforementioned codes for two different degree distributions of the parallel concatenated LDGM scheme. The curves associated to the original CSS codes are also included.

Similar to the results for the i.i.d. $X/Z$ channel, Figure \ref{parallel_dep} portrays how the $R_Q=0.255$ non-CSS codes display performance very close to that of the $R_Q=\frac{1}{4}$ CSS schemes. This phenomenon can be explained for both the depolarizing and the i.i.d. $X/Z$ channels by the nature of the non-CSS construction, which adds a very small number of edges and removes very few syndrome nodes from the original CSS factor graph.  As in the case of the $X/Z$ channel model, the $q=500$, $R_Q=\frac{1}{4}$ non-CSS codes designed utilizing method 1 also outperform the CSS code. 

An important observation, which is reflected in Figure \ref{parallel_dep}, is that the results are consistent regardless of the degrees $(y_2,z_2)$ that are chosen. This is significant, since increasing $y_2$ and $z_2$ enables us to lower the error-floor of the QLDGM code. As shown in \cite{jgf3} and \cite{jgf4} for CSS codes, we can see from Figure \ref{parallel_dep} that selecting $\mathrm{P}_2$ with larger degrees lowers the error floor but worsens the decoding threshold.

\begin{figure}[p!]
\centering

  \includegraphics[width=\linewidth,height=4in]{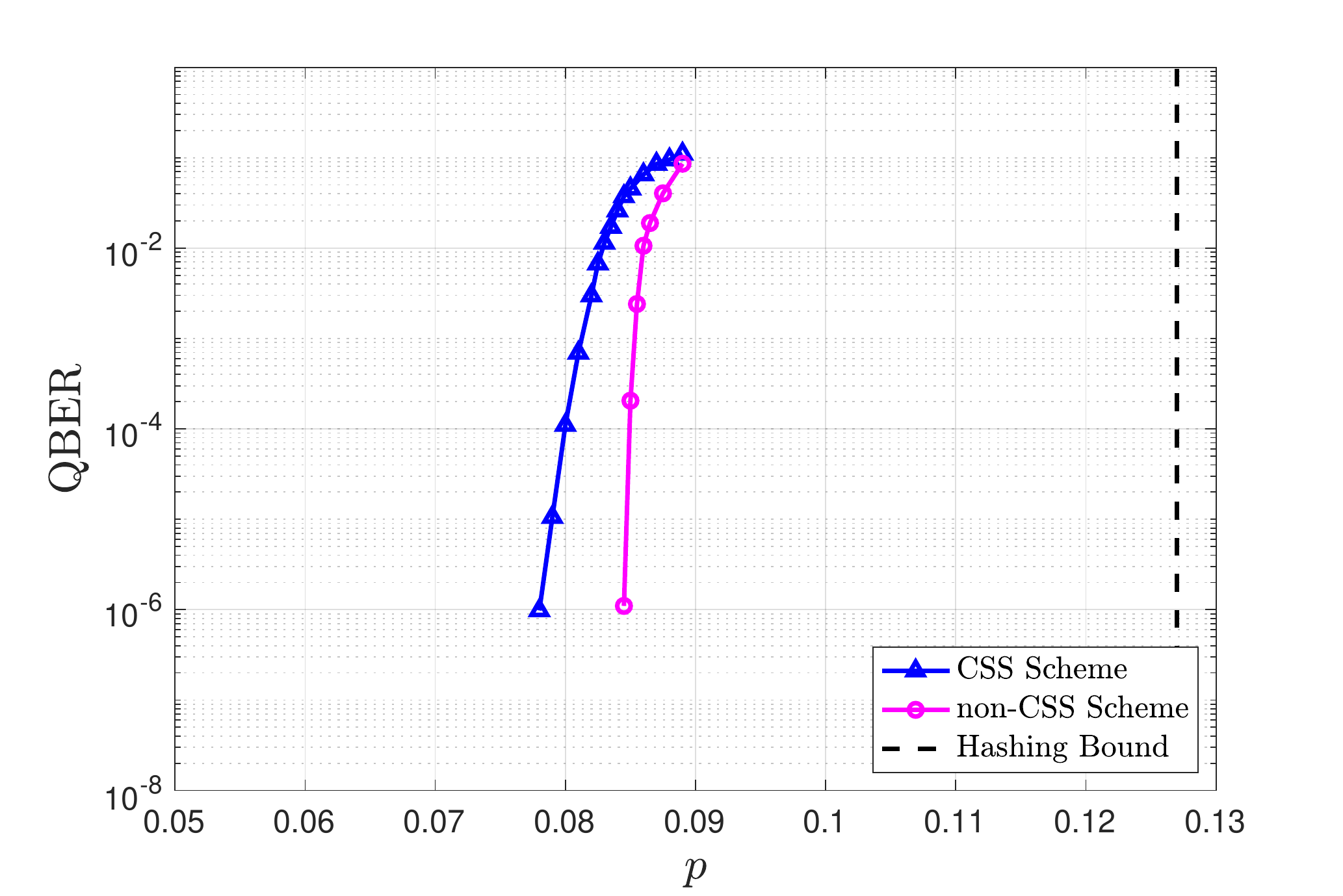}

\caption{\normalsize{Simulated QBER for the best $R=\frac{1}{4}$ CSS QLDGM code in \cite{jgf4} and the best non-CSS QLDGM code designed in this paper. The Hashing bound is also shown. The codes are based on the parallel concatenation of regular LDGM codes with degree distribution $\mathrm{P}[(8, 8);(8, 160)]$. $p$ is the depolarizing probability.}}
\label{bestqber}
\end{figure}

\subsubsection{Distance to the theoretical limit}

The most effective way to characterize the performance improvement of our proposed non-CSS codes is to measure their gap with respect to the Hashing bound. For this comparison, we will employ the design parameters that yield the best possible code. Such a scheme is obtained by using the first construction method with $t=5000$, $q=500$, $\mathrm{M}(3;1,11.04)$, and a parallel concatenated LDGM code of degree distribution $\mathrm{P}[(8, 8);(8, 160)]$.

Figure \ref{bestqber} depicts the performance of this non-CSS scheme as well as that of the CSS code used as a starting point for the design, which is the best CSS code proposed in \cite{jgf4}. The Hashing bound for $R_Q = \frac{1}{4}$ is also shown.

We compute the distance to the Hashing bound $\delta$ as defined in \eref{eq:hash_distance}, knowing that the the noise limit for $R_Q=\frac{1}{4}$ is $p^* \approx 0.127$, and taking $p_\text{CSS} = 0.0825$ and $p_\text{non-CSS} = 0.0865$ as the depolarizing probabilities at which the CSS and non-CSS codes enter the waterfall region, respectively. This yields $\delta_\text{CSS} = 1.873 $ dB and $\delta_\text{non-CSS} = 1.668 $ dB. In other words, the non-CSS scheme is about $0.2$ dB closer to the Hashing bound. Thus, in terms of overall performance over the depolarizing channel, the non-CSS codes proposed in this article outperform existing CSS techniques.

\subsubsection{Comparison with existing QLDPC schemes}

\begin{figure}[H]
\centering
\includegraphics[width=\linewidth, height=4in]{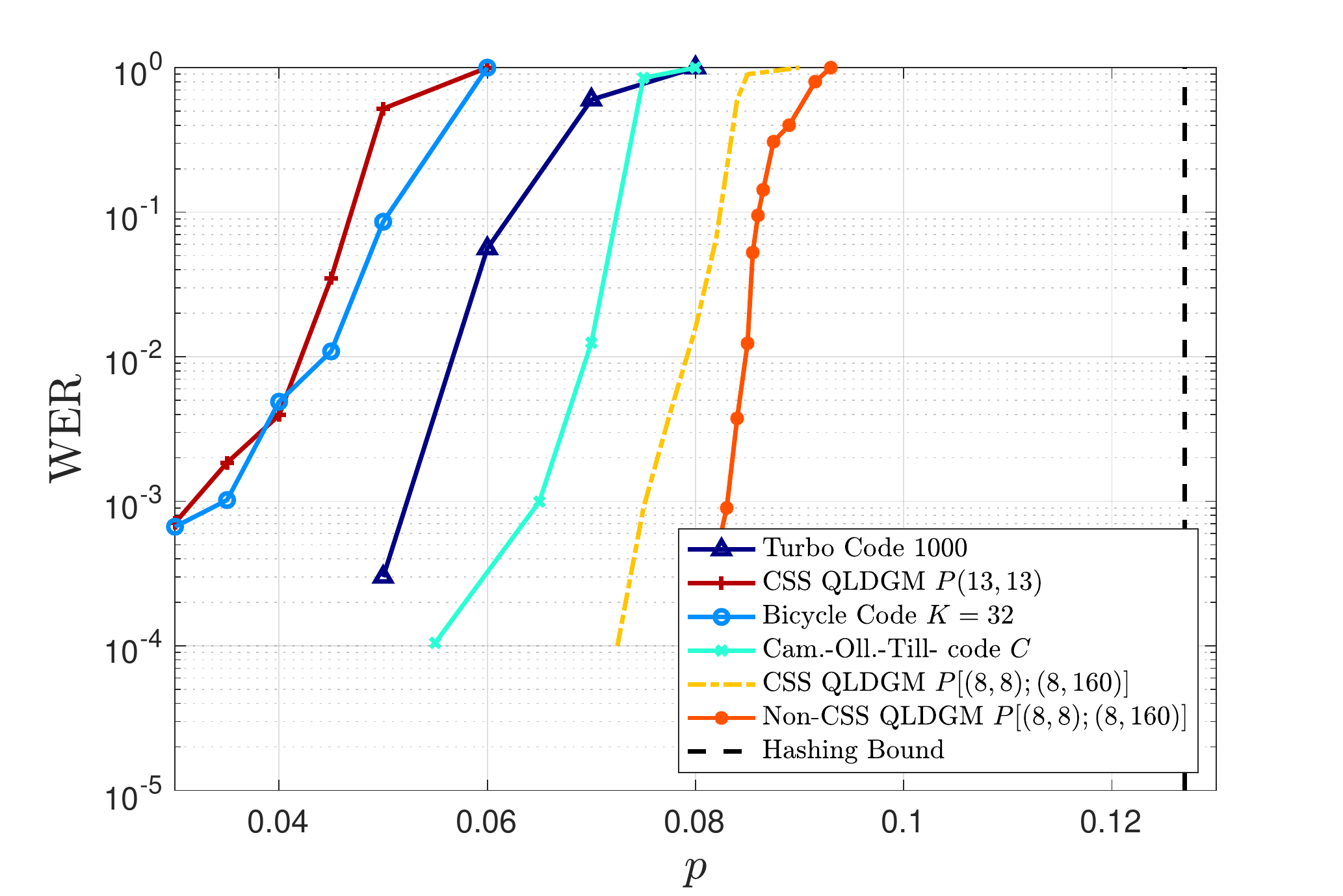}
\caption{\normalsize{Word error rate for various quantum codes when they are applied over the depolarizing channel. The Hashing bound is also included. $p$ is the depolarizing probability.}}
\label{finalcomp}
\end{figure}

We close out this section by studying how our proposed non-CSS codes measure up against other QLDPC schemes in the literature. For this purpose, we conduct two different comparisons. We begin by comparing the codes in Figure \ref{bestqber} to other quantum codes of rate $\frac{1}{4}$ and then use a second comparison to study how decoders capable of exploiting the correlation between $X$ and $Z$ operators match up to our non-CSS construction. The first comparison is shown in Figure \ref{finalcomp}, which includes the performance for the following codes:

\begin{itemize}
    \item The CSS QLDGM code based on a single regular LDGM code from \cite{jgf2}. The degree distribution of the underlying classical LDGM code is $\mathrm{P}(13,13)$. The block length of the code is $19014$.
    \item The $K=32$ bicycle code of block length $19014$ introduced by MacKay et al. in \cite{bicycle}. 
    \item The quantum serial turbo Code of \cite{QTC}, with block length $4000$.
    \item The non-CSS concatenated code (Code $C$) from \cite{nonCSS1}, with block length $138240$. 
\end{itemize}

Figure \ref{finalcomp} shows that our proposed non-CSS QLDGM codes outperform existing quantum turbo codes and previously proposed quantum LDPC codes. 

\subsubsection*{Comparison to improved CSS decoding strategies}

In Chapter $5$ we mentioned that CSS decoders capable of improving performance by exploiting the correlation that exists between $X$ and $Z$ operators over the depolarizing channel have been studied in the literature. Some of the earliest work on this topic was conducted in \cite{degen3}, where a set of modified BP decoders for CSS codes that reintroduce $X/Z$ correlations were proposed. The most notable of these decoding schemes is known as the random perturbation decoder. In \cite{mod-BP}, further work on the topic of modified BP decoders for CSS codes was conducted and two novel CSS decoders were presented: the adjusted decoder and the augmented decoder. 

The adjusted decoder attempts to reintroduce the correlations between $X$ and $Z$ operators, neglected by a standard binary BP decoder, by adjusting prior probabilities. The augmented decoder operates by appending a specific subset of rows $\mathrm{H}_\delta$ from the original PCM $\mathbf{H}$ into an exact copy of that matrix, resulting in an augmented PCM $\mathbf{H}_A = \begin{pmatrix} \mathbf{H} \\ \mathrm{H}_\delta \end{pmatrix} $ which is associated to a larger factor graph over which the decoding algorithm is run. Also in \cite{mod-BP}, the adjusted and augmentation techniques are combined to form a new CSS decoder known as the combined decoder. An in-depth look at some of these modified decoding strategies is provided in Appendix \ref{app:modified}.

In Figure \ref{qbers}, we compare the performance of the random perturbation decoder, as well as that of the adjusted, augmented, and combined decoders, to that of our proposed non-CSS codes when they are applied over the depolarizing channel. The underlying LDGM code and CSS configuration is the same as in most of our previous simulations: the classical code is defined by the parallel concatenation $\mathrm{P}[(8,8);(3,60)]$ and the CSS construction is achieved using $\mathrm{M}(3;1,8.72)$ and $t = 4361$. Performance of a typical CSS decoder (separate decoding of $X$ and $Z$ operators) is also included in Figure \ref{qbers}. The non-CSS scheme is obtained from the CSS code by
using the first design method and setting $q = 500$, which is
the parameter configuration that produced the best results in
our earlier simulations. Both the CSS code and the non-CSS
configuration have rate $R_Q = \frac{1}{4}$.

\begin{figure}[h!]
\centering

  \includegraphics[width=\linewidth,height=4in]{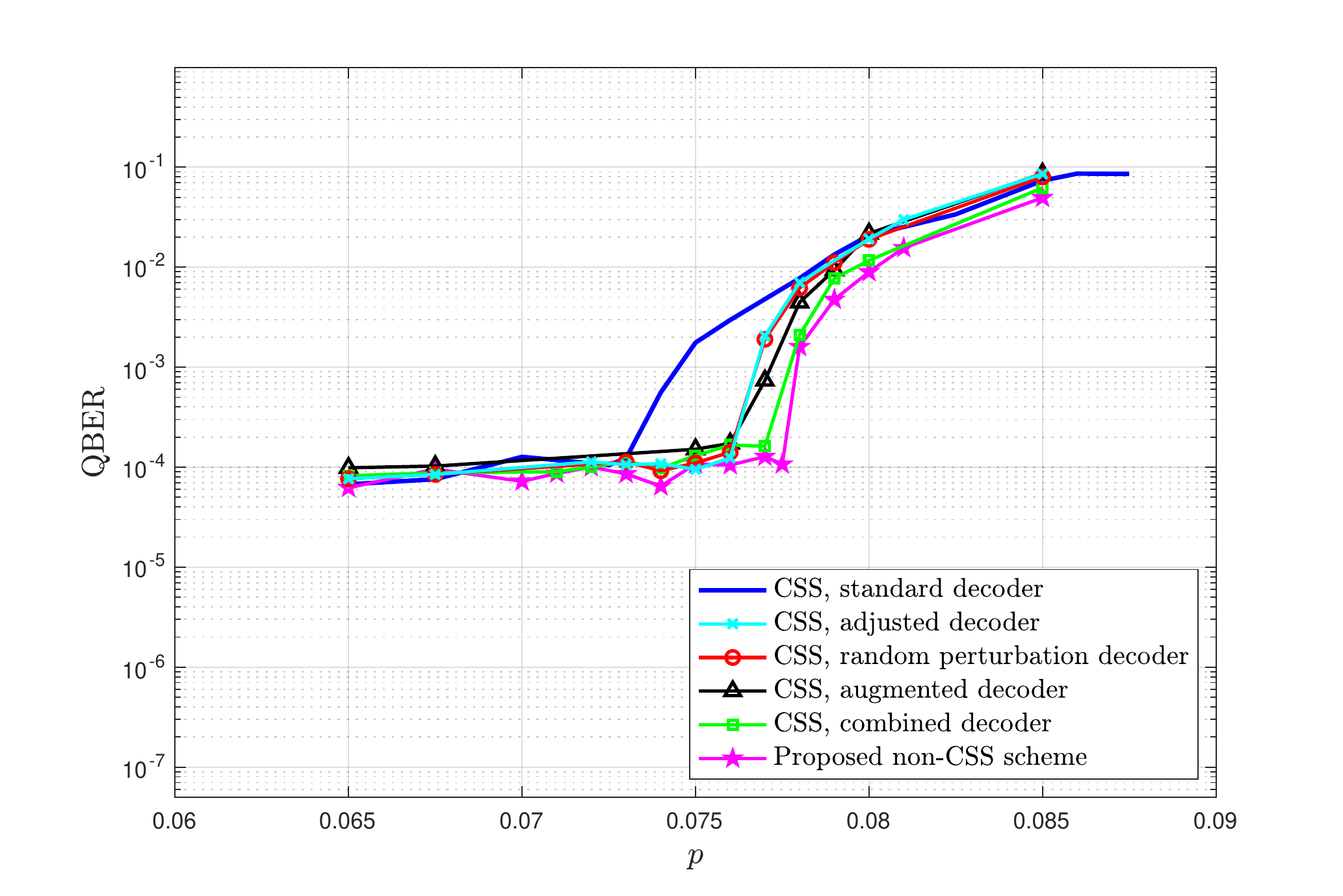}

\caption{\normalsize{Simulated QBER for different types of modified CSS BP decoders over the depolarizing channel. Results for the separate decoding of $X$ and $Z$ errors (standard CSS decoding) and for our proposed non-CSS scheme are also included. $p$ is the depolarizing probability of the depolarizing channel.}}
\label{qbers}
\end{figure}  

As shown in Figure \ref{qbers}, all of the modified CSS BP decoders yield performance improvements when compared to the generic CSS decoding scheme. Nonetheless, they are all outperformed by our proposed non-CSS construction, despite the close proximity between the QBER curves of the combined decoder and the non-CSS scheme. Importantly, we must note that although these modified decoders achieve performance close to that of our proposed scheme, they require specific modifications to the decoding algorithm that result in higher decoding complexity. This increase in decoding complexity arises from the fact that the modified CSS schemes require the execution of a standard decoder prior to their application. As mentioned in section \ref{sec:heuristics}, most modified decoding strategies rely on the failure of a standard decoder to then apply modifications to the factor graph according to the channel error estimate produced by the failed standard decoder. Once these modifications have been made (either to the a priori probabilities or to the factor graph itself) decoding is re-attempted. In some instances, decoding must be re-attempted multiple times before a correct syndrome estimation is produced. Moreover, the augmented decoder and the combined decoder operate over larger factor graphs, which further increases decoding complexity.

Since our proposed construction is decoded by running the SPA over the corresponding factor graph, decoding never has to be re-attempted, and so the complexity of our decoder is essentially the same as that of a standard CSS decoder. Therefore, it is clear that the scheme proposed in this chapter outperforms the aforementioned CSS decoders while displaying a lower decoding complexity.

\section{Chapter summary}

This chapter has introduced a technique to design non-CSS quantum codes based on the use of the generator and parity check matrices of LDGM codes. The proposed methods are based on modifying the upper layer of the decoding graph in CSS QLDGM constructions. The simplicity of the proposed scheme ensures that the high degree of flexibility in the choice of the quantum rate and the block length for the CSS code utilized as a starting point is translated to the non-CSS design. Compared to quantum CSS codes based on the use of  LDGM codes, the proposed non-CSS scheme is $0.2$ dB closer to the Hashing bound in the depolarizing channel and outperforms all other existing quantum codes of comparable complexity.
\clearemptydoublepage
\chapter{Performance of QLDPC codes over Pauli channels} \label{chapter7}

\epigraph{\textit{``The future can ever promise but one thing and one thing only: surprises''}}{\textbf{Steven Erikson}.}

\noindent\hrulefill

In Chapter \ref{chapter6} we studied and analyzed the performance of quantum codes over the depolarizing channel. However, despite serving as a practical abstraction that is useful to design and evaluate the performance of quantum codes, the depolarizing channel model we have worked with up to this point does not always provide an accurate representation of practical communication schemes or the real behaviour of quantum devices. A good example of this is the scenario known as channel mismatch, which describes the event in which the value of the depolarizing probability is not known prior to decoding. Notice how, up to this point, a perfect channel knowledge assumption was implicit in our simulations, i.e, we considered that the decoder knew the precise value of the depolarizing probability prior to decoding. Clearly, the depolarizing channel model that we have considered previously does not suffice to represent the channel mismatch scenario. Another instance in which a different channel model becomes necessary is when seeking to represent the behaviour of realistic quantum devices. Many of these devices behave asymmetrically: the probability of a phase-flip taking place is actually orders of magnitude higher than the probability of a bit-flip; a type of behaviour that cannot be appropriately described using the depolarizing channel. For these reasons, in this chapter we introduce different Pauli channel models that are appropriate to represent realistic scenarios that the depolarizing channel cannot re-enact. We focus on the phenomenon of channel mismatch in section \ref{sec:mismatch} and then analyze the performance of CSS codes over asymmetric Pauli channels in \ref{sec:asym}.

\section{Performance of non-CSS QLDPC codes over the misidentified depolarizing channel} \label{sec:mismatch}

In the previous chapter we introduced a non-CSS inspired LDGM-based strategy that outperforms all other CSS and non-CSS codes of similar complexity over the depolarizing channel. We did so, as is common throughout the literature of QEC, under the tacit premise that perfect knowledge of the quantum channel in question was available. In reality, such a scenario is highly unlikely, and the depolarizing probability must somehow be estimated prior to decoding. This makes it relevant to study how the behaviour of QEC codes can change in terms of the existing information about the quantum channel. 

The phenomenon of channel mismatch in the context of QEC was first considered in \cite{first-impact}, where the authors studied the impact of decoding QLDPC codes with imperfect channel knowledge over the depolarizing channel. In \cite{current}, the same authors designed an improved decoding strategy for QLDPC codes when only an estimate of the channel depolarizing probability was available. This last method made use of quantum channel identification, a technique that requires the introduction of a probe (a known quantum state) into the quantum channel and the subsequent measurement of the channel output state to produce an accurate estimate of the depolarizing probability. Quantum channel identification procedures typically require additional qubits and result in a latency increase, making the error correction procedure increasingly cumbersome. For this reason, researchers have since focused on the design of methodologies that can minimize this overhead (by avoiding the use of channel identification) while yielding performance similar to the perfect channel knowledge scenario. Such techniques have already been explored and successfully demonstrated for QTCs. In \cite{josu2} the authors derive a so-called on-line depolarizing probability estimation technique that yields similar performance to that obtained when using the same QTCs with perfect channel information but without the need for additional resources. In light of this outcome and because an equivalent strategy has not been proposed for QLDPC codes, in this section we derive a similar on-line estimation procedure to that of \cite{josu2} that can be applied to sparse quantum codes. 

\subsection{Quantum Channel Identification} \label{sec:identification}

A common assumption in the field of QEC is that perfect knowledge of the quantum channel under consideration is available prior to decoding. In reality, this information cannot be readily obtained, and estimates of the corresponding quantum channel parameter must be derived. In the context of the depolarizing channel, this means that the decoder must be provided with an estimate of the channel depolarizing probability prior to the execution of the decoding process. If the estimated value of the depolarizing probability $\hat{p}$ is different to its actual value $p$, i.e., $\hat{p} \neq p$, we encounter the scenario known as \textit{channel mismatch}, which typically results in the degradation of the performance of QEC codes (most codes performs worse at the channel noise value $\hat{p}$ than at the real value $p$). 

To study the sensitivity of the non-CSS QLDGM scheme derived in the previous chapter to the channel mismatch phenomenon, we consider a scenario where $\hat{p}$ is varied while the true depolarizing probability remains fixed. We conduct these simulations using the $R_Q = \frac{1}{4}$ $q = 500$ non-CSS strategy derived based on the first methodology presented in section \ref{sec:first} and that employs an underlying parallel-concatenated LDGM code with degree distribution $P[(8,8)(3,60)]$. We have chosen the parallel concatenation with the smallest degrees tested in Chapter \ref{chapter6} in order to ease simulation requirements. As was mentioned in the previous chapter, the $q=500$ $M'_d$ matrix is obtained from two identical matrices defined by the configuration $M(3;1,8.72)$ and $t = 4361$. 

\begin{figure}[h!]
\centering
  \includegraphics[width=\linewidth,height=3.5in]{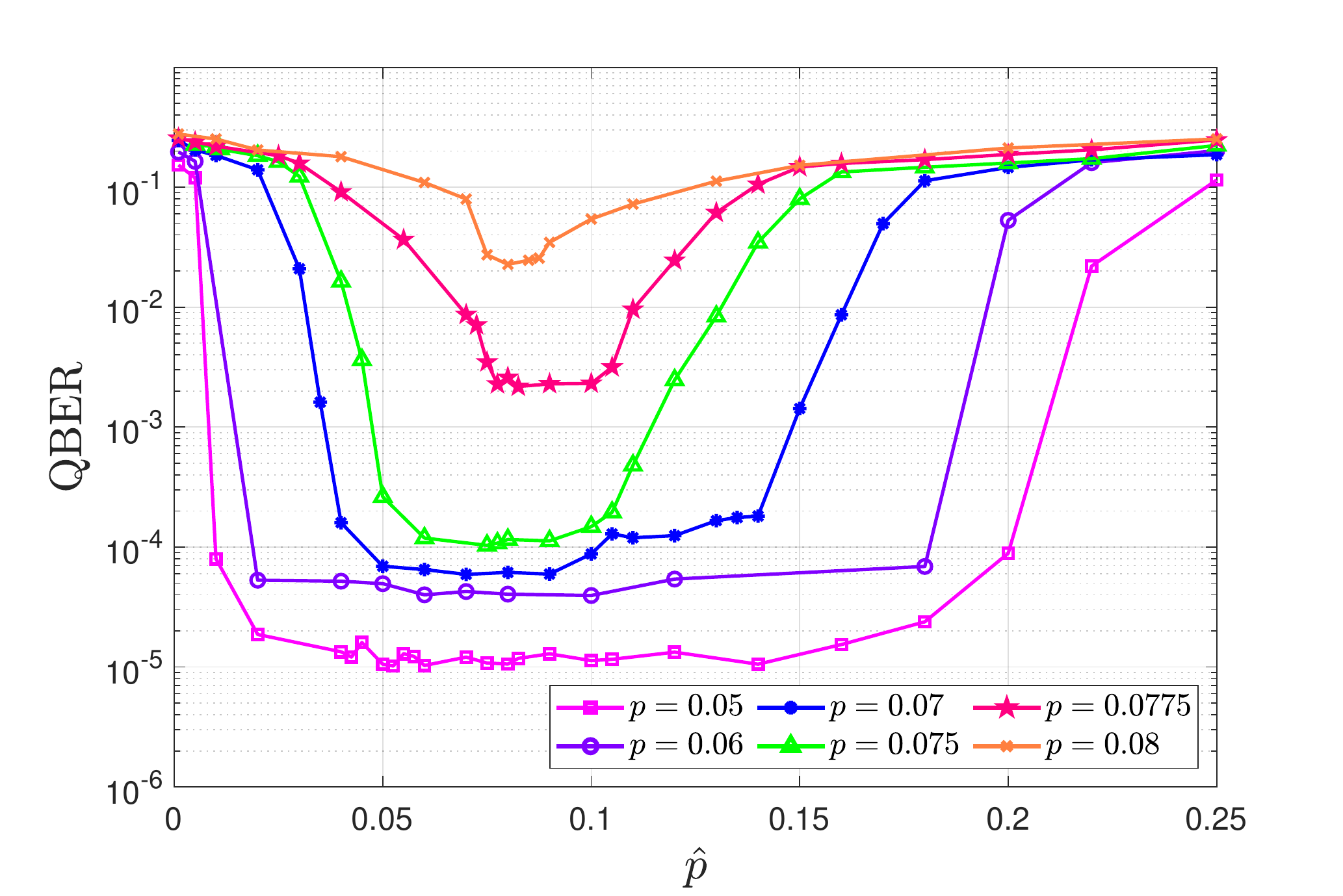}
\caption{Simulated QBER as a function of the estimated depolarizing probability $\hat{p}$ when the true depolarizing probability $p$ is fixed.}
\label{est_mismatch}
\end{figure}

The results of these simulations are shown in terms of the QBER in Figure \ref{est_mismatch}. The values of $p$ have been chosen so that they gradually get closer to the waterfall region\footnote{The decoding threshold or waterfall region of an error correction code is the region (value of $p$) where a sharp drop in the error rate takes place.} of the code ($p= [0.05, 0.06, 0.07, 0.075, 0.0775, 0.08]$), as this allows us to study the sensitivity of the code to the accuracy of the estimate $\hat{p}$ when the actual value of the depolarizing probability is varied.

As shown in Figure \ref{est_mismatch}, for smaller values of $p$ ($p\leq0.075$), the less accurate $\hat{p}$ needs to be to attain similar performance to when perfect channel knowledge is available (when $\hat{p} = p$). On the contrary, for larger values of the depolarizing probability that are closer to the waterfall region of the code ($p > 0.075$), higher accuracy of the estimate $\hat{p}$ will be necessary to achieve the best possible performance. This is reflected by the decrease in the width of the flat regions of the QBER curves as $p$ is increased, where the flat region is defined as the part of the curve where the QBER is not significantly degraded \cite{josu2}. This reduction in the size of the flat region indicates that the precision of the estimate $\hat{p}$ becomes increasingly important as $p$ grows. For instance, if instead of estimating the value of the depolarizing probability we fix $\hat{p}$ to $p
^*=0.127$, which corresponds to the Hashing limit for a $R_Q=\frac{1}{4}$ quantum code, Figure \ref{est_mismatch} shows that for small values of $p$ the resulting performance would be very close to that of a scheme with perfect knowledge of the depolarizing probability. However, as larger values of $p$ are considered, performance deteriorates further and further from the perfect channel knowledge scenario. For a noise level of $p = 0.075$, the QBER is $10^{-4}$ if $\hat{p}\approx0.075$ but degrades to $10^{-2}$ if $\hat{p}=0.0127$. It is clear from these results that techniques capable of providing good estimates of $\hat{p}$ are necessary when facing a channel mismatch scenario. This is discussed in the following subsections, where two different estimation methodologies than can be used to obtain $\hat{p}$ are presented.

\subsubsection{Off-line Estimation Method} 

In quantum channel identification, estimates of the parameter that governs the behaviour of a specific quantum channel $\xi(p)$ are obtained by exposing a known quantum state $\sigma$, referred to as the probe, to the effects of said channel. Performing quantum measurements on the corrupted quantum probe, described by its density matrix $\rho_o(p)$, yields classical information from which an estimation of the parameter $p$ can be obtained. Numerous experimental schemes have been devised to perform quantum channel identification: we can use unentangled quantum probes, probes entangled with ancilla qubits, or even probes entangled with other probes. Each of these strategies provides a unique set of advantages and disadvantages. However, because analyzing the performance of these schemes is outside the scope of this dissertation, in what follows we simply assume that an estimation set-up capable of obtaining the information-theoretical optimal performance is available.

Optimal estimation of the depolarizing probability of the depolarizing channel has previously been studied by making use of a metric known as the quantum Fisher information \cite{fissh1}. The quantum Fisher information of $p$ is given by $$ J(p) = \text{Tr}[\rho_o(p)\mathrm{\hat{L}}^2(p)],$$ where $\text{Tr}[\rho_o(p)\mathrm{\hat{L}}^2(p)]$ represents the trace of the matrix $[\rho_o(p)\mathrm{\hat{L}}^2(p)]$, $\rho_o(p)$ is the quantum state at the output of the channel, and $\mathrm{\hat{L}}(p)$ is the symmetric logarithm derivative defined implicitly as $$ \frac{\partial\rho_o(p)}{\partial p} = \frac{1}{2}[\mathrm{\hat{L}}(p)\rho_o(p) + \rho_o(p)\mathrm{\hat{L}}(p)].$$

Since estimations of $p$ are dependent on statistically distributed quantum measurements obtained from $\rho_o(p)$, the estimate of the depolarizing probability $\hat{p}$ will be a random variable. Therefore, quantum channel identification comes down to selecting a procedure that provides the most accurate values of $p$. This is analogous to finding a method that minimizes the variance of the estimation $E\{(\hat{p}-p)^2\}$, assuming that the estimator is unbiased, $E\{\hat{p}\}=p$. In this context, the best possible performance of any estimator will be given by the quantum Cramér-Rao bound \cite{cram1, lastcite}, which states that the variance of the best possible estimator is bounded by
$$ \text{var}(\hat{p}) \geq \frac{1}{n_mJ(p)} = \frac{1}{J_{n_m}(p)} ,$$ where $J_{n_m}(p) = n_mJ(p)$ defines the overall Fisher information for $n_m$ independent quantum measurements \cite{lastcite} and $J(p)$ denotes the Fisher information of $p$ for a single measurement. As mentioned previously, because we operate under the assumption that our estimator attains the information-theoretical optimal performance, the variance of our estimator will be given by the quantum Cramér-Rao bound.

Based on the results shown in Figure \ref{est_mismatch}, where we studied the QBER in terms of the mismatched depolarizing probability $\hat{p}$, we can now compute the average $\tilde{\text{QBER}}(p)$ with regard to the real depolarizing probability of the channel $p$. This can be done as is shown in \eref{eq:est_QBER}, where $P(\hat{p})$ is the probability density function of our optimal estimator. 

\begin{equation} \label{eq:est_QBER}
    \tilde{\text{QBER}}(p) = \int \text{QBER}(\hat{p})P(\hat{p})d\hat{p}.
\end{equation}

As in \cite{josu2}, we assume that $P(\hat{p})$ is the truncated normal distribution defined between $a$ and $b$, with mean $\mu$, and variance $\frac{1}{J_{n_m}(p)}$. Once again, recall that the variance of $P(\hat{p})$ is chosen as the inverse of the asymptotically achievable Fisher information because we assume that we have the best possible quantum channel identification method at our disposal. The overall Fisher information $J_{n_m}(p)$ will vary in terms of the type of selected quantum probe. Although other works have studied the impact of using different types of probes \cite{josu2}, for the sake of simplicity, herein we will only consider the use of unentangled pure states as channel probes. In this case, the overall Fisher information is given by $ J_{n_m}(p) = n_m\big(\frac{9}{8p(3-2p)}\big)$ \cite{probes}.  

\begin{figure}[H]
\centering
  \includegraphics[width=\linewidth,height=4in]{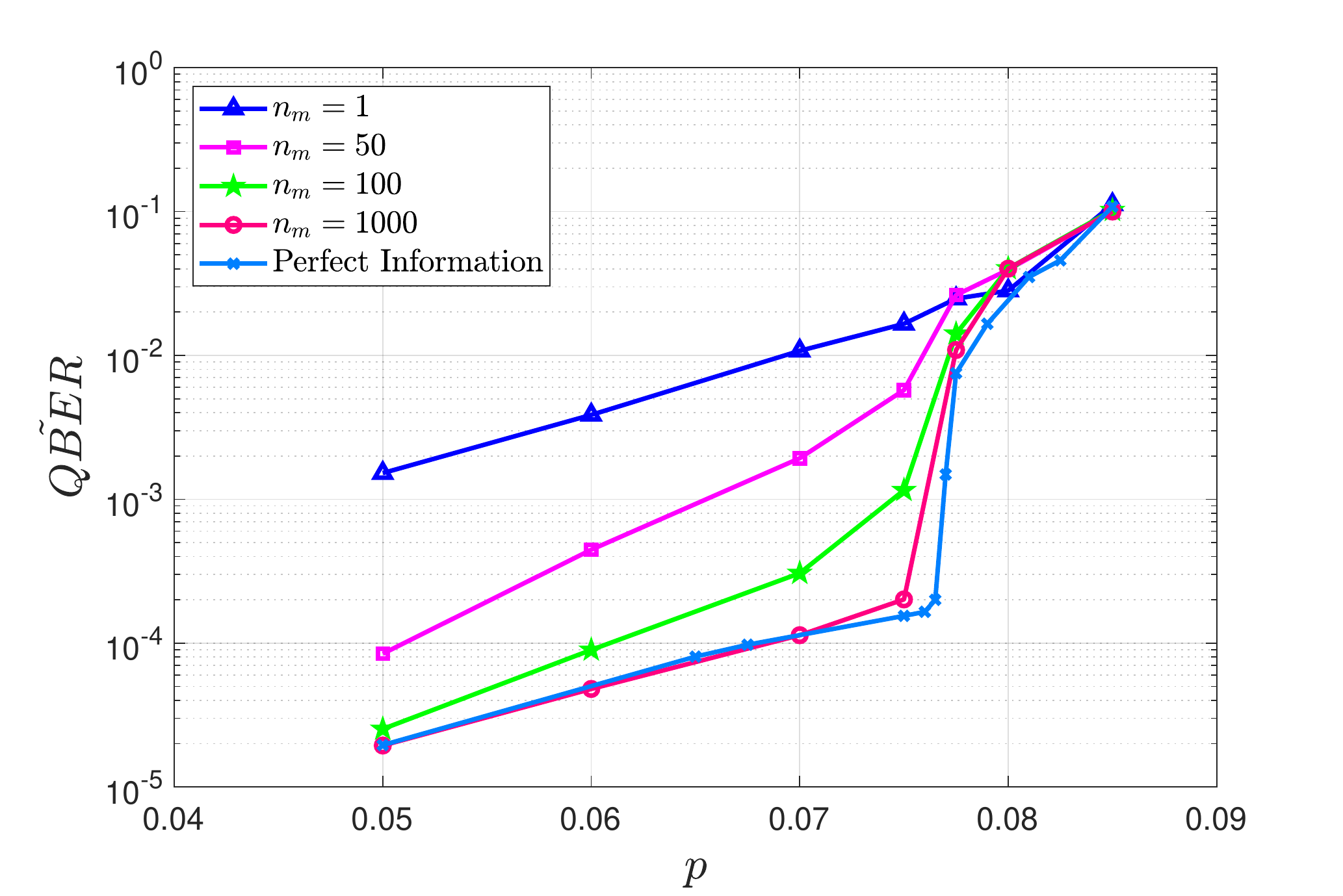}
\caption{Average QBER in terms of $p$ when the number of used probes $n_m$ is varied.}
\label{offline}
\end{figure}

Figure \ref{offline} shows the result of computing $\tilde{\text{QBER}}(p)$ for the non-CSS QLDGM code considered previously as a function of the number of channel probes $n_m$. In \cite{josu2}, the off-line estimation protocol achieves the same performance as the perfect channel information case when $n_m \approx 1000$. The results shown in Figure \ref{offline} indicate that convergence is faster for our codes, since performance close to the perfect information case is obtained for $n_m \approx 100$. As is also shown in \cite{josu2}, convergence to the perfect information case may be further improved by using maximally entangled pairs as probes (EPR pairs instead of pure states), i.e, less probes than when using unentangled pure states will be required. 

Regardless of the type of quantum probe, the main handicap of off-line estimation protocols is that if the channel varies for every transmitted block, the overall rate of the QLDPC code that is being used will be severely reduced. Although this reduction in rate is asymptotically negligible for constant channels, it represents a significant drawback when using this estimation method in rapidly varying quantum channels \cite{superconducting-josu}.

\subsubsection{Online Estimation Method} \label{sec:online}

In similar fashion to what is done in \cite{josu2} for QTCs, we apply slight modifications to the generic syndrome SPA iterative QLDPC decoder that allow us to estimate the depolarizing probability while decoding is taking place. This on-line estimation scheme does not require quantum channel identification, meaning that rate reduction is avoided regardless of the type of quantum channel under consideration, be it constant or block-to-block time varying. 

In section \ref{construction} we saw how decoding of a non-CSS QLDGM code is performed by running the SPA over the factor graph associated to the equation $\mathbf{w} = \mathbf{H}_Q \odot \mathbf{e}$, where $\mathbf{w}$ is the measured syndrome, $\mathbf{H}_Q$ is the QPCM of the non-CSS code\footnote{Recall that to reap the benefits of the non-CSS structure, the decoding algorithm must be run over the complete factor graph representation of the matrix product that defines the QPCM: $\mathrm{M}'_d\times \mathbf{H}_\text{CSS}$. Decoding on the factor graph representation of the final matrix, $\mathbf{H}_Q$, results in worse performance.}, and $\mathbf{e}$ is the symplectic representation of the error pattern induced by the quantum channel. The decoding objective is to find the most likely estimate of the channel error from the observed syndrome, i.e, the decoder must find the most likely estimate of the channel error, $\hat{\mathbf{E}}$, such that the estimated syndrome $\hat{\mathbf{w}} = \mathbf{H}_Q\odot \hat{\mathbf{e}} = (\mathrm{M}'_d \times \mathbf{H}_\text{CSS}) \odot \hat{\mathbf{e}}$, is equal to the observed syndrome $\mathbf{w}$, where $\hat{\mathbf{e}}$ is the symplectic representation of $\hat{\mathbf{E}}$. 

Against this backdrop, the online estimation decoding process works as follows:

\begin{enumerate}
    \item First, the SPA is initialized using a ``flooding" schedule in which the lower layer nodes of the factor graph transmit messages upwards in a layer-by-layer sequential manner until the top-most nodes are reached. These messages are based on an initial estimate of the depolarizing probability of the channel $\hat{p}^{(1)}$, which is used to compute the a priori log likelihood ratios of the algorithm.
    \item Once information gets to the top layer, we say that the graph has been ``flooded" and decoding can actually begin. Decoding then proceeds using a reversed schedule, in which, starting from the top-most syndrome nodes, messages are exchanged downwards and layer-by-layer until the bottom-most nodes are reached. The messages transmitted by the syndrome nodes are computed considering information of the measured syndrome $\mathbf{w}$. 
    \item When two messages have been transmitted over every edge of the factor graph, an iteration of the decoding algorithm has been completed. At the end of each iteration, an estimate of the symplectic representation of the error pattern $\hat{\mathbf{e}}$ is produced and used to compute $\hat{\mathbf{w}}$.
    \item  If $\mathbf{w} = \hat{\mathbf{w}}$, then the algorithm has finished. If $\mathbf{w} \neq \hat{\mathbf{w}}$, the algorithm continues until it finds a matching syndrome or until a maximum number of iterations is reached. We can obtain an estimate of the depolarizing probability at each iteration $j$ by assessing the number of $X$, $Y$, and $Z$ operators present in the estimated error pattern\footnote{Recall that we can obtain the Pauli operator representation from the symplectic representation by applying the inverse isomorphism $\beta^{-1}$ to $\hat{\mathbf{e}}$ (see section \ref{sec:stabs}).} $\hat{\mathbf{E}}$ and dividing them by the block length of the code. This can be expressed as

\begin{equation}
    \hat{p}^{(j)} = 1 - \frac{1}{N}\sum_{i=1}^{N}P^{(j)}(\mathrm{\hat{E}}_i = I|\hat{\mathbf{w}}),
\end{equation}

\noindent where $N$ is the block length of the code, $I$ is the identity operator, $\mathrm{\hat{E}}_i$ is the $i$-th component of the estimated error pattern $\mathbf{\hat{E}}$, and $P^{(j)}(\mathrm{\hat{E}}_i = I|\hat{\mathbf{w}})$ is the probability at iteration $j$ that the $i$-th component of the estimated error pattern is equal to the identity operator conditioned on the estimated syndrome. 

    \item Once $\hat{p}^{(j)}$ is obtained, it is used as the depolarizing probability to compute the necessary sum-product messages in the following iteration. Given the iterative nature of the decoding algorithm, we expect that each successive estimate $\hat{p}^{(j)}$ will get closer to the actual value of $p$, leading to better decoding performance. 

\end{enumerate}

\begin{figure}[htp]
\centering
  \includegraphics[width=\linewidth,height=4in]{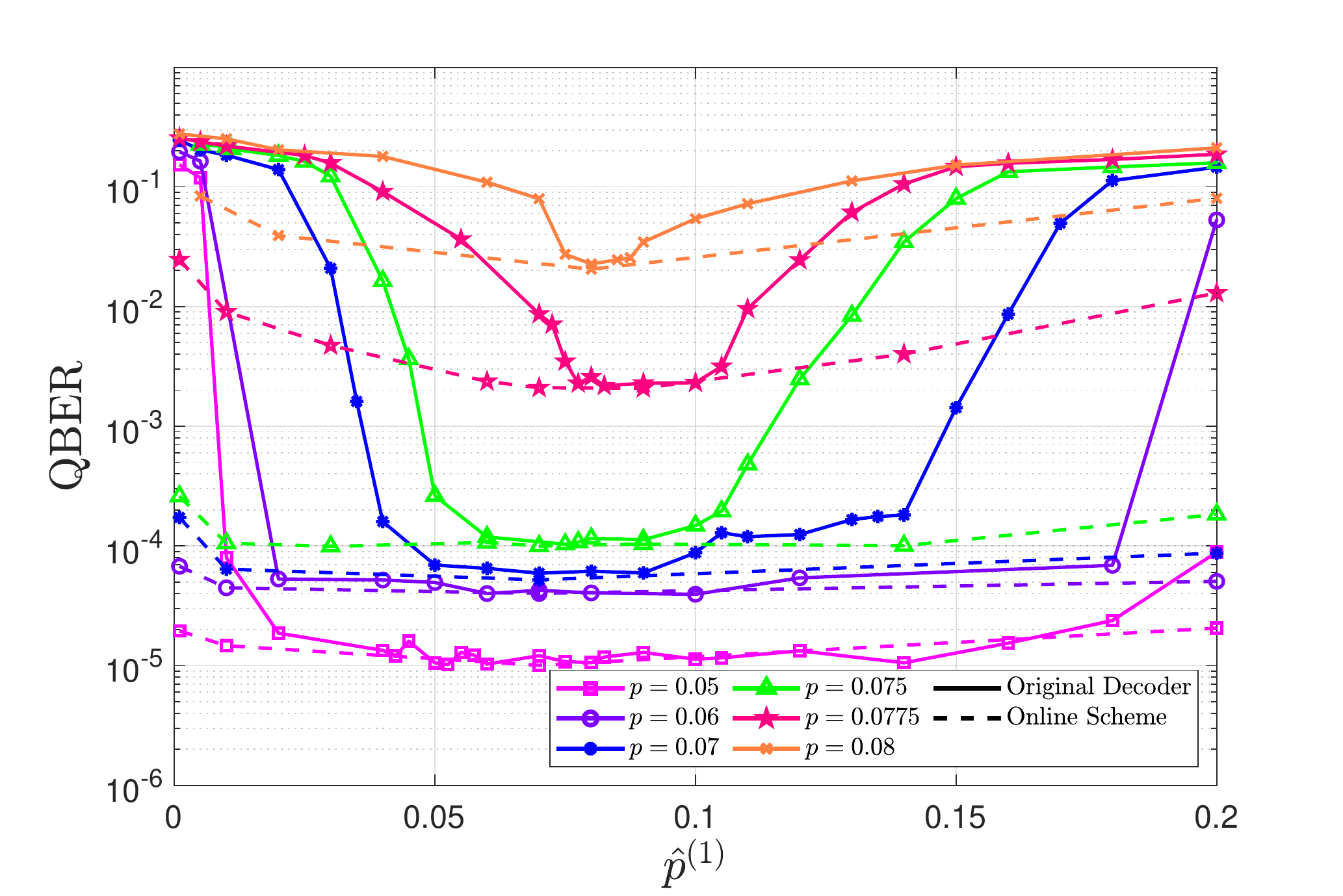}
\caption{Simulated QBER as a function of the initial estimate of the depolarizing probability of the channel $\hat{p}^{(1)}$. The continuous lines are associated to the original iterative QLDPC decoder which uses $\hat{p}^{(1)}$ for every iteration as if it were the true depolarizing probability. The dashed lines are obtained when the modified iterative decoder that uses the on-line estimation method is applied.}
\label{online}
\end{figure}

A matter that we have yet to discuss is how an appropriate value for the initial estimate of the depolarizing probability $\hat{p}^{(1)}$ can be defined. It is possible that this initialization might affect the convergence of the estimates $\hat{p}^{(j)}$ to $p$, which may have an impact on decoder performance. Thus, we must conduct an analysis in which $\hat{p}^{(1)}$ is varied while $p$ remains fixed. The results of this analysis are shown in Figure \ref{online}, where each dashed curve corresponds to a different value of the true depolarizing probability $p$. The curves associated to the original decoder, which where previously shown in Figure \ref{est_mismatch}, are also included in Figure \ref{online} as continuous lines. 

Upon closer inspection of Figure \ref{online}, we can see that the performance of the on-line estimation method is similar to that of the perfect information scenario regardless of the value of $\hat{p}^{(1)}$. In fact, the modified on-line decoder significantly outperforms the original mismatched decoder, as is reflected by the flatter appearance of its performance curves. Figure \ref{online} also shows how the sensitivity of the modified on-line decoder to the initial estimate $\hat{p}^{(1)}$ increases as $p$ grows. This is reflected in a reduction of the flatness of the on-line curves as $p$ increases, which is most noticeable for $p=0.0775$ and $p=0.08$. 

Once the depolarizing probability is higher than a certain threshold we enter the waterfall region of the code, where qubit errors occur with much higher probability than in the error-floor region. Even though the on-line method computes $\hat{p}^{(j)}$ during every iteration, our simulations results show that for sufficiently large values of $p$, if $\hat{p}^{(1)}$ is either too small or too large, the convergence of $\hat{p}^{(j)}$ to $p$ is weakened and performance is hindered. This happens due to the combination of two factors: On the one hand, performance of the code is worse outside of the error floor region, and so the estimated error patterns are much more likely to have errors. On the other, small or large enough values of $\hat{p}^{(1)}$ make initial estimates of the error pattern contain either not enough or too many $X$, $Y$, and $Z$ operators, which can corrupt the values of $P^{(j)}(\hat{\mathrm{E}}_i = I|\hat{\mathbf{w}})$ to the point that subsequent estimates $\hat{p}^{(j)}$ become increasingly inaccurate. This does not occur in the error floor region, where $\hat{p}^{(j)}$ converges to $p$ regardless of the value of $\hat{p}^{(1)}$. 

Ideally, we would like to define the value of $\hat{p}^{(1)}$ for which performance with the on-line estimation method is optimal. If we look at the curves corresponding to $p = 0.0775$ and $p=0.08$, we can see performance is significantly degraded when $\hat{p}^{(1)} \leq 0.05$ or $\hat{p}^{(1)} \geq 0.13$. For the $R_Q= \frac{1}{4}$ code under consideration, the hashing limit is $p^* = 0.127$, which falls within the range $0.05 < p^* < 0.13$. Thus, performance on par with the perfect channel information scenario can be obtained with the on-line estimation method, regardless of the actual value of the depolarizing probability and without any additional resources or reductions in code rate, by setting $\hat{p}^{(1)} = p^*$.

\section{Design of asymmetric QLDGM codes} \label{sec:asym}

Throughout this chapter we have mentioned numerous times that most of the research in the field of QEC considers the symmetric instance of the general Pauli channel, commonly referred to as the depolarizing channel. However, realistic quantum devices often exhibit asymmetric behaviour, where the probability of a phase-flip taking place is orders of magnitude higher than the probability that a bit-flip occurs \cite{josurev, josuconf, asymmetry-1}. This asymmetry stems from the nature of the physical mechanisms that define the behaviour of the materials that practical quantum processors are built from. These mechanisms are essentially governed by the single-qubit relaxation and dephasing times, which define the likelihood that a quantum device suffers bit-flips and phase-flips. Generally, relaxation causes both bit-flips and phase-flips, while dephasing only leads to phase-flip errors. In real quantum processors, the relaxation time can sometimes be orders of magnitude larger than the dephasing time. This difference in relaxation and dephasing times gives rise to the aforementioned asymmetric behaviour, where bit-flip errors are much less likely to occur than phase-flips. 

Clearly, the depolarizing channel cannot be used to represent this type of behaviour. However, it can be accurately modelled using the general Pauli channel \cite{josurev, model-asymmetry-1, model-asymmetry-2, model-asymmetry-3, model-asymmetry-4, model-asymmetry-5}. Given the difference in the likelihood of bit-errors and phase-errors, it stands to reason that the QEC schemes built for this asymmetric channel must somehow exploit this asymmetry to achieve good performance. This has previously been studied in the context of QTCs in \cite{asymmetry-2}, where the authors introduce an EXIT-chart based methodology to design QTCs specifically for the general Pauli channel. This work was later extended in \cite{josuconf}, where an online estimation protocol to decode QTCs over general Pauli channels was proposed. Because performance of the QTCs was shown to vary as a function of the asymmetry of the channel, these results speak to the merit of constructing a coding scheme tailored to the particular asymmetric characteristics of the quantum channel in question.

For this reason, in this section we study the performance of quantum CSS LDGM codes when they are applied over a general Pauli channel. We show how although they are not the best known codes for the depolarizing channel, their simplicity allows for them to be almost seamlessly adapted to the general Pauli channel. Based on this result, we introduce a simple yet effective method to derive CSS QLDGM codes that perform well over channels with varying degrees of asymmetry. Such a strategy is necessary because CSS codes designed for the depolarizing channel perform poorly over its asymmetric counterpart. To the extent of our knowledge and at the time of writing, the research on designing quantum codes specifically for asymmetric quantum channels is quite limited \cite{short-asym}, especially when compared to results regarding the depolarizing channel. Thus, the work included herein represents one of the first attempts at designing QLDPC codes specifically for asymmetric quantum channels.

\subsection{Realistic Pauli Channel model} \label{sec:asymmetry-model}

Having established that the materials that make up quantum processors cause these devices to behave in an asymmetric manner, we must now build a model that can re-enact this particular behaviour. Asymmetry in the behaviour of quantum devices is embodied by the probability of a bit-flip $p_x$ being orders of magnitude smaller than the probability of a phase-flip $p_z$. We know from our previous discussions regarding the general Pauli channel (see Chapters \ref{chapter3} and \ref{chapter6}) that it is completely characterized by the probabilities $p_x, p_y$, and $p_z$. Because it allows us to vary these probabilities, the Pauli channel can be adapted to accurately represent the asymmetric behaviour exhibited by realistic quantum devices. To derive this so-called realistic Pauli channel model, we must first establish the asymmetric relationship between the probabilities $p_x$, $p_y$, $p_z$, and $p$. This can be done by introducing the parameter $\alpha$, known as the channel's ratio of asymmetry \cite{asymmetry-2}, which represents the ratio of the phase-flip probability and the bit-flip probability as \cite{alpha1, alpha2, alpha3}

\begin{equation} \label{eq:alpha}
\alpha = \frac{p_z}{p_x} = 1 + 2\frac{e^{\frac{-t}{T_1}} - e^{(\frac{-t}{2T_1}-\frac{2t}{T_2})}}{1-e^{\frac{-t}{T_1}}} \,,
\end{equation}

where $T_1$ is the relaxation time, $T_2$ represents the dephasing time, and $t$ is the coherent operation duration of a physical quantum gate \cite{time}. In \cite{josurev, alpha1} expressions for $p_x$, $p_y$, and $p_z$ are given in \eref{eq:bit-phase-probs}, where the bit-flip probability and bit-and-phase-flip probability can be considered to be equal.
\begin{equation}\label{eq:bit-phase-probs}
\begin{gathered} 
p_x = p_y = \frac{1-e^{\frac{-t}{T_1}}}{4} \\
p_z = \frac{1}{2} - p_x - \frac{e^{\frac{-t}{T_2}}}{2} .
\end{gathered}
\end{equation}

If the coherent operation duration $t$ is assumed to be reasonably short, i.e, $t << T_1$ \cite{alpha1}, then from \eref{eq:alpha} the ratio of the phase and bit-flip probabilities can be approximated by $ \alpha \approx 2\frac{T_1}{T_2} - 1 $. In consequence, this model allows us to completely determine the values of $p_z$, $p_x$, and $p_y$ from $\alpha$ and $p$. Common values for the ratio of asymmetry are given in \cite{josurev, josuconf, asymmetry-2}, with most materials used to build quantum devices having $\alpha = [10^2, 10^4, 10^6]$. Notice that, if we select $\alpha = 1$, we obtain the depolarizing channel model that is considered in most circumstances and that satisfies $p_x = p_y = p_z = \frac{p}{3}$. Although we have not mentioned it explicitly, the case of $\alpha = 1$ has been observed for specific types of devices, meaning that the depolarizing channel can sometimes also provide a realistic representation of the behaviour of quantum machines.

\subsection{Performance comparison with other QLDPC codes} \label{sec:initial-comp}

From our discussion in Chapter \ref{chapter6} (see section \ref{sec:CSS}), we know that designing LDGM-based CSS quantum codes is a fairly straightforward procedure. In fact, all that is needed to build the factor graphs of such codes is a set of four matrices: the generator and parity check matrices of a classical LDGM code, $\tilde{\mathrm{G}}$ and $\tilde{\mathrm{H}}$, and the matrices $\mathrm{M}_1$ and $\mathrm{M}_2$ described in theorem \ref{theo:css}. These matrices define code parameters such as the rate $R_Q$ or distance of the code. In Chapter \ref{chapter6} we designed these matrices for optimal performance over the depolarizing channel, but given how simple it is to modify them, it is likely that we can also construct them for good performance under different channel requirements. As will be shown in the following subsection, this comes in handy when designing codes for the asymmetric Pauli channel.

\begin{figure}[!h]
\centering
\hbox{\hspace*{-1.5cm}\includegraphics[width = 1.25\linewidth,height=4in]{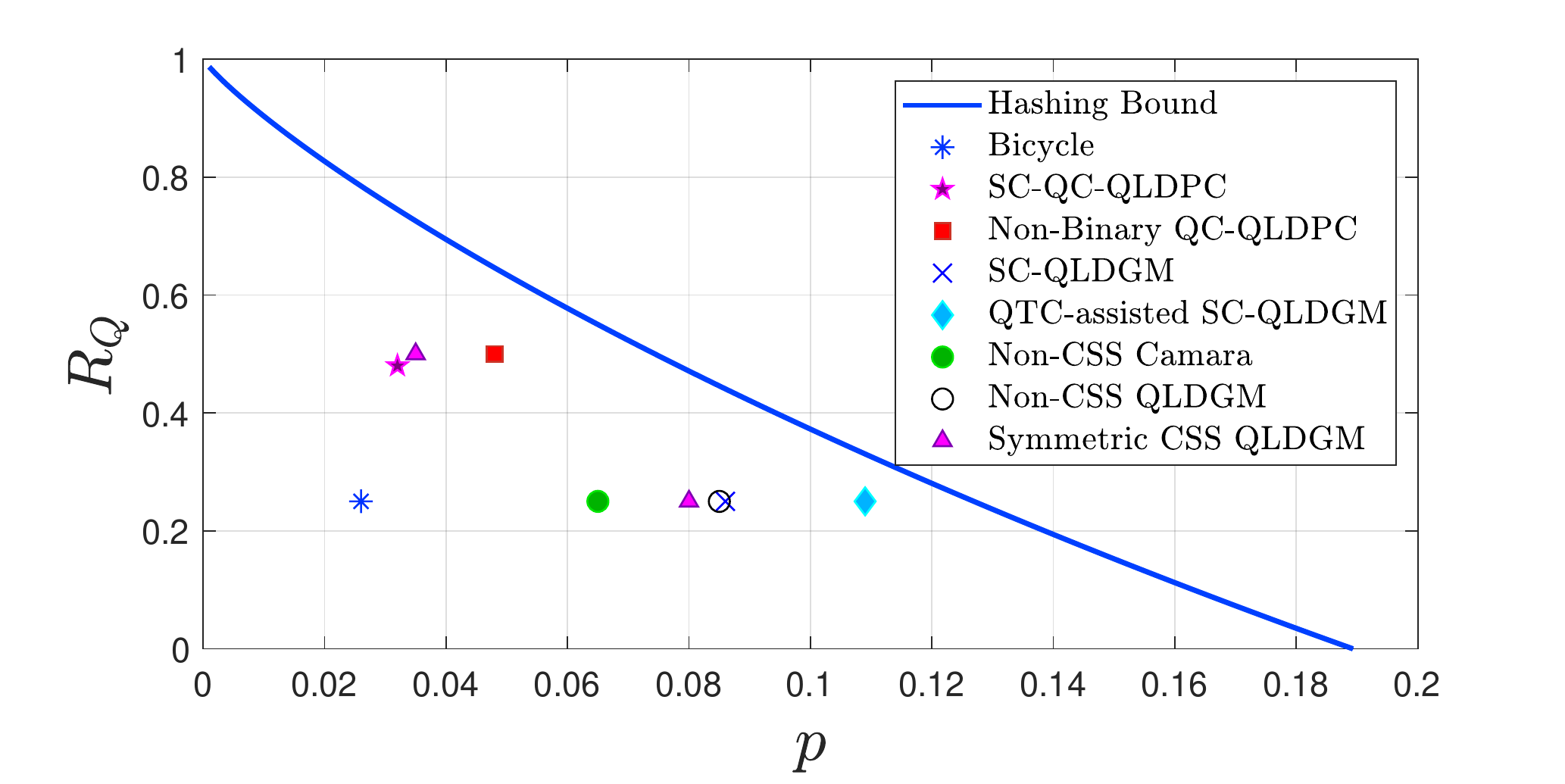}}
\caption{\normalsize{Achievable quantum coding rate at a WER of $10^{-3}$ for different types of QLDPC codes.}}

    \label{comparison-Sym}
\end{figure}

Since CSS QLDGM codes were originally designed to operate over the depolarizing channel, it is important to analyze how they perform over the depolarizing channel when compared to other existing QLDPC codes prior to modifying them for performance over different quantum channels. Although we mentioned them in Chapter \ref{chapter6}, we have not yet analyzed the performance of CSS QLDGM codes in detail. For this purpose, in Figure \ref{comparison-Sym} we show the highest possible coding rate at which various QLDPC codes that have appeared in the literature can achieve a WER of $10^{-3}$ over the depolarizing channel. This figure compares the performance of the following codes:

\begin{enumerate}

    \item Symmetric CSS QLDGM code based on the structure $\mathrm{M}(3; 1, 11.04)$ and $\mathrm{P}[(8,8);(8,160)]$ with block length $n=19014$ and rate $R_Q = \frac{1}{2}$.
    \item Symmetric CSS QLDGM code based on the structure $\mathrm{M}(3; 1, 11.04)$ and $\mathrm{P}[(8,8);(8,160)]$ with block length $n=19014$ and rate $R_Q = \frac{1}{4}$.
    \item The $K=32$ bicycle code with block length $n=19014$ and rate $R_Q = \frac{1}{4}$ proposed in \cite{bicycle}.
    \item The Spatially Coupled (SC) Quasi-Cyclic (QC) QLDPC code of rate $R_Q = 0.49$ and block length $n=181000$ given in \cite{hag2}.
    \item The non-binary QC-QLDPC GF($2^{10}$) code of rate $R_Q = \frac{1}{2}$ and block length $n=20560$ proposed by Kasai et al. in \cite{hag3, hag4}.
    \item The SC-QLDGM code of $R_Q = \frac{1}{4}$ and block length $n=76800$ proposed in \cite{nonCSS2}.
    \item The QTC-assisted SC-QLDGM code of $R_Q = \frac{1}{4}$ and block length $n=821760$ of \cite{CSSbound}.
    \item The non-CSS concatenated code (code C) of $R_Q = \frac{1}{4}$ and block length $n=138240$ of \cite{nonCSS1}.
    \item The $t= 5000$, $q = 500$, $\mathrm{M}(3; 1, 11.04)$, $\mathrm{P}[(8,8);(8,160)]$ non-CSS QLDGM code of $R_Q = \frac{1}{4}$ and block length $n=19014$ of \cite{patrick} that was introduced in Chapter \ref{chapter6}.
\end{enumerate}

As can be seen in Figure \ref{comparison-Sym}, at both of the considered rates, the symmetric CSS QLDGM code is outperformed by some of the other codes. At a rate of $R_Q = \frac{1}{4}$, the symmetric CSS QLDGM code is beaten by the non-CSS implementation of \cite{patrick} (as was shown in the previous chapter), the SC-QLDGM code of \cite{nonCSS2}, and the QTC-assisted SC-QLDGM code of \cite{Til}. However, this comes as no surprise, since all three of these codes take a symmetric CSS QLDGM code as their starting point and then modify it (by changing the factor graph or combining them with a QTC) with the purpose of improving performance. For a rate of $R_Q = \frac{1}{2}$, the symmetric CSS QLDGM code is once again outperformed by the non-binary QC-QLDPC GF($2^{10}$) code of \cite{hag2, hag3}. We can further expand this comparison by looking at the distance of each code to the Hashing bound. This can be done as in Chapter \ref{chapter6} by computing 
\begin{equation} \label{eq:distance-hashing}
\delta = 10\log_{10}\bigg(\frac{p^*}{p}\bigg) ,
\end{equation}
where $p^*$ is the noise limit of the depolarizing channel for a specific quantum rate $R_Q$, and $p$ is the highest depolarizing probability at which the code in question
can operate with a WER of $10^{-3}$.

At a rate of $R_Q = \frac{1}{2}$, the QC-QLDPC code of \cite{hag3, hag4} is $\delta_{\text{QC-QLDPC}} = 1.9$ dB away from the Hashing bound. At this same rate, the distance for the symmetric CSS QLDGM code is $\delta_{\text{CSS-QLDGM}} = 2.86$ dB. These values make it clear that when $R_Q = \frac{1}{2}$, QC-QLDPC codes significantly outperform CSS QLDGM codes. In contrast, at a rate of $R_Q = \frac{1}{4}$, the symmetric CSS QLDGM code is $\delta_{\text{CSS-QLDGM}} = 1.95$ dB away from the Hashing bound. Once more, despite the improvement in performance at a this lower rate, the non-CSS and the SC-QLDGM codes of \cite{patrick} and \cite{nonCSS2} which exhibit approximately the same distance to the Hashing Bound $\delta_{\text{non-CSS}} \approx \delta_{\text{SC-QLDGM}} = 1.69$ dB, outperform the CSS QLDGM code. However, it is worth mentioning that at this rate the difference in performance is notably less significant than when $R_Q = \frac{1}{2}$. This means that the performance of CSS QLDGM codes is better at a lower rate, which implies one of two things: that a different construction strategy must be employed when constructing higher rate CSS QLDGM codes, similar to what is done classically in \cite{eurasip}, or that they should not be used for error correction when high rates are necessary.

At this point, it should be noted that these improvements in performance come at the expense of the complexity of the error correction schemes. This is especially true for the best known code for $R_Q = \frac{1}{4}$, the QTC-assisted SC-QLDGM code of \cite{CSSbound}, which requires a QTC and a large block length in order to get closer to the Hashing bound. Herein lies the main appeal of CSS QLDGM codes, because although they are slightly worse than the state-of-the art codes at $R_Q = \frac{1}{4}$, the simplicity with which their design parameters can be manipulated allows for them to be seamlessly adapted to different channels. Doing so for the other codes included in this discussion is a much more difficult task, since their increased complexity does not allow direct modifications like those permitted by CSS QLDGM codes. For this reason, and knowing that performance of CSS QLDGM codes over the depolarizing channel is acceptable at $R_Q = \frac{1}{4}$, we use these codes in the following section as the basis to design asymmetric CSS QLDGM codes for a more realistic Pauli channel model. 

\subsection{Asymmetric QLDGM CSS codes}

Over Pauli channels that model practical quantum devices, a phase-flip is generally much more likely to occur than a bit-flip. Therefore, it is reasonable to assume that for a quantum error correction scheme to be optimal for this type of channel, it must be capable of appropriately exploiting the channel's asymmetry. To be more precise, attaining the best performance over such channels requires a more complex strategy (by modifying the code construction) than just adapting\footnote{The simplest decoding strategy for an asymmetric channel would be to use a symmetric CSS QLDGM code in which the a priori probabilites of each individual decoder are the bit and phase flip probabilities instead of $\frac{2p}{3}$.} the decoding strategy employed over the depolarizing channel by modifying the flip probabilities according to the new channel model. This means that decoding a symmetric QLDGM CSS scheme by feeding the a priori bit and phase error probabilities $f_m^1 = p_x$ and $f_m^2=p_z$ of an asymmetric channel to the corresponding bit and phase error decoders, while certainly an improvement to decoding over an asymmetric channel based on the mismatched probability of the original i.i.d. channel $f_m = \frac{2p}{3}$, will not result in noticeable performance improvements. This is shown in the following section, which also portrays the significant improvement yielded by asymmetric CSS schemes that tailor specifically to the Pauli channel model for asymmetry. 

\subsubsection{Adaptation of symmetric CSS QLDGM codes to the Pauli channel}

\begin{figure}[!h]
\centering
\includegraphics[width=\linewidth,height=3.5in]{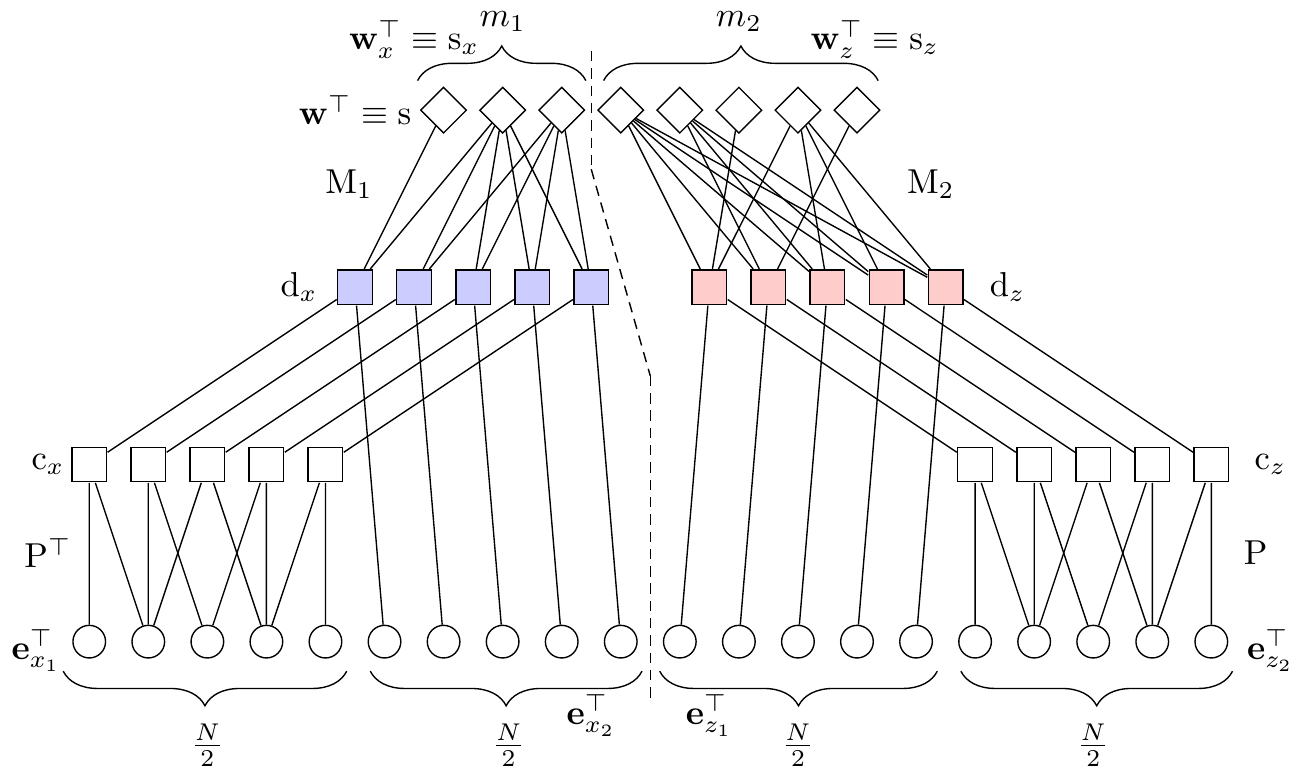}
\caption{\normalsize{Decoding graph for an asymmetric QLDGM CSS scheme. The dashed line is included to emphasize the separation of the two constituent subgraphs. The leftmost subgraph decodes the $X$ errors while the one on the right decodes the $Z$ errors. We have assumed that $n_1 = n_2 = \frac{N}{2}$, and $m = m_1 + m_2$.}}

    \label{asym_CSS}
\end{figure}

The QLDGM CSS scheme introduced in section \ref{sec:CSS} can be adapted to an asymmetric channel by increasing the number of syndrome nodes used to decode the $Z$ operators and decreasing the number of syndrome nodes used to decode the $X$ operators. In this way, the decoder can take advantage of the channel's asymmetric behaviour and improve the performance of the error correcting scheme. The factor graph of a QLDGM CSS code tailored to the Pauli channel model for asymmetry is shown in Figure \ref{asym_CSS}.

Despite how intuitive the idea appears, it is worth discussing why utilizing more syndrome nodes to decode $Z$ operators and less syndrome nodes to decode $X$ operators is beneficial when the considered channel is asymmetric. The asymmetry-integrating Pauli channel model causes phase-flips ($Z$ errors) with much higher probability than bit-flips or bit-and-phase-flips ($X$ and $Y$ errors, respectively). Thus, the symplectic error representation of a pattern induced by this asymmetric channel will have a much higher number of non-zero elements in its $\mathbf{e}_z$ string than in its $\mathbf{e}_x$ string. In contrast, when $\mathbf{e}$ is produced by a depolarizing channel, $\mathbf{e}_x$ and $\mathbf{e}_z$ will, on average, have the same number of non-zero entries. This presence of similar amounts of $X$ and $Z$ errors in error sequences produced by the depolarizing channel is the reason why the decoding graph of Figure \ref{fullCSS} uses the same amount of syndrome information to decode the $\mathbf{e}_x$ and $\mathbf{e}_z$ nodes. However, such a graph will more than likely not be optimal for an asymmetric scenario in which the distribution of non-zero entries over the length $N$ constituent strings of the symplectic representation of an error pattern is not equal like for the depolarizing channel.

Let $\mathbf{e}^{\text{dep}} = (\mathbf{e}^{\text{dep}}_x|\mathbf{e}^{\text{dep}}_z)$ and $\mathbf{e}^{\text{asym}} = (\mathbf{e}^{\text{asym}}_x|\mathbf{e}^{\text{asym}}_z)$ denote the symplectic representations of two error patterns induced by a depolarizing channel and a Pauli channel that models a realistic quantum device, respectively. Let us also define the operator $\sigma(\mathbf{a})$, which computes the number of non-zero entries in a binary string $\mathbf{a}$. Finally, assume that the asymmetry coefficient of the asymmetric Pauli channel in question satisfies $\alpha \geq 10^2$. For the same value of $p$, $\sigma(\mathbf{e}^{\text{dep}}) \approx \sigma(\mathbf{e}^{\text{asym}})$. However, while $\sigma(\mathbf{e}^{\text{dep}}_x) \approx \sigma(\mathbf{e}^{\text{dep}}_z)$, the same does not occur for the asymmetric channel, $\sigma(\mathbf{e}^{\text{asym}}_z) \ggg \sigma(\mathbf{e}^{\text{asym}}_x)$. Additionally, $\sigma(\mathbf{e}^{\text{asym}}_z) \gg \sigma(\mathbf{e}^{\text{dep}}_z)$ and $\sigma(\mathbf{e}^{\text{asym}}_x) \ll \sigma(\mathbf{e}^{\text{dep}}_x)$. In consequence, it is quite obvious that a decoder tasked with decoding error patterns induced by a general Pauli channel will benefit from an uneven decoding graph in which more syndrome information is employed to decode $\mathbf{e}_z$ and less syndrome nodes are utilized to decode $\mathbf{e}_x$. 

The design of such a decoding graph, an example of which is shown in Figure \ref{asym_CSS}, gives rise to a whole new set of questions. The first and most significant one is how can the optimum values for $m_1$ and $m_2$ be determined, where $m_1$ and $m_2$ denote the number of syndrome nodes used to decode the $X$ and $Z$ operators, respectively. It is evident that $m_2 > m_1$, with said difference growing larger as the asymmetry of the channel increases. Ideally, we would like to devise a mathematical formulation from which the values of $m_2$ and $m_1$ that yield the best possible performance could be obtained. 

Another important matter, of which little insight is possessed, is which configuration of the $\mathrm{M}_1$ and $\mathrm{M}_2$ matrices will yield the best results.
However, establishing which values of $(y;1,x)$ and $t$ for each of these matrices is optimal, further augments the complexity of the asymmetric design procedure when compared to the symmetric scenario. This increase in complexity is caused by the fact that exploiting the asymmetry of the channel requires $\mathrm{M}_1 \neq \mathrm{M}_2$, which allows for myriads of different configurations in terms of the values chosen for $m_i$, $y_i$, $x_i$, and $t_i$, where $i=1,2$. Finding the optimum configuration through a brute force search requires such a plethora of simulations that the issue becomes computationally intractable. 

In order to simplify our search for these matrices, we recover the design methodology used in Chapter \ref{chapter6} (that of \cite{jgf3, jgf4, jgf5}) to construct symmetric CSS codes. This procedure reduces the complexity of the construction of the $\mathrm{M}_1$ and $\mathrm{M}_2$ matrices by considering $\frac{m}{2} = m_1 = m_2$ and $\mathrm{M}_1 = \mathrm{M}_2 = \mathrm{M}$. This means that instead of building two different matrices, the same matrix $\mathrm{M}$ is used to define the upper layers of both CSS subgraphs. 

The methodology to construct $[\mathrm{M}(y;1,x)]_{\frac{m}{2}\times \frac{N}{2}}$ begins by defining the values of $m$, the total number of syndrome nodes of the decoding graph, and $N$, the block length of the code. $N$ is chosen to be sufficiently large so as to ensure that the code will possess good error correcting capabilities, while $m$ is selected to guarantee that the code has the desired quantum rate, which for these CSS QLDGM codes is given by $R_Q = \frac{N-m}{N}$. Once again, since these codes are built for the symmetric Pauli channel, $\frac{m}{2}$ syndrome nodes are assigned to each CSS subgraph. Following this, $y$ is set as a natural number to make sure that the $\mathrm{d}$ nodes of the CSS subgraphs have the same number of edges, and the number of $\mathrm{s}_A$ nodes is chosen as $t \leq \frac{m}{2}$. Finally, $x$ is obtained from the following equation:

\begin{equation} \label{eq:M_equation}
\bigg(\frac{m}{2}-t\bigg)x + t = y\frac{N}{2}.
\end{equation}

In \cite{jgf3, jgf4, jgf5}, where $R_Q = \frac{1}{4}$ codes with $N = 19014$ and $m = 14262$ are considered, the configurations of $[y,x,t]$ that achieved the best performance were: $[3,8.72,4161]$ and $[3,11.04,5000]$. These configurations were also shown to be slightly dependant on the characteristics of the underlying parallel-concatenated classical LDGM code.

Finding the combination of the parameters involved in \eref{eq:M_equation} that produces the code with best possible performance is no easy task. Nonetheless, as was done in the previous chapter, certain assumptions can be made in order to reduce the complexity of this endeavour. To begin with, we know that $N$ and $m$ are fixed in order to define the desired rate of the code. If $N$ is appropriately chosen (it is large enough to guarantee good error correction potential of the code), this simplifies matters and reduces the number of parameters from \eref{eq:M_equation} that must be studied. Now, recall that $y$ must be set as a natural number to ensure the regularity of the $\mathrm{d}$ nodes. In \cite{jgf3, jgf4, jgf5} results showed that only a single value of this parameter yielded positive outcomes\footnote{Choosing $y=2$ resulted in too little syndrome information being propagated throughout the graph and applying $y=4$ resulted in worse and slower decoding due to the large amount of messages exchanged over the graph.}, $y = 3$. Given that the codes we will construct in this chapter are based on the structures proposed in \cite{jgf3, jgf4, jgf5}, it is reasonable to adopt the same value for $y$ in our constructions. In consequence, this results in $N$, $m$, and $y$ being fixed to specific values, implying that the only parameters in equation \eref{eq:M_equation} that can actually be modified are $t$, the number of $\mathrm{s}_A$ nodes, and $x$, the degree of the $\mathrm{s}_B$ nodes. Several insights regarding the value of these parameters can be obtained based on our previous discussions and the aforementioned equation:

\begin{itemize}
    \item For large values of $x$, the reliability of the messages transmitted by the $\mathrm{s}_B$ nodes in the decoding process is significantly reduced. This occurs because when nodes have many edges in SPA-based decoding, the messages that are considered in the computations of each of these nodes are numerous enough to have an ``averaging" effect and reduce the impact of any one given message. Naturally, this should hinder the overall performance of the code.
    \item As the values of $x$ grow, given that the RHS of \eref{eq:M_equation} is fixed, the value of $t$ will also be larger. Note that this is intuitive: the more degree-1 syndrome nodes that there are (the larger the value of $t$), the larger the degree of the remaining $\mathrm{s}_B$ nodes (the larger the value of $x$) will be because the degree $y$ of the lower layer $\mathrm{d}$ nodes must remain the same, which can only be guaranteed by adding more edges to the $\mathrm{s}_B$ nodes. Growth in the value of $t$ should have a positive impact on performance, as having more $\mathrm{s}_A$ nodes in the decoding graph means that more ``perfect" information from these degree-$1$ syndrome nodes will be transmitted to the lower layer nodes in the initial decoding iterations.
\end{itemize}

Against this backdrop, it seems likely that the optimum values of $x$ and $t$ will be dictated by a trade-off between these two effects and that the choice of $x$ will be inherently linked to the choice of $t$. As was mentioned previously, this search for the best combination of $t$ and $x$ is further complicated by the fact that our goal is to design asymmetric CSS codes. This is more difficult than building CSS QLDGM codes for the depolarizing channel because the matrices $\mathrm{M}_1$ and $\mathrm{M}_2$ must now be different in order to exploit the asymmetry of the channel. Because more syndrome nodes are used to decode phase flips and less syndrome nodes are used to decode bit flips ($m_1 \neq m_2$), the $x$ and $t$ parameters corresponding to each of these matrices must now be optimized. Therefore, to appropriately design LDGM-based CSS codes for the Pauli channel model presented in section \ref{sec:asymmetry-model}, we have to adapt equation \eref{eq:M_equation} into the new version shown below

\begin{equation} \label{eq:M_a_equation}
\bigg(\frac{m_i}{2}-t_i\bigg)x_i + t_i = y_i\frac{N}{2}, 
\end{equation}

where $i = 1, 2$. Essentially, two designs have to be optimized instead of one: we must now find the configurations of $\mathrm{M}_{{m_1}\times\frac{N}{2}}(y_1;1,x_1)$ and $t_1$ and $\mathrm{M}_{{m_2}\times\frac{N}{2}}(y_2;1,x_2)$ and $t_2$ that yield the CSS codes with the best performance. Recall that the rate of the scheme will remain fixed since the sum $m = m_1 + m_2$ does not change. The demands of this process are discussed in the following section, where we show how in reality, most of the parameter optimization is only needed for one of these matrices.

\subsection{Simulations}\label{sec:results}

We will now study the performance of the proposed asymmetric CSS scheme over the Pauli channel model for asymmetry. First, we perform simulations to analyze the behaviour of a variety of asymmetric schemes over a Pauli channel with a specific degree of asymmetry and compare these results to the performance of a symmetric CSS code when it is applied over that same channel. Based on these results, we proceed by studying those schemes that yield the best performance and narrowing down the search for the optimum size and configuration of the $\mathrm{M}_1$ and $\mathrm{M}_2$ matrices. Then, we propose a methodology to design asymmetric CSS QLDGM codes based on the asymmetry coefficient of the channel. Finally, we compare the performance of the proposed asymmetric schemes to the theoretical limits of the Pauli channel and study how they measure up against other codes that are found in the literature.

\subsubsection{Performance over the asymmetric Pauli Channel} \label{sec:degrade}

Realistic Pauli channel models for quantum devices induce phase-flips ($Z$ errors) with much higher probability than bit-flips ($X$ errors) \cite{josurev}. Asymmetric CSS QLDGM schemes can exploit this phenomenon by utilizing more syndrome nodes to decode $Z$ errors and employing less syndrome information to decode $X$ errors. Given that the only design guideline we possess to begin this analysis is that $m_2$ should be larger than $m_1$, we start by fixing the asymmetry coefficient of the channel to $\alpha = 10^2$, and simulating different configurations of the proposed asymmetric CSS scheme. For comparison purposes, we also simulate a symmetric CSS code over the Pauli channel with $\alpha = 10^2$. We select the value $\alpha = 10^2$ because it is the smallest out of the set of realistic values for the asymmetric coefficient provided in \cite{josurev, asymmetry-2}. Performance of the proposed schemes for channels with other degrees of asymmetry is studied in the last part of this section.

\begin{figure}[!h]
\centering
\includegraphics[width = \linewidth,height=4in]{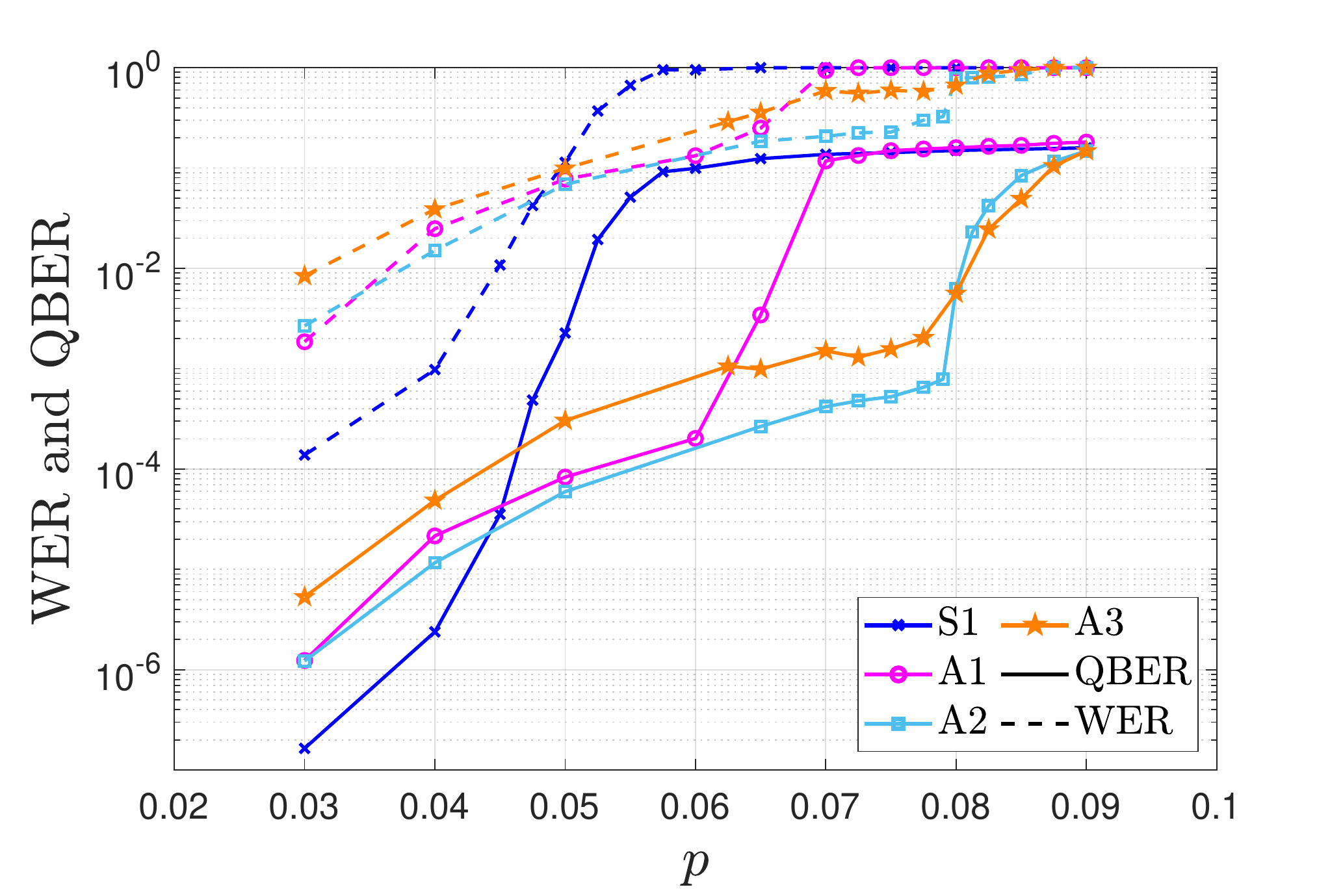}
\caption{\normalsize{Simulated WER and QBER for different CSS QLDGM schemes. $p$ represents the gross flip probability of the Pauli channel with asymmetry coefficient $\alpha = 10^2$. }}

    \label{results}
\end{figure}

For our simulations, we build codes of quantum rate $R_Q = \frac{1}{4}$ and block length $N=19014$ that encode $k = 4752$ qubits into $N$ qubits. The pseudorandom matrix $\mathrm{P}$ of the underlying LDGM code has size $9507\times9507$, has the same degree distribution as its transpose $\mathrm{P}^\top$, and corresponds to a rate $\frac{1}{2}$ classical irregular LDGM code. The irregular LDGM code is designed via the parallel concatenation of two regular LDGM codes. As was done in Chapter \ref{chapter6}, we use the particular concatenation $\mathrm{P}[(8,8)(3,60)]$ because of its relatively small number of degrees, which reduces simulation time substantially. Once the optimum $\mathrm{M}_1$ and $\mathrm{M}_2$ configuration has been found, a parallel concatenated LDGM code of larger degrees can be used to improve performance. Figure \ref{results} shows the performance of the simulated schemes and table \ref{data} outlines the details of each specific design. The results are depicted using the QBER and the WER.

 \begin{table}[h!]

    \centering\caption{\normalsize{Parameter values and configurations of the CSS codes simulated over an Pauli channel with $\alpha = 10^2$. The results of these simulations are shown in Figure \ref{results}. }}
    \vspace{0.15mm}\begin{tabular}{ccccc}
    \toprule 
    Code Type &\# &$[m_1, t_1, x_1, y_1]$ &$[m_2, t_2, x_2, y_2]$\\
    \midrule
    Symmetric  &S1 &$[7131, 4161, 8.22, 3]$ &$[7131, 4161, 8.22, 3]$ \\
    Asymmetric &A1 &$[6262, 3000, 7.82, 3]$ &$[8000, 5100, 6.33, 3]$\\
    Asymmetric &A2  &$[3262, 1500, 12.49, 3]$ &$[11000, 8497, 8, 3]$  \\
    Asymmetric &A3 &$[1262, 750, 54.24, 3]$ &$[13000, 9507, 6.9, 3]$ 
    \\\bottomrule
    \label{data}
    \end{tabular}
    \end{table}

As was expected, the results shown in Figure \ref{results} portray how the asymmetric CSS schemes outperform the symmetric CSS code over the Pauli channel with $\alpha = 10^2$. This can be appreciated by observing how the QBER and WER curves of the symmetric CSS code (code S1) enter the waterfall region and experience significant degradation at a substantially lower value of $p$ than the asymmetric codes.

 The simulation outcomes depicted in Figure \ref{results} also serve to provide insight regarding the performance determining factors of our asymmetric schemes. For instance, defining an increasingly unbalanced configuration of the upper layer of the decoding graph by selecting larger values of $m_2$ and decreasing the values of $m_1$ appears to have a positive impact on our construction. This is reflected by the betterment in QBER and WER results of the codes of table \ref{data} as $m_2$ grows. However, the performance curves of code A3, which has the largest value of $m_2$, exhibit a higher error floor than all the other simulated codes. This may occur because selecting such a large value for $m_2$ reduces the number of syndrome nodes leftover to decode the $X$ operators to such an extent that the corresponding decoder, despite the low probability of bit-flips over the channel, sees an inevitable increase in its error rate. At the same time, it may also be that for such a small value of $m_1$, the choice of $[t_1, x_1, y_1]$ is so critical that inappropriate selection of these values degrades performance significantly. Thus, although large values of $m_2$ provide more syndrome information to decode the $Z$ operators and improve the ability of the code to correct phase-flips, they come at the expense of using less information to correct bit-flips (values of $m_1$ that are too low), which results in increased error floors and worse overall performance if the bit-flip decoder is not correctly designed.

Another aspect of the proposed asymmetric CSS scheme that is integral to its performance and which was discussed in the previous section, is the relationship between the degree of the $\mathrm{s}_B$ nodes of the CSS decoding subgraphs, denoted by $x_i$, and the number of degree-1 syndrome nodes $t_i$, where $i=1,2$. The impact of this relationship, and specifically its aforementioned trade-off nature, is significant, as it ties into the selection of $m_i$. For the symmetric CSS schemes of \cite{jgf3, jgf4} the best performance was obtained for codes that utilized $\mathrm{M}_1 = \mathrm{M}_2 = \mathrm{M}(3;1,11.02)_{7131\times9507}$ and $t=5000$. Let us assume that the optimal degree of $x$ for matrix $\mathrm{M}$ of a symmetric CSS code will still be optimal for the $\mathrm{M}_1$ and $\mathrm{M}_2$ matrices used to build each of the subgraphs of an asymmetric CSS code. In reality, achieving a configuration where $x_i = 11.02$ for a scenario in which $\mathrm{M}_1 \neq \mathrm{M}_2$ will not always be possible. If we revisit equation \eref{eq:M_a_equation} we can understand why this happens. Once the quantum rate of the code has been selected, aside from $t_i$, the only other parameter we can modify is $y_i$. Recall, that in \cite{jgf3, jgf4}, results showed that only a single value of this parameter yielded good outcomes, $y = 3$. Thus, since $N$, $m_i$, and $y_i$ are fixed and $t_i$ is bounded\footnote{The parameter $t_i$ can theoretically be as large as $m_i$. However, since each decoding subgraph only has $\frac{N}{2}$ $\mathrm{d}$ nodes, it is does not make sense to choose $t_i > \frac{N}{2}$ since with $t_i = \frac{N}{2}$ all the $\mathrm{d}$ nodes are already $\mathrm{d}_a$ nodes (they receive perfect syndrome information).} by $\frac{N}{2}$, it will not always be possible to build matrices that have $x_i = 11.02$. 

We illustrate this with an example: Introducing $N=19014$, $y_2=3$, $m_2=11500$ and the maximum possible value of $t_2 = 9507$ into \eref{eq:M_a_equation}, we obtain $x_2 = 9.54$. Since $t_2$ cannot be increased further, a scheme with this parameter configuration will have a maximum $\mathrm{s}_B$ node degree of $x_2 = 9.54$. In consequence, it becomes apparent that our choice of $m_1$ and $m_2$ also affects the values of $x_1$ and $x_2$. 

Although this entire design procedure for $\mathrm{M}_1$ and $\mathrm{M}_2$ may seem overwhelming, this last remark regarding the value of $x_i$ is actually a positive outcome. If we can show that performance of the asymmetric schemes is optimized for a specific value of $x_i$, then the asymmetric CSS code design procedure can be reduced to finding the values of $m_i$ and $t_i$ that produce this particular value of $x_i$. 

The last detail worthy of mention related to the results shown in Figure \ref{results} is that, aside from code A3, whose decoding errors are overwhelmingly caused by $X$ operators, the entirety of the decoding errors of all the other asymmetric codes are caused by phase-flips. Although not surprising given the nature of an asymmetric Pauli channel, it would be ideal to design an asymmetric scheme in which errors are equally distributed, as is the case over the depolarizing channel. In other words, we would like the $X$ and $Z$ operator decoders of our asymmetric CSS codes to fail with similar rates, instead of all the decoding errors being attributed to one of them. 

\subsubsection*{Impact of CSS decoding} \label{CSSdecoding}

Generally, CSS codes are decoded separately by executing the SPA over each of the subgraphs of the overall CSS factor graph. Over the depolarizing channel, given the equal likelihood of $X$ and $Z$ error events, the error contributions of each individual decoder to the overall code are essentially identical. Over the asymmetric channel, however, maintenance of this separate decoding policy results in each of the CSS decoders impinging on the error correcting capabilities of the overall CSS code in a different manner. In what follows we will study the impact of each of the individual decoders on the performance of the asymmetric CSS code by simulating different configurations. Throughout this discussion we will use the terms $X$ operator decoder and $Z$ operator decoder to refer to the decoding process that unfolds over the CSS subgraphs associated to the $X$ and $Z$ operators, respectively. Note that we are slightly abusing the concept of a decoder since what we actually simulate are different subgraph configurations, not different decoders.

\begin{figure}[!h]
\centering
\includegraphics[width = \linewidth,height=4in]{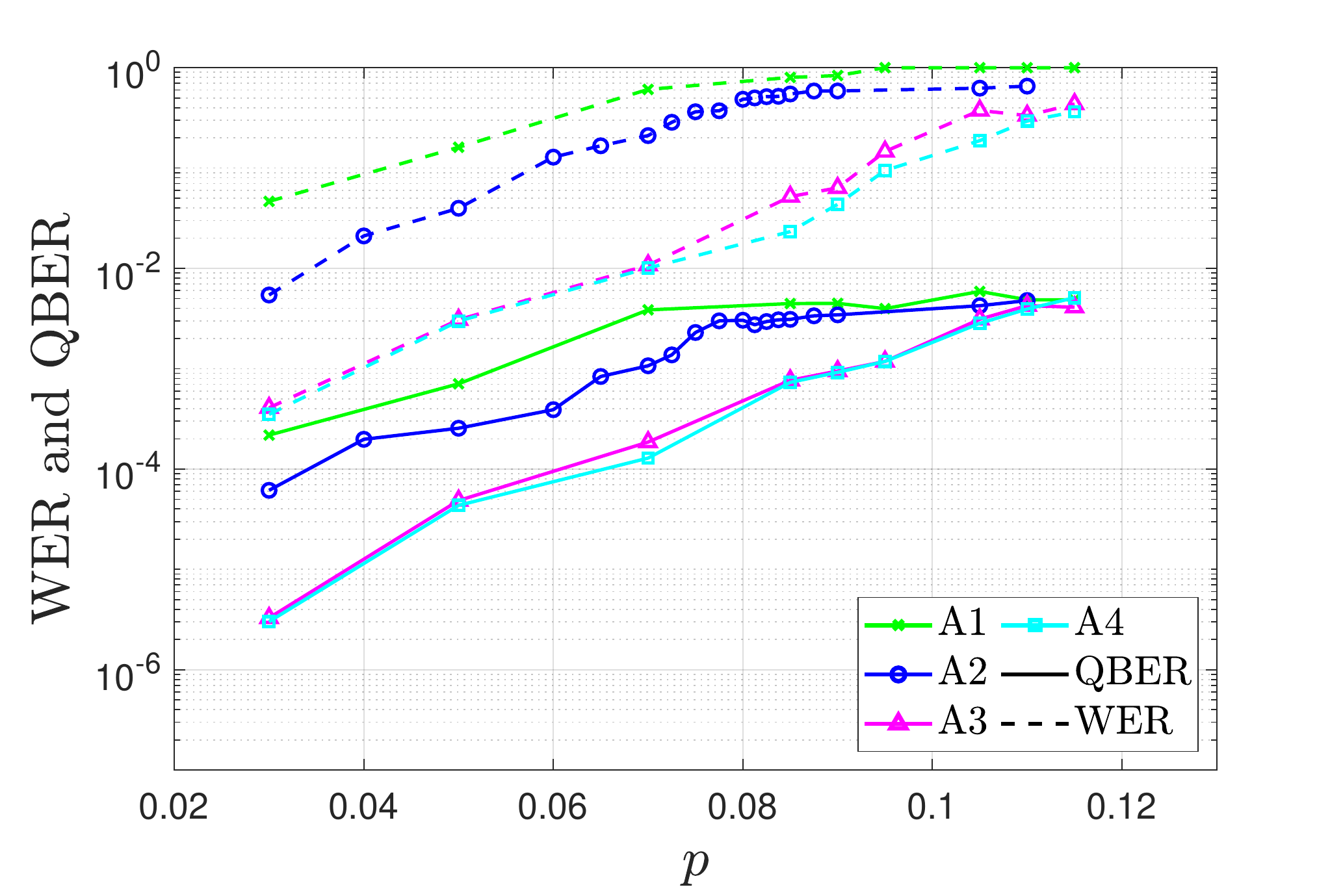}
\caption{\normalsize{Simulated WER and QBER for different $X$ operator decoders. $p$ represents the gross flip probability of the Pauli channel with asymmetry coefficient $\alpha = 10^2$. }}

    \label{Xdocoder}
\end{figure}

Let us first consider the $X$ operator decoder. This decoder can cause an increment in the error floor of the code if the values of $m_1$ and $[y_1, x_1, t_1]$ are not chosen correctly. For sufficiently large values of $m_1 < \frac{N}{2}$, due to the low likelihood of $X$ errors, the bit-flip decoder performs well regardless of the values of $[y_1, x_1, t_1]$, but when $m_1$ becomes too small, performance of the decoder is only acceptable if $[y_1, x_1, t_1]$ are chosen appropriately. This can be seen in Figure \ref{Xdocoder}, where the QBER and WER curves of $X$ operator decoders with $m_1 = 1262$ and different configurations of $[y_1, x_1, t_1]$ are shown. The complete characteristics of these $X$ decoders are detailed in table \ref{dataX}.

 \begin{table}[h!]

    \centering\caption{\normalsize{Parameter values and configurations of $X$ operator decoders of asymmetric CSS codes simulated over a Pauli channel with $\alpha = 10^2$. The results of these simulations are shown in Figure \ref{Xdocoder}. }}
    \vspace{0.15mm}\begin{tabular}{ccccc}
    \toprule 
    Decoder &$m_1$ &$t_1$ &$x_1$ &$y_1$\\
    \midrule
    A1 &$1262$ &$900$ &$76.3$ &$3$ \\
    A2 &$1262$ &$750$ &$54.2$ &$3$\\
    A3 &$1262$ &$300$ &$29.3$ &$3$\\
    A4 &$1262$ &$100$ &$24.4$ &$3$\\
    \bottomrule
    \label{dataX}
    \end{tabular}
    \end{table}

\newpage
As is shown in Figure \ref{Xdocoder}, performance of the $X$ decoders varies significantly depending on the values of $[y_1, x_1, t_1]$. Decoders A1 and A2 are substantially worse than decoders A3 and A4. They mainly differ in the values of the parameters $x_1$ and $t_1$, which are much larger for decoders A1 and A2 than for decoders A3 and A4. This implies that the performance of $X$ decoders with smaller values of $m_1$ will be better when lower values of $x_1$ are chosen by using less degree-1 syndrome nodes in the decoding graph, i.e, selecting smaller values for $t_1$. This is logical, since as was mentioned in the previous section, despite the fact that selecting larger values of $t_1$ will increase the amount of exact information transmitted from the upper layer nodes during initial iterations, it will also make the degree of the $s_B$ nodes so large that the impact of these ``perfect" messages might be mitigated and message passing may not operate successfully. For a decoder with a small value of $m_1$ ($m_1 = 1262$), if $[y_1, x_1, t_1]$ are chosen correctly, performance of the $X$ decoder is excellent, with its QBER and WER curves increasing in an almost linear fashion as functions of $p$ (codes A3 and A4 of Figure \ref{Xdocoder}). We will later see how the performance curves of such aptly built $X$ decoders can sometimes be orders of magnitude below those of the corresponding $Z$ decoder. On the contrary, if the selected configuration of $[y_1, x_1, t_1]$ yields a value of $x_1$ that is too large, performance of the decoder will be degraded enough to cause an increase in the error floor of the overall CSS code.

\begin{figure}[!h]
\centering
\includegraphics[width = \linewidth,height=4in]{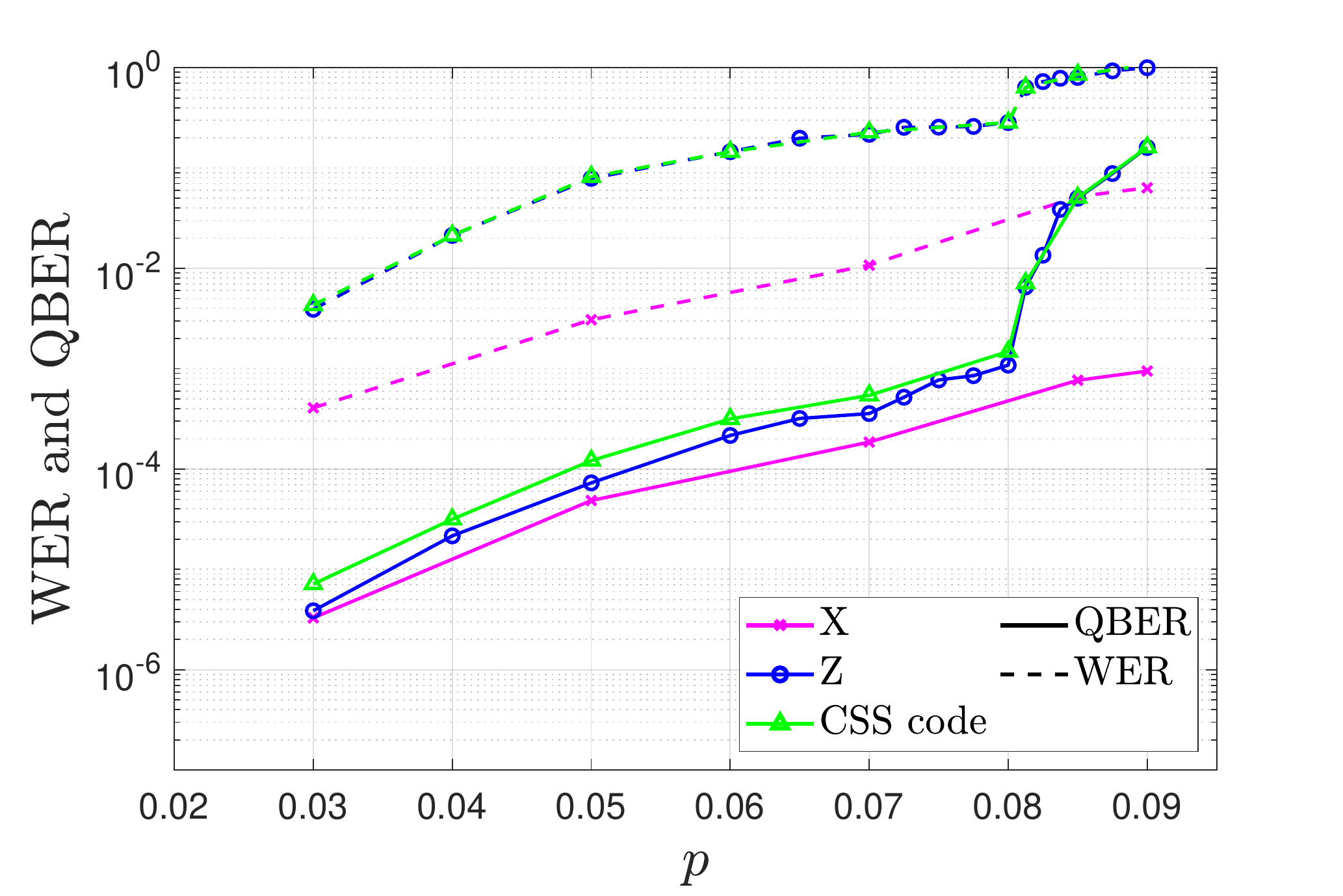}
\caption{\normalsize{Simulated WER and QBER for the constituent decoders of a CSS code. $p$ represents the gross flip probability of the Pauli channel with asymmetry coefficient $\alpha = 10^2$. }}

    \label{resultsXZ}
\end{figure}

The $Z$ operator decoder faces the daunting task of correcting the much more frequent $Z$ errors. Considering the previous discussion regarding the $X$ operator decoders, if we assume that we have an appropriately designed $X$ operator decoder, we hypothesize that the performance of the CSS code as a whole will be majorly determined by the quality of its phase-flip decoder. To evaluate this hypothesis we study the performance curves shown in Figure \ref{resultsXZ}, which correspond to the $Z$ decoder of code A3 of table \ref{data}, the $X$ decoder A3 of table \ref{dataX}, and the the CSS code that arises when using these two decoders simultaneously.\footnote{The performance curves of the CSS code in Figure \ref{resultsXZ} have been simulated. Nonetheless, summing the WER/QBER of each constituent CSS decoder (the $X$ and $Z$ operator decoders) is also a valid method to obtain the performance curves of the overall code.}

Figure  \ref{resultsXZ} shows how the error contribution of the X decoder to the QBER and WER curves of the overall CSS code is almost negligible when compared to the $Z$ decoder. This occurs because the $X$ decoder is correctly designed (the design parameters $m_1$ and $[y_1, x_1, t_1]$ have been selected appropriately), contrary to the $X$ decoder of code A3 in Figure \ref{results}. Additionally, the results of Figure \ref{resultsXZ} show  that the performance curves of the CSS code and its $Z$ operator decoder are very similar, especially at the decoding threshold\footnote{Recall that the decoding threshold or waterfall region of an error correction code is the region where a sharp drop in the error rate takes place.}. Two factors play a role in defining when the CSS code reaches its decoding threshold. The first, which will be discussed later on in this section, is the block length of the code itself. The second, is the design of the individual CSS decoders of the code. The abrupt increase in the error rate for the $Z$ decoder (both for the WER and the QBER) while the $X$ decoder maintains performance at the error floor, proves that for our proposed scheme, the decoding threshold is essentially defined by the $Z$ operator decoder. Therefore, this confirms our hypothesis that the performance of an asymmetric CSS code, if it is aptly designed (the values of $m_1$ and $[y_1, x_1, t_1]$ are chosen correctly), will be determined by the error correcting capabilities of its phase-flip decoder.

In short, this implies that the design of the best possible decoder for the proposed asymmetric CSS scheme can be approached through the separate optimization of its constituent $Z$ and $X$ decoders. This can be done by conducting simulations of different configurations of $[y_2, x_2, t_2]$ and $m_2$ that allow a sufficiently large value of $m_1$ or configuration of $[y_1, x_1, t_1]$ for which the bit-flip decoder exhibits few errors. In this manner, even if an equal distribution of $X$ and $Z$ errors is not obtained, we avoid the increased error-floor associated to bad $X$ operator decoders while optimum performance, in terms of at what noise levels the waterfall region is entered, is achieved. Additionally, this serves to simplify matters, since by having shown that the performance of these asymmetric CSS codes is overwhelmingly determined by the quality of the phase flip decoder (assuming the the bit flip decoder is aptly built), we can now focus only on optimizing the parameter configuration of a single decoder. With this goal in mind, in the sequel we simulate different configurations of $m_2$ and $[y_2, x_2, t_2]$ and determine which one results in the best performance.

\subsubsection{Optimization of the $Z$ decoder}

In the previous subsection we showed that the performance of these asymmetric CSS codes is determined by the behaviour of its constituent decoders. A faulty $X$ operator decoder can degrade performance by raising the error floor of the code, while its decoding threshold can change depending on the quality of the $Z$ operator decoder.  Having previously established that if the $X$ operator decoder is designed appropriately performance is completely determined by the $Z$ operator decoder, we now conduct simulations for various $Z$ operator decoders in an attempt to discover the optimum values of $m_2$ and $[y_2, x_2, t_2]$. 

\begin{table}[h!]

    \centering
    \caption{\normalsize{Parameter values and configurations of $Z$ operator decoders of asymmetric CSS codes simulated over a Pauli channel with $\alpha = 10^2$. The results of these simulations are shown in Figure \ref{Z_decoders}. }}
    \vspace{0.15mm}\begin{tabular}{ccccc}
    \toprule 
    Decoder &$m_2$ &$t_2$ &$x_2$ &$y_2$\\
    \midrule
    Z1 &$12262$ &$8198$ &$5$ &$3$ \\
    Z2 &$12262$ &$9010$ &$6$ &$3$ \\
    Z3 &$12262$ &$9507$ &$6.9$ &$3$ \\
    Z4 &$11000$ &$8497$ &$8$ &$3$\\
    Z5 &$11000$ &$8810$ &$9$ &$3$\\
    Z6 &$11000$ &$9060$ &$10$ &$3$\\
    \bottomrule
    \label{dataZ}
    \end{tabular}
    \end{table}

\begin{figure}[!h]
\centering
\includegraphics[width = \linewidth,height=4in]{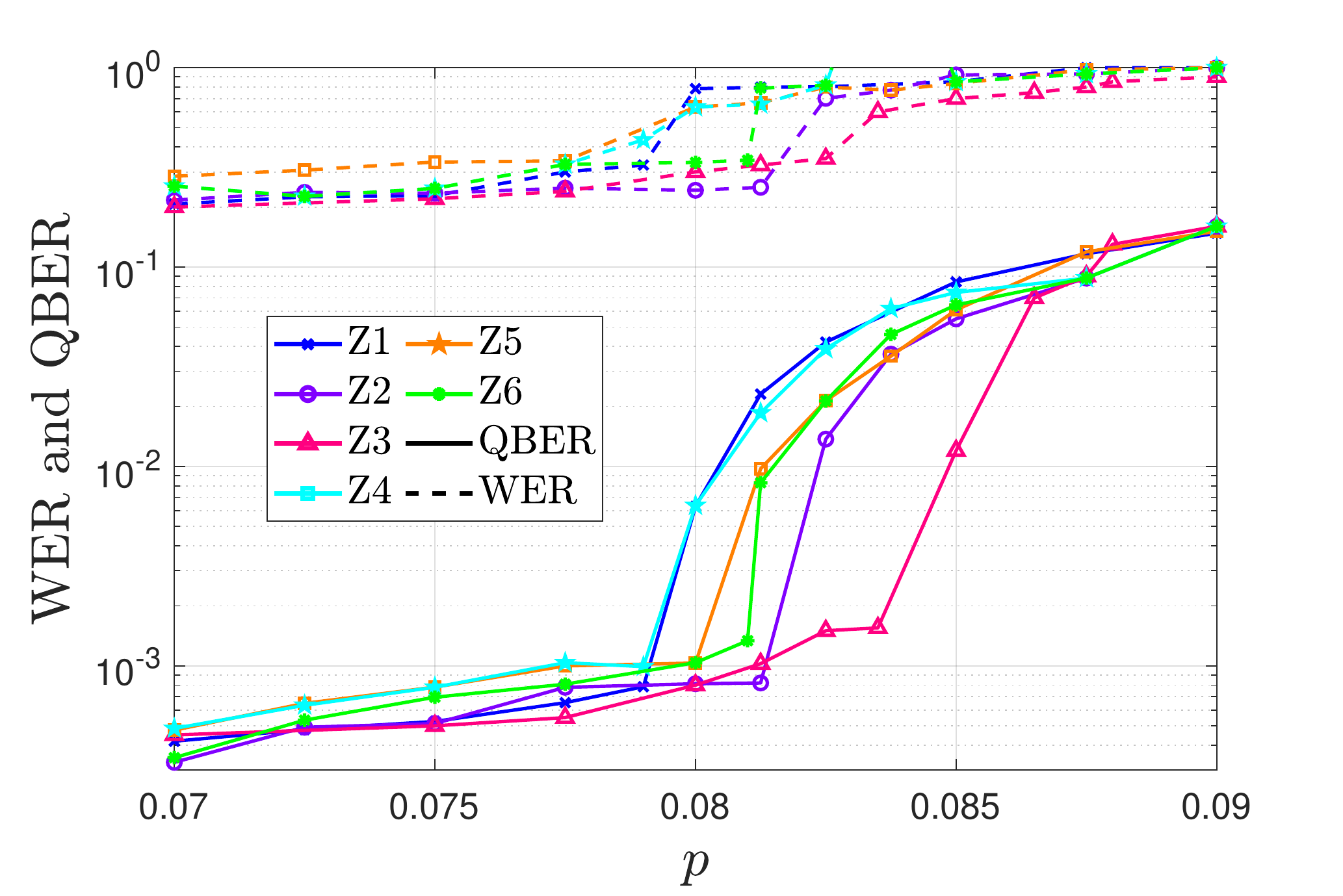}
\caption{\normalsize{Simulated WER and QBER for different CSS QLDGM schemes. $p$ represents the gross flip probability of the Pauli channel with asymmetry coefficient $\alpha = 10^2$. }}

    \label{Z_decoders}
\end{figure}

Given the flexibility the design of these $Z$ error decoders allows, it is important to provide structure to the simulation process. Earlier in this paper we mentioned that if $x_2$ could be shown to be a good indicator for the performance of the overall scheme, the design process could be reduced to simply finding the parameter configuration that would yield the optimum value of $x_2$. The value of $x_2$ is representative of its relationship with the parameter $t_2$, given that growth or reduction in one of these parameters will have the same effect on the other. In fact, when all the other parameters are fixed, the only way we have to modify the value of $x_2$ is by changing $t_2$. At the same time, the parameter $m_2$ is intricately related to the value of $x_2$, which means that the relationship between these parameters will also play a part in the performance of the $Z$ decoder. In an attempt to verify how good a performance indicator $x_2$ is and the nature of the relationship between this parameter and $m_2$, we test the Z decoder schemes detailed in table \ref{dataZ}. These decoders differ in unit increments of $x_2$ while the value of $m_2$ is maintained equal as long as the design process permits\footnote{Recall that for a specific value of $m_2$, given that $t_2 \leq \frac{N}{2}$, there is a maximum value of $x_2$ that can be obtained.}. The performance curves of these decoders are portrayed in Figure \ref{Z_decoders}.

The results shown in Figure \ref{Z_decoders} shed light on the relationship between $m_2$ and $x_2$, as well as how these parameters are linked to the performance of the decoder. For starters, consider decoders Z1, Z2, and Z3. All three of them have the same value of $m_2$, but by selecting larger values of $t_2$, each scheme attains a higher value of $x_2$, with $x_2 = 6.9 \approx 7$ representing the largest possible value of the parameter for $m_2 = 12262$. The performance curves of these decoders show how the waterfall region is entered for subsequently higher values of $p$ (the decoding threshold improves) as $x_2$ grows. For instance, code Z1 which has $x_2 = 5$, enters the waterfall region at roughly $p = 0.081$. In contrast, code Z3 which has $x_2 \approx 7$, enters the waterfall region at approximately $p = 0.085$. The same trend of performance improvement as $x_2$ becomes larger can be observed by looking at decoders Z4, Z5 and Z6. All three decoders have the same value of $m_2$ and the performance curves of decoder Z6, which has the largest value of $x_2$, are slightly better than those of its counterparts. In consequence, this outcome proves that for a given value of $m_2$ the decoder that will attain the best performance will be the one for which the value of $x_2$ is maximized. Notice that maximizing $x_2$ also means maximizing $t_2$, which, in this particular instance means that the negative effects associated to having larger degree $\mathrm{s}_B$ nodes ($x_2$ is maximized) are outweighed by the positive impact of having the largest possible amount of $\mathrm{s}_A$ nodes ($t_2$ is maximized). This is the exact opposite of what happened in the previous subsection when studying the bit flip decoder, where maximizing the parameter $x$ yielded worse results.

Let us now compare the decoders of Table \ref{dataZ} in terms of their value of $m_2$. The results of Figure \ref{Z_decoders} show that performance of the decoder improves as the value of $m_2$ becomes larger. In fact, even though decoder Z6 has the highest value of $x_2$, it is outperformed by both decoders Z2 and Z3. These results present a new conundrum to which we must now give answer: which decoders will perform better, those with larger values of $m_2$ and lower values of $x_2$ or those with larger values of $x_2$ and smaller values of $m_2$? We can also formulate this question as: how is the trade-off relationship between $x_2$ and $t_2$ affected by the value of $m_2$? To analyze this, we simulate the Z decoders shown in table \ref{finalZ_sims}. 

\begin{figure}[!htp]
\centering
\includegraphics[width = \linewidth,height=4in]{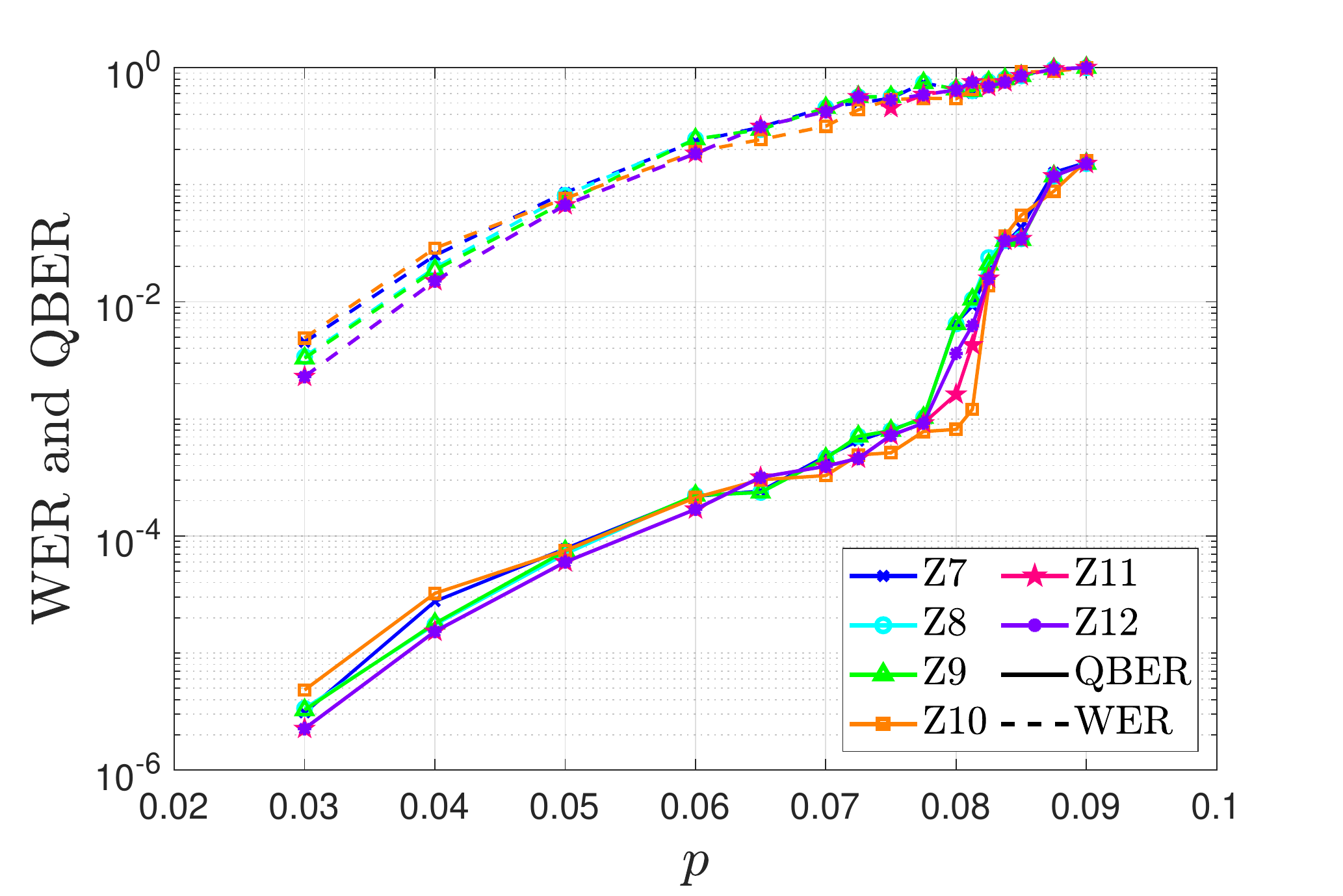}
\caption{\normalsize{Simulated WER and QBER for the decoders of table \ref{finalZ_sims}. $p$ represents the gross flip probability of the Pauli channel with asymmetry coefficient $\alpha = 10^2$. }}

    \label{opt_Z_decs}
\end{figure}

Figure \ref{opt_Z_decs} portrays the simulation results for the $Z$ decoders of table \ref{finalZ_sims}. Performance is similar for all the simulated decoders, with decoder Z10 having the best decoding threshold. In terms of the relationship between $x_2$ and $m_2$, the curves shown in Figure \ref{opt_Z_decs} show that, up to a certain point, better performance is obtained by maximizing $m_2$ over $x_2$. In other words, increasing the number of syndrome nodes used to decode $Z$ errors is more important, within certain limits, than trying to obtain the largest value of $x_2$. This is reflected by decoder Z10, which has $m_2 = 12262$, outperforming decoders Z11 and Z12, which have $m_2 = 12676$ and $m_2 = 13262$. Therefore, the best performance of the proposed scheme over the asymmetric channel with $\alpha = 10^2$ is obtained by setting $m_2 = 12262$ and maximizing $t_2$ ($t_2=9507$ in this case) so that the largest possible value of $x_2$ is obtained for the selected $m_2$ value. This outcome tells us that increasing $m_2$ beyond a certain value has a negative impact on performance, despite the maximization of $t_2$ and $x_2$.

A plausible cause for this is that when $m_2 > 12262$, there is an increased number of $\mathrm{s}_B$ nodes and a reduced percentage of degree-1 $\mathrm{s}_A$ nodes (since $t_2$ is bounded). A smaller percentage of $\mathrm{s}_A$ nodes means that a lower amount of perfect information will be propagated during initial decoding interations, which coupled with the increased number of $\mathrm{s}_B$ nodes, explains the degradation in the performance of the message passing decoding algorithm for the $m_2 > 12262$ schemes. Nonetheless, the performance of all the decoders shown in Figure \ref{opt_Z_decs} differs by such a small margin that we can confidently state the following: \textit{The best or near-best configuration of the proposed asymmetric CSS schemes is obtained by selecting $m_2 = \beta m$, where $\beta \in (0,1)$, that allows a sufficiently large value of $m_1 = (1-\beta)m$ for the $X$ operator decoder to function well. In terms of doping, $t_2$ should be maximized as $t_2 = \frac{N}{2}$, and $t_1 = 0.3m_2$.}

\begin{table}[h!]

    \centering\caption{\normalsize{Parameter values and configurations of $Z$ operator decoders of asymmetric CSS codes simulated over a Pauli channel with $\alpha = 10^2$. The results of these simulations are shown in Figure \ref{opt_Z_decs}. }}
    \vspace{0.15mm}\begin{tabular}{ccccc}
    \toprule 
    Decoder &$m_2$ &$t_2$ &$x_2$ &$y_2$\\
    \midrule
    Z7 &$10410$ &$9507$ &$21$ &$3$ \\
    Z8 &$10600$ &$8972$ &$12$ &$3$ \\
    Z9 &$11232$ &$9507$ &$11$ &$3$ \\
    Z10 &$12262$ &$9507$ &$6.9$ &$3$ \\
    Z11 &$12676$ &$9507$ &$6$ &$3$\\
    Z12 &$13262$ &$9507$ &$5$ &$3$\\
    \bottomrule
    \label{finalZ_sims}
    \end{tabular}
    \end{table}

 For $\alpha = 10^2$, from the decoders of table \ref{finalZ_sims} we can ascertain that setting $0.73 \leq \beta \leq 0.9$ results in good performance, provided that $x_2$ is maximized by setting $t_2 = \frac{N}{2}$ once $m_2$ is chosen. We expect the value of the parameter $\beta$ to vary with the degree of asymmetry of the channel, becoming larger as $\alpha$ grows. In terms of the value of $t_1$, the results of Figure \ref{Xdocoder} show that setting $30\%$ of the $m_1$ nodes to be degree-1 syndrome nodes produces the best results. 
We test the validity of these statements for channels with larger degrees of asymmetry in the final subsection of this chapter. Prior to doing so, we show how the error floor of the CSS codes can be reduced by increasing the degrees of the underlying classical LDGM code, and we also show how the decoding threshold of our schemes can be improved by selecting larger values for the block length.

\subsubsection*{Error floor reduction}

Most of the CSS and non-CSS QLDGM schemes we have seen throughout this dissertation are based on an underlying classical irregular LDGM code constructed through the parallel concatenation of two regular LDGM codes. The motivation behind such a construction is that the error floor of a single regular LDGM code can be substantially reduced when it is parallel-concatenated with another regular LDGM code of much higher degree. This results in an irregular configuration in which using a second LDGM code with larger degrees serves to lower the error floor of the whole scheme. However, as has been observed in \cite{patrick}, the use of larger degrees in the second code of the concatenation may worsen the decoding threshold. To study this behaviour in the context of asymmetric CSS codes we conduct simulations in which the optimum configuration of the CSS decoders devised for the Pauli channel with $\alpha = 10^2$ is employed: $[m_1, t_1, x_1, y_1] = [2000, 700, 11.03, 3]$ and $[m_2, t_2, x_2, y_2] = [12262, 9507, 6.9, 3]$. 

\begin{figure}[!h]
\centering
\includegraphics[width = \linewidth,height=4in]{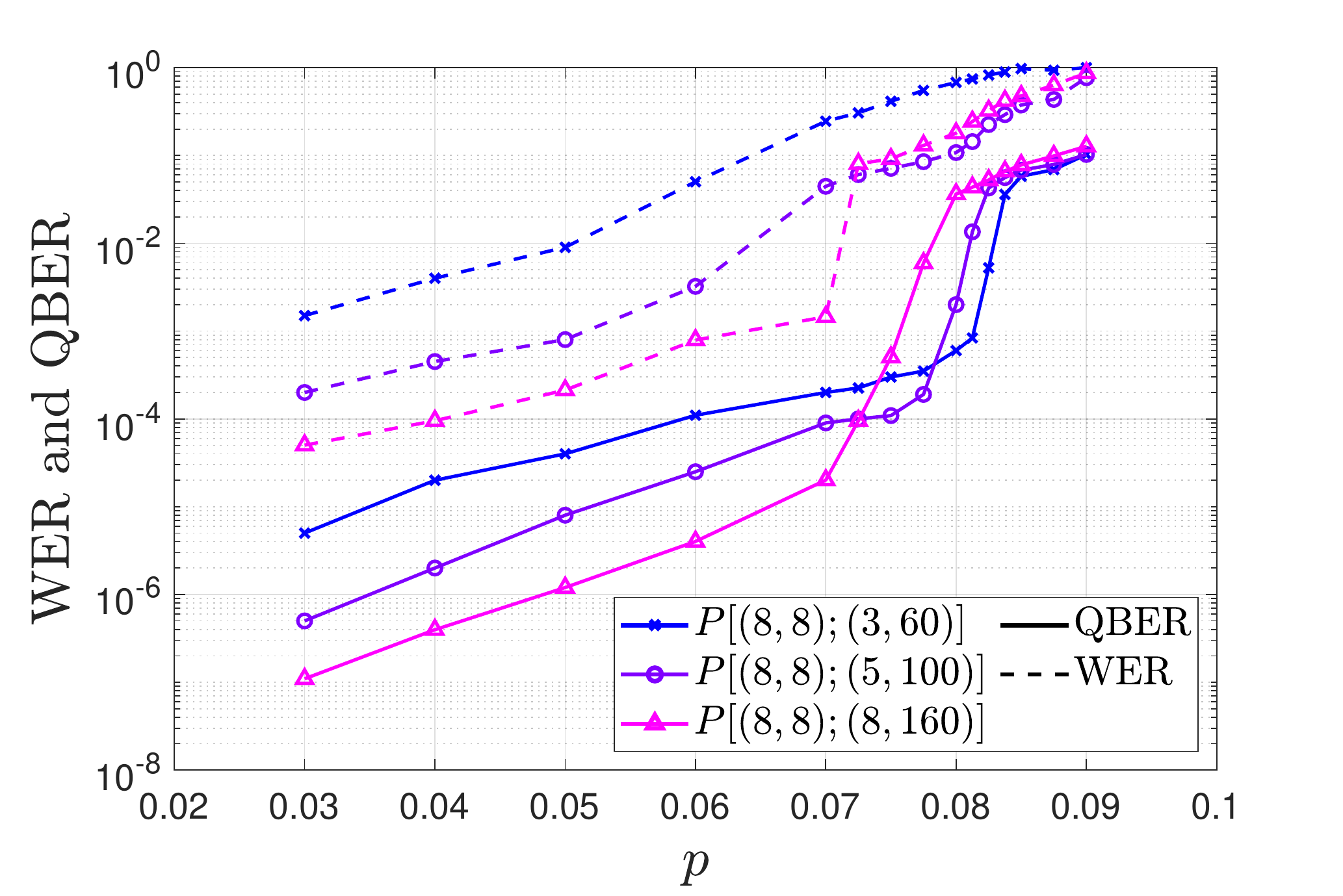}
\caption{Simulated WER and QBER for  asymmetric CSS schemes with $[m_1, t_1, x_1, y_1] = [12262, 9507, 6.9, 3]$, $[m_2, t_2, x_2, y_2] = [2000, 700, 11.03, 3]$. Different degrees of the underlying irregular LDGM code have been tested. $p$ represents the gross flip probability of the Pauli channel with asymmetry coefficient $\alpha = 10^2$.}

    \label{opt_dec_degs1}
\end{figure}

\begin{figure}[!h]
\centering
\includegraphics[width = \linewidth,height=4in]{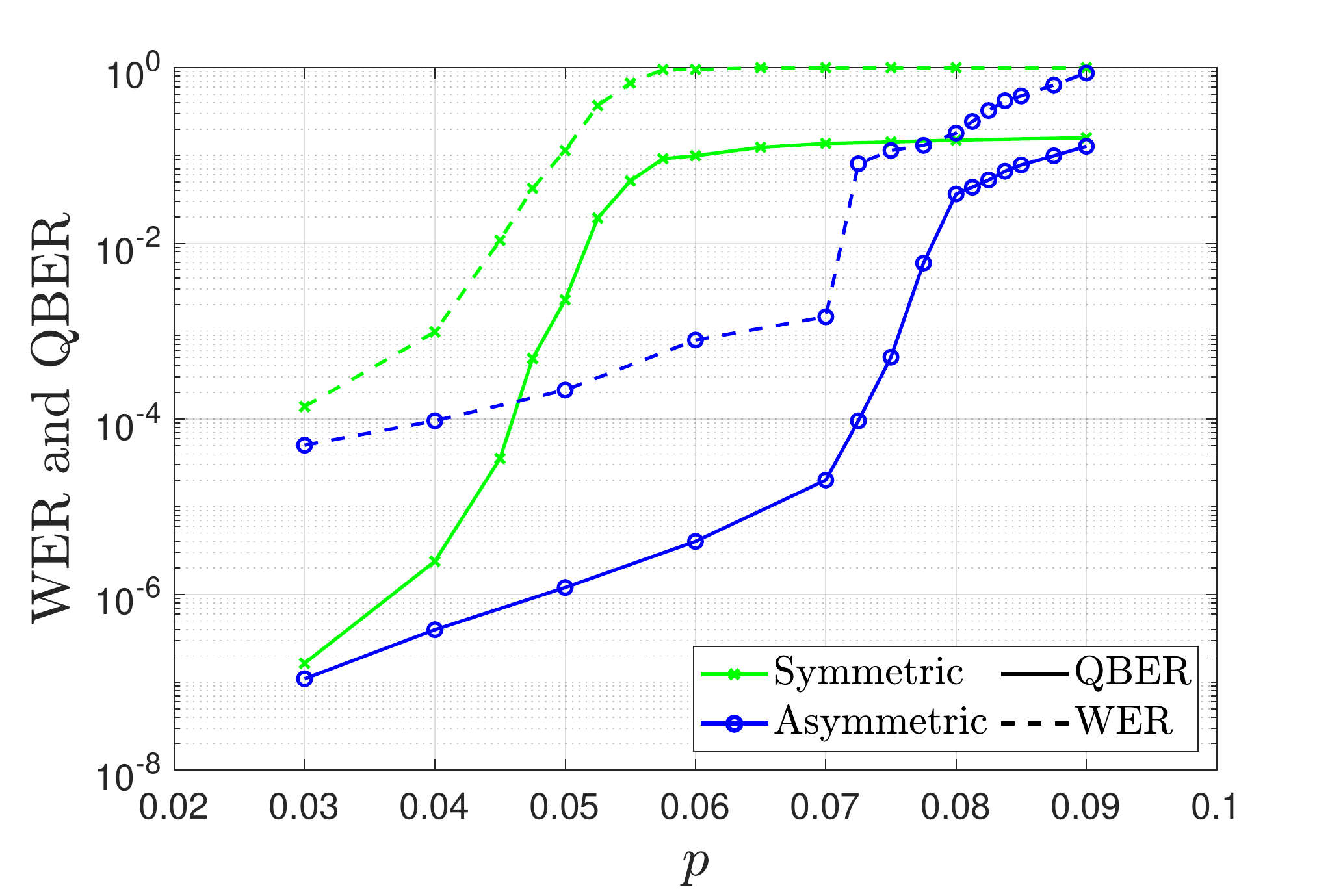}
\caption{Simulated WER and QBER for  an asymmetric CSS scheme with $[m_1, t_1, x_1, y_1] = [12262, 9507, 6.9, 3]$, $[m_2, t_2, x_2, y_2] = [2000, 700, 11.03, 3]$, and a symmetric CSS scheme of the same block length ($N=19014$). The degree of the underlying LDGM code is the same for both schemes $\mathrm{P}[(8,8);(8,160)]$. $p$ represents the gross flip probability of the Pauli channel with asymmetry coefficient $\alpha = 10^2$.}

    \label{opt_dec_degs2}
\end{figure}

As is done in \cite{patrick}, we utilize the irregular LDGM codes described by the concatenations $\mathrm{P}[(8,8);(3,60)]$, $\mathrm{P}[(8,8);(5,100)]$ and $\mathrm{P}[(8,8);(8,160)]$. The simulation results are shown in Figures \ref{opt_dec_degs1} and \ref{opt_dec_degs2}, where the best symmetric scheme of \cite{jgf3}, \cite{jgf4} is included for comparison purposes. 

It is easy to see from Figure \ref{opt_dec_degs1} that increasing the degrees of the underlying irregular LDGM code lowers the overall error floor of the CSS code. Moreover, these results also show that as the degrees of the second regular LDGM code used in the parallel concatenation become larger, in accordance with what has been shown throughout the literature, the decoding threshold of the scheme (more visible in terms of the QBER) begins to deteriorate. Despite the slight worsening of the decoding threshold, the error floor yielded by the CSS code that uses $P[(8,8);(8,160)]$ (the irregular LDGM code with the largest degrees) is orders of magnitude better than for the other simulated concatenations, both in terms of the WER and the QBER. Hence, as occurs for quantum LDGM-based CSS codes designed for the depolarizing channel, using irregular LDGM codes of larger degrees is a valuable technique to improve the performance of codes designed for the general Pauli channel.

Additionally, we include Figure \ref{opt_dec_degs2} to showcase the improvements provided by designing CSS codes specifically for the Pauli channel with $\alpha = 10^2$. In this figure, the performance curves of the best symmetric CSS QLDGM scheme in the literature are compared to those of our best asymmetric CSS QLDGM code. Both constructions are based on the parallel concatenation described by $P[(8,8);(8,160)]$ and both have block length $N = 19014$. Consider the decoding threshold of the symmetric CSS scheme: the code enters the waterfall region at approximately $p_{\text{sym}} = 0.0525$. The asymmetric scheme enters the waterfall region at $p_{\text{asym}} = 0.08$, which when compared to the symmetric code, is equivalent to an improvement of approximately 41 \%. 

\subsubsection*{Decoding threshold improvements}

We know from the work of Shannon \cite{Shannon} that the error rate of error correction codes vanishes asymptotically as their block length grows. Naturally, this means that real error correction codes will never achieve this performance as infinite block lengths cannot be employed in practice \cite{jgf5}. However, this also means that increasing the value of $N$ may lead to performance improvements. 

\begin{figure}[!h]
\centering
\includegraphics[width = \linewidth,height=4in]{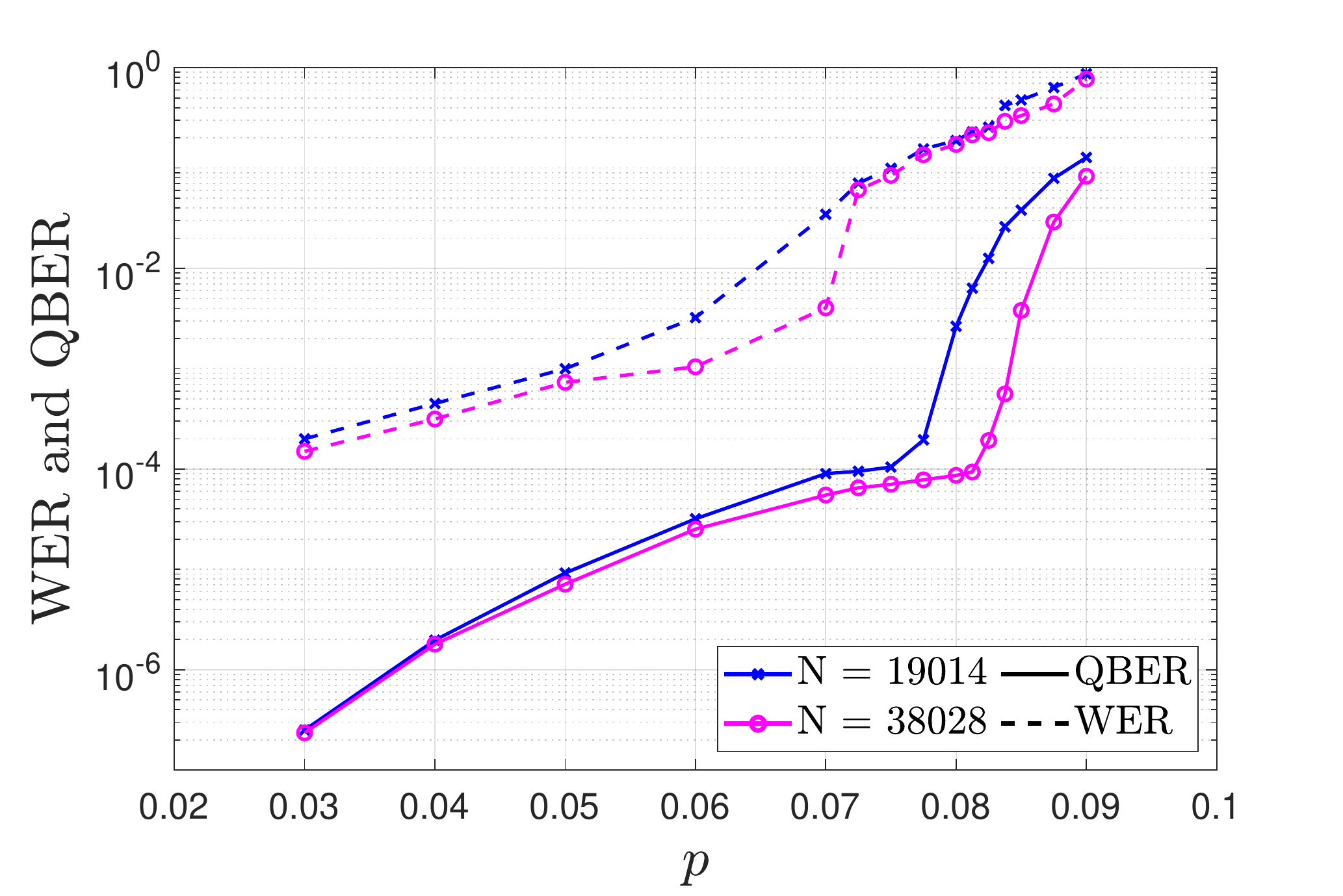}
\caption{Simulated WER and QBER for two asymmetric CSS codes with block lengths $N = 19014$ and $N = 39028$. $p$ represents the gross flip probability of the Pauli channel with asymmetry coefficient $\alpha = 10^2$. }

    \label{largerBlock}
\end{figure}

We close out this subsection by showing how an augmentation of the block length of our asymmetric scheme results in an improvement of its decoding threshold. For this purpose we compare the code with $[m_1, t_1, x_1, y_1] = [2000, 700, 11.03, 3]$ and $[m_2, t_2, x_2, y_2] = [12262, 9507, 6.9, 3]$ that uses the concatenation $P[(8,8);(5,100)]$ to its equivalent when the block length is doubled ($N = 2\times19014 = 38028$), i.e $[m_1, t_1, x_1, y_1] = [4000, 1400, 11.03, 3]$ and $[m_2, t_2, x_2, y_2] = [24524, 19014, 6.9, 3]$. Once again, the considered channel is the Pauli channel with $\alpha = 10^2$. The results are shown in Figure \ref{largerBlock}, where the betterment of the decoding threshold associated to a larger block size can be clearly observed.  

\subsubsection{Simulations and adaptation to other asymmetric parameter values}

\begin{figure*}[!htp]
	\centering
	\subfloat[ \label{eeee}]{%
		\includegraphics[width=.48\textwidth, height = 2.25in]{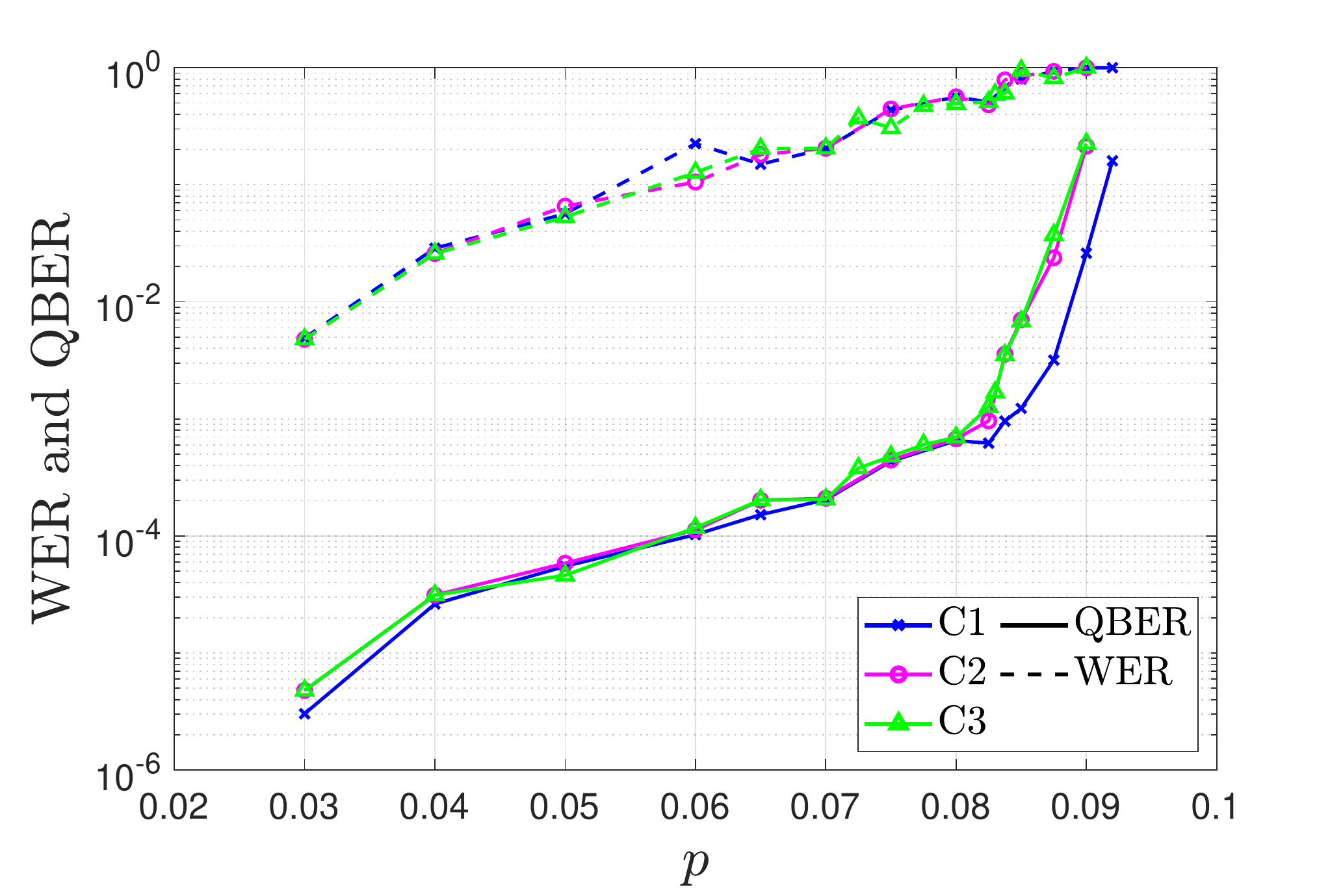}\label{fig:res_A}
	} \hfill
	\subfloat[ \label{rrrrr}]{%
		\includegraphics[width=.48\textwidth, height = 2.25in]{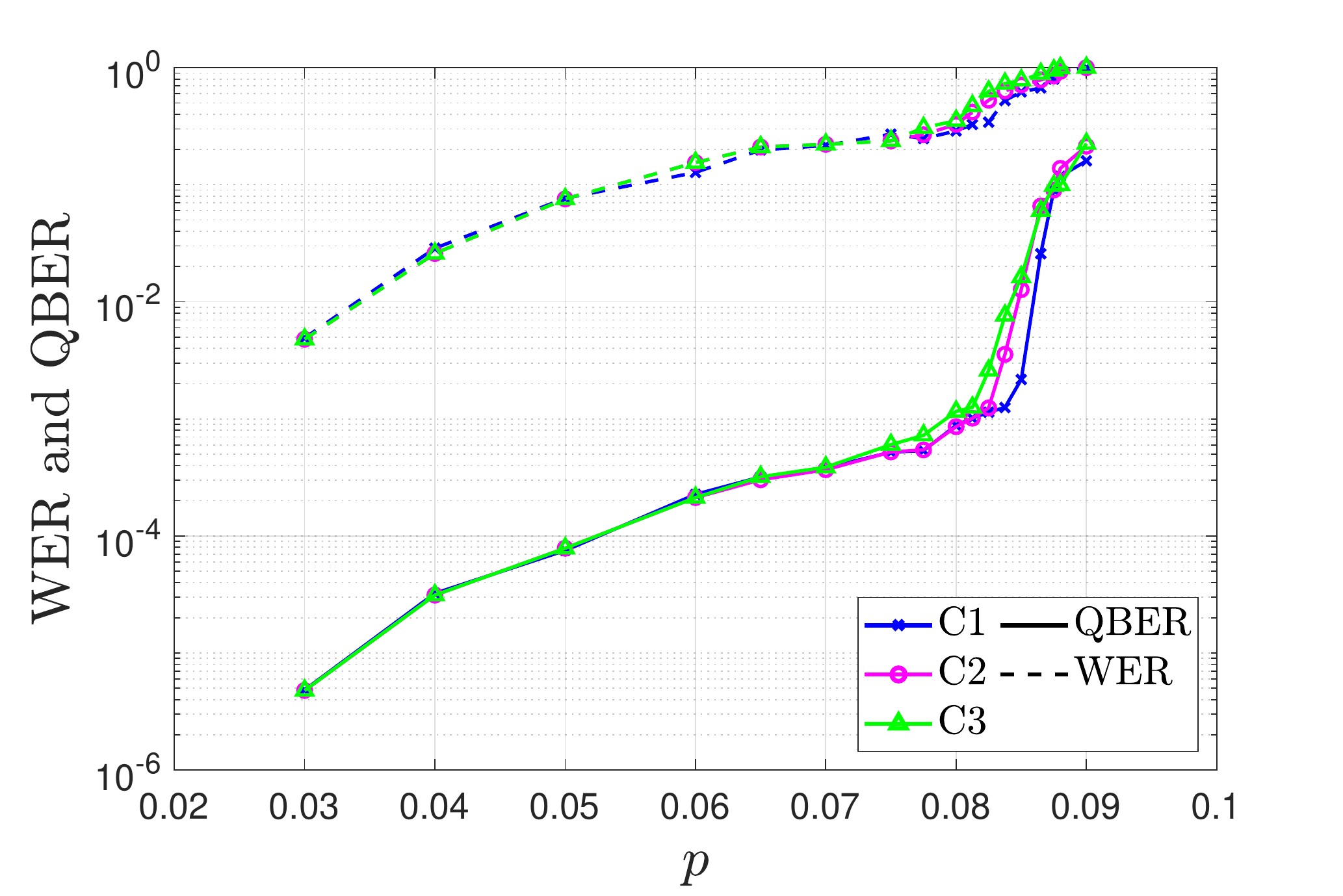} \label{fig:res_B}
	}\hfill
	\subfloat[ \label{aaaaaa}]{%
		\includegraphics[width=.5\textwidth, height = 2.25in]{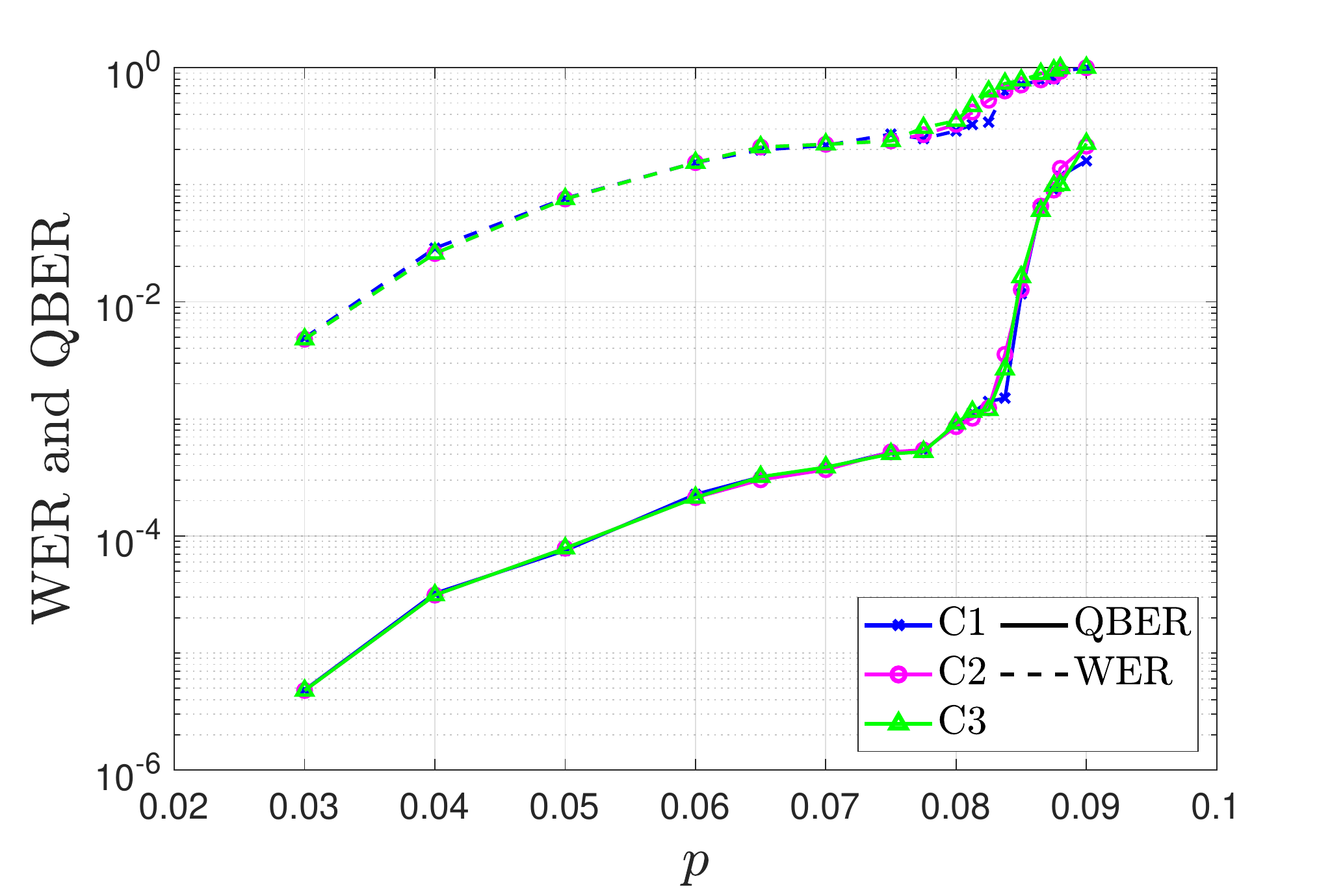} \label{fig:res_C}
	} 
	\caption{Simulated WER and QBER for the asymmetric CSS schemes of table \ref{finaltable}: \textbf{(a)} $p$ represents the gross flip probability of the Pauli channel with an asymmetry coefficient $\alpha = 10$. \textbf{(b)} $p$ represents the gross flip probability of the Pauli channel with asymmetry coefficient $\alpha = 10^4$. \textbf{(c)} $p$ represents the gross flip probability of the Pauli channel with asymmetry coefficient $\alpha = 10^6$.} \label{perf}
\end{figure*}

A matter that has yet to be discussed is the behaviour of our scheme over Pauli channels with different degrees of asymmetry. In \cite{josurev, josuconf, asymmetry-2}, the values of the asymmetry coefficient $\alpha = [1, 10^2, 10^4, 10^6]$ are said to provide a realistic representation of practical quantum devices. Earlier in this work, we predicted that for larger degrees of asymmetry our proposed schemes would benefit from allowing more syndrome information to be used to decode $Z$ errors (increasing the value of $m_2$). In a similar manner, this implies that for smaller degrees of asymmetry, the asymmetric CSS codes should provide more syndrome information to the $X$ decoder. Essentially, asymmetric CSS QLDGM schemes should have larger $Z$ operator decoding subgraphs and smaller $X$ operator decoding subgraphs as the parameter $\alpha$ grows. To verify this hypothesis, we simulate different asymmetric CSS codes for the asymmetry coefficients $\alpha = [10, 10^4, 10^6]$. The particular configurations of the considered CSS codes are shown in table \ref{finaltable}, while the performance of these codes over the Pauli channels with asymmetry coefficients $\alpha = [10, 10^4, 10^6]$ is shown in Figure \ref{perf}. 

\begin{table}[h!]

    \centering\caption{\normalsize{Parameter values and configurations of the CSS codes simulated over Pauli channels with $\alpha = [10, 10^4, 10^6]$. The results of these simulations are shown in Figure \ref{perf}. }}
    \vspace{0.15mm}\begin{tabular}{ccccc}
    \toprule 
    Code \# &$[m_1, t_1, x_1, y_1]$ &$[m_2, t_2, x_2, y_2]$\\
    \midrule
    C1 &$[2000, 700, 11.03, 3]$ &$[12262, 9507, 6.9, 3]$  \\
    C2 &$[1262, 300, 29.3, 3]$ &$[12676, 9507, 6, 3]$  \\
    C3 &$[1000, 100, 31.5, 3]$ &$[13262, 9507, 5, 3]$  \\
    \bottomrule
    \label{finaltable}
    \end{tabular}
    \end{table}

These results prove that our hypothesis is correct. For $\alpha = 10$, code C1 outperforms C2 and C3, but as alpha grows, the performance of C1 becomes increasingly degraded while that of C2 and C3 improves. This outcome is consistent with our initial hypothesis because for the smallest value of $\alpha$ that we have simulated, $\alpha = 10$, the code with the best performance is C1 which has the smallest value of $m_2$ (number of syndrome nodes used by the $Z$ decoder), whereas for $\alpha = 10^6$ codes C2 and C3, which have larger values of $m_2$, overtake code C1. This behaviour is also congruous with the fact that the hashing bound of a Pauli channel increases as this channel becomes more asymmetric \cite{asymmetry-2}.

However, the results of Figure \ref{perf} are somewhat surprising in the sense that, even for the most asymmetric instance that we have simulated $\alpha = 10^6$, the performance improvement provided by the more asymmetric codes, C2 and C3, is relatively small. In fact, for $\alpha = 10^4$ codes C2 and C3, which have larger values of $m_2$, do not outperform the optimal scheme (code C1) that was derived for a Pauli channel with  $\alpha = 10^2$. Let us discuss why this happens.

It is clear from our initial simulation results (Figure \ref{results}) that asymmetric CSS schemes in which more syndrome information is used to decode $Z$ operators perform better over a general Pauli channel than symmetric CSS codes. The extent to which $m_2$ must be increased has been thoroughly discussed throughout this section and has been shown to also be dependant on other characteristics of the asymmetric CSS construction. The entirety of this analysis has been performed considering a Pauli channel with $\alpha = 10^2$. We can now use the knowledge we have obtained for channels with $\alpha = 10^2$ to try and understand what happens over Pauli channels with $\alpha = 10^4$ and $\alpha = 10^6$. 

In reality, the only difference between these channels lies in the error probabilities of the $X$ and $Z$ operators. For instance, if we assume a channel gross flip probability of $p = 0.075$, we will go from $p_z = 0.0735$ and $p_x= 7\times10^{-4}$ when $\alpha = 10^2$, to $p_z \approx 0.075 $ and $p_x \approx 7 \times 10^{-6}$ when $\alpha = 10^4$, to $p_z \approx 0.075 $ and $p_x \approx 7 \times 10^{-8}$ when $\alpha = 10^6$. These changes in the values of $p_z$ and $p_x$ as $\alpha$ grows, show why in Figure \ref{perf}, the optimum code for $\alpha = 10^2$ performs as well as C2 and C3 when $\alpha = 10^4$ and $\alpha = 10^6$. The change in $p_z$ when going from $\alpha = 10^2$ to $\alpha = 10^4$ or $\alpha = 10^6$ is too subtle to appreciate improvements\footnote{It may even slightly impinge on performance due to imperfect configuration of the other parameters of the scheme.} when increasing the value of $m_2$ from $12262$ to $m_2 = 12676$ and $m_2 = 13262$. In stark contrast, when we go from $\alpha = 1$ to $\alpha = 10^2$, i.e, we compare our schemes to symmetric CSS codes, the improvements in performance when increasing $m_2$ are evident throughout the results provided in this paper. This happens because the change in $p_z$ and $p_x$ is substantial enough when going from $\alpha = 1$ to $\alpha = 10^2$, that changing $m_2 = 7131$ to $m_2 = 12262$ results in a palpable boost in performance. This also coincides with what is shown in \cite{asymmetry-2}, where the capacity of a Pauli channel grows sharply when the asymmetry coefficient goes from $\alpha = 1$ to $\alpha = 10^2$, but only a marginal capacity improvement is observed when $\alpha$ increases further beyond $10^2$. Thus, the main reason for the almost negligible improvements that C2 and C3 provide with regard to C1 when $\alpha > 10^2$ stems from the fact that the change in the nature of the asymmetric channel when increasing $\alpha$ beyond $10^2$ is too small to allow us to appreciate improvements in performance when increasing $m_2$ for the selected block length.

An interesting future research problem will be to study whether more drastic performance improvements can be obtained by increasing the block length of the code, as it may be that for a sufficiently large value of $N$ (for the same quantum rate, increasing the block length implies an increase in the total number of syndrome nodes), for channels with values of $\alpha > 10^2$, the improvements provided by codes with larger values of $m_2$ when compared to the optimum scheme for the channel with $\alpha = 10^2$ will be more noticeable. 

\subsubsection{Distance to the Hashing bound of the Pauli channel model for asymmetry}

We close this section by benchmarking the performance of our proposed schemes against the theoretical limits of the Pauli channel. The Hashing bound for a Pauli channel with asymmetry coefficient $\alpha$ \cite{josuconf} can be computed as

\begin{align*}
C_Q(p,\alpha) = 1 + (1-p)\log_2(1-p)\ + \ &\bigg(\frac{2p}{\alpha + 2}\bigg)\log_2\bigg(\frac{p}{\alpha + 2}\bigg)\\ + \ \bigg(\frac{\alpha p}{\alpha + 2}\bigg)\log_2\bigg(\frac{\alpha p}{\alpha + 2}\bigg).
\end{align*}

\begin{figure}[!h]
\centering
\hbox{\includegraphics[width = \linewidth,height=4in]{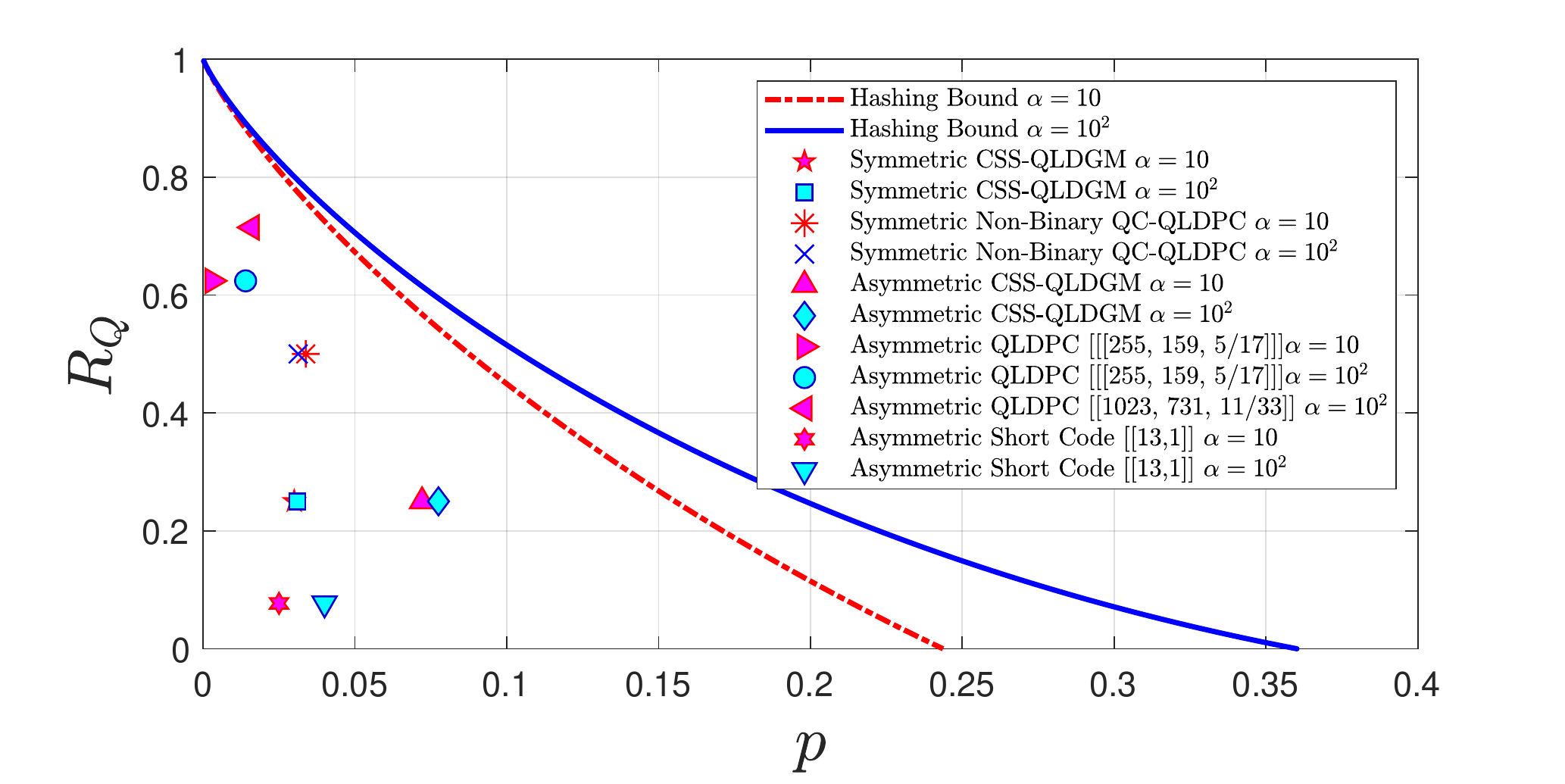}}
\caption{\normalsize{Achievable coding rate at a WER of $10^{-3}$ for various QLDPC codes over Pauli channels with $\alpha = 10$ and $\alpha = 10^2$. }}

    \label{asym-comps}
\end{figure}

We can use this expression to assess the distance to the Hashing bound for a specific rate and asymmetry coefficient. This is reflected in Figure \ref{asym-comps}, where the Hashing bounds for two Pauli channels with asymmetry coefficients $\alpha = 10$ and $\alpha = 10^2$ are shown alongside the points at which the best $R_Q = \frac{1}{4}$ asymmetric schemes\footnote{These schemes are defined by the parameters $[m_1, t_1, x_1, y_1] = [4000, 1400, 11.03, 3]$ and $[m_2, t_2, x_2, y_2] = [24524, 19014, 6.9, 3]$, $N = 38028$, and $P[(8,8);(8,160)]$.} designed in the previous sections for each of these channels can function with WER $=10^{-3}$. As was done earlier in section \ref{sec:initial-comp}, in Figure \ref{asym-comps} we also show the highest possible coding rate at which the asymmetric codes that have been proposed in the literature can function with WER $=10^{-3}$ over each of these asymmetric channels. These codes are:

\begin{enumerate}
    \item The [[$255$, $159$, $\frac{5}{17}$]] asymmetric QLDPC code of rate $R_Q \approx 0.624$ introduced in \cite{klap}.
    \item The  [[$1023$, $731$, $\frac{11}{33}$]] asymmetric QLDPC code of rate $R_Q \approx 0.714$ introduced in \cite{alpha1}.
    \item The [[$13,1$]] asymmetric short code of rate $R_Q \approx 0.077$ introduced in \cite{short-asym}.
    
\end{enumerate}

To provide further context, we also include the coding rates that can be achieved while maintaining WER $=10^{-3}$ by the symmetric CSS QLDGM codes with $N = 19014$ of \cite{jgf3, jgf4} and the non-binary QC-QLDPC codes with $N = 20560$ of \cite{hag2, hag3} over the Pauli channels with asymmetry coefficients $\alpha = 10$ and $\alpha = 10^2$. The results for the symmetric CSS QLDGM codes have been obtained via Monte Carlo simulations (see Appendix \ref{app:sims}). The results for the codes of \cite{hag2, hag3} have been derived as follows.

As is mentioned in section \ref{sec:asymmetry-model}, the most commonly employed noise model in the literature of CSS codes \cite{bicycle, jgf3, jgf4, hag2, hag3} approximates the action of a depolarizing channel by means of two independent BSCs with marginal bit-flip probabilities $f_m = \frac{2p}{3}$. Given the fact that each constituent code of a CSS scheme is decoded separately, the use of this model simplifies the simulation process because it only requires one of the constituent codes to be executed in most cases. This can be done because the error rate of the CSS code over the complete channel is computed as the sum of the error rates of each constituent code over each separate BSC, and considering the fact that each BSC will have the same flip probability, it will be possible to compute the performance of the scheme over the overall channel by simply obtaining the error rate of one of the constituent codes and summing it to itself.

We can use this framework to estimate the performance over the general Pauli channel model by adjusting the flip probabilities of the separate BSCs as $f_m^{x} = \frac{2p}{\alpha + 2}$ and $f_m^{z} = \frac{p(\alpha + 1)}{\alpha + 2}$ \cite{alpha1, klap}. This means that each BSC serves as an $X$ and $Z$ error channel, respectively. Against this backdrop, we can compute the individual flip probabilities that each constituent code of the symmetric $R_Q = \frac{1}{2}$ code of \cite{hag2, hag3} would have to function at by substituting the value of $p$ into the expressions that have been given for $f_m^{x}$ and $f_m^{z}$. For instance, over a Pauli channel with $\alpha = 10$, for the same value of $f_m$ at which the code performs with WER = $10^{-3}$ over the depolarizing channel, its $X$ error decoder will now have to operate at $f_m^x \approx 0.0077$ while the $Z$ operator decoder has to function at $f_m^z \approx 0.0426$. These flip probabilities can then be used to obtain the WER of each constituent code of the non-binary QC-QLDPC CSS code from the results given in \cite{hag2, hag3}. This yields $\text{WER}_x \ll 10^{-5}$ and $\text{WER}_z \approx 10^{-1}$, where $\text{WER}_x$ and $\text{WER}_z$ denote the WER for the $X$ and $Z$ operator decoders of the CSS code, respectively. Finally, we can add these error probabilities to obtain the overall WER over the complete asymmetric quantum channel\footnote{Note that this procedure would be valid to approximate the performance of any symmetric CSS code over a Pauli channel with asymmetry coefficient $\alpha$.},  which in this particular case would be $\text{WER}_{\text{non-bin-QC-QLDPC}} = \text{WER}_z + \text{WER}_x \approx 10^{-1}$. Analogously, since $\alpha \geq 10$, $p_z$ will be much larger than $p_x$ and we will be able to compute the depolarizing probability at which the code will have WER = $10^{-3}$ over the general Pauli channel by solving for $p$ in $f_m^{\text{dep}} \approx f_m^{z} = \frac{p(\alpha + 1)}{\alpha + 2}$, where $f_m^{\text{dep}}$ is the flip probability at which the code performs with WER = $10^{-3}$ over the i.i.d. $X/Z$ channel model.

This discussion proves that a CSS code designed for a symmetric channel over which the probability distribution for $X$ and $Z$ errors is identical will be unable to yield the same performance over an asymmetric channel. This is reflected in Figure \ref{asym-comps}, where a coding rate $R_Q = \frac{1}{2}$, which is achievable with the codes of \cite{hag2, hag3} over the depolarizing channel for a value of $p \approx 0.0465$, can now only be achieved for $p \approx 0.0338$ and $p \approx 0.0313$ when the corresponding asymmetry coefficient of the general Pauli channel is $\alpha = 10$ and $\alpha = 10^2$, respectively. The same phenomenon was exhibited by the symmetric CSS QLDGM codes in section \ref{sec:degrade}, where performance was shown to be substantially degraded over the general Pauli channel.

As was done previously for the comparison over the depolarizing channel, we can use the distance to the Hashing bound\footnote{We denote the distance to the Hashing bound of a Pauli channel with asymmetry coefficient $\alpha$ by $\delta^\alpha$. The superscript $\alpha$ is introduced to distinguish this distance measure from $\delta$, the measure used for the depolarizing channel.}, computed as shown in \eref{eq:distance-hashing}, to analyze the quality of the strategies shown in Figure \ref{asym-comps}. For instance, for the Pauli channel with asymmetry coefficient $\alpha = 10$, the proposed asymmetric CSS QLDGM scheme exhibits a distance to the Hashing bound of $\delta^{\alpha}_{\text{asym-CSS}} = 3.1$ dB. In the case of the non-binary QC-QLDPC codes of \cite{hag2, hag3} applied to this same channel, the distance to the Hashing bound is $\delta^{\alpha}_{\text{non-bin-QC-QLDPC}} = 4.2$ dB. This distance is $\delta^{\alpha}_{\text{sym-CSS}} = 6.66$ dB for the symmetric CSS QLDGM codes. Clearly, these outcomes showcase the improvements provided by building CSS designs specifically for the Pauli channel model for asymmetry. However, given that other families of QLDPC codes like those of \cite{CSSbound, hag2, hag3} exhibit excellent performance over the depolarizing channel, it is likely that asymmetric adaptations of these codes may surpass the performance we obtain over asymmetric channels with the asymmetric CSS QLDGM codes proposed herein. Nonetheless, given the increased complexity of these error correction strategies, optimizing them for asymmetric quantum channels will be more complex than the schemes proposed in this work. 

All in all, the discussion provided in this section along with the results that are portrayed in Figure \ref{asym-comps} prove that the best symmetric CSS codes of \cite{hag2, hag3} and asymmetric codes that can be found in the literature are outperformed over asymmetric Pauli channels by the codes we have proposed in this chapter.  

\section{Chapter summary}\label{sec:conclusion}

In this chapter we have introduced a technique to design CSS quantum codes based on the use of the generator and parity check matrices of LDGM codes specifically for the general Pauli channel. The proposed methods are based on simple modifications to the upper layer of the decoding graph of a symmetric CSS QLDGM code designed for the depolarizing channel. For the block length used in this article, an asymmetric CSS code has been found for practical Pauli channels with different values of the asymmetry coefficient. Additionally, we have shown how for larger block lengths, the proposed asymmetric CSS codes can be further optimized based on the asymmetry coefficient $\alpha$ by increasing the block length of the code and the value of $m_2$ according to the guidelines provided in the paper. Over Pauli channels with $\alpha = 10$ and $\alpha = 10^2$, the proposed schemes are closer to the theoretical limit than other existing asymmetric codes and the best codes designed for the depolarizing channel.  
\clearemptydoublepage
\chapter{QTCs and Quantum Channels} 

\epigraph{\textit{``Somewhere, something incredible is waiting to be known''}}{\textbf{Carl Sagan}.}

\noindent\hrulefill

This chapter serves as a summary of the main results of other QEC-related publications that the author has been involved in during this PhD thesis. The chapter is divided into four different sections, each one dedicated to a different research topic:

\begin{itemize}
    \item Section \ref{sec:turbo} studies the design of decoding protocols to improve the peformance of QTCs over realistic Pauli channels (those that include asymmetry). It reviews the primary contents and takeaways of \cite{josuconf}: J. Etxezarreta Martinez, P. Fuentes, P. M. Crespo, and J. Garcia-Fr\'ias,`` Pauli channel online estimation protocol for quantum turbo codes,'' \textit{IEEE International Conference on Quantum Computing and Engineering (QCE20)}, 2020. doi: 10.1109/QCE49297.2020.00023.
    \item Section \ref{sec:twirl} discusses the derivation of classically-tractable quantum channel models that can be used to approximate and reenact the effects of decoherence. It provides a succint description of the motivation and results of \cite{josurev}: J. Etxezarreta Martinez, P. Fuentes, P. M. Crespo, and J. Garcia-Frias, ``Approximating Decoherence Processes for the Design and Simulation of Quantum Error Correction Codes in Classical Computers,'' \textit{IEEE Access}, vol. 8, pp. 172623-172643, 2020. doi: 10.1109/ACCESS.2020.3025619.
    \item Section \ref{sec:TV} introduces the idea of time-varying quantum channel models, which serves as an overview of the work conducted in \cite{superconducting-josu}: J. Etxezarreta Martinez, P. Fuentes, P. M. Crespo, and J. Garcia-Frias, ``Time-varying quantum channel models for superconducting qubits,'' \textit{npj Quantum Inf.}, vol. 7, no. 115, 2021. doi: 10.1038/s41534-021-00448-5.
    \item Section \ref{sec:TV-2} further explores the concept of time-varying quantum channel models by studying the asymptotical error correction limits of these channels. This serves as a summary of the work conducted in \cite{josuarxiv}: J. Etxezarreta Martinez, P. Fuentes, P. M. Crespo, and J. Garcia-Frias, ``Quantum outage probability for time-varying quantum channels,'' submitted to \textit{Phys. Rev. A}, 2021. arXiv: 2108.13701. 
\end{itemize}

Before moving onward, it must be mentioned that the contents shown herein are meant only as a cursory overview. Readers should refer to Josu Etxezarreta Martinez's PhD dissertation (first-author of this research) or the journal articles themselves for a complete discussion regarding these findings.

\section{Online Estimation Protocol for QTCs over Pauli channels} \label{sec:turbo}

As was shown previously in Chapter \ref{chapter7}, the assumption that information regarding the noise level of a quantum channel is available prior to decoding does not generally hold in real QEC scenarios. In practice, quantum error correction strategies face the phenomenon of channel mismatch, where the noise level of the channel must be estimated prior to decoding. Additionally, we also mentioned how the noise suffered by some quantum devices does not usually exhibit the symmetry described by the depolarizing channel. Instead, general Pauli channels must be employed to provide an accurate representation of the real behaviour of these quantum devices. This means that, although QEC techniques are generally evaluated using the depolarizing channel under the assumption of perfect channel knowledge, in some instances, QEC codes must actually be studied (and achieve good performance) over asymmetric quantum channels where prior knowledge of channel parameters is unavailable.

We tackle this issue in \cite{josuconf}, where we address the channel estimation problem for realistic quantum channels when using QTCs. This involves applying the online estimation protocol for QTCs derived for the depolarizing channel in \cite{josu2} to decode over Pauli channels that accurately represent the asymmetric behaviour of realistic quantum devices. This online estimation protocol achieves excellent performance without the need for offline channel estimation, which, as we also discussed in Chapter \ref{chapter7}, requires additional resources and reduces the rate of the QEC strategy. 

\begin{figure}[!h]
\centering
\hbox{\includegraphics[width = \linewidth,height=4in]{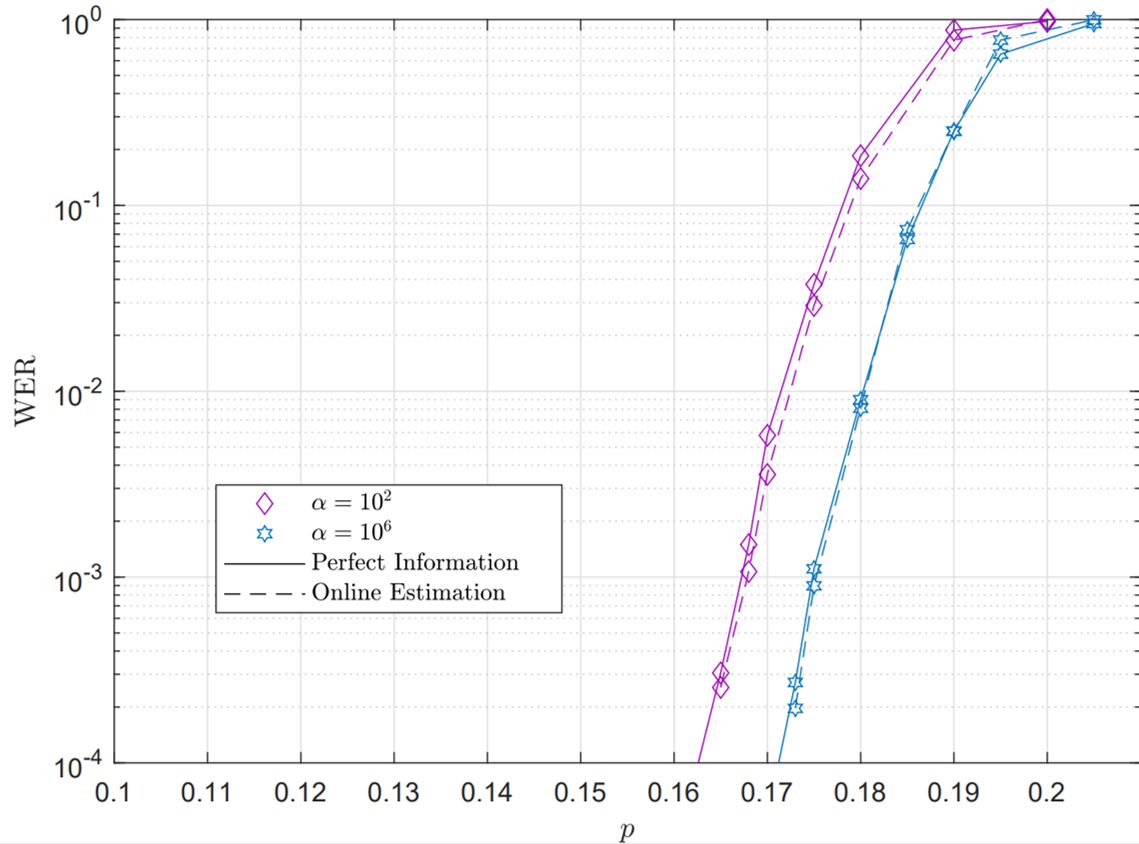}}
\caption{Performance of a QTC over Pauli channels with asymmetry coefficient $\alpha = 10^2$ and $\alpha = 10^6$ under the perfect channel knowledge assumption and when using the online estimation strategy.}

    \label{fig:results_josu1}
\end{figure}

In \cite{josuconf}, we apply the aforementioned online estimation technique for QTCs to the Pauli channel model for asymmetry shown in section \ref{sec:asymmetry-model}. This is achieved by modifying the online estimator to account for the asymmetry coefficient of the channel and produce estimates of its value. The estimates of the channel asymmetry coefficient can then be used to compute the individual probabilities $p_x$ and $p_z$ and begin the decoding process. As is shown in Figure \ref{fig:results_josu1}, the online estimation technique that accounts for asymmetry attains the same performance as that obtained under a perfect channel knowledge scenario. This means that, as was the case for the depolarizing channel, the online technique achieves excellent performance over asymmetric quantum channels without the need for offline channel estimation. 

\section{Approximating decoherence processes for the design of QECCs on classical computers} \label{sec:twirl}

The design and optimization of QEC codes requires the use of an error model, typically in the form of a quantum channel, that accurately represents the decoherence processes that affect quantum information. Based on these error models, appropriate strategies to combat
the effects of decoherence can be derived. In order for these QECCs to be applicable in realistic quantum devices, quantum channels should capture the defining
characteristics of the physical processes that result in qubit decoherence. From what we have seen throughout this dissertation, we know that the depolarizing channel is the most widespread decoherence model used to evaluate the error correcting abilities of QECC families. This decoherence model is especially useful because it satisfies the Gottesman-Knill theorem (see section \ref{sec:channels}), which makes it possible to efficiently simulate it on a classical computer. The reduced qubit-count and limited accessibility of modern quantum computers makes it implausible to simulate decoherence models via quantum means, which implies that having a classically tractable model for decoherence is quite a momentous result. Thus, classical resources remain an invaluable tool for the design of advanced QECCs, even those that will be used beyond the Noisy Intermediate-Scale Quantum (NISQ) era\footnote{The NISQ
era, a term coined by famous physicist John Preskill, makes reference to the time when quantum computers will be able to perform tasks that are out of the reach of classical computers, but will
still be too small (in qubit number) to provide fault-tolerant implementations of quantum algorithms.} \cite{34-preskill}.

The motivation behind our work in \cite{josurev} is to provide a clear description of how the physical processes that lead to the corruption of quantum information (decoherence processes) can be approximated so that they are tractable using classical resources. For this purpose, our work surveys the techniques that have been presented in the literature to approximate general quantum channels by the widely used Pauli channels. We commence with a review of the decoherence processes that compromise the integrity of quantum information by studying the mathematical description/abstraction of these processes into quantum channels. Then, we apply the technique known as \textit{twirling} \cite{twirl1, twirl2, twirl3}, which enables us to obtain classically tractable approximations of those quantum channels and allows us to derive the asymmetric Pauli channel and the depolarizing channel. During this process we also prove that any error correction methods designed for the twirled approximated channels will also be valid for the more realistic original channels. Furthermore, we provide an extensive discussion regarding the Pauli channel and its symmetric instance, the depolarizing channel, and show why these approximations are so widely used for QECC construction/simulation with classical resources. Additionally, even though the memoryless version is almost always considered, we also discuss the emergence and impact of memory effects on those channels \cite{mem1, mem2}. Finally, we present the way in which twirled approximations are implemented on a classical computer and discuss how they can be employed to simulate the performance of QECCs.

All in all, the contributions of our work in \cite{josurev} can be summarized as follows. Primarily, we explain in a clear manner how it is possible to construct and evaluate certain families of quantum error correction codes without the explicit need for a quantum computer (the error models and quantum codes can be entirely simulated in the classical domain). Moreover, we describe how classical computers can be used to simulate the performance of the aforementioned quantum error correcting schemes when they are exposed to the effects of general decoherence models. The key here is that decoherence models can be approximated as Pauli channels, which can be implemented with classical resources, and that the performance of quantum error correcting codes over the approximated channels is equivalent to that obtained for the original decoherence model.

\section{Time-varying quantum channel models for superconducting qubits} \label{sec:TV}

Error models that accurately describe the decoherence processes that corrupt quantum information are a necessity when seeking to construct appropriate quantum error correcting codes. The best known quantum channel models for decoherence are the amplitude damping channel $\xi_{AD}$ and the combined amplitude and phase damping channel $\xi_{APD}$. However, such channels cannot be efficiently simulated in a classical computer for qubit counts that exceed a small limit. As was mentioned in the previous section, the quantum information theory technique known as twirling allows us to approximate $\xi_{AD}$ and $\xi_{APD}$ by the classically tractable family of Pauli channels. It is important to note that the dynamics of the original quantum channels depend on the qubit relaxation $T_1$ and dephasing times $T_2$. Naturally, this dependence will be inherited by the Pauli channel family obtained by twirling the original quantum channels. Throughout the literature, $T_1$ and $T_2$ are considered to be fixed parameters, i.e, that they do not fluctuate over time, which implies that the noise dynamics suffered by the qubits in a quantum device are identical for each quantum information processing task.

Unfortunately, the assumption that $T_1$ and $T_2$ are time invariant has been disproven in recent experimental studies that analyzed the stability of qubits in superconducting quantum processors \cite{decoherenceBenchmarking,fluctAPS,fluctApp,temperature}. The results pesented in these studies show that the relaxation and dephasing times of superconducting qubits can vary by up to $50\%$ of the mean value of the sample data, which strongly suggests that the dynamics of the decoherence effects change drastically as functions of time. Therefore, if we wish to assess the performance of QEC codes accurately, quantum channel models must somehow incorporate the time-variations suffered by the parameters that define their dynamics. 

\begin{figure}[!h]
\centering
	\subfloat[ \label{rste}]{%
		\includegraphics[width=.75\textwidth, height = 2.6in]{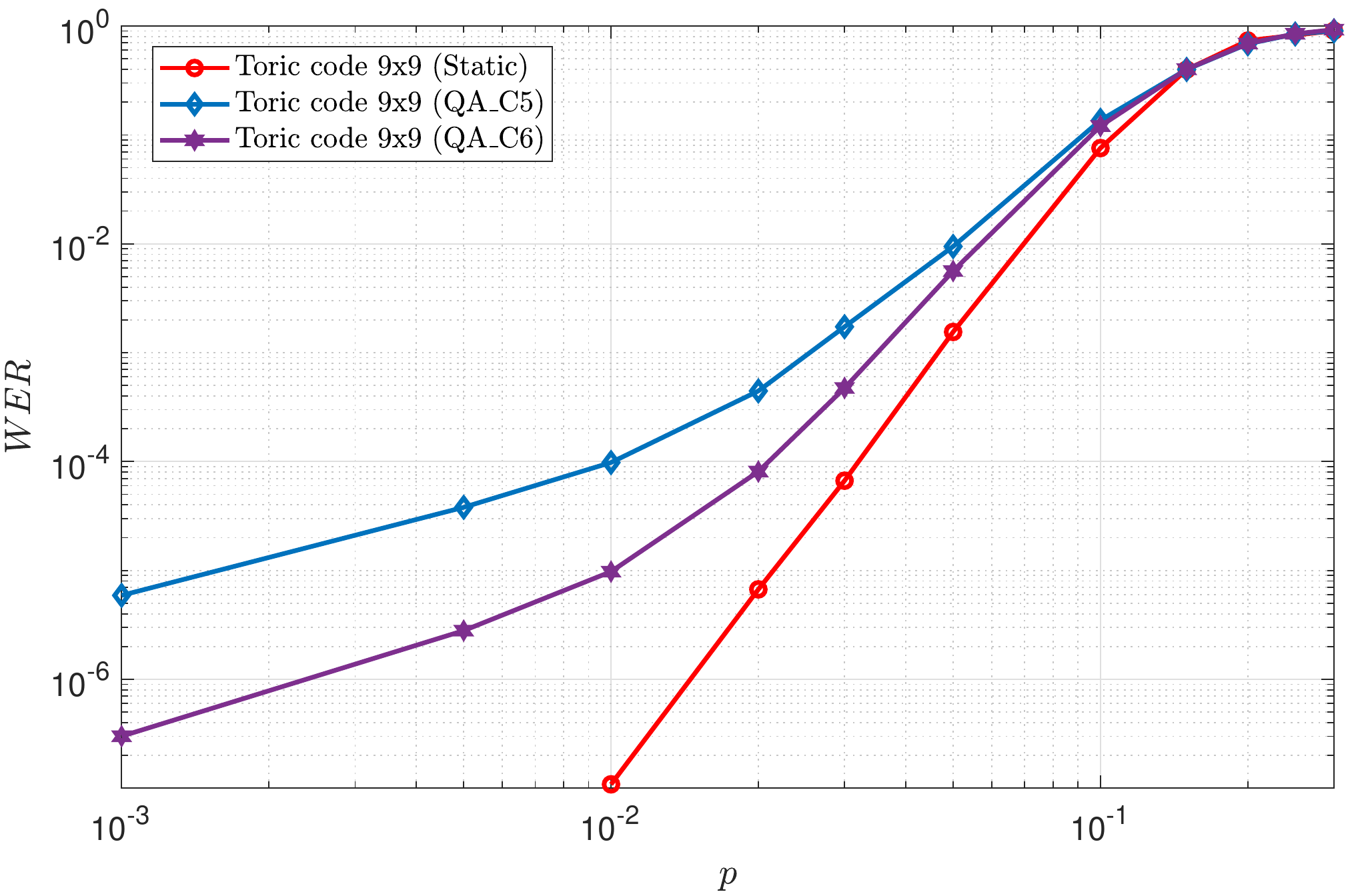} \label{fig:res_BA}
	}\hfill
	\subfloat[ \label{aaadwf}]{%
		\includegraphics[width=.75\textwidth, height = 2.6in]{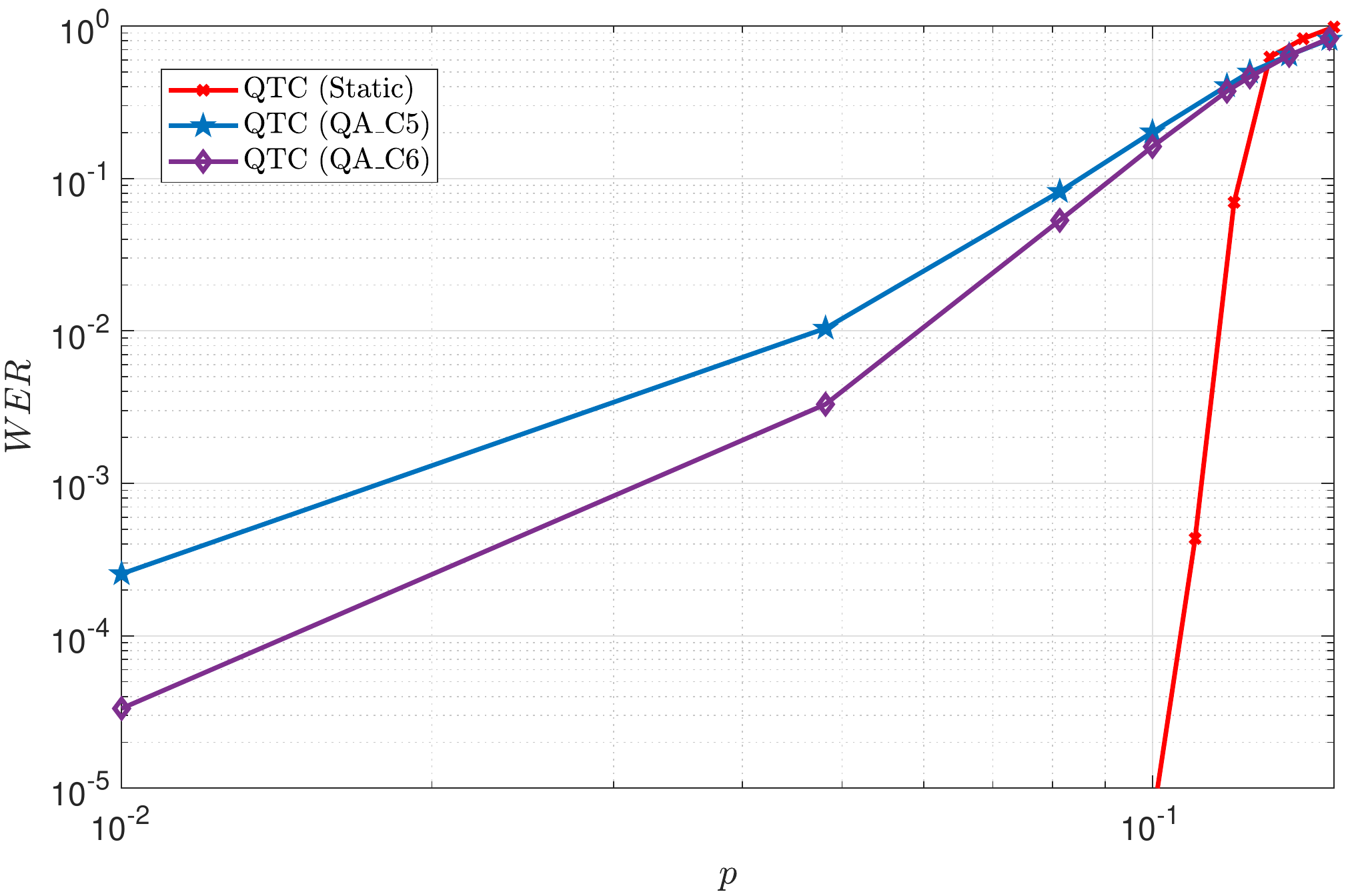} \label{fig:res_CA}
	} 
\caption{Performance of the $d \times d$ Kitaev Toric codes of \cite{toric} with $d = 9$ and the QTC of \cite{josuconf} over a static quantum channel and two different time-varying quantum channels. \textbf{(a)} $9\times 9$ Toric code. \textbf{(b)} QTC.}

    \label{fig:results_josu2}
\end{figure}

In our work \cite{superconducting-josu}, we amalgamate the findings of \cite{decoherenceBenchmarking,fluctAPS,fluctApp,temperature} with the existing models for quantum noise, which culminates in the proposal of Time-Varying Quantum Channels (TVQCs) $\xi(\omega,t)$ for superconducting qubits. The dynamics of these models are defined by the random processes $T_1(\omega,t)$ and $T_2(\omega,t)$ whose behaviour is described by the decoherence governing parameters of existing superconducting quantum computers. Against this backdrop, we use a metric known as the diamond norm distance\footnote{The diamond norm $||\xi_1 - \xi_2||_\diamond$ provides a measure of how different two quantum channels are, where $\xi_1$ and $\xi_2$ denote two quantum channels.} \cite{diamondNat,FanoDiamond} to assess the difference between the commonly employed static channels and our proposed TVQCs. The results shown in \cite{superconducting-josu} prove that neglecting the fluctuating nature of the relaxation and dephasing times in the construction of noise models provides an unrealistic (overly optimistic) portrayal of decoherence effects. This can also be seen in Figure \ref{fig:results_josu2}, included to provide context, which shows how the performance of QEC codes deteriorates when the time fluctuations of the decoherence parameters are accounted for in the quantum channel model (note how the WER is significantly worse over the time varying channels, QA\_C5 and QA\_C6, than over the static channel).

In summary, the primary takeaway from our work in \cite{superconducting-josu} is that TVQC models are necessary to construct and appropriately simulate/predict the performance of QECCs. Additionally, despite the fact that the derived time-varying channel models are based on the statistical experimental characterization of the parameters $T_1$ and $T_2$ of \cite{decoherenceBenchmarking,fluctAPS,fluctApp}, they are also applicable to any superconducting quantum processor whose decoherence parameters exhibit slow fluctuations. In fact, the model is actually also applicable to any quantum-coherent two-level system that presents similar time dependencies, regardless of its physical implementation.

\section{Quantum outage probability for time-varying quantum channels
} \label{sec:TV-2}

TVQCs allow us to account for the experimentally validated time-variance of the relaxation and dephasing times of superconducting qubits and enable us to study and predict the performance of QEC codes with higher precision than with previously existing channel models. However, because the TVQC framework includes the time dependence of $T_1$ and $T_2$, the asymptotical error correction limits associated to time-varying  channels will be different to the limits of those channels that assume time independence of the aforementioned decoherence defining parameters. In \cite{josuarxiv}, we take on the challenge of studying the asymptotical limits of error correction in the TVQC paradigm. To do so, we draw from the well-known classical scenario of slow/block fading and introduce the concepts of the quantum outage probability and the quantum hashing outage probability of a TVQC. 

In the context of classical communications, we know from the Shannon channel coding theorem \cite{Shannon} that it will be possible to achieve reliable communications with a code of rate $R_c$ over an arbitrary channel if the realization of the channel capacity $C(\omega)$ is larger than $R_c$. On the contrary, when for a particular realization of the channel capacity we have $C < R_c$, communication at a low probability of error will not be possible. Thus, we refer to the probability that communications fail when transmitting a codeword with rate $R$ as the outage probability and we define it as 

$$ p^c_\text{out}(R_c) = \{\omega \in \Omega: C(\omega) < R_c\}. $$

We can define the quantum outage probability for a TVQC by following this same reasoning. Knowing that the
qubit relaxation and dephasing times are modelled as random variables\footnote{They are actually modelled as random processes, i.e, $T_1(\omega, t)$ and $T_2(\omega, t)$, but because the coherence time of a qubit is generally longer than the processing time of a quantum algorithm, realizations of these random processes can be considered to remain constant during each processing round and can thus be modelled as random variables $T_1(\omega)$ and $T_2(\omega)$ \cite{superconducting-josu}.}, $T_1(\omega)$ and $T_2(\omega)$, realizations of the relaxation and dephasing times will result in a realization of the TVQC in question, and consequently, in a specific value for the quantum channel capacity $C_Q$. Reminiscent of the classical scenario, if the realization of the decoherence parameters leads to a channel capacity that is lower than the quantum coding rate, $R_Q$, then the error probability (the QBER or the WER) will not vanish asymptotically with the block length, independently of the selected QEC code. For such realizations, the channel can be said to be in outage. Therefore, we can write

$$ p^Q_\text{out}(R_c) = \{\omega \in \Omega: C_Q(\omega) < R_Q\}. $$

The above expression tells us that, with probability $p^Q_\text{out}(R_c)$, the capacity of the channel will be lower than the rate of the quantum code and so reliable communication will not be possible. Conversely, with probability $1-p^Q_\text{out}(R_c)$ reliable communication over the TVQC will be possible. Thus, the quantum outage probability of a TVQC is defined as the asymptotically achievable error rate of a QEC code with quantum rate $R_Q$ that operates over said TVQC. 

\begin{figure}[!h]
\centering
\hbox{\includegraphics[width = \linewidth,height=3.75in]{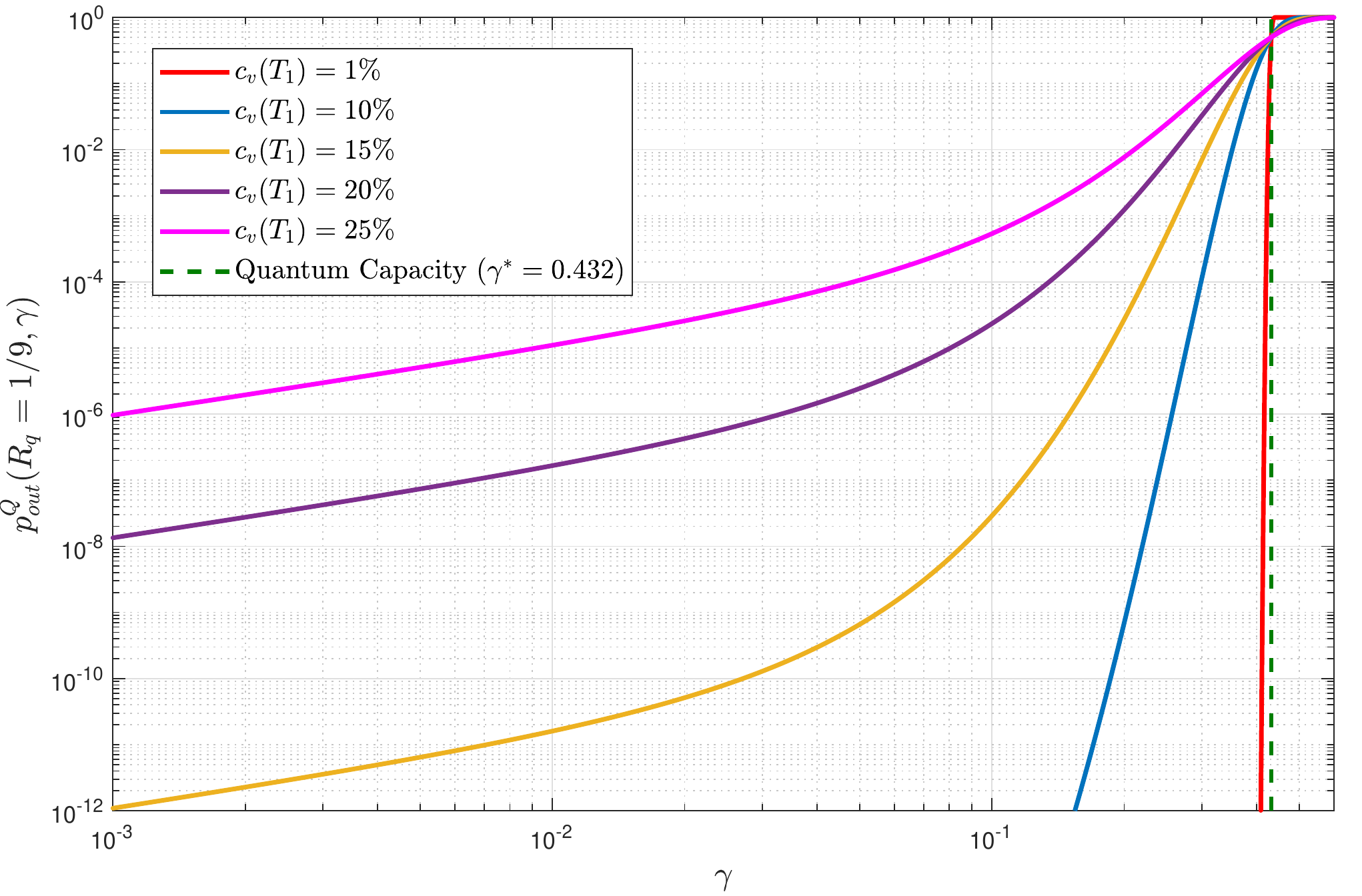}}
\caption{Quantum outage probability of the TVAD channel. TVADs with $c_v(T1) = \{1, 10, 15, 20, 25\}$ assuming a rate $R_Q = 1/9$ have been considered for the simulations.}

    \label{fig:last_fig}
\end{figure}

In \cite{Shannon} these ideas are extended and, based on the results of \cite{superconducting-josu, decoherenceBenchmarking}, closed-form expressions for known TVQCs, such as the time-varying amplitude damping channel (TVAD), the time-varying amplitude damping Pauli twirl approximated channel (TVADPTA), and the time-varying amplitude damping Clifford twirl approximated channel (TVADCTA), are provided. This work also thoroughly analyzes the quantum outage probability and quantum hashing outage probabilities of the aforementioned TVQCs in different scenarios. The work conducted in \cite{Shannon} concludes by simulating the performance of QTCs over the considered channels and benchmarking them against the derived information theoretic limits. An example of this can be seen in Figure \ref{fig:last_fig}, which plots the quantum outage probability against the damping parameter $\gamma$ of the TVAD channel\footnote{In the AD channel the damping parameter defines the noise dynamics of the channel similarly to how the depolarizing probability determines the noise level of the depolarizing channel.}. This figure clearly shows how the probability of the channel being in outage increases with the variability\footnote{By variability we refer to how much realizations of $T_1$ can differ compared to its mean. This is given by the coefficient of variation $c_v = \frac{\sigma}{\mu}$, where $\mu$ and $\sigma$, are the mean and variance of the random variable in question.} of $T_1(\omega)$.

\clearemptydoublepage
\chapter{Conclusion and Future Work} 

\epigraph{\textit{``Science never solves a problem without creating ten more''}}{\textbf{George Bernard Shaw}.}

\noindent\hrulefill

This thesis set out with the objective to ponder and analyze the following two aspects of sparse QEC codes:

\begin{itemize}
    \item how to improve the performance of CSS and non-CSS QLDPC codes in various different scenarios.
    \item how to characterize the manifestation of the ``quantum-exclusive'' phenomenon known as degeneracy and its impact on the performance of sparse quantum codes.
\end{itemize}
 
With these goals in mind, the dissertation began with an introduction to the realms of quantum computing, quantum information, and classical error correction. This preliminary part of the thesis commenced in Chapter $2$, where we showed how quantum computing can be used to address problems that cannot be solved efficiently on classical machines and we presented various different technologies that are currently being employed to build quantum processors. Then, in Chapter $3$, we presented background concepts on quantum information and classical error correction, such as quantum channels and the factor graph representation of classical codes, that are critical to follow the rest of our discourse. 

Following this, we entered the part of the thesis that focuses on understanding degeneracy and its relationship to sparse quantum codes. In Chapter $4$ we presented a group theoretic interpretation of stabilizer-based error correction that allowed us to accurately define degeneracy and classify its effects on the decoding algorithm applied to sparse quantum codes. This chapter also included a numerical example that helped to further portray what degeneracy is and how it is involved in the decoding process of stabilizer codes. Based on this new perspective, in Chapter $5$ we studied the ways in which the performance of sparse quantum codes has been evaluated in the literature, and we showed how this can sometimes lead to an overestimation of the error rate. Furthermore, we discussed previously existing methods to compute the logical error rate of sparse quantum codes and we also proposed our own classically-inspired strategy for this same purpose. We closed this chapter by discussing the most relevant improved decoding strategies for sparse quantum codes that have appeared in the literature.

After Chapter $5$, we entered the second major part of the thesis, which concerned itself with the analysis and optimization of the performance of LDGM-based quantum codes in a variety of different scenarios. We started with the derivation of a non-CSS construction in Chapter 6, where we showed how it is possible to improve the performance of known CSS LDGM-based codes by rearranging the structure of their corresponding factor graphs. We also showed how the proposed non-CSS codes outperformed other existing codes of comparable complexity. We followed this up by studying the impact that the channel mismatch phenomenon has on this family of non-CSS LDGM-based codes in Chapter $7$. This analysis showed that having no prior knowledge of the channel led to a worsening of the performance of our codes. To face this, we derived an online estimation strategy similar to one that was previously used for QTCs. We finished this chapter by studying the performance of CSS LDGM-based codes in the context of asymmetric quantum channels, which have been shown to represent specific quantum devices with more accuracy than the depolarizing channel. In the final chapter of the dissertation, Chapter $8$, we provided a brief overview of other QEC-related works that the author has been involved in during his time as a PhD student. 

All in all, in this thesis we have seen how degeneracy is intricately related to the performance of quantum codes and we have studied and improved the performance of a specific family of sparse quantum codes in different communication scenarios. At the same time, we have also seen how there is ample room for growth, progress, and (obviously) work, in both of these areas. Consequently, we close this chapter (and the dissertation itself) by entertaining some of the questions and potential research topics that our work herein has left unanswered.

\textbf{$\bullet$ Design of degenerate sparse quantum codes.}

Having shown that degenerate errors do not actually alter the codespace of a stabilizer code and that they do not result in decoding mistakes, it is clear that degeneracy can have a positive impact on the performance of sparse quantum codes. Thus, it is likely that constructing quantum codes with a maximized probability of suffering degenerate errors will lead to improvements in performance. An added benefit of such a strategy is the fact that the performance gain would come at no cost to the decoding complexity, which cannot be said for other performance improvement methods. In the literature, most of the research related to the exploitation of degeneracy has focused on the decoder. The best example of this can be found in \cite{exploiting} were the performance of particular sparse quantum codes is improved by building a decoder that can take advantage of degeneracy. However, this decoder-focused approach differs significantly from our suggestion of specifically designing codes to be degenerate. For this reason, it may be difficult to find inspiration in the literature and define a starting point for our idea. Based on intuition, one could begin by studying the coset structure of a short stabilizer code using the degenerate error detecting methods discussed in Chapter $5$ and then attempting to re-organize it so that the lowest weight operators all fall within the coset with highest probability. If successful, we could then try to generalize this approach to larger codes.

\textbf{$\bullet$ Improvement and design of modified decoding strategies.}

The idea of improved decoding strategies for sparse quantum codes has been mentioned numerous times throughout this thesis. In Chapter $5$ we saw how these methods can be used to reduce the number of end-to-end errors with different syndromes, albeit at the cost of an increase in the decoding complexity. However, some of these schemes can only be applied to CSS quantum codes. Therefore, it is likely that the adaptation and application of these strategies to the non-CSS codes derived in Chapter $6$ will lead to performance improvements. Furthermore, because the construction of a ``degenerate'' decoder for sparse quantum codes remains an open problem in QEC, another potential line of work would involve seeking ways in which to modify the SPA decoding algorithm of sparse quantum codes to account for the degeneracy phenomenon.

\textbf{$\bullet$ Construction \& analysis of QEC codes on real quantum computers.}

A topic that we have rarely mentioned in this dissertation is the construction of practical error correction schemes on existing quantum computers. This is mostly due to the fact that sparse quantum codes generally require large block lengths (over $1000$ qubits) to provide noticeable improvements in performance; a prohibitive requirement, since currently existing quantum computers barely exceed a count of $100$ physical qubits. In other words, practical QEC strategies for the present must function for significantly smaller block lengths than those we have considered here. This opens up various different research options that remain somewhat unexplored in the literature. One possibility would be to study and optimize shorter block length sparse code constructions and then seek to derive stabilizer encoding circuits from their corresponding QPCMs. Despite the suboptimal performance of shorter block length sparse quantum codes, deriving a valid method to obtain the encoding circuit from the corresponding QPCM would bridge an important gap between theoretical and practical QEC. QEC codes are physically implemented using quantum circuits and a strategy capable of deriving them from QPCMs represents an important advancement for both present and future QEC methods. Following this, and conditioned by the number of available qubits, these encoding circuits could be simulated on cloud-accessible quantum computers such as IBM-Q. Another possibility would be to look at these same problems from the perspective of different quantum code families, like surface codes, instead of sparse quantum codes.

\textbf{$\bullet$ Sparse quantum codes over TVQCs.}

In Chapter $8$ we saw that TVQCs provide a more accurate portrayal of the decoherence-induced noise experienced by superconducting qubits. Based on what we have seen in this dissertation, it would be interesting to benchmark the performance of CSS and non-CSS QLDGM codes over these channels and to study them based on the view provided by the channel outage probability.

\clearemptydoublepage

\def\chaptername{APPENDIX}
\appendix
\begin{appendices}

	\chapter{Syndrome-based Decoding of LDPC codes} \label{app:spa}

Sum-product based decoding of an LDPC code can be easily understood by describing how the procedure unfolds over the factor graph associated to the PCM of the LDPC code under consideration. Assume that the receiver has obtained the syndrome $\mathbf{s} = \mathbf{H}_c\mathbf{x}^\top$, where $\mathbf{H}_c$ is the (sparse) PCM of the LDPC code and $\mathbf{x}$ is the channel output. Aside from the syndrome, an SPA-based decoder requires information regarding the communication channel. It is commonplace in many classical and quantum coding scenarios to assume that the receiver knows the probability distribution of the channel, $\mathrm{P}(\mathbf{e})$. In reality, practical communication systems must somehow perform channel estimation to estimate the noise level of the channel. This has been studied in both the classical domain \cite{class-mismatch1, class-mismatch2, class-mismatch3} and in the quantum paradigm \cite{josu2, josuconf, first-impact, current, patrick2}, and we also discuss it in depth in Chapter \ref{chapter7}. For the sake of simplicity, in what follows we will simply assume that the probability distribution of the channel is known to the receiver. Under this premise, we can now address how the sum product algorithm is executed over the corresponding factor graph. As mentioned in Chapter \ref{chapter3}, the SPA attempts to estimate the symbol-wise most likely channel error by computing the solution to
\begin{align} \label{eq:bwMLapp}
\begin{split} 
\hat{\mathrm{e}}_{j}^{\text{SW}} = \argmax_{\mathrm{e}} P(\mathrm{e}_j = \mathrm{e}|\mathbf{z}) \\ = \argmax_{\mathrm{e}_j \in \{0,1\}} \sum_{\mathrm{e}_1,\ldots,\mathrm{e}_{j-1},\mathrm{e}_{j+1},\ldots,\mathrm{e}_N} P(\mathrm{e}_1, \ldots, \mathrm{e}_N|\mathbf{z}).
\end{split}
\end{align} To do so, the algorithm operates by exchanging messages between the variable nodes and the parity check nodes of the factor graph associated to the PCM of the LDPC code in question. Assuming that the graph has no cycles, the solution provided by the SPA agrees with the solution of (\ref{eq:bwMLapp}).

In order to simplify the description of the SPA, assume that the selected LDPC code is binary and that communications take place over a Binary Symmetric Channel (BSC) with crossover probability\footnote{A bit transmitted over a BSC with crossover probability $q$ will be flipped with probability $q$ and will remain unchanged with probability $1-q$.} $q$. In this context, the sum product algorithm can be summarized into the following set of sequential steps: 

\begin{enumerate}
    \item \textbf{Initialization:} In order to begin, the algorithm must be initialized with channel information. This is done by computing the message pair ($m_{{\mathrm{e}_i}\rightarrow{\mathrm{c}_j}}^0, m_{{\mathrm{e}_i}\rightarrow{\mathrm{c}_j}}^1$) for each variable node on the factor graph, where $\mathrm{e}_i$ denotes the $i$-th variable node\footnote{Since each variable node is associated to a component of the error vector, we extend our notation (regular lower case romans) to denote the variable nodes.}, $\mathrm{c}_j$ denotes the $j$-th variable node, and the subscript $\mathrm{e}_i\rightarrow \mathrm{c}_j$ denotes the message transmitted from variable node $\mathrm{e}_i$ to check node $\mathrm{c}_j$. In this notation, the superscripts represent the value of the $i$-th component of the error sequence, i.e., a message $m_{{\mathrm{e}_i}\rightarrow{\mathrm{c}_j}}^x$ can be understood as the 
   ``belief'' that variable node $\mathrm{e}_i$ sends to parity check node $\mathrm{c}_j$ regarding the certainty it has of being equal to $\mathrm{x} = \{0,1\}$. During this initialization phase the receiver only knows the distribution probability of the channel, and so these beliefs are simply equal to the BSC flip probabilities:
   
   \begin{align} \label{eq:beliefs}
\begin{split} 
m_{{\mathrm{e}_i}\rightarrow{\mathrm{c}_j}}^0 &= P_\text{BSC}(\mathrm{e}_i=0) = 1 - q, \\m_{{\mathrm{e}_i}\rightarrow{\mathrm{c}_j}}^1 &= P_\text{BSC}(\mathrm{e}_i=1) = q.
\end{split}
\end{align}
   
   As code block lengths increase, the computation of these message pairs for each node of the factor graph can become increasingly cumbersome. For this reason, it is common to use Log-Likelihood Ratios (\textit{llrs}). By defining channel \textit{llr}s during this initialization process, each variable node need only compute and send a single message to the check nodes it is connected to. As a result, the algorithm computes and provides each variable node $\mathrm{e}_i$ with its corresponding channel \textit{llr} $$l_\text{ch}(\mathrm{e}_i) = \log\bigg(\frac{m_{{\mathrm{e}_i}\rightarrow{\mathrm{c}_j}}^0}{m_{{\mathrm{e}_i}\rightarrow{\mathrm{c}_j}}^1}\bigg) = \log\bigg(\frac{1-q}{q}\bigg).$$ Subsequently, each variable node
   transmits the message $\mu^1_{{\mathrm{e}_i}\rightarrow{\mathrm{c}_j}} = l_\text{ch}(\mathrm{e}_i)$, to all of the parity check nodes it is attached to. The superscript now denotes that this exchange occurs in the first algorithm iteration, which is logical given that it takes place during the initialization step. This process is shown in Figure \ref{fig:spainit}.
   
\begin{figure}[!ht]
\centering
\includegraphics[width=0.99\columnwidth,  height=2.25in]{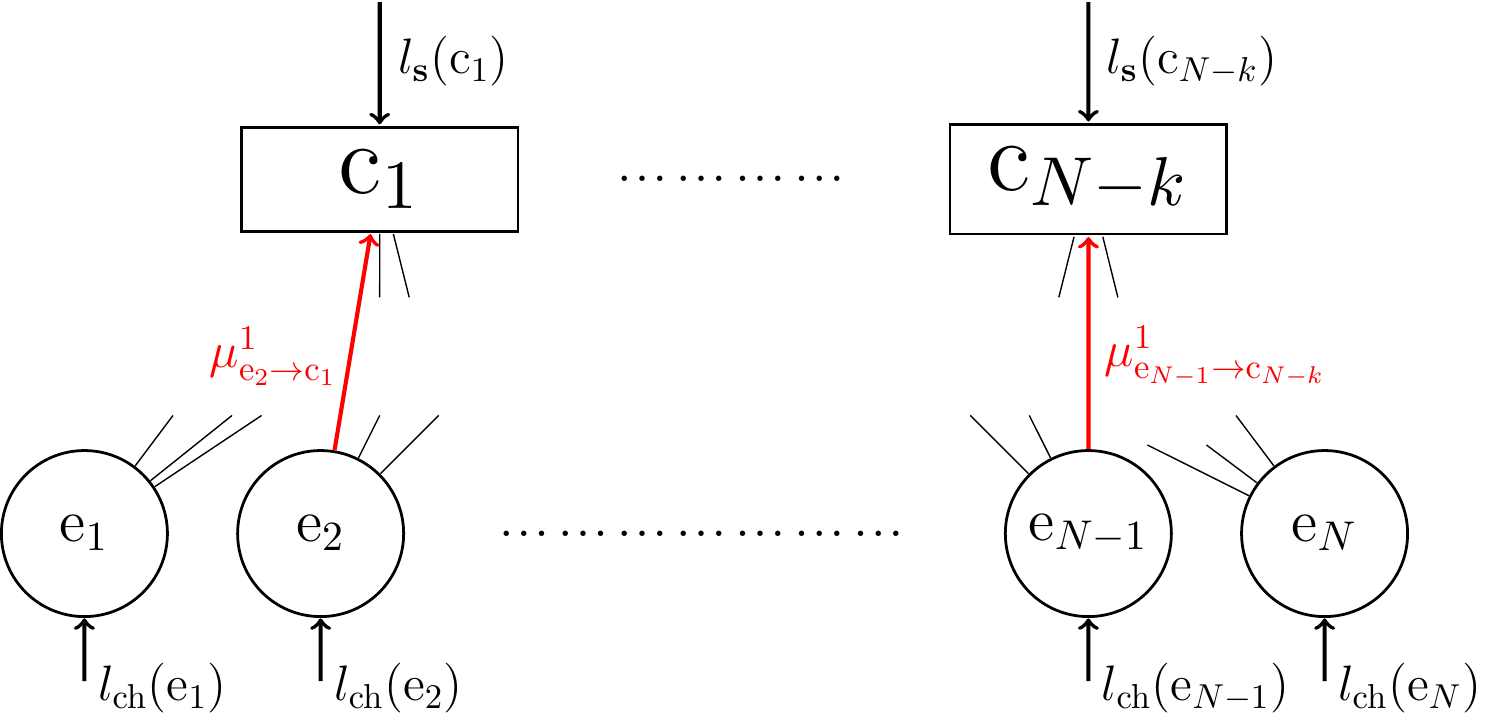}
    \caption{Initialization of the sum-product algorithm and transmission of the first messages from variable nodes to parity check nodes.}
    \label{fig:spainit}
\end{figure}

\item \textbf{Parity check node to variable node messaging:} Aside from the channel \textit{llr}s that they receive from variable nodes, each parity check node $\mathrm{c}_i$ is also associated to a component $\mathrm{s}_i$ of the obtained syndrome $\mathbf{s} = [\mathrm{s}_1\ldots \mathrm{s}_{N-k}]^\top  \in \mathbb{F}_2^{N-k}$.  The procedure is similar to the one employed for the channel messages: the binary value $\mathrm{s}_i \in [0,1]$ is transformed into an \textit{llr} as $$l_\mathbf{s}(\mathrm{c}_i) = \log\bigg(\frac{P(\mathrm{s}_i = 0)}{P(\mathrm{s}_i = 1)}\bigg)$$ where $l_\mathbf{s}(\mathrm{c}_i) = -\infty$ if $\mathrm{s}_i = 1$ or $l_\mathbf{s}(\mathrm{c}_i) = \infty$ if $\mathrm{s}_i = 0$. The association of syndrome \textit{llr}s to parity check nodes is also shown in Figure \ref{fig:spainit}. In practical implementations of the algorithm, these syndrome likelihoods are chosen as finite approximations in order to guarantee numerical stability: $l_\mathbf{s}(\mathrm{c}_i) \approx -\infty$ and $l_\mathbf{s}(\mathrm{c}_i) \approx \infty$. Parity check nodes combine this syndrome knowledge with the information they receive from variable nodes and compute new messages which they then relay back to the variable nodes. These messages are computed according to the SPA tanh rule \cite{spa, BP, tanh}, and consider all incoming variable node messages except those received over the edge they are to be transmitted along, as required by the SPA message update rule \cite{spa, BP}. Essentially, this comes down to computing 

\begin{equation} \label{eq:check} \mu_{{\mathrm{c}_i}\rightarrow{\mathrm{e}_j}}
^1= 2\text{atanh}\prod_{k=1}
^{\Xi-1}\tanh\bigg(\frac{\mu^1_{{\mathrm{e}_k}\rightarrow{\mathrm{c}_i}}}{2}\bigg)\tanh\bigg(\frac {l_\mathbf{s}(\mathrm{c}_i)}{2}\bigg), \end{equation}

where $\Xi$ represents the number of edges of each parity check node. The product considers up to $\Xi - 1$
messages since the message $\mu^1_{{\mathrm{e}_j}\rightarrow{\mathrm{c}_i}}$ received on the edge $\mu_{{\mathrm{c}_i}\rightarrow{\mathrm{e}_j}}
^1$ will be transmitted over, is not considered. Once more, the superscript $1$ denotes that these messages are exchanged in the first algorithm iteration or decoding round. This procedure is portrayed over a factor graph in Figure \ref{fig:step2}.

       \begin{figure}[!ht]
\centering
\includegraphics[width=0.99\columnwidth,  height=1.75in]{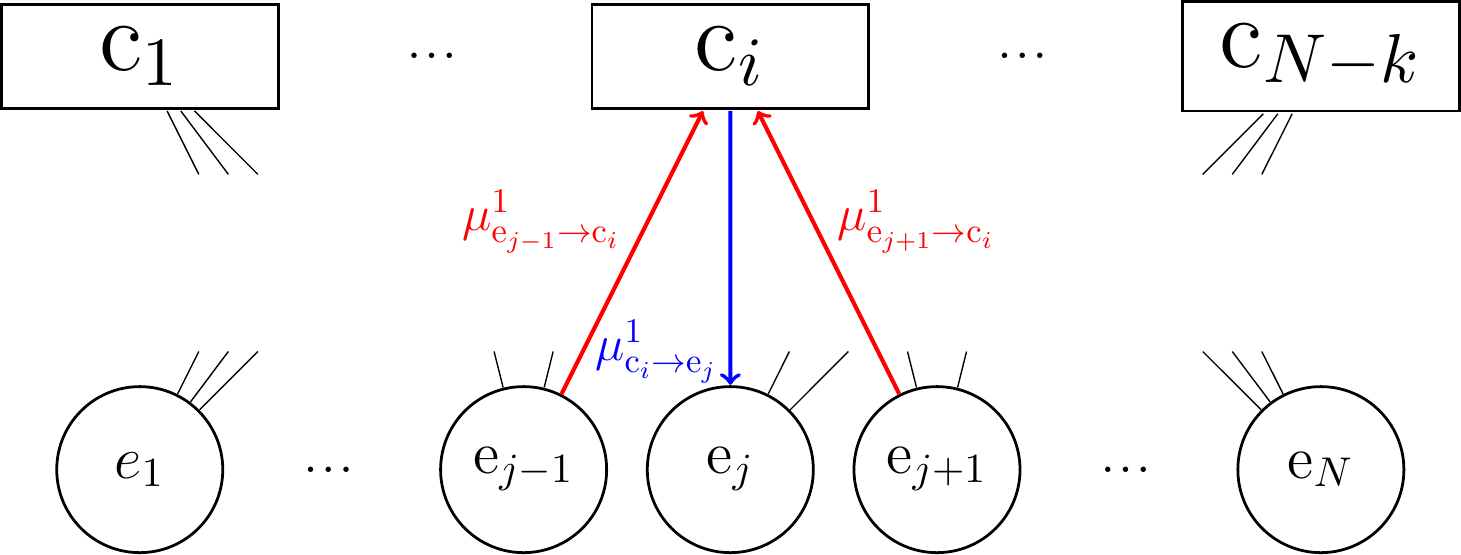}
    \caption{Message exchange between parity check nodes and variable nodes during the first decoding iteration.}
    \label{fig:step2}
\end{figure}
 
 \item \textbf{Computation of the marginal probabilities:} This stage of the decoding process is reached once the message pair $(\mu^1_{{\mathrm{e}_j}\rightarrow{\mathrm{c}_i}}, \mu_{{\mathrm{c}_i}\rightarrow{\mathrm{e}_j}}
^1)$ has been exchanged over every edge of the factor graph, where $i=1,\ldots,N-k$ and $j=1,\ldots,N$. The exchange of these two messages over the factor graph edges also symbolizes the conclusion of the first sum product decoding iteration. At this point, the algorithm must determine whether it has estimated the correct error sequence or if subsequent decoding iterations will be necessary. In order to produce an estimate of the error sequence, the decoder needs to obtain the marginal probabilities $P(\mathrm{e}_j|\mathbf{s})$, where we have assumed $\mathrm{e}_j \in \mathbb{F}_2$ and $j=1,\ldots,N$. These marginal probabilities are calculated by each variable node $\mathrm{e}_j$, in  \textit{llr} form and at the end of each decoding iteration\footnote{Subsequent algorithm steps operate identically for every decoding iteration, hence it is logical to describe them for an arbitrary iteration.} $t$, via the following computation:

$$ l^t_\text{ap}(\mathrm{e}_j) = l_\text{ch}(\mathrm{e}_j) + \sum_{k=1}^\sigma \mu_{{\mathrm{c}_k}\rightarrow{\mathrm{e}_j}}^t,$$ where $\sigma$ denotes the degree of the variable nodes, i.e., the summation considers all incoming check node messages to each variable node. 

\item \textbf{Stop/Continue criterion:} Once every variable node has obtained its corresponding \textit{a posteriori} log-likelihood ratio $l^t_\text{ap}(\mathrm{e}_j)$, the symbol-wise most likely error sequence $\hat{\mathbf{e}}^{\text{SW},t}$ can be obtained. Recall that $\hat{\mathbf{e}}^{\text{SW},t} = [\hat{\mathrm{e}}^{\text{SW},t}_1\ldots\hat{\mathrm{e}}^{\text{SW},t}_N]$, where $\hat{\mathrm{e}}^\text{BW}_j$ is given by (\ref{eq:bwMLapp}), ergo $\hat{\mathrm{e}}^{\text{SW},t}_j$ is derived by  making a hard decision\footnote{This is as simple as estimating $\hat{\mathrm{e}}^\text{SW}_j = 0$ if $l^t_\text{ap}(\mathrm{e}_j) \geq 0$ and $\hat{\mathrm{e}}^\text{SW}_j = 1$ otherwise.} on $l^t_\text{ap}(\mathrm{e}_j)$. Once $\hat{\mathbf{e}}^{\text{SW},t}$ has been calculated, the decoder computes the product $\hat{\mathbf{s}}_t = (\hat{\mathbf{e}}^{\text{SW},t})\mathbf{H}_c^\top$, which yields the syndrome associated to the most likely error sequence. $\hat{\mathbf{s}}_t$ is critical in dictating what is left of the decoding process: if $\hat{\mathbf{s}}_t = \mathbf{s}$, the correct solution has been found and the procedure halts. However, if $\hat{\mathbf{s}}_t \neq \mathbf{s}$ another decoding round will be executed. This procedure is repeated until $\hat{\mathbf{s}}_t = \mathbf{s}$ or a maximum number of decoding iterations $T$ is reached.

\item \textbf{Node messaging in posterior iterations:} In order to complete this description of SPA decoding, the manner in which messages are exchanged over the factor graph after the first decoding iteration ($t=1$) must be explained. Given that the computation of marginal probabilities and the stop/continue criterion have been discussed in terms of an arbitrary decoding iteration $t$, the same must be done for the messaging between check nodes and variable nodes, and vice-versa. If we assume that in the previous decoding round $t-1$ an incorrect solution was obtained, then during the subsequent iteration $t$, each variable node $\mathrm{e}_j$ will compute and transmit the following message over each of its edges:

$$ \mu_{{\mathrm{e}_j}\rightarrow{\mathrm{c}_i}}
^t= l_\text{ch}(\mathrm{e}_i) + \sum_{k=1}^{\sigma-1} \mu_{{\mathrm{c}_k}\rightarrow{\mathrm{e}_j}}^{t-1},$$

where $\sigma$ denotes the degree of the variable nodes, and so messages are computed according to the sum-product update rule: disregarding the message received in the previous iteration through the edge over which transmission of $\mu_{{\mathrm{e}_j}\rightarrow{\mathrm{c}_i}}
^t$ is to occur. This is shown in Figure \ref{fig:step3}.

    \begin{figure}[!ht]
\centering
\includegraphics[width=0.99\columnwidth,  height=1.5in]{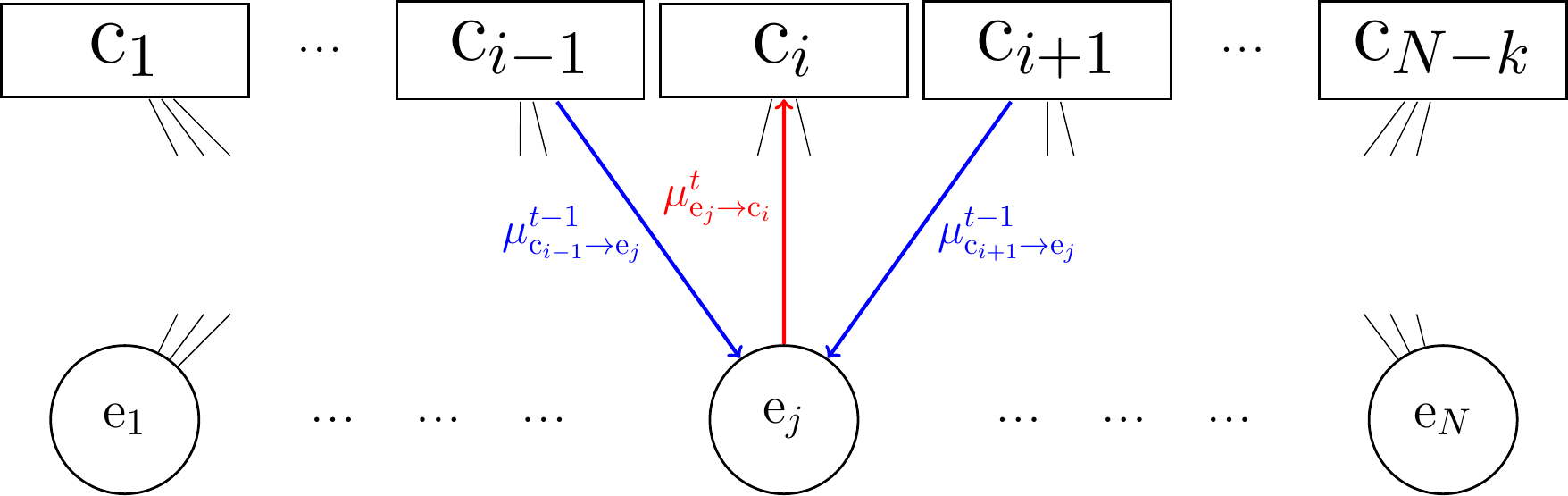}
    \caption{Message exchange between variable nodes and parity check nodes during decoding iteration $t$.}
    \label{fig:step3}
\end{figure}

 In terms of messages transmitted from parity check nodes to variable nodes during iteration $t$, the messages are computed according to the expression shown in (\ref{eq:check}). This exchange of messages is shown in Figure \ref{fig:step2}. Notice that said figure and expression (\ref{eq:check}) are a representation of the first decoding iteration: $t=1$. Hence for subsequent iterations messages should actually be denoted by the superscript $t$.

\end{enumerate}

\chapter{Improved Decoding Strategies for QLDPC codes} \label{app:modified}

In what follows we provide a detailed summary of the most relevant modified/improved QLDPC decoding strategies that have appeared in the literature.

\section{Correlation Exploiting Decoder}

The correlation exploiting decoder of \cite{jgf2} implements the SPA modifications originally suggested in \cite{bicycle} and represents one of the earliest attempts of an improved decoding strategy for CSS QLDPC codes. Although they are discussed in Chapter $6$, in case readers have come directly from Chapter $5$ we will also present CSS codes here. CSS codes are a subclass of stabilizer codes that are built from existing classical codes in such a manner that the fulfilment of the symplectic criterion is guaranteed. The QPCM of a generic CSS code is given by 

\begin{equation} \label{CSS}
\mathbf{H}_Q = (\mathrm{H}_x|\mathrm{H}_z) =
\begin{pmatrix} \mathrm{H}_x^{'} &0 \\
    0 &\mathrm{H}_z^{'} \end{pmatrix} , 
\end{equation}

where $\mathrm{H}_x = \begin{pmatrix} \mathrm{H}_x^{'} \\ 0 \end{pmatrix}$ and $\mathrm{H}_z = \begin{pmatrix} 0 \\ \mathrm{H}_z^{'} \end{pmatrix}$. In this construction, $\mathrm{H}_x'$ and $\mathrm{H}_z'$ are the parity check matrices of two classical LDPC codes $C_1$ and $C_2$, respectively, where each matrix is used to correct either bit-flips or phase-flips. The classical codes are chosen so that $C_2^\perp \subseteq C_1$, where $C_2^\perp$ is the dual of the classical LDPC code $C_2$. This design constraint,  generally referred to as the \textit{CSS condition}, reduces the expression of the symplectic criterion shown in \eref{eq:symplec} to $\mathrm{H}_z^{'}\mathrm{H}_x^{'T}=0$. Due to the particular structure of a CSS QPCM (\ref{CSS}), CSS codes can and are almost always decoded by means of two separate binary BP decoders. This is made possible because $\mathbf{w} = \mathbf{H}_Q \odot \mathbf{e} = \mathrm{H}_z\mathbf{e}_x^\top \oplus \mathrm{H}_x\mathbf{e}_z^\top = [\mathrm{H}'_z\mathbf{e}_x^\top \  \mathrm{H}'_x\mathbf{e}_z^\top]= [\mathbf{w}_x \ \mathbf{w}_z]$, where $\mathbf{H}_C$ is given in (\ref{CSS}), $\mathbf{w}_x = \mathrm{H}'_z\mathbf{e}_x^\top$ and $\mathbf{w}_z = \mathrm{H}'_x\mathbf{e}_z^\top$, respectively. Thus, the syndrome that is obtained when using a CSS code is of the form $\mathbf{w}=[\mathbf{w}_x \ \mathbf{w}_z]$, i.e, half of it contains information strictly about bit flips and the other half only provides knowledge regarding phase flips. In this way, two separate decoders can be used simultaneously to produce the error estimates $\hat{\mathbf{e}}_x$ and $\hat{\mathbf{e}}_z$, and so obtain $\hat{\mathbf{w}}_x = \mathrm{H}'_z\hat{\mathbf{e}}_x^\top$ and $\hat{\mathbf{w}}_z = \mathrm{H}'_x\hat{\mathbf{e}}_z^\top$, where these are estimates of $\mathbf{w}_x$ and $\mathbf{w}_z$, respectively. 
 
 Despite being convenient, a decoding stratagem based on two independent binary BP decoders will ignore the correlations that exist between $X$ and $Z$ errors over the depolarizing channel due to its separate processing of bit and phase flips. The correlation exploiting decoder circumvents this issue by iteratively exchanging messages between the $X$ and $Z$ operator decoders. After each iteration, the a priori probability of each $X$ or $Z$ variable node\footnote{In this case, variable nodes represent qubits.} $q$ is updated based on the decoding outcome of the opposite decoder, i.e, the variable nodes associated to the $X$ operators, which we will denote as $\mathbf{e}_x$, are updated based on the results obtained by the $Z$ operator decoder and viceversa. More explicitly, during each decoding iteration $t$, the $X$ decoder transmits the message $p^t(\mathbf{E}_{z,q} = 1)$ to the $Z$ decoder and the $Z$ decoder relays the message $p^t(\mathbf{E}_{x,q} = 1)$ to the $X$ decoder. These messages are given by 
 
 \begin{equation}\label{correl}
\begin{aligned}
&p^t(\mathbf{E}_{z,q} = 1) = p^t(\mathbf{E}_{x,q} = 1)p(\mathbf{E}_{z,q}=1|\mathbf{E}_{x,q}=1) \\
&+ (1-p^t(\mathbf{E}_{x,q} = 1))p_q(\mathbf{E}_{z,q}=1|\mathbf{E}_{x,q}=0), \\
&p^t(\mathbf{E}_{x,q} = 1) = p_q^t(\mathbf{E}_{z,q} = 1)p(\mathbf{E}_{x,q}=1|\mathbf{E}_{z,q}=1)\\
&+ (1-p^t(\mathbf{E}_{z,q} = 1))p(\mathbf{E}_{x,q}=1|\mathbf{E}_{z,q}=0),
\end{aligned}
\end{equation}
 
where $p^t(\mathbf{E}_{x,q} = 1)$ and $p^t(\mathbf{E}_{z,q} = 1)$ denote the probability of an $X$ or $Z$ error taking place on the $q$-th variable node of each corresponding decoder during that same iteration $t$, and the conditional probabilities are obtained from the joint distribution of $X$ and $Z$ operators over the depolarizing channel shown in table \ref{tab:joint}. The messages $p^t(\mathbf{E}_{z,q} = 1)$ and $p^t(\mathbf{E}_{x,q} = 1)$ are used as the a priori probabilities for the $q$-th $Z$ and $X$ variable nodes in the $(t+1)$-th decoding round. In \cite{jgf2}, this correlation exploiting decoder is shown to yield small performance improvements, foreshadowing the potential that degeneracy-exploiting decoding strategies have to improve the performance of QLDPC codes. 

\begin{table}
\centering
    \caption{Joint Distribution of $X$ and $Z$ operators over the depolarizing channel. $p$ is the depolarizing probability of the channel.}
    \label{tab:joint}
    \setlength{\tabcolsep}{6pt}
    \begin{tabular}{|p{45pt}|p{45pt}|p{45pt}|}
        \hline
\multicolumn{1}{|c|}{$P(\mathbf{E}_x, \mathbf{E}_z)$}
                    &  $\mathbf{E}_x = 0$ & $\mathbf{E}_x = 1$\\[0.1cm]
        \hline
$\mathbf{E}_z = 0$  & $1-p$ & $\frac{p}{3}$\\[0.1cm]
 \hline
$\mathbf{E}_z = 1$  & $\frac{p}{3}$ & $\frac{p}{3}$\\[0.1cm]
        \hline
    \end{tabular}
\end{table}

\section{Freezing Decoder}

The ``freezing'' technique is the first and most simple method proposed in \cite{degen3}. The procedure works by selecting a random frustrated check node from the original factor graph and tampering with the prior probabilities of one of the variable nodes connected to it. Frustrated check nodes are those whose associated real syndrome component $\mathrm{w}_c$ does not match with the same estimated syndrome component $\hat{\mathrm{w}}_c$, where $\mathbf{w} = [\mathrm{w}_1,\ldots,\mathrm{w}_{N-k}]$, $\hat{\mathbf{w}}$ is obtained from the symplectic product $\mathbf{H}_Q \odot \hat{\mathbf{e}}$, $\mathbf{H}_Q$ is the QPCM of the code, and $\hat{\mathbf{e}}$ denotes the symplectic representation of the estimate of the error pattern produced by a standard BP decoder. 

Once a frustrated check node has been chosen, the prior probability $p(\mathbf{E}_q)$ or channel llr $l_\text{ch}(\mathbf{E}_q)$ of one of the variable nodes $\mathbf{e}_q$ connected to the aforementioned check node is 
``frozen'' to a value $\delta_q$. The decoder is re-run based on this new prior probability for a fixed number of iterations. If the correct syndrome estimate is found during this second decoding phase then the technique has been successful. If not, then another of the variable nodes connected to the frustrated check node is ``frozen'', the previous check node is restored and decoding is reattempted. If one of these combinations results in a non-frustrated check node but not the appropriate syndrome, another frustrated check node is selected, a variable node connected to it is frozen at random, and the decoding process is re-run. Given the myriads of frozen and non-frozen node combinations that are possible with this technique, if the correct syndrome estimate is not found in the first few decoding reattempts, the increase in decoding complexity can become quite substantial.

\section{Random Perturbation Decoder}

This technique is similar to the ``freezing" decoder in that it also works with the prior probabilities of qubits related to frustrated check nodes. It consists in identifying all of the frustrated check nodes of the factor graph and applying the following perturbations (up to normalization) to the prior probabilities of each variable node $q$ connected to these check nodes:

\begin{equation}\label{random-pert}
\begin{aligned}
&p(\mathbf{E}_q = \mathrm{I}) \rightarrow p(\mathbf{E}_q = I) \\
&p(\mathbf{E}_q = \mathrm{X}) \rightarrow (1+\delta_x)p(\mathbf{E}_q = \mathrm{X}) \\
&p(\mathbf{E}_q = \mathrm{Y}) \rightarrow (1+\delta_y)p(\mathbf{E}_q = \mathrm{Y})\\
&p(\mathbf{E}_q = I) \rightarrow (1+\delta_z)p(\mathbf{E}_q = \mathrm{Z}),
\end{aligned}
\end{equation}

where $\delta_x, \delta_y$, and $\delta_z$ are random variables uniformly distributed over the range $[0,\delta]$ for a fixed $\delta$. Notice that this technique is strictly designed for a $\text{GF(4)}$ decoder. A similar strategy for binary decoders, known as the adjusted decoder, is shown later on. The primary goal of this method is to resolve symmetric degeneracy errors (see section \ref{sec:end-to-end}) by perturbing the prior probabilities used by the qubit-wise BP decoder. Symmetric degeneracy errors occur because degenerate errors of equal weight are completely symmetric under qubit-wise BP decoding, and so the only way to resolve them is by breaking this symmetry. This requires the perturbations to be random, since an equal increment in the probabilities would cause the issue to persist. In \cite{degen3}, a random perturbation decoder is shown to be successful in solving instances in which symmetric degeneracy errors have taken place.

\section{Collision Decoder}

Decoding based on collisions is the last technique introduced in \cite{degen3}. This strategy is used in conjunction with either of the previously discussed methodologies (freezing or random perturbation) to have a more structured approach to fixing symmetric degeneracy errors. Once the traditional BP decoder has failed, the collision decoder begins by identifying a pair of ``colliding'' check nodes, which are two unsatisfied check nodes that share some variable nodes. Assuming that colliding pairs of check nodes occur as a result of errors on their shared variable nodes, the decoder then chooses to either apply random perturbations to those shared nodes or to freeze one of them. 

\section{Enhanced Feedback Decoder}

The authors of \cite{degen3} mention that despite the performance improvements provided by their proposed decoders, all of the errors that arise in their simulations are still end-to-end different syndrome errors (no degenerate or identical syndrome errors occur). This means that all of the errors that happen when using these modified decoding strategies can be attributed to the decoders themselves, which implies that further improvements can be made to these modified decoding strategies and a higher number of those end-to-end errors with different syndromes might possibly be corrected. This is done in \cite{efb}, where the authors, following the observations made in \cite{degen3},  present an enhanced feedback BP iterative decoding algorithm that provides useful information to the
BP decoder based on exploiting not only the syndrome but also the stabilizer elements themselves. 

As do most of the other modified decoding approaches, this method uses a conventional BP decoder as its default decoding tool until an end-to-end different syndrome error is found. Once this happens, a frustrated check node $\mathrm{c}_j$ and a variable node $q$ that share an edge on the factor graph are chosen at random. If $\mathrm{w}_j=1$ and $\hat{\mathrm{w}}_j=0$, where $\mathrm{w}_j$ represents the $j$-th component of the error syndrome $\mathbf{w} = [\mathrm{w}_1,\ldots,\mathrm{w}_{N-k}]$ and $\hat{\mathrm{w}}_j$ represents the $j$-th component of the estimated error syndrome $\mathbf{\hat{w}} = [\hat{\mathrm{w}}_1,\ldots,\hat{\mathrm{w}}_{N-k}]$, it is clear that the real error $\mathbf{E}$ anticommutes with the $j$-th stabilizer generator $\mathbf{S}_j$ while the estimated error $\hat{\mathbf{E}}$ commutes with $\mathbf{S}_j$. The enhanced feedback decoder uses this information to modify the prior probabilities of the variable node $q$ (as it connects to the check node $\mathrm{c}_j$ associated to the syndrome component $\mathrm{w}_j$) in order to make an anticommuting error more likely than the trivial error that the decoder generally leans towards (the decoder is biased towards estimating the identity operator). Thus, in the case that $\mathrm{w}_j=1$ and $\hat{\mathrm{w}}_j=0$, the prior probabilities are changed to

\begin{equation}\label{efb1}
p(\mathbf{E}_q = \gamma) \rightarrow 
	\begin{cases}
		\frac{p}{2}  & \mbox{if } \gamma = I \ \text{or}  \ \gamma = \mathbf{S}_{j}^{q}, \\
		1-\frac{p}{2} & \text{otherwise},
	\end{cases}
\end{equation}
where $\mathbf{S}_{j}^{q}$ is the $q$-th component of the $j$-th stabilizer generator and $p$ is the depolarizing probability of the channel. If the inverse scenario is encountered, i.e, $\mathrm{w}_j=0$ and $\hat{\mathrm{w}}_j=1$, then the prior probabilities are changed to 
\begin{equation}\label{efb2}
p(\mathbf{E}_q = \gamma) \rightarrow 
	\begin{cases}
		1 - \frac{p}{2}  & \mbox{if } \gamma = I \ \text{or}  \ \gamma = \mathbf{S}_{j}^{q}, \\
		\frac{p}{2} & \text{otherwise}.
	\end{cases}
\end{equation}

For instance, if $\mathbf{S}_{j}^{q} = X$ when $\mathrm{w}_j=1$ and $\hat{\mathrm{w}}_j=0$, then we would have the adjusted prior probabilities $ p(\mathbf{E}_q= I) = P(\mathbf{E}_q=X) = \frac{p}{2} $ and $p(\mathbf{E}_q= Z) = P(\mathbf{E}_q=Y) = 1 - \frac{p}{2} $. Whereas if $\mathrm{w}_j=0$ and $\hat{\mathrm{w}}_j=1$ we would have $ p(\mathbf{E}_q= I) = P(\mathbf{E}_q = X) = 1-\frac{p}{2} $ and $p(\mathbf{E}_q= Z) = P(\mathbf{E}_q=Y) = \frac{p}{2} $.

After the prior probabilities of the variable node are adjusted, decoding is reattempted for a fixed number of iterations. If the algorithm halts during this process, i.e, $\mathbf{w} = \hat{\mathbf{w}}$, then the correct syndrome has been found. If the check node $\mathrm{c}_j$ is still frustrated after this process then the prior probabilities of variable node $i$ are restored and the prior probabilities of a different variable node $i'$ connected to $\mathrm{c}_j$ are modified following equations (\ref{efb1}) and (\ref{efb2}). This procedure runs until $\mathrm{c}_j$ is no longer frustrated or until decoding based on modified prior probabilities has been attempted with all the variable nodes connected to $\mathrm{c}_j$. If $\mathrm{c}_j$ is no longer frustrated and $\mathbf{w} = \hat{\mathbf{w}}$ then decoding is complete. However, if $\mathrm{c}_j$ is no longer frustrated but $\mathbf{w} \neq \hat{\mathbf{w}}$, or all the qubits that connect to $\mathrm{c}_j$ have been exhausted, then a different frustrated check node $\mathrm{c}_j'$ is selected and the process begins anew. It is obvious that the complexity increase resulting from this technique will be heavily dependent on the efficacy of the prior probability modifications, i.e, how many frustrated check nodes must it tamper with prior to the correct syndrome being found. In \cite{efb}, the enhanced feedback decoder is shown to significantly outperform the random perturbation decoder of \cite{degen3} in terms of both QBER performance and decoding efficiency. In fact, when compared to a standard BP decoder, the work of \cite{efb} shows that the enhanced feedback technique is approximately $10$ times better\footnote{It is also worth noting that the authors of \cite{efb} assess their decoding performance using the physical error rate; they do not account for the presence of degeneracy. This means that the enhanced feedback decoder may actually exhibit better performance than is reflected in \cite{efb}.} while only requiring a $30\%$ increase in the number of decoding iterations.

\section{Supernode Decoder}

In \cite{qldpc15}, a modified GF($4$) decoder for dual-containing CSS codes, also referred to as homogeneous CSS codes, is proposed. A CSS code is said to be dual-containing if $\mathrm{H}_z^{'} = \mathrm{H}_x^{'}$. These codes are useful because $\mathrm{H}_z^{'}\mathrm{H}_x^{'T}= \mathrm{H}_z^{'}\mathrm{H}_z^{'T} = 0$ is always fulfilled. The main drawback of these dual-containing CSS codes is that their performance under SPA-based decoding is hindered by their large amount of length-4 cycles. To make matters worse, these cycles become even more prevalent when the error correction scheme is quaternary. Quaternary quantum codes are built based on the Pauli-to-GF($4$) isomorphism, which similarly to the mapping shown in (\ref{eq:Paulimapping}), serves to map the single qubit Pauli operators to equivalent $4$-ary symbols as \begin{equation}\label{eq:Quatmapping}
I \rightarrow 0, \
Z \rightarrow 1, \
X \rightarrow \omega, \
Y \rightarrow \bar{\omega}.
\end{equation}

Using this mapping, a quaternary $m \times N$ CSS QPCM $\mathbf{H}_Q$ can be constructed as: 
\begin{equation}\label{quatmat}
\mathbf{H}_Q = \begin{pmatrix}
\mathrm{H}'_x \\
\omega \mathrm{H}'_z \\
\end{pmatrix},
\end{equation}
where $\mathrm{H}_z^{'}$, and $\mathrm{H}_x^{'}$ are the same as in (\ref{CSS}). Herein lies the reason why the length-$4$ cycle issue is further aggravated for homogeneous quaternary CSS codes. Since $\mathrm{H}_z^{'} = \mathrm{H}_x^{'}$, the same two classical PCMs are used to build $\mathbf{H}_Q$, which forces the $i$-th and $(i+\frac{m}{2})$-th rows to completely overlap and leads to an increase in the number of the aforementioned cycles. 

Fortunately, the supernode decoder is capable of eliminating these specific cycles (those that are generated because of the homogeneous CSS structure of $\mathbf{H}_Q$). It achieves this by means of a modified Tanner graph where the $i$-th and $(i+\frac{m}{2})$-th check nodes become the $i$-th supernode, which automatically halves the number of length-$4$ cycles of the original factor graph. Naturally, such a change requires that the message exchange between the resulting supernodes and the original variable nodes be updated so that both of the original checks, $\mathrm{c}_i$ and $\mathrm{c}_{i+\frac{m}{2}}$, are still satisfied. Since the $i$-th and $(i+\frac{m}{2})$-th rows of $\mathbf{H}_Q$ fulfill $\mathrm{H}_i = \omega \mathrm{H}_{i+\frac{m}{2}}$, the $i$-th and $(i+\frac{m}{2})$-th syndrome components will be identically related, $\mathrm{w}_i = \omega \mathrm{w}_{i+\frac{m}{2}}$. Thus, the computation of the check-to-variable node messages changes only in that an updated syndrome $\bar{\mathrm{w}}_i = \mathrm{w}_i + \omega \mathrm{w}_{i+\frac{m}{2}} \in $ GF($4$) is used, where $i = 1,\ldots,\frac{m}{2}$. Essentially, the factor graph is reduced in size without the computation of the check node messages becoming more complex, which implies that the supernode decoder also represents an improvement in terms of decoding complexity when compared to the previously discussed heuristic methods. 

In \cite{qldpc15}, this supernode decoder is shown to exhibit  superior WER performance and lower decoding complexity when compared to existing state-of-the-art decoding techniques. Additionally, given that some short cycles are still present in the supernode factor graph (those that arise because of the symplectic criterion), the authors of said work also apply the Uniformly-Reweighted Belief Propagation (URW-BP) strategy of \cite{URW1, URW2} to their supernode decoder in order to alleviate the impact of the remaining short cycles. The authors also mention that the supernode decoder may be seamlessly amalgamated with other heuristic methods to improve performance even further. This is done in \cite{mod-BP}, where a set of novel improved decoding strategies are proposed. The reader is referred to \cite{qldpc15, imagra, URW2, imagra-nonbin} for a rigorous discussion on the specifics of quaternary decoding and its increased nuance.
 
 \section{Adjusted Decoder}
 
 The adjusted decoder is the first improved decoding technique that is proposed in \cite{mod-BP}. A similar decoder is derived and shown to improve the performance of surface codes in \cite{adj-surf}. Reminiscent of the correlation exploiting decoder, the goal of the adjusted decoder is to reintroduce the correlations between $X$ and $Z$ errors that are present in the depolarizing channel but are ignored when decoding a CSS code with a generic binary CSS decoder.
 
  The adjusted decoder tackles this issue by adding the correlations neglected by the separate binary decoders via ``adjusting'' the a priori channel probabilities of specific variable nodes of the factor graph. As with most of the previously discussed heuristic methods, the procedure begins by attempting to decode with a generic binary CSS decoder comprised of two separate decoders, one to decode bit-flips and the other to decode phase-flips. If decoding is successful or if both estimates $\hat{\mathbf{w}}_x \neq \mathbf{w}_x$ and $\hat{\mathbf{w}}_z \neq \mathbf{w}_z$, then the process halts. However, if one of the estimates is correct, then decoding is reattempted for the incorrect component based on a set of adjusted probabilities. For instance, if $\hat{\mathbf{w}}_x = \mathbf{w}_x$ but $\hat{\mathbf{w}}_z \neq \mathbf{w}_z$, then the a priori probabilities of the $Z$ operator variable nodes are modified to:
 
 \begin{equation}\label{adjusted_1}
p(\mathbf{e}_{z,q} = 1) \rightarrow 
	\begin{cases}
		\frac{p_y}{p_x+p_y}  & \mbox{if } \hat{\mathbf{E}}_{x,q} = 1, \\
		\frac{p_z}{1-(p_x+p_y)} & \text{otherwise},
	\end{cases}
\end{equation}
 
where the subscript $q$ denotes the $q$-th variable node of the $Z$ decoder and $\mathbf{e}$ denotes the symplectic representation of the error operator in question. If instead $\hat{\mathbf{w}}_z = \mathbf{w}_z$ but $\hat{\mathbf{w}}_x \neq \mathbf{w}_x$, then the adjustment is the following:

 \begin{equation}\label{adjusted_2}
p(\mathbf{e}_{x,q} = 1) \rightarrow 
	\begin{cases}
		\frac{p_y}{p_y+p_z}  & \mbox{if } \hat{\mathbf{E}}_{z,q} = 1, \\
		\frac{p_x}{1-(p_y+p_z)} & \text{otherwise},
	\end{cases}
\end{equation}

where the subscript $q$ denotes the $q$-th variable node of the $X$ decoder.

 \section{Augmented Decoder}
 
 The second modified decoding technique proposed in \cite{mod-BP} is the augmented decoder. It is based on the  decoder proposed in \cite{class-aug} for classical binary LDPC codes and serves to improve the performance of both binary and quaternary BP decoders. As with most of the previously discussed methodologies, decoding is initially attempted with a standard BP decoder. If this is unsuccessful, decoding is reattempted with an ``augmented'' PCM $$\mathbf{H}_A = \begin{pmatrix}
\mathbf{H}_Q \\
\mathbf{H}_\delta \\
\end{pmatrix},$$ where $\mathbf{H}_\delta$ denotes a subset of rows chosen at random from the original QPCM $\mathbf{H}_Q$. The size of this subset is determined by the parameter $\delta$, which is known as the augmentation density. Decoding based on $\mathbf{H}_A$ also requires that the syndrome be augmented accordingly, $\mathbf{w}_A =[\mathbf{w} \ \mathbf{w}_\delta]$, where $\mathbf{w}$ is the measured syndrome and $\mathbf{w}_\delta$ represents the syndrome values associated to the rows of $\textbf{H}_Q$ that comprise $\mathbf{H}_\delta$. If the initial decoding reattempt based on $\mathbf{H}_A$ is not successful, the process is repeated until either $\hat{\mathbf{w}}=\mathbf{w}$ or a predefined maximum number of attempts is reached. In the particular case of binary CSS decoders, both the $X$ decoder PCM and the $Z$ decoder PCM, as well as the syndromes $\mathbf{w}_x$ and $\mathbf{w}_z$, have to be augmented. 

Since decoding directly over the augmented matrix $\mathbf{H}_A$ conduces to an increase in complexity (larger matrices and syndromes are required), an equivalent decoding alternative capable of mitigating the impact of the augmentation strategy on the complexity of the scheme is also proposed in \cite{mod-BP}. Instead of using the augmented matrix, this method operates by modifying the marginal probabilities of the SPA based on a binary function whose entries represent if a check node from the factor graph has been duplicated when performing the augmentation. In terms of performance, augmented $\text{GF}(4)$ decoders and augmented supernode decoders have been shown to outperform random perturbation and enhanced feedback decoders when used to decode dual-containing CSS codes \cite{mod-BP}. When applied to non-dual-containing CSS codes and non-CSS codes, the augmented decoder was shown to perform similarly to random perturbation and enhanced feedback decoders.  

 \section{Combined Decoder}

The last technique proposed in \cite{mod-BP} is a CSS decoder that combines both the adjusted and augmented decoding methods. Initially, a standard binary CSS decoder is executed. If both $\hat{\mathbf{w}}_x$ and $\hat{\mathbf{w}}_z$ do not match with the measured syndromes, then decoding is reattempted for the $X$ operators using the augmentation technique. If this is unsuccessful after a fixed number of attempts, the augmented decoder is used for the $Z$ operators. If both of the estimated syndromes still do not match, then the procedure is halted. However, if we obtain $\hat{\mathbf{w}}_x = \mathbf{w}_x$ or $\hat{\mathbf{w}}_z = \mathbf{w}_z$, either from the initial standard binary decoding or due to one of the augmented modifications, then decoding is reattempted for the remaining failed component by means of adjusted a priori probabilities. If this is unsuccessful, the last resort is to reattempt decoding for the failed component using the augmentation technique and maintaining the adjusted probabilities. In \cite{mod-BP}, the combined decoder is shown to outperform random perturbation and enhanced feedback decoding techniques when applied to dual-containing CSS codes. When used to decode non-dual containing CSS codes, the combined decoder outperforms a regular GF($4$) decoder but performs slightly worse than quaternary modified decoders (enhanced feedback, augmented and random perturbation).

 \section{Ordered Statistics Decoder}
 
 The Ordered Statistics Decoding (OSD) technique of \cite{class-osd1} is a well known classical decoding strategy that improves performance at the expense of the decoding complexity. In \cite{osd1}, the OSD algorithm is applied to the quantum domain as a post-processing methodology to improve performance whenever the traditional BP decoder fails. The algorithm uses the soft-decisions made by the traditional BP decoder to re-arrange the qubits in descending order of their total probability of error. This serves to sort the qubits in terms of their reliability, which is then used to make hard decisions on the qubits for which the traditional BP decoder has the most certainty. Next, assuming that the decoder makes correct hard decisions on $N-\mu$ qubits, the OSD-decoder flips the $\mu$ most unreliable qubits of the traditional BP error estimate in order to find an error estimate that produces a matching syndrome. If multiple correction operators are found, the error sequence of minimum weight is chosen as the correct estimate of the error pattern. In some cases, the initial number of unreliable qubits that is used by the OSD algorithm is insufficient to produce an acceptable estimate of the error. If so, the qubits are reprocessed by flipping an increased number of unreliable qubits until an appropriate estimate is found or until a predefined number of processing rounds is reached. If one processing round is sufficient to find a matching syndrome the algorithm is known as order-$0$ or OSD-$0$. If subsequent processing rounds are required, the algorithm is referred to as OSD-$i$ or order $i$ OSD, where $i$ denotes the number of processing rounds.
 
 Given its particular structure and provided that sufficient decoding time is permitted, the OSD decoder will always recover an error pattern that maps to the correct syndrome. This stands in stark contrast to the generic SPA decoder and all of the aforementioned improved decoding techniques, which may sometimes yield an incorrect syndrome regardless of the allowed decoding time. 
 In \cite{osd1, osd2} the OSD decoder is shown to vastly improve the performance of a generic SPA decoder, regardless of the type of employed QLDPC code. In fact, the OSD decoding algorithm is compatible with non-CSS decoding, which if applied to existing non-CSS decoders, would serve to further improve their performance \cite{patrick, nonCSS1}. Also in \cite{osd1}, the order 0 version of the algorithm is shown to surpass the enhanced feedback, augmented, and random perturbation decoders. 
 
 \section{Refined Belief Propagation Decoding}
 
 Recently, in \cite{refined}, methods to improve the performance and simplify the complexity of a GF($4$) decoder for QLDPC codes have been proposed. Despite the fact that this decoder is not specifically designed to correct end-to-end errors with different syndromes, it is relevant to our discussion. First off, it represents a leap forward for quantum quaternary decoding, which in terms of performance is a better decoding strategy to handle end-to-end errors with different syndromes than the binary BP decoder. The authors of \cite{refined} show that their refined BP quaternary decoder has the same decoding complexity as a standard binary BP decoder. This symbolizes a major reduction in the complexity of quaternary decoding, since the decoding algorithm of a generic GF($4$) decoder runs 16 times slower than the binary equivalent when decoding a quantum stabilizer code. Another reason that makes the decoding strategy of \cite{refined} relevant is that this modified GF($4$) decoder is compatible with all of the previously discussed improved decoding methods, which implies that performance of QLDPC codes could become orders of magnitude better while only paying a small price in the form of a more complex decoder. 
 
 The reduction in the decoding complexity of the refined BP decoder is achieved by passing single-valued messages over the factor graph instead of the multivalued messages that are typically exchanged in quaternary decoding. The use of single-valued messages in a GF($4$) decoder is made possible by the insight that check node messages are more indicative of the commutation status of an error component and the corresponding stabilizer generator than the actual type of Pauli operator acting on said component. This stems from the fact that a quantum syndrome, whose components are directly related to the check nodes of a QLDPC factor graph, is a binary vector that represents the commutation status of each error component with the stabilizer generators.  
 
Aside from the use of single-valued messages, two additional useful strategies to improve the performance of quantum BP decoders while keeping their complexity low are detailed in \cite{refined}. The first one involves modifying the message passing schedule of the decoder from a parallel message passing schedule to a serial one. The most common implementation of the SPA or BP algorithm is based on a parallel schedule, where messages of the same kind are exchanged simultaneously over the factor graph during each iteration, i.e, all the variable node messages are transmitted to the check nodes at the same time instant $t_1$ and all the check node messages are transmitted to the variable nodes at the same time instant $t_2$. A less common but possibly better messaging schedule is the serial implementation of the BP algorithm. In \cite{serialBP1, serialBP2}, it is shown to  improve the convergence behavior when the underlying Tanner graph has many short cycles at no increase in decoding complexity. If messages are updated in this manner, the process operates in a top-down sequential fashion, where the first variable node receives messages from its  neighbouring check nodes, to who it then replies with its own messages, who then move on to the next variable node, and so on. 

The second strategy involves adjusting the magnitude of the messages exchanged over the factor graph, since they can easily become overestimated \cite{norm1,norm2,norm3}. Such an adjustment can be performed via message normalization or offset, which can be applied seamlessly to the refined BP decoder given the complexity reduction that is attained by migrating to single-valued message exchanges. 
 
 The results shown in \cite{refined} show that performance can be significantly improved by using a serial BP algorithm and normalization techniques. The authors of this work mention that further improvements may be possible by improved exploitation of quantum degeneracy, either by stacking their decoder with any of the heuristic methods we have explained if complexity increments are affordable\footnote{Decoding efficiency is primordial in the quantum paradigm, since the coherence of quantum states decays rapidly.}, or by devising new techniques. This is discussed at length in \cite{exploiting}.
 
 \chapter{Monte Carlo Simulations} \label{app:sims}

 Monte Carlo simulations, also known as the Monte Carlo Method or multiple probability simulations, are a mathematical technique that can be used to model the probability of different outcomes in a process that cannot easily be predicted due to the intervention of random variables. It is a technique used to understand the impact of risk and uncertainty in prediction and forecasting models. A simple example of a Monte Carlo Simulation is to consider calculating the probability of rolling two standard dice. There are 36 combinations of dice rolls. Based on this, you can manually compute the probability of a particular outcome. Using a Monte Carlo Simulation, you can simulate rolling the dice 10,000 times (or more) to achieve more accurate predictions.

In this dissertation we employ the Monte Carlo Method to assess the performance of various families of quantum codes. The method provides us with guidelines that ensure that the simulations we conduct are accurate, which ultimately allows us to draw valid conclusions from the results these simulations provide. Although they vary depending on the quantum channel model under consideration, all our simulations operate based on the following principle:

\begin{enumerate}
    \item First, an $N$-qubit operator $\mathbf{A}\in\mathcal{\overline{G}}_N$ is generated according to the probability distribution of the particular quantum channel that is being considered. This probability distribution will be different for each of the quantum channels (i.i.d. $X/Z$ channel, depolarizing channel, asymmetric channel) discussed in this dissertation.
    \item Next, the syndrome $\mathbf{w}$ associated to the operator $\mathbf{A}$ is computed and fed to the decoder so that the decoding process can begin. 
    \item Once the decoding process has finished, the decoding estimate $\mathbf{\hat{A}}$ and the real operator $\mathbf{A}$ are compared.
    \item After this comparison, a new simulation round begins by going back to step $1$ and generating another $N$-qubit operator.
\end{enumerate}

In terms of the comparison that is performed in step $3$ different performance assessment metrics or figures of merit can be obtained. Throughout this dissertation we employ the WER and the QBER as the figures of merit. The WER represents the probability that at least one qubit of the estimated operator is incorrect, i.e, it compares the entire operator instead of its constituent single-qubit operators. Thus, the WER will consider that the scheme has been successful whenever $\mathbf{\hat{A}}\star\mathbf{A} \in \mathcal{\overline{Z}(\overline{S})}$ and that an error has occurred whenever this does not hold\footnote{Recall that comparison up to a stabilizer element computes the logical error rate, which is the appropriate way of assessing the performance of degenerate quantum codes.}. On the other hand, the QBER compares the individual single-qubit operators that make up $\mathbf{A}$ and $\mathbf{\hat{A}}$, so it represents the fraction of qubits that experience an error. This means that it is computed by checking if $\mathrm{A}_i = \mathrm{\hat{A}}_i, \forall i=1,\ldots,N$, where a qubit error occurs whenever the equality does not hold.  

To ensure that the WER and QBER results of our simulations are accurate, we invoke the rules of Monte Carlo simulations. This requires that we follow the guideline shown below to select the number of necessary simulation rounds\footnote{Simulation rounds are also sometimes referred to as blocks due to the similarity with classical communications in which each simulation round represents the transmission of a block of information bits.} $\mathrm{N}_\text{rounds}$ \cite{MonteCarlo}:

\begin{equation} \label{eq:sims_1000}
    \mathrm{N}_\text{rounds} = \frac{100}{\text{WER}}.
\end{equation} 

If we assume that the observed error events are independent, the above rule of thumb yields the following 95 \% confidence interval:

$$ P(0.8\tilde{\text{WER}} \leq \text{WER} \leq 1.25\tilde{\text{WER}}) = 0.95,$$

where $\tilde{\text{WER}}$ represents the empirically estimated value of the WER. 

For example, in order to have high confidence (with a $95\%$ confidence level) regarding the performance of an arbitrary quantum code at a WER of $10^{-3}$, this rule tells us that we would need to conduct $\mathrm{N}_\text{rounds}=100000$ simulation rounds. In summary, conducting our simulations based on the principle of \eref{eq:sims_1000} ensures that the corresponding results will be statistically representative and that they will accurately portray the performance of the quantum codes that are being considered.

\end{appendices}
\clearemptydoublepage

\def\bibname{References}
\addcontentsline{toc}{chapter}{\protect\numberline{}References}



\end{document}